\documentclass{article}
\usepackage{epsfig}
\usepackage{subfigure}
\usepackage{amssymb}
\usepackage{authblk}
\usepackage{multicol}
\usepackage{rotfloat}

\usepackage[breaklinks=true,colorlinks=true,linkcolor=blue]{hyperref}
\begin{document}
\begin{center}
{\Huge Hall A Annual Report \\ 2013}
\end{center}
\vspace{2cm}
\begin{center}
\includegraphics[width=0.95\textwidth, angle = 0.]{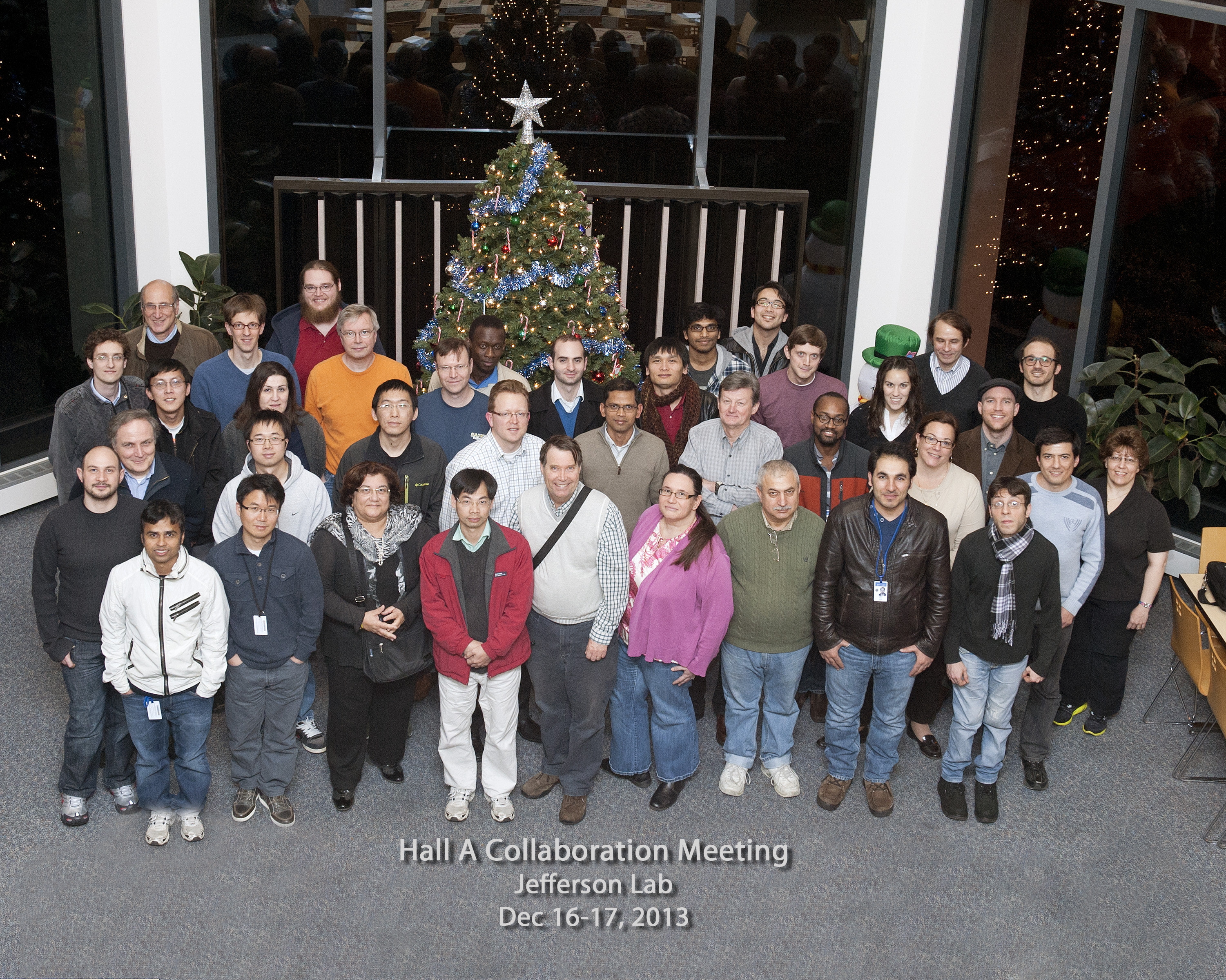}
\end{center}
\vspace{2cm}
\begin{center}
{\Large Edited by Mark M Dalton}
\end{center}
\title{Hall A Annual Report 2013}

\author[1]{M.M.~Dalton} 
\author[2]{K.~Allada} 
\author[3]{K.~Aniol}
\author[4]{W.~Boeglin}
%\author[1]{A.~Camsonne} 
\author[1]{E.~Chudakov} 
\author[5]{M.~Cummings}
\author[6]{D.~Flay}
\author[7]{M.~Friedman}
\author[8]{O.~Glamazdin} 
\author[1]{J.~Gomez}
\author[1]{C.~Keppel}
\author[4]{H.P.~Khanal}
\author[9]{R.~Lindgren}
\author[10]{E.~Long}
\author[1]{R.~Michaels}
\author[11]{M.~Mihovilovi\v{c}}
\author[12]{C.~Mu\~noz~Camacho}
%\author[1]{S.~Nanda} 
\author[8]{R.~Pomatsalyuk}
\author[13]{S.~Riordan}
\author[11]{S.~\v{S}irca}
\author[9]{C.~Smith}
\author[1,10]{P.~Solvignon}
\author[6]{N.F.~Sparveris}
\author[8]{V.~Vereshchaka}
\author[14]{X.~Yan} 
\author[5]{Z.~Ye}
\author[14]{Y.X.~Zhao}
\author[]{the Jefferson Lab Hall A Collaboration}

\affil[1]{Thomas Jefferson National Accelerator Facility, Newport
News, VA 23606, USA}
\affil[2]{Massachusetts Institute of Technology, Cambridge, MA 02139,
USA}
\affil[3]{California State University Los Angeles, Los Angeles, CA 90032, USA}
\affil[4]{Florida International University. Miami, FL 33199, USA}
\affil[5]{College of William and Mary, Williamsburg, VA 23187, USA}
\affil[6]{Temple University, Philadelphia, PA 19122}
\affil[7]{Racah Institute of Physics, Hebrew University of Jerusalem,
Givat Ram 91904, Israel}
\affil[8]{National Science Center Kharkov Institute of Physics and Technology,
Kharkov 61108, Ukraine}
\affil[9]{University of Virginia, Charlottesville, VA 22901, USA}
\affil[10]{University of New Hampshire, Durham, NH 03824, USA}
\affil[11]{University of Ljubljana, SI-1000 Ljubljana, Slovenia}
\affil[12]{Institut de Physique Nucleaire d'Orsay, IN2P3, BP 1,
91406 Orsay, France}
\affil[13]{University of Massachusetts, Amherst, MA 01003, USA}
\affil[14]{University of Science and Technology of China, Hefei
230026, People's Republic of China}

\renewcommand\Authands{ and }
\maketitle
\newpage
\tableofcontents
\newpage
\listoffigures
\newpage

\section{Introduction}

\begin{center}
contributed by C. Keppel.
\end{center}

The year 2013 was one of transition to making the 12 GeV upgrade to the Continuous Electron Beam Accelerator Facility a reality, and anticipation of seeing the first beam to an experiment back in Hall A in 2014. 
The year has been dedicated largely to preparing for two experiments: E12-06-114, a measurement of deeply virtual Compton scattering (DVCS), and E12-07-108, a measurement of the proton magnetic form factor G$_M^p$. These two experiments will be the first to receive beam in the 12 GeV era. Preparations for E12-07-108 and E12-06-114 have involved significant detector upgrades to both HRS’s, a hydrogen target with an improved cell design, and other complimentary equipment. The DVCS collaboration has been hard at work preparing the stand-alone calorimeter necessary to their experiment. Requisite hall beamline modifications for the higher energy beam have included the Moller and Compton polarimeters, higher field mapping of the Hall A arc, a new raster system, and also reviving the Unser and BCMs for precision charge measurement. 

The 12 GeV scientific plans for the hall consist of many compelling experiments to utilize the standard Hall A equipment, some with slight modifications, in conjunction with the higher energy beam. Two require a $^3$H target, one to measure the F$_2^n$/ F$_2^p$ structure function ratio at large x, and one to continue the successful Hall A studies of  short range correlation phenomena. This target and associated systems are in design currently for a run after the E12-07-108 and E12-06-114 experiments. Beyond experiments that will utilize the standard Hall A equipment are ambitious plans involving multiple new experiment installations. 

This year brought construction commencement of one of these larger scale installation experiments, the Super Bigbite Spectrometer (SBS) program. The SBS project consists of a set of three form factor experiments centered around somewhat common equipment and new experimental capabilities. First activities to begin this program have included re-design of a magnet from the Brookhaven National Laboratory, delivered this year to JLab, the successful completion of pre-research and development of GEM tracking detectors, and a host of scientific development activities including detector construction projects such as prototyping a standalone hadron calorimeter, data acquisition upgrades, and refined physics projections. 
Work has continued effectively as well on many other fronts, including infrastructure improvements in data acquisition, offline analysis, and core hall capabilities. Technical preparations are underway for planned experiments such as PREX-II, APEX, A$_1^n$ and others. A new experiment, CREX, was approved to measure the neutron radius of $^{48}$Ca. 

Moreover, there has been active engagement in analyses of past experiments. Here, twelve new publications related to Hall A experiments were authored by members of the Hall A collaboration, and two new Hall A related doctoral theses were successfully defended. Excitingly, the first APS Topical Group on Hadron Physics Dissertation Award was given in 2013 to Dr. Jin Huang, whose dissertation research was on a Hall A measurement of double spin asymmetries in charged pion production from a transversely polarized $^3$He target.

In all, this has been a year of achievements coupled with anticipation and diligent preparations. It is a joy and a privilege to work with the Hall A staff and user community, and I am looking forward to successfully entering into the 12 GeV era together. Please accept my many, many thanks to you all for your expert, industrious, innovative contributions to the hall. I look forward to welcoming the higher energy beam into Hall A with you!

\clearpage

\section{General Hall Developments}

%\documentclass{article}
%\usepackage{epsfig}
%\setlength{\textwidth}{6.5in}
%\setlength{\oddsidemargin}{0in}
%\setlength{\evensidemargin}{0in}
%\setlength{\textheight}{9in}
%\setlength{\topmargin}{0in}
%\setlength{\headheight}{0in}
%\setlength{\headsep}{0in}

%\begin{document}

%\section{General Hall Developments}

\subsection{Status of the Hall A M\o{}ller Polarimeter DAQ }
\label{sec:moller_status}

\begin{center}
$^1$O.~Glamazdin, $^2$E.~Chudakov, $^2$J.~Gomez, $^1$R.~Pomatsalyuk, $^1$V.~Vereshchaka \\
$^1$National Science Center Kharkov Institute of Physics and Technology, Kharkov 61108, Ukraine  \\
$^2$Thomas Jefferson National Accelerator Facility, Newport News, VA23606, USA  \\
\end{center}

\subsubsection{M\o{}ller polarimeter} 
\label{polarimeter}

\begin{figure}[hbt]
   \center{\epsfig{figure=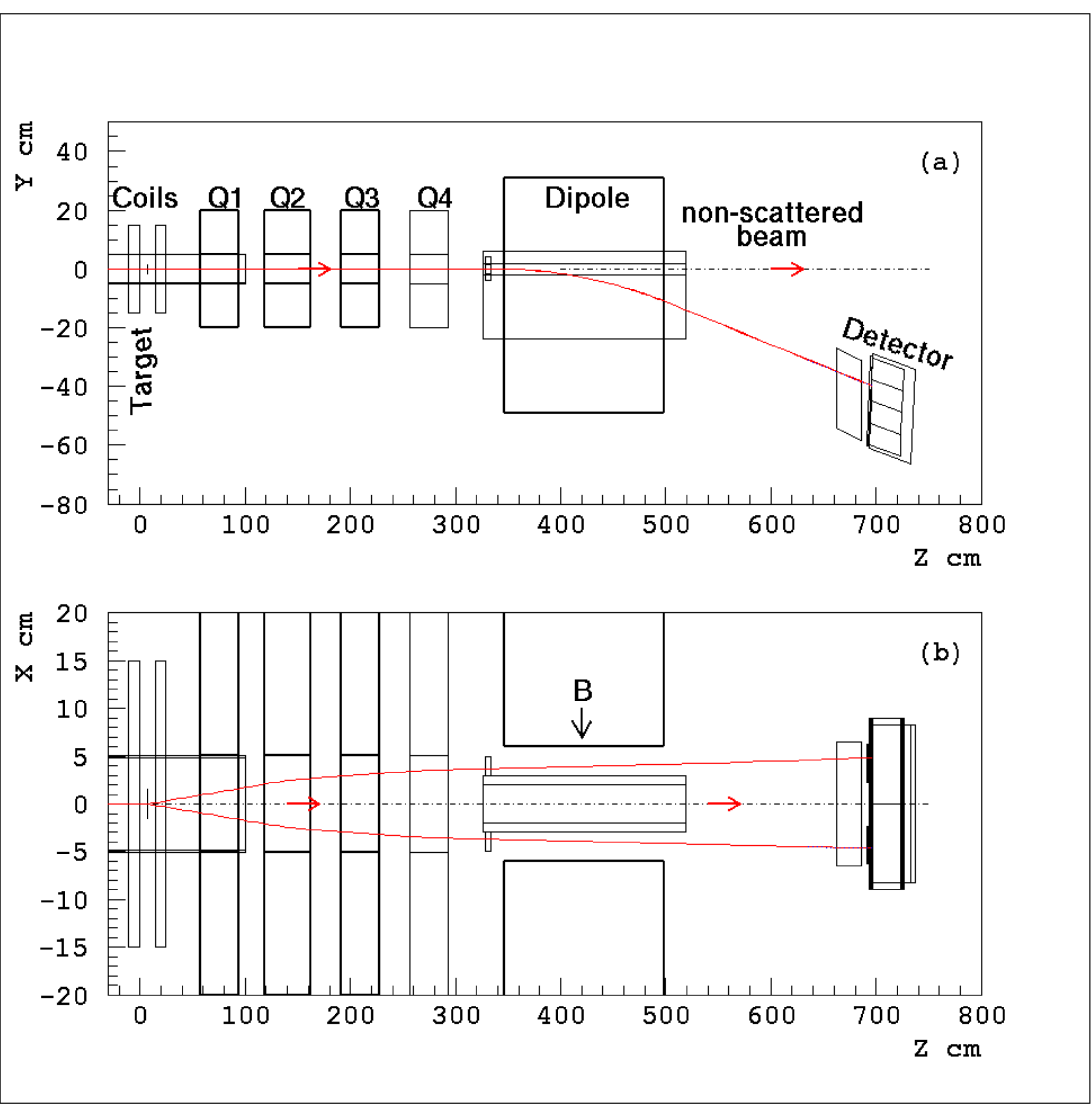,width=12.0cm}}
   \caption{Layout of the Hall A M\o{}ller polarimeter. a) side view, b) top view. }
\label{fig:layout}
\end{figure}

The Hall A M\o{}ller polarimeter is used to measure a beam polarization for  Hall A experiments with polarized electron beam. The polarimeter consists of polarized electrons target, a magnetic spectrometer and a detector (see Fig.~\ref{fig:layout}). M\o{}ller electrons appear as a result of a polarized electron beam scattering on a polarized electron target. Scattered M\o{}ller electrons  are analyzed by the magnetic spectrometer. The spectrometer consists of four quadrupole (Q1, Q2, Q3, Q4) and one dipole magnets. Scattered electrons are focused by the quadrupole magnets in the horizontal plane at the entrance gap of the dipole magnet. The dipole magnet deflects 
M\o{}ller electrons down to detector.

The electron detector consists of two total absorption calorimeters, which allow to detect 
M\o{}ller events in coincidences. Each calorimeter has four channels.  There is a four channels aperture detector made of plastic scintillator attached to the face panel of each calorimeter. The typical rate of the detector during measurements is 100~-~200~kHz in one arm and 50~kHz in coincidences. The event rate in the coincidence mode for different signs of the electron beam polarization and known values of the target polarization and the detector analyzing power allow to calculate the measured polarization of the electron beam.

The Hall A M\o{}ller polarimeter has been upgraded in accordance with the CEBAF upgrade plan to 
the beam energy range $1\div11$~GeV. Description and status of the Mo\o{}ller polarimeter after the upgrade can be found in \cite{report-2012}.

The Hall A M\o{}ller polarimeter has two data acquisition and processing systems:

\begin{itemize}
  \item {} Old system is based on CAMAC, VME, NIM modules;
  \item {} New system is based on VME module flash-ADC F-250 designed in Jefferson Lab.
\end{itemize}

Both DAQs are used simultaneously to measure the electron beam polarization. The old DAQ, in operation since 1998, is fully functional but may not be repairable in case of malfunction, as the system modules are not in stock and are not manufactured anymore. The new DAQ based on flash-ADC, which is in operation since 2010, is more precise and provides more detailed data analysis. However, it currently requires more careful adjustment and further improvement.

\subsubsection{The old DAQ}
\label{sec:old_DAQ}

\begin{figure}[hbt]
   \center{\epsfig{figure=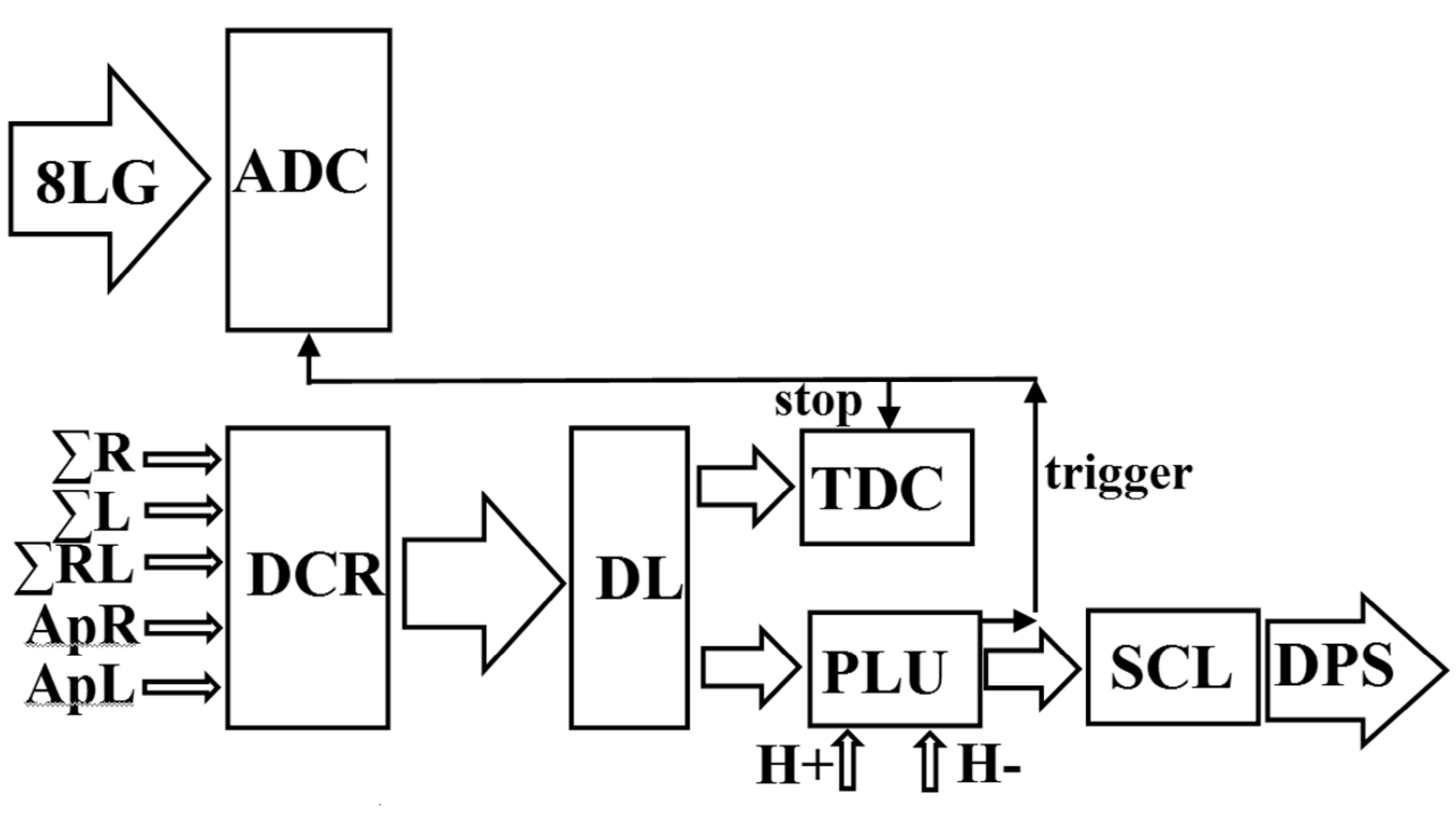,width=12.0cm}}
   \caption[Scheme of the old M{\o}ller DAQ.]{Scheme of the old M\o{}ller DAQ. 8LG - signals of 8 blocks (calorimeter), ADC - analog to digital converter, $\Sigma$L - sum of 4 left blocks (left calorimeter), $\Sigma$R - sum of 4 right blocks, $\Sigma$LR - sum of 8 calorimeter blocks, ApL - signal from left aperture counter, ApR - signal from right aperture counter, DCR - discriminator, DL - delay lines, TDC time to digital converter, PLU - programmable logic unit, H+, H- -signals of ''+'' and ''-'' helicity, SCl - scalers, DPS - data processing system. }
\label{fig:old_daq} 
\end{figure}

The old system is fully functional for both (''high magnetic field'' and ''low magnetic field'') polarimeter targets. It is time-tested and well-studied. However, it has a low event recording speed, the system modules occupy several crates, and there is a large number of inter-unit connections and cables which reduces the system reliability. Moreover, some system modules went out of production and are not replaceable in case of malfunction. The old DAQ layout is shown in Fig.~\ref{fig:old_daq}. The CODA system remotely controls the data acquisition system and data recording from the M\o{}ller polarimeter ~\cite{CODA}.

A few electronic modules of the old DAQ were replaced with more modern ones with higher bandwidth. The main goals of the upgrade are:

\begin{itemize}
  \item {} To replace programmable logic module (PLM) which is no longer available;
  \item {} To increase bandwidth (up to 200~MHz);
  \item	{} To decrease readout time from ADC and TDC modules;
  \item {} To decrease the number of crates required for the DAQ and to eliminate CAMAC as an outdated and slow standard.
\end{itemize}

\begin{figure}[hbt]
   \center{\epsfig{figure=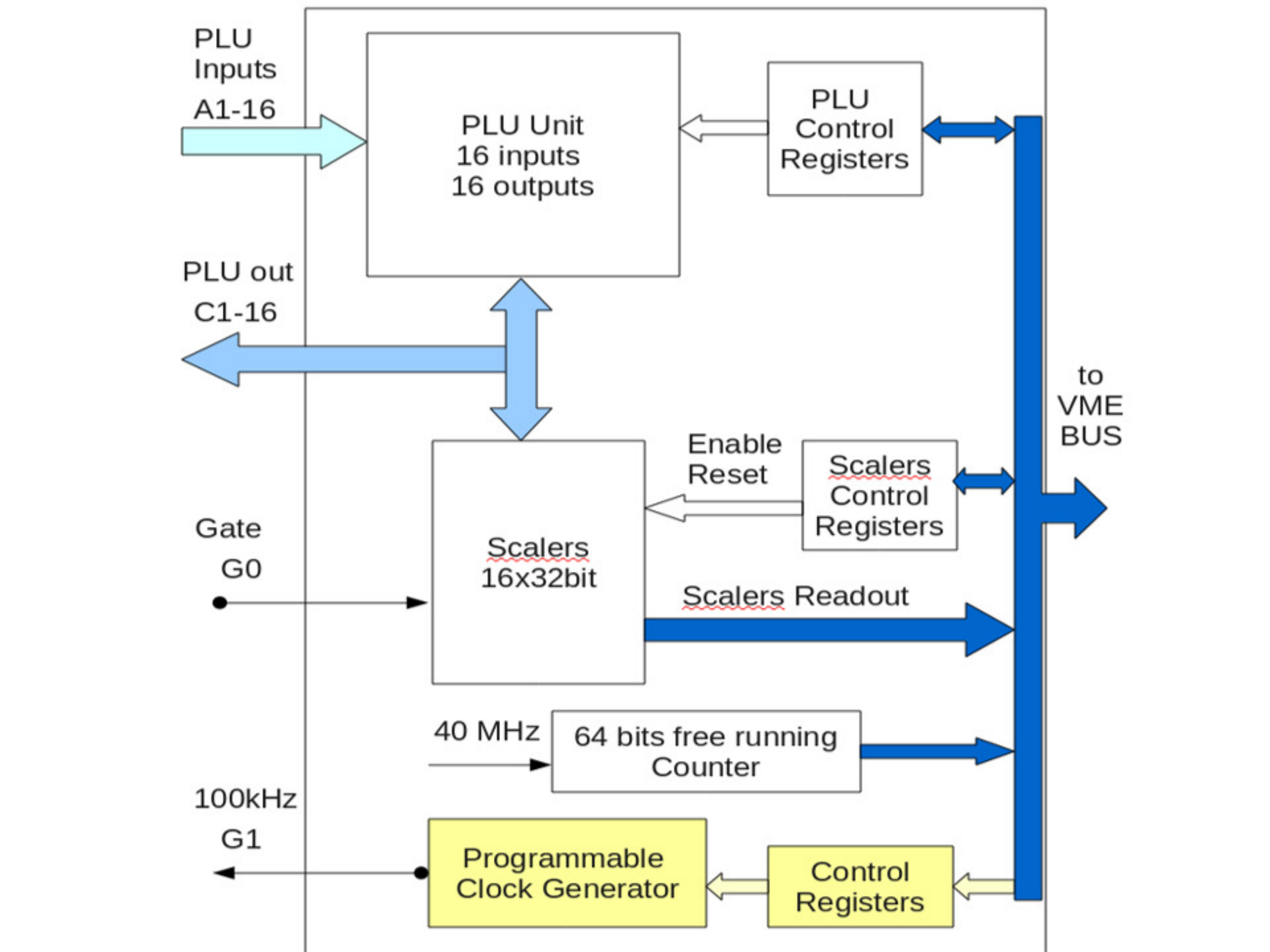,width=12.0cm}}
   \caption{Diagram of programmable logic unit based on module CAEN V1495. }
\label{fig:new_plu}
\end{figure}

The list of replaced modules is:
\begin{itemize}
   \item To increase bandwidth:
     \begin{enumerate} 
       \item PLM LeCroy-2365 (frequency $<$75~MHz, crate CAMAC) was replaced with PLU based on CAEN V1495 
             (frequency 200~MHz, crate VME) (see Fig.\ref{fig:new_plu});
       \item Discriminator Ortec-TD8000 (frequency $<$150~MHz, crate CAMAC) was replaced with discriminator 
             P/S~708 (frequency $<$300~MHz, crate NIM).
      \end{enumerate}
   \item To reduce readout time:
      \begin{enumerate}
         \item ADC LeCroy 2249A (12 channels, crate CAMAC) was replaced with ADC CAEN V792 
              (32 channels, crate VME);
         \item TDC LeCroy 2229 (crate CAMAC) was replaced with TDC CAEN V1190B (64 channels, 
               0.1~ns, crate VME).
      \end{enumerate}
\end{itemize}

All new modules have been tested, and new software and the analyzer for the upgraded DAQ are under development. 

It is planned to use the old DAQ after reconstruction at least until the system based on flash-ADC is fully functional for operation with two targets of the 
M\o{}ller polarimeter. It should be stressed that the new DAQ (based on FADC) was created for operation with the new target (''high magnetic field'') and its software is not intended for operation with the old target (''low magnetic field''). As a result, if the M\o{}ller polarimeter is operating with the old target, the old DAQ system is mandatory whereas the new one is optional. Moreover, simultaneous use of the two systems allows to study the polarimeter systematic errors.

\subsubsection{The new DAQ based on FADC}
\label{sec:new_DAQ}

\begin{figure}[hbt]
   \center{\epsfig{figure=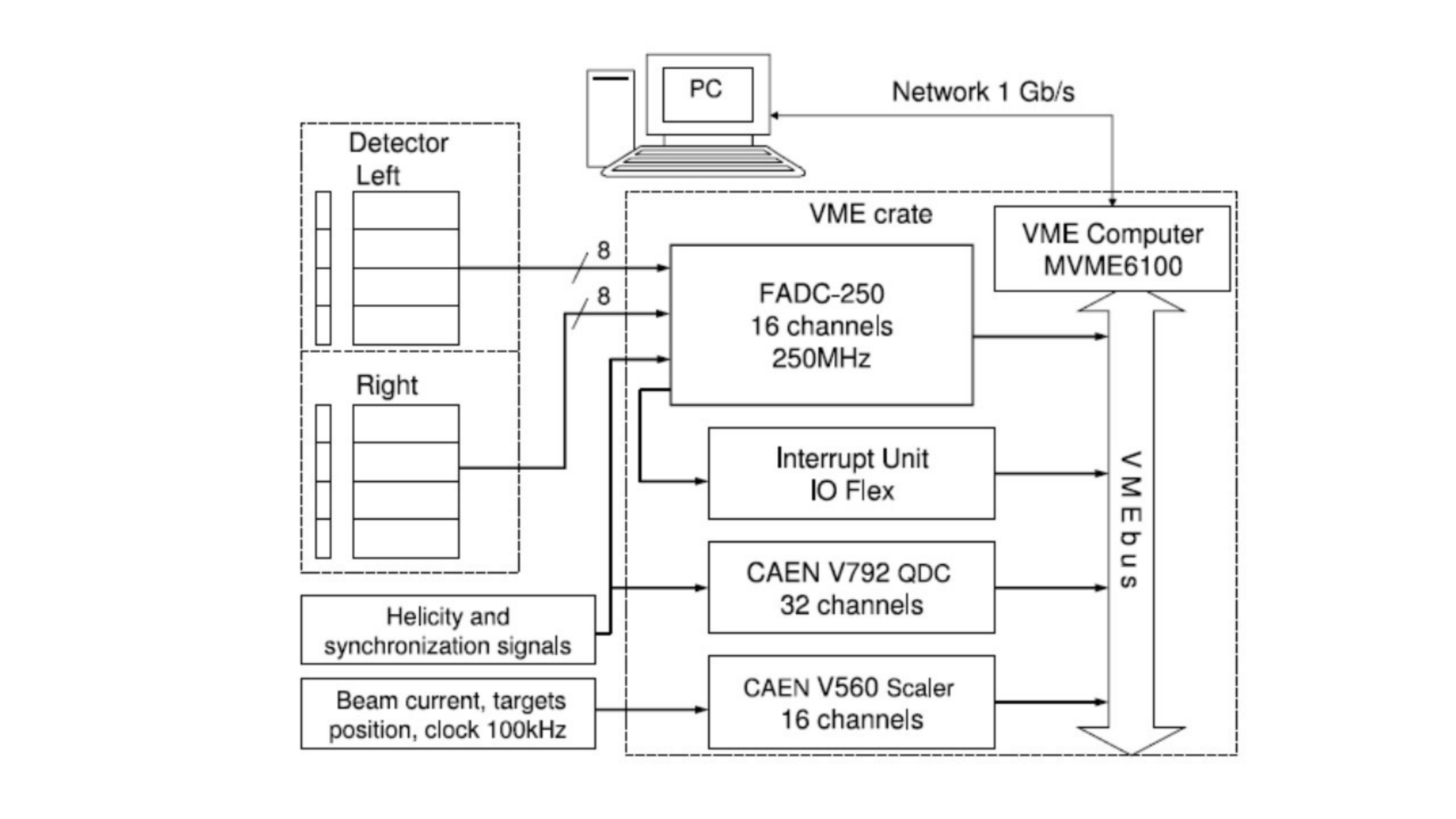,width=17.0cm}}
   \caption{New M\o{}ller polarimeter DAQ based on FADC. }
\label{fig:new_daq}
\end{figure}

One of the main goals for implementing the new DAQ is to reduce systematic error of the polarization measurement with the Hall A M\o{}ller polarimeter. One of the M\o{}ller polarimeter systematic error components is a dead time of event record system. One of the ways to reduce the dead time is to increase DAQ event rate registration and record speed.  The programmable module flash-ADC F250, designed in Jefferson Lab, with algorithm for processing and recording events from the polarimeter ~\cite{review} was chosen for the new fast DAQ. A block diagram of the data acquisition system on the basis of flash-ADC is shown in  Fig.~\ref{fig:new_daq}.

The new data acquisition system consists of:
   \begin{itemize}
      \item Flash-ADC F250;
      \item Interrupt unit IO Flex;
      \item Additional QDC module CAEN V792;
      \item Scalers module CAEN V560;
      \item VME controller MVME6100;
      \item NIM crate and modules for ECL/NIM levels conversion (not shown in the figure);
      \item Gigabit Ethernet network;
      \item Controlling PC.
   \end{itemize}

Electronic modules are mounted in the VME crate located in the Hall A behind a shielding wall.  QDC module ADC V792 is used with the flash-ADC for supplementary record of polarization sign signals and synchronization signals. Scalers module V560 is used for recording of the beam current, target position relating to the beam and the signals of the 100~kHz reference generator.

\begin{figure}[hbt]
   \center{\epsfig{figure=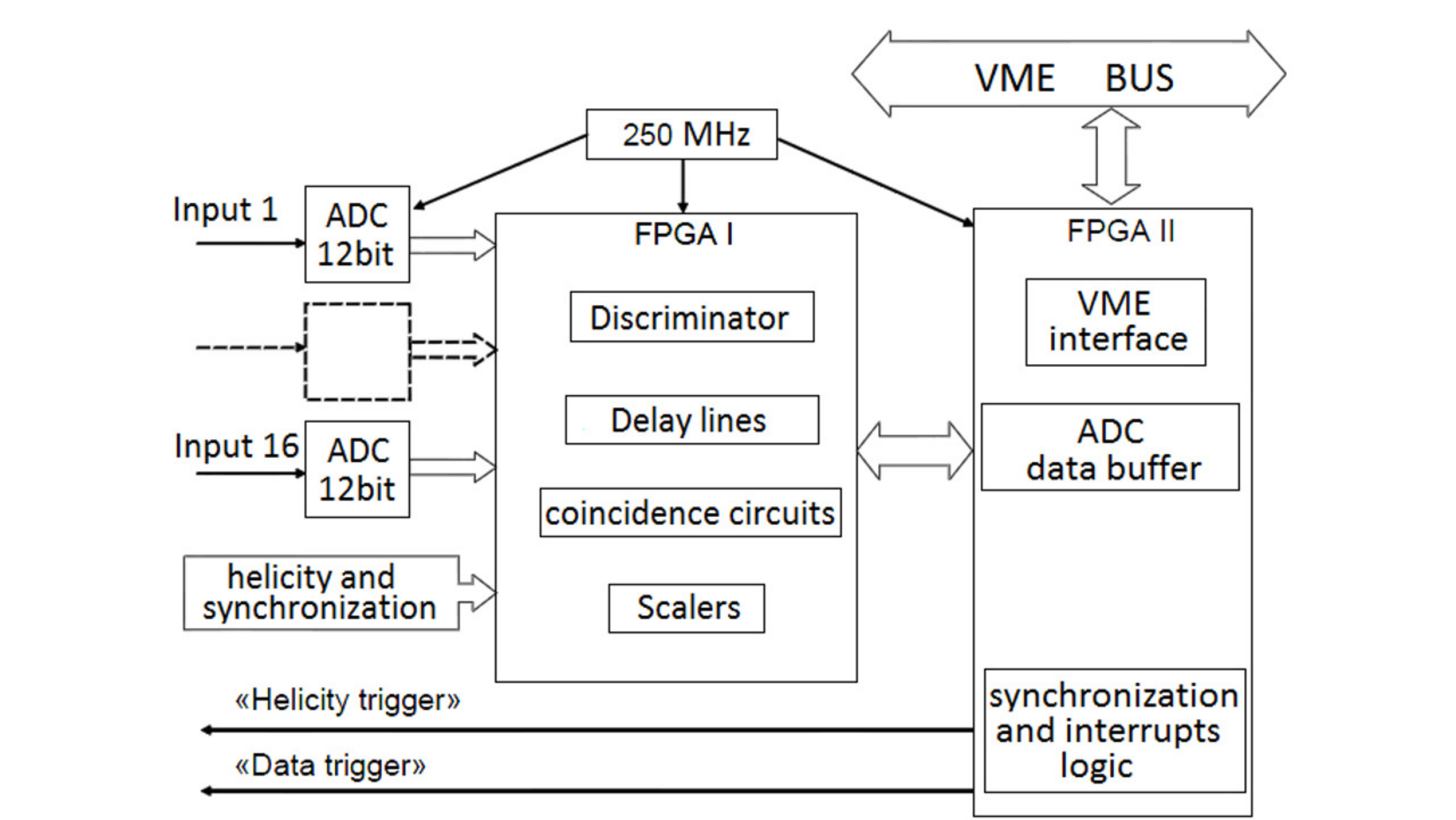,width=15.0cm}}
   \caption{Flash ADC module. }
\label{fig:fadc}
\end{figure}

The module flash-ADC is an integration of 16-channel 12-bit ADC of conveyor type with the conversion frequency 250~MHz and the programmable logic array FPGA in one VME unit. Analog signals with amplitude up to +1~Volt and duration $30\div35$~ns from the detector PMTs are sent to the module input and are digitized in the ADC. The programmable logic module allows to construct the whole logic circuit for event recording and to process digitized signals from the ADC directly in the module. Functions of the discriminators, delay lines, coincidence circuits, scalers and event recording logic are programmed in the FPGA (see Fig.~\ref{fig:fadc}).

After digitizing in the ADC, signals from the detector are summed for each calorimeter arm. The summed signal is selected in the discriminator. If the level of the summed signal exceeds the preset discriminating threshold, the event (digitized signals of all channels) is recorded to inner data buffer of the module and the logic signals are formed at the inputs of the coincidence circuits. Parameters of ADC operation are programmed and uploaded to the module at the CODA start.

The data acquisition system based on flash-ADC generates 2 types of triggers (events):

\begin{itemize}
   \item \textit{Helicity trigger} -when values of scalers are read;
   \item \textit{Data trigger} - when inner buffer is filled with data from the ADC.
\end{itemize}

Upon interrupt signals, the data of inner scalers and ADC buffer are read into the common data flow of the CODA system and are transmitted via network to the controlling computer. On the PC, the data are written into files for further analysis and processing. This system allows to detect and to record data flow at the speed up to 50~MB/s in the coincidence mode which corresponds to the event rate in the arms $\sim$~160kHz.

\begin{figure}[hbt]
   \center{\epsfig{figure=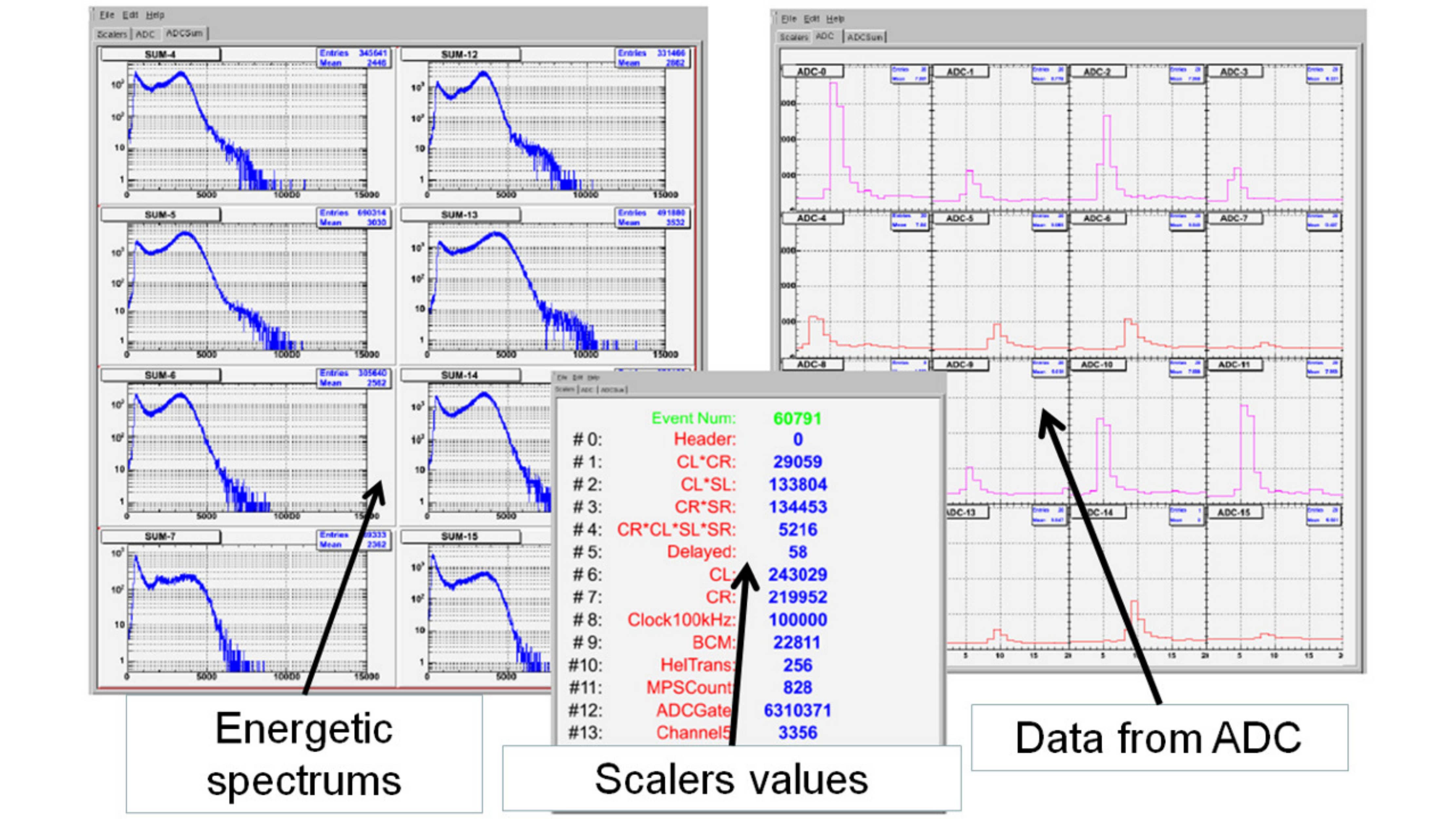,width=17.0cm}}
   \caption{Data from on-line monitoring program. }
\label{fig:online}
\end{figure}

%The software package includes the program for data acquisition and on-line monitoring, and programs for off-line data  processing. Monitoring program allows to control the quality of incoming information by displaying current values of coincidence scalers, digitized analog signals from each detector unit and amplitude spectra of signals from the detector. Fig.~\ref{fig:online} shows an example of displayed information by the on-line monitoring program. The programs for off-line data processing allows to convert data files from CODA into ROOT, to process and analyze data, and to obtain the results of the beam polarization measurement.

The software package includes the program for data acquisition and on-line monitoring, and programs for off-line data  processing. Monitoring program allows to control the quality of incoming information by displaying current values of coincidence scalers, digitized analog signals from each detector unit and amplitude spectra of signals from the detector. Fig.~\ref{fig:online} shows an example of displayed information by the on-line monitoring program. The programs for off-line data processing allows to convert data files from CODA into ROOT~\cite{ROOT}, to process and analyze data, and to obtain the results of the beam polarization measurement.

\begin{figure}[hbt]
  \center{\epsfig{figure=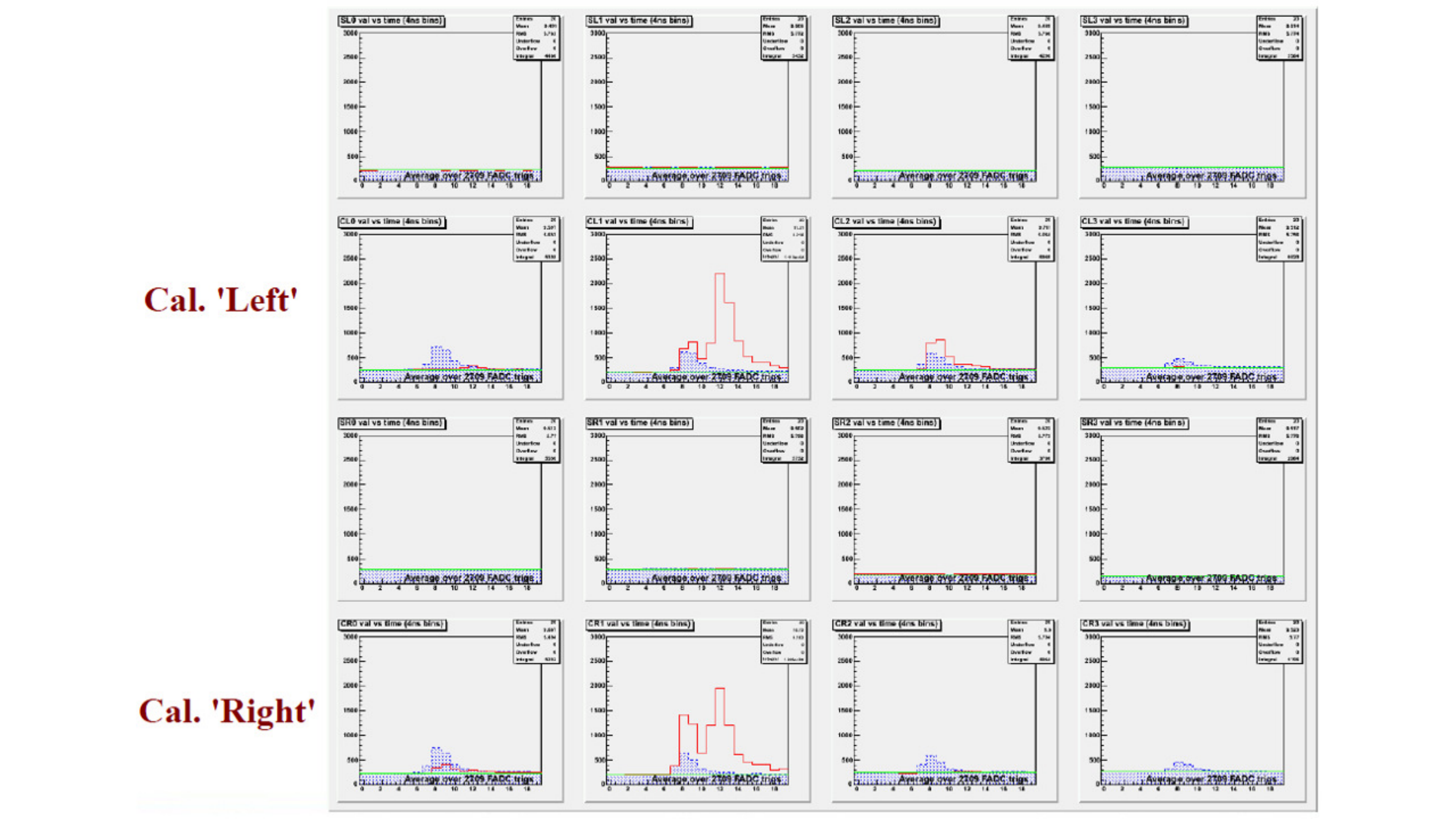,width=17.0cm}}
  \caption{An example of data trigger (pile-up events).}
\label{fig:pileup}
\end{figure}

Ability of the flash ADC to record every single event from the detector (data trigger) allows to study systematic errors. Analysis of this information should help to improve the polarimeter GEANT model, to increase the accuracy of measuring average analyzing power, and to take into account such effects as ``pile-up events'' (see Fig.~\ref{fig:pileup}), Levchuk effect, etc. During operation of the flash-ADC data trigger some problems were found. Due to these problems this new instrument cannot be used in full. The bugs and errors have to be eliminated, and development of the software for the data analysis has to be completed.

\subsubsection{Comparison of two DAQs results}
\label{subsec:compar}

\begin{figure}[hbt]
  \center{\epsfig{figure=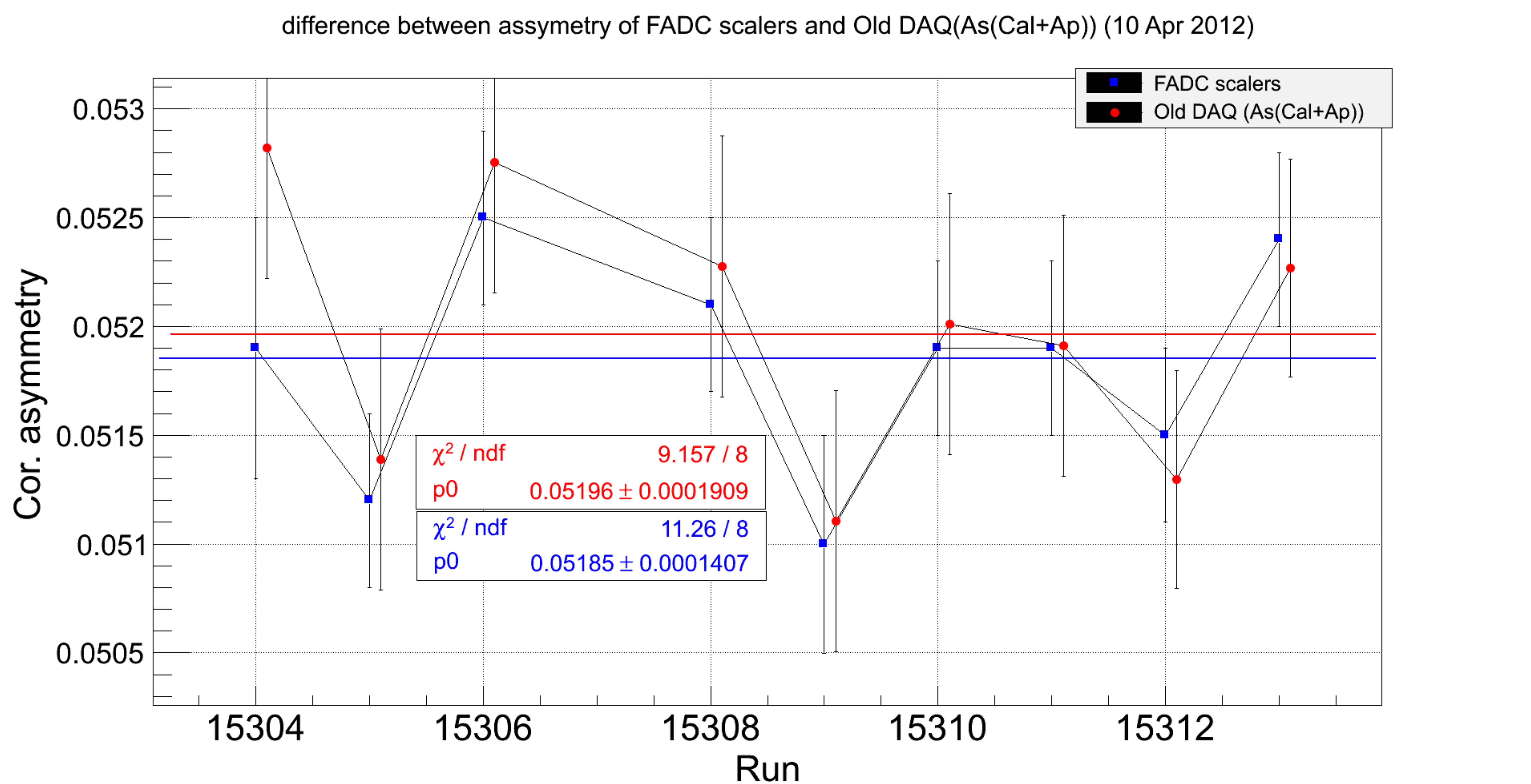,width=15.0cm}}
  \caption{Result of asymmetry measurement with two M\o{}ller polarimeter DAQs. }
\label{fig:compar}
\end{figure}

Fig.~\ref{fig:compar} shows the results of comparison of the asymmetry values measured by both data acquisition systems.

Blue dots show the measurement result with the new DAQ system based on flash-ADC, and red dots show the measurement result with using the old DAQ system. The discrepancy between two DAQs results is not beyond the statistical error.

\subsubsection{Conclusion}
\label{sec:conclusion}

The Hall A M\o{}ller polarimeter has two data acquisition and processing systems operating simultaneously. The systems are based on different element bases with different types of triggers (events).

The old DAQ is fully functional. After the upgrade a few modules will be replaced.

%The new system based on high-speed multichannel flash-ADC has been developed and tested. This system allows to record events with the rate up to 160~kHz in coincidence mode, and allows data acquisition rate up to 50~MB/s without significant increase of the system dead time. It allows to increase the accuracy of measurement of the electron beam polarization with the Hall A M\o{}ller polarimeter ~\cite{review}. The new system is more accurate but currently requires more accurate adjustment and further improvement.

The new system based on high-speed multichannel flash-ADC has been developed and tested. This system allows to record events with the rate up to 160~kHz in coincidence mode, and allows data acquisition rate up to 50~MB/s without significant increase of the system dead time. It allows to increase the accuracy of measurement of the electron beam polarization with the Hall A M\o{}ller polarimeter ~\cite{an_report}. The new system is more accurate but currently requires more accurate adjustment and further improvement.

%\end{document}

\clearpage

%\documentclass{article}
%\usepackage{epsfig,multirow}
%\setlength{\textwidth}{6.5in}
%\setlength{\oddsidemargin}{0in}
%\setlength{\evensidemargin}{0in}
%\setlength{\textheight}{9in}
%\setlength{\topmargin}{0in}
%\setlength{\headheight}{0in}
%\setlength{\headsep}{0in}

%\newcommand{\pvdisqsqI}{1.121}
%\newcommand{\pvdisqsqII}{1.925}

%\begin{document}

%\section{Equipment}                          

\subsection{Compton DAQ}
\label{sec:compton_daq}

\begin{center}
Upgrade of Compton Polarimeter DAQ:  Progress and Plans
\end{center}

\begin{center}
R. Michaels, A. Camsonne, S. Nanda \\
contributed by R. Michaels
\end{center}

In this contribution, I describe the progress and plans to
upgrade the Compton Polarimeter DAQ.  The project can also be
viewed as a test for the new pipelining electronics 
needed for some future projects like SOLID.  
The Hall A Compton Polarimeter consists of an
integrating-mode DAQ and a counting-mode DAQ for
the detectors, plus a slow DAQ for auxiliary signals
such as laser state and beam current.  The two 
modes have different advantages and different systematics.
The integrating mode was developed largely by the 
Carnegie Mellon group prior to the three
parity experiments that ran in 2009-2010, and we don't
foresee modifying this for now.  The counting-mode
DAQ had been developed in 1998 by Saclay, and has served
us well, but it needs to be upgraded.

\begin{figure}
\includegraphics[width=1.05\textwidth]{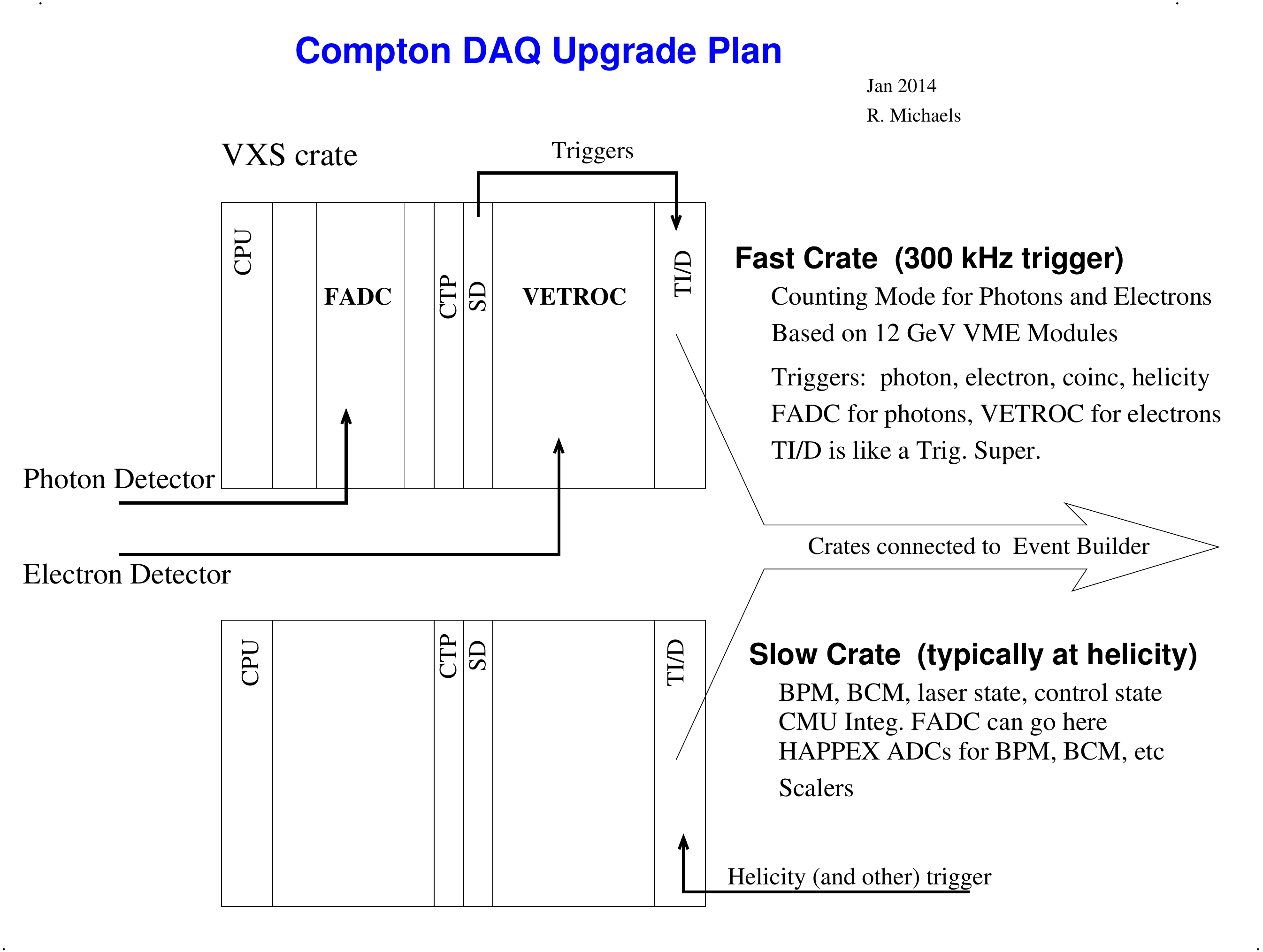}
\caption{Proposed upgrade of the Compton DAQ}
\label{fig:comptondaq}
\end{figure}

The plan for the upgrade and some details about the
progress are kept at this URL:

\url{http://hallaweb.jlab.org/equipment/daq/compton}

The photon detector will be read out by the the JLab FADC.
The electron detector will require a new board, which is being developed
by the JLab DAQ Group and the JLab Fast Electronics Group.
This new board is called the ``VETROC'' and it will replace
the existing ``ETROC''.  (ETROC means electron trigger readout
card, and the new ``V'' implies VXS-based).  
The FADC and VETROC are part of a ``fast''
DAQ which handles triggers, as shown in fig~\ref{fig:comptondaq}.
The triggers will be either
singles triggers from the two detectors or coincidence
triggers.  These electronics boards are based on the 12 GeV
VME modules being used in hall D, for example.

The two main areas of progress in 2013 were : 1) a test
stand was setup to test the FADC and several useful things were
learned about the performance, the deadtime correction, and
systematic errors in the extraction of asymmetries; and
2) a detailed specification of the VETROC was written
and the JLab DAQ Group and JLab Fast Electronics Group
have agreed to build a prototype; indeed, the board
may have broad application at the lab as a flexible trigger module.

The FADC test stand consists of a set of NIM electronics, 
a random high-rate pulser, 
and the VME crate with the FADC and associated DAQ boards 
to test the ability to measure photon signals at 
up to 500 kHz with a known
and variable helicity-correlated asymmetry of
1 to 10\% (note, the Compton asymmetry is 3\%), where the
helicity is provided by the JLab Helicity Generator Board. 
The tests up to now can be summarized as follows.
The fastest method for event-mode readout is to run
the FADC in ``pulse-integral'' mode, whereby a pulse
produces a trigger if it is above a threshold, and
the FADC board provides both the time of the trigger
and the integrated signal, which is integrated over
a time window that extends by a programmable amount 
before the trigger time to an amount after; a window
of about 80 nsec was usually used.  The data are readout
in a multiblock mode.  Since the FADC
is pipelined, there is zero deadtime for 
rates below a critical frequency $f_c$.  The value of
$f_c$ depends on the setup parameters -- mainly the number
of enabled channels and the integration window -- and
was 300 kHz for our setup.  Below $f_c$ the asymmetries
extracted from the data agreed with expectation to better
than 100 ppm.  Above $f_c$ there is a significant deadtime
which, so far, we cannot correct; however, we've learned
that this deadtime affects the data in contiguous time bins
or ``chunks'' of time; the majority of helicity windows remain
deadtime-free if $f \le$ 500 kHz.  Above 500 kHz the
data become less useful as one approaches 100\% deadtime.  
The ``chunks'' of data affected by deadtime can be flagged 
and eliminated, leaving a clean sample of data whose 
asymmetry agrees with expectation.

The near-term goal is to commission this counting-mode 
photon DAQ with the beam, and run it in conjunction with 
the existing integrating-mode DAQ.  Meanwhile, the VETROC
prototype will be available in about one year and we will
test it in the aforementioned DAQ test stand.

%\end{document}

\clearpage

\section{Summaries of Experimental Activities}

%\documentclass[11 pt]{report}
%\usepackage[pdftex]{graphicx}
%\usepackage{float}
%\usepackage{amsmath}
%\usepackage{multirow}
%\restylefloat{figure}
%\usepackage{subfig}
%\usepackage[nosort]{cite}
%\usepackage[top = 1.0in, bottom=1.25in,left = 1.5in, right= 1.0 in]{geometry}
%\usepackage{setspace}
%\usepackage[compact]{titlesec}
%\titlespacing{\subsubsection}{0pt}{0.1ex}{0ex}
%\setlength{\parindent}{0pt}
%\setlength{\parskip}{1ex plus 0.5ex minus 0.2ex}
%\begin{document}
%\begin{spacing}{1}
%\title{}{ The Electro-disintegration of Deuteron at High Four-momentum Transfer}
%\author{Hari P Khanal \and  W. U. Boeglin}

\subsection[E01-020: (e, e'p) Studies of the Deutron at High $Q^2$]{E01-020: (e, e'p) Studies of the Deutron at High $Q^2$} \label{sect:e01020}

%\begin{center}
%(e, e'p) Studies of the Deutron at High $Q^2$
%\end{center}
\begin{center} 
Hari P Khanal and  W. Boeglin \\
Spokespersons: W. Boeglin, E.Voutier, M.Jones, A.Klein, P.E.Ulmar, \\
The Hall A Collabration.
\end{center}

\subsubsection {Introduction}
The deuteron is the simplest nuclear system consisting of a single  proton and a single neutron. Exclusive electron scattering from the deuteron is an efficient way of probing the dynamics of two nucleons at very short space-time distance. The simplest model of deuteron electro-disintegration is the Plane Wave Impulse Approximation (PWIA) in which the proton ejected by the virtual photon does not further interact with  the recoiling neutron and is described by a plane wave. In this approximation, the missing momentum  $\vec{p_{m}} = \vec{q} - \vec{p_{f}}$, where $\vec{p_{f}}$ is the momentum of the outgoing proton and $\vec{q}$ is the 3-momentum transfer, represents the momentum of the recoiling nucleon and is opposite to the initial momentum $\vec{p_i}$ of the struck nucleon $\vec{p_i} = - \vec{p_m}$. However the out-going proton can interact strongly with the recoiling neutron thereby destroying the simple relationship between $\vec{p_m}$ and $\vec{p_i}$ (final state interaction, FSI).
 
The main focus of the analysis consists of the determination of the $d(e,e'p)n$ cross section as a function of (I) missing momenta for fixed angles between the momentum transfer and the momentum of the recoiling nucleon ($\theta_{nq}$) and (II) as a function of $\theta_{nq}$ for fixed values of $p_m$. State of the art calculations~\cite{Sabine1,FMGSS95,FSS96,Jesch2001,Ciofi08,treeview,JW08,JW09,Kap05,La05,JW10,noredepn}
and recent results have indicated that at high energies ($\approx$ 1 GeV) in the final n-p system  the strength of FSI depend strongly on the recoiling neutron angle.

The $d(e,e'p)n$ differential cross section in one photon exchange  is  written in the following way
  \begin{equation}
  \frac{d^5 \sigma}{d \omega d \Omega _{e} d \Omega _{p}} =Kf\sigma_{Mott}(\nu_{L}R_{L} +\nu_{T}R_{T} + \nu_{TT}R_{TT}cos(2\phi) + \nu_{LT}R_{LT}cos(\phi)).
  \end{equation}
   Where $R_{i}$ and $R_{ij}$ are the nuclear  response functions, $\nu_{i}$ and $\nu_{ij}$ are the coefficients which depend only on the electron kinematics  and $\sigma_{Mott}$ is the Mott cross section.
  In  PWIA, the differential cross section can be  factorized to 
  \begin{equation}
  \frac{d^5 \sigma}{d \omega d \Omega _{e} d \Omega _{p}} =K\sigma_{ep} \rho(p_{m}).
  \end{equation}

Where K is a  kinematic factor, $\sigma_{ep}$ is the off-shell electron-proton cross section  and  $\rho(p_{m})$ is the momentum distribution, which describes the probability of finding the nucleon with initial  momentum $-\vec{p}_{m}$ in the ground state of the deuteron. 

\subsubsection {Experimental Overview}
Data have been taken during  experiment  E01020  in Hall-A at Jefferson Lab at energies of  2.83 , 4.7  and  5.0 GeV  for $Q^2$ values of 0.8, 2.1 and 3.5  (GeV/c)$^2$, respectively. The Hall A cryogenic target system provided a 15 cm long liquid deuterium target at beam currents ranging from 1  to 100 $\mu A$.  The scattered electrons and out-going protons were detected by  the two 4 GeV/c High Resolution Spectrometers(HRS).  In each HRS, the timing information was provided by two scintillator planes. A pair of  Vertical Drift Chambers (VDC) provided the tracking information in each spectrometer and electrons were identified using the gas Cherenkov detector. $d(e,e'p)n$ events have been selected by a time of flight cut between the spectrometers, a cut on missing energy and a cut on the reconstructed vertex location from each spectrometer.

\subsubsection {Analysis Status}
First results at $ Q^2 = 3.5 $ (GeV/c)$^2 $ have already been published in \cite{werner1} while the analysis of the $Q^2 = 0.8$ and 2.1 (GeV/c)$^2 $ data is nearing completion. For the lower $Q^2$ data detector calibrations, beam position determinations, detector efficiency determinations, radiative correction and phase space calculations have been completed. Differential cross sections for the angular distribution and for the momentum distribution at both $ Q^2 = 0.8$ and  2.1  (GeV/c)$^2$ have been determined. We are currently working on the estimation of the systematics uncertainties and  an extraction of $A_{LT}$. 

\paragraph{Momentum Distribution}

\begin{figure}[hbt]
\begin{center}
\includegraphics[height=3.in]{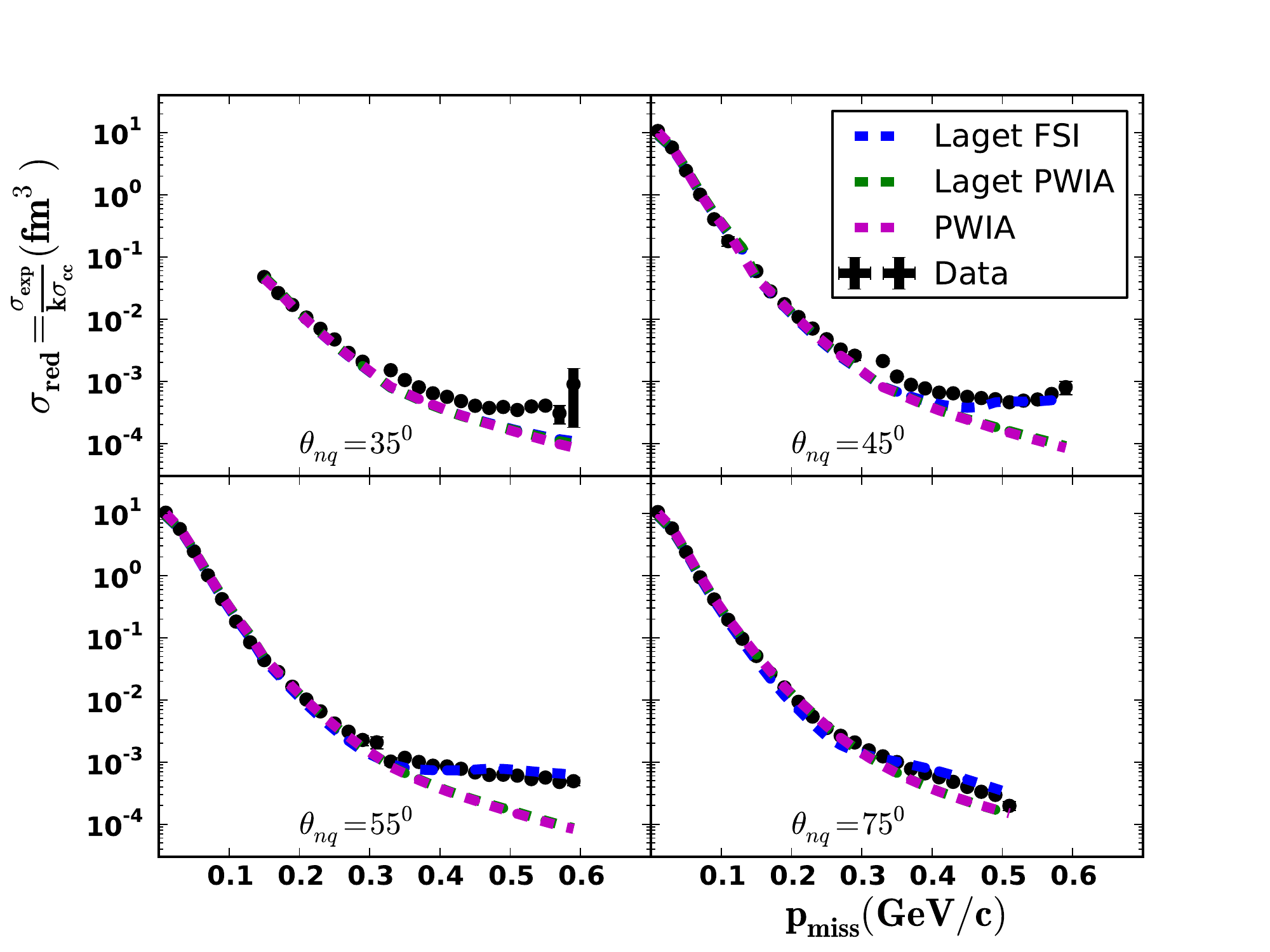}
\caption {{\label{fig1}}The momentum distribution at low $ Q^2 = 0.8 $ (GeV/c)$^2$ for different recoil angles $\theta_{nq}$} 
\end{center}
\end{figure}

\begin{figure}
\begin{center}
\includegraphics[height=3.5in]{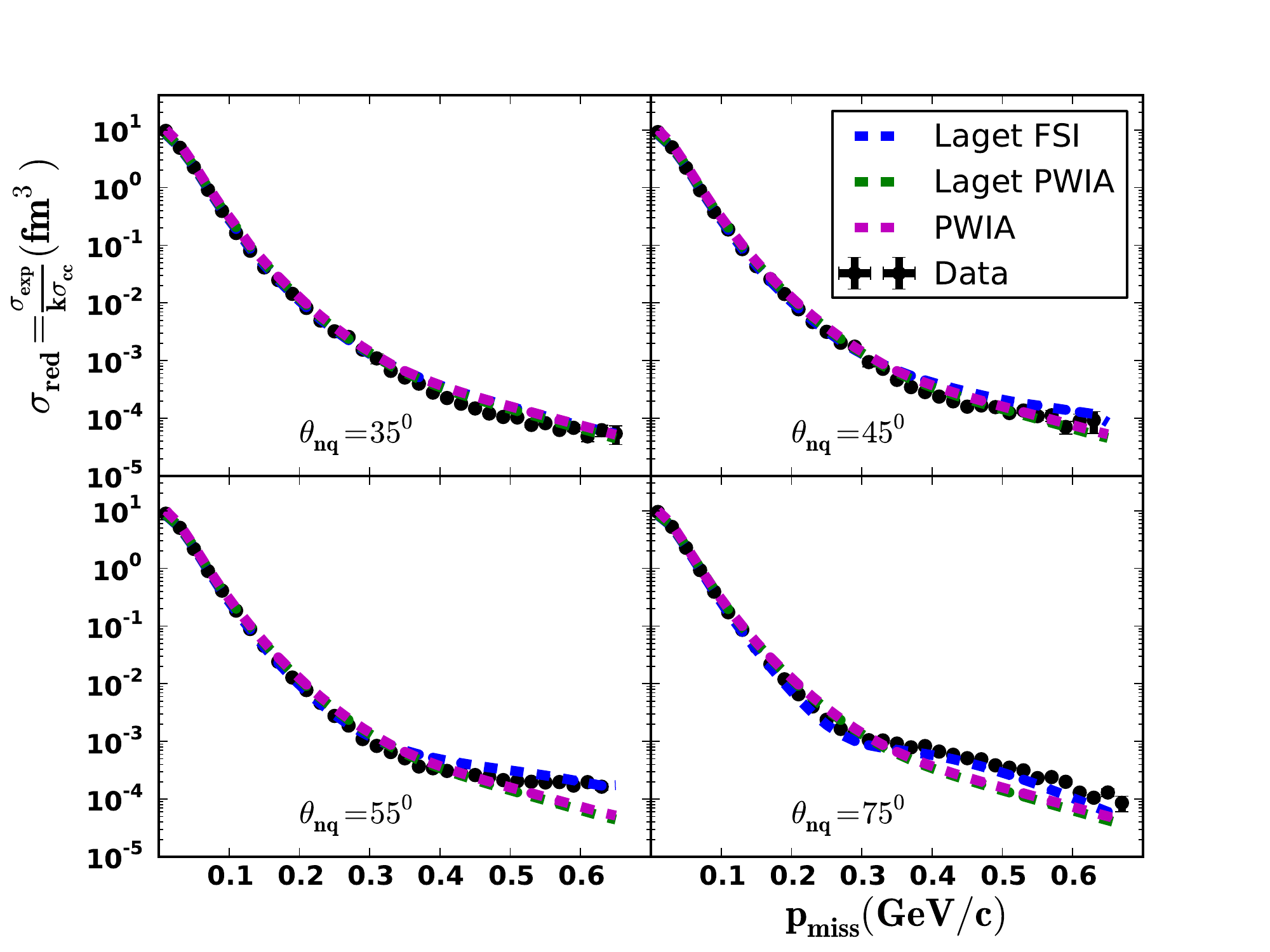}
\caption {{\label{fig2}}The momentum distribution at high $ Q^2 = 2.1 $ (GeV/c)$^2$ for different recoil angles $\theta_{nq}$} 
\end{center}
\end{figure}

\begin{figure}
 \begin{center}
\includegraphics[height=3.5in]{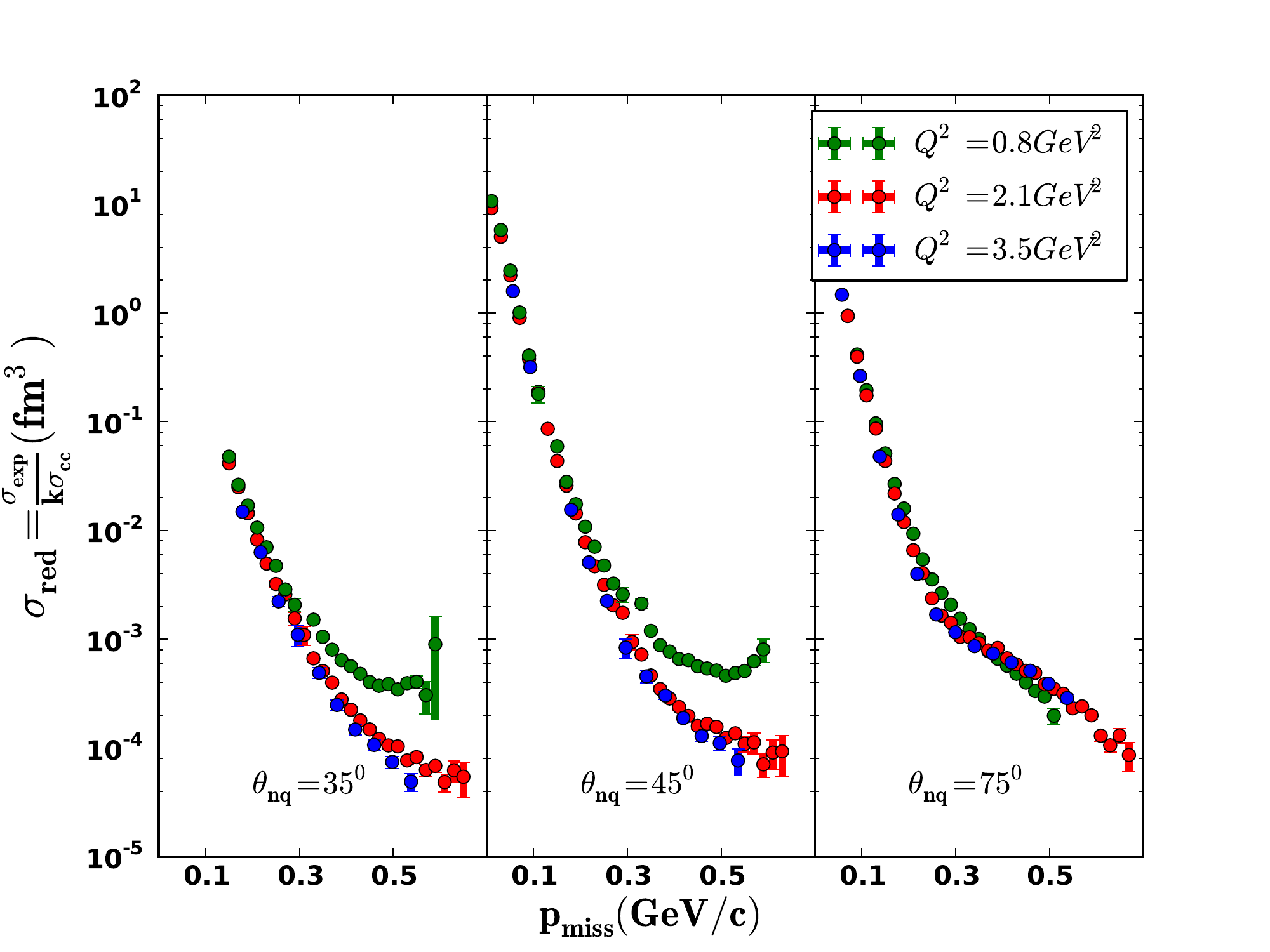}
\caption {{\label{fig3}} Comparison of the experimental momentum distributions for $ Q^2$ = 0.8  (green),  2.1 (red), and 3.5 (GeV/c)$^2$ \cite{werner1} (blue) for different values of $\theta_{nq}$  } 
\end{center}
\end{figure}

If there were no FSI the momentum distribution could be extracted from the measured cross section by simply dividing them by $K\sigma_{ep}$. In reality FSI are always present to a certain degree and this ratio is referred to as the reduced cross section.
Figs.~\ref{fig1} and ~\ref{fig2} show the reduced cross section  
 as a function of the missing momentum for a set of four, fixed recoil angles and for $Q^2 = 0.8$  and  $Q^2 = 2.1$ (GeV/c)$^2$. The $\theta_{nq}$ bin width of each setting of recoil angle is $\pm 5^\circ$ and the missing momentum bin width is $\pm 10$MeV/c. The experimental reduced cross section has been compared to a calculated one with and without FSI. At low $Q^2$ FSI start to contribute significantly for missing momenta above 0.2 - 0.3 GeV/c for all angles. In contrast at higher $ Q^2 $ as shown in Fig.~\ref{fig2}, FSI dominates the cross section only around  $\theta_{nq} = 75 ^0 $. Fig.~\ref{fig3} shows a comparison of the experimental reduced cross sections (momentum distributions)  at $Q^2 = 0.8, 2.1$ and $3.5$ (GeV/c)$^2$. All distributions  agree with each other in the low missing momentum region.

\paragraph {Angular Distribution} 

\begin{figure}
\begin{center}
\includegraphics[height=3.5in]{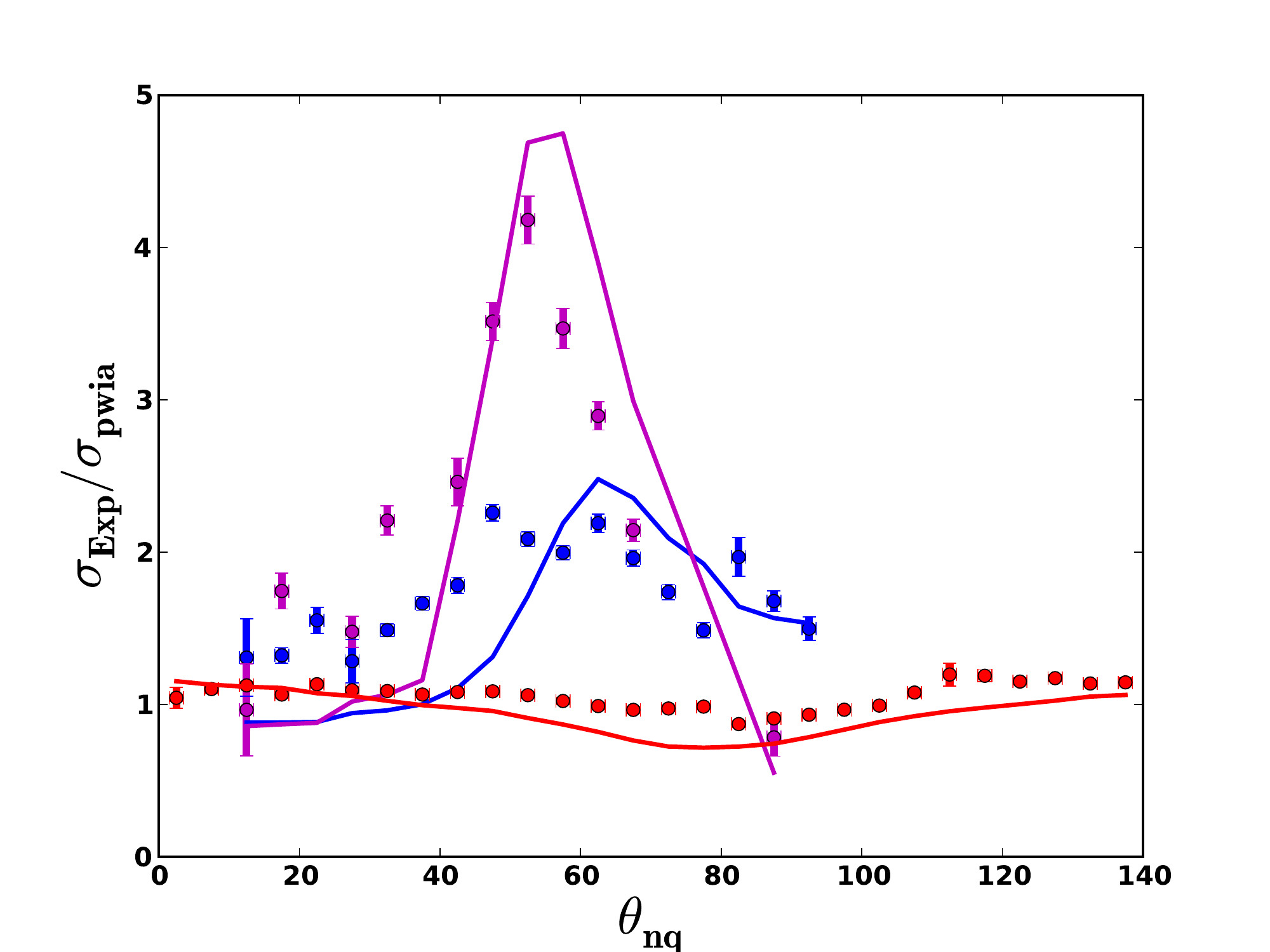}
\caption {{\label{fig4}}$R$ as a function of $\theta_{nq}$ for $ Q^2 = 0.8 $(GeV/c)$^2$. Red: $p_{m} = 0.2 $GeV/c , blue: $p_{m} = $ 0.4 GeV/c and  magenta: $p_{m} = 0.5 $GeV/c . The corresponding sold lines represent calculations using Laget's model including FSI } 
\end{center}
\end{figure}

\begin{figure}
\begin{center}
\includegraphics[height=3.5in]{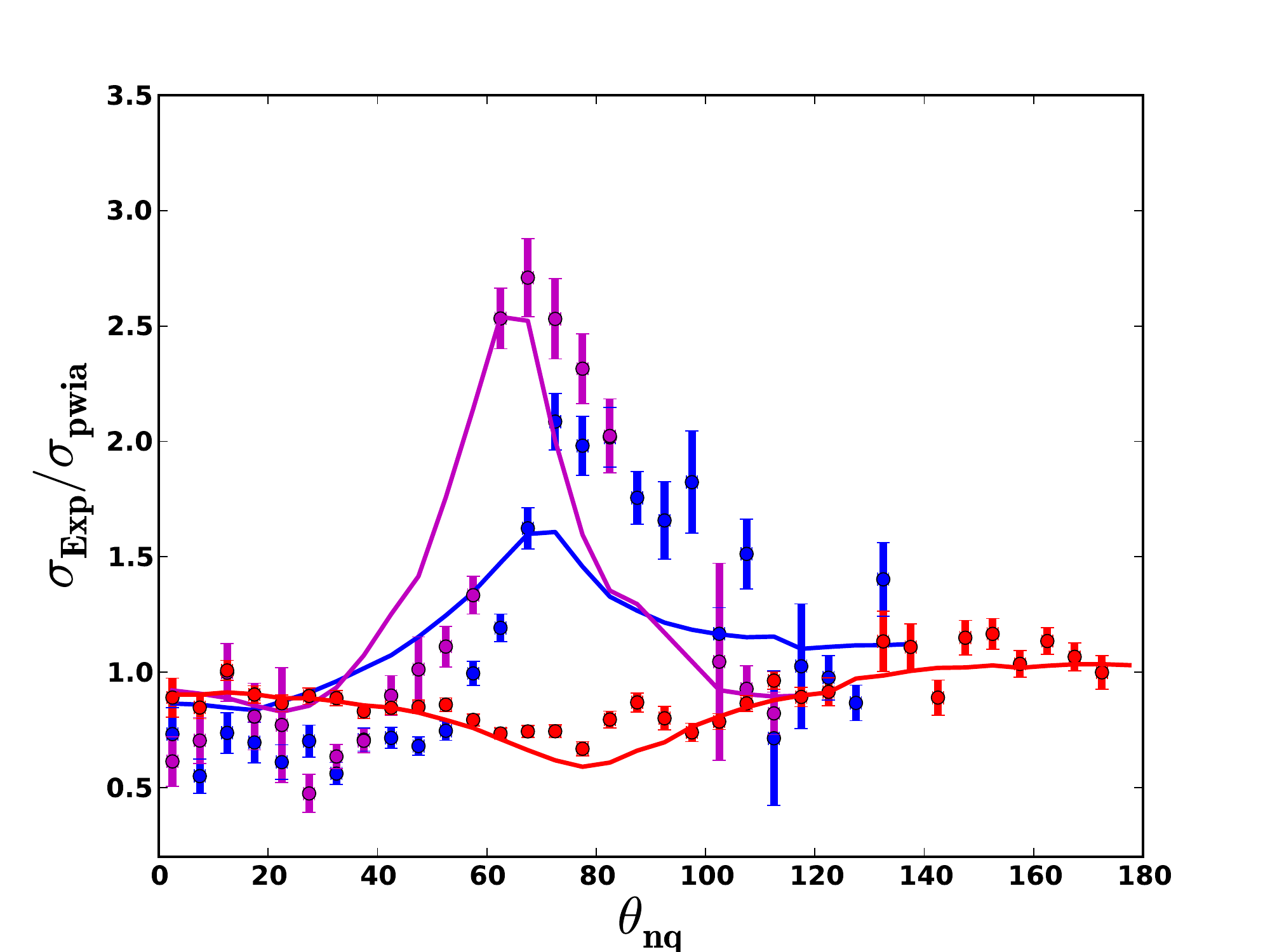}
\caption {{\label{fig5}}Like Fig.~\ref{fig4} for $Q^2 = 2.1$ (GeV/c)$^2$}
\end{center}
\end{figure}

In order to study the angular dependence of FSI contributions we determined the ratio $ R =\sigma_{exp}/\sigma_{pwia}$ of the experimental cross section ($\sigma_{exp}$) to the PWIA cross section ($\sigma_{pwia}$). If there were no FSI and the momentum distribution used correctly described the deuteron structure then $R = 1$ independent of $\theta_{nq}$ would be found. The experimental values of $R$ are shown for missing momenta $p_{m}= 0.2, 0.4$ and 0.5 GeV/c  and at $ Q^2 = 0.8 $ and 2.1  (GeV)$^2$ in Figs.~\ref{fig4} and \ref{fig5}.

At low $ Q^2$, the distributions are quite  broad with large FSI contributions even at small angles $\theta_{nq}<40^\circ$ and missing momenta of $p_m = 0.4$ and 0.5 GeV/c. Only small fluctuations around 1 are found for $p_m = 0.2 $ GeV/c. 
 
In contrast at $ Q^2 = 2.1  $(GeV)$^2$ $R$ has a well defined peak at around 75$^\circ$ as shown in Fig.\ref{fig5}.  At missing momentum $p_{m} = 0.2$ GeV/c  $R$ is reduced by about 30\% at $\theta_{nq}$ around $75^\circ$. For $p_{m} = $ 0.4 GeV/c and   $p_{m} = $ 0.5 GeV/c  R increases at around $75^\circ$ by factor 2.5 and 3.0 respectively. The angular dependence of R clearly indicates that FSI between the two final state nucleons at high missing momenta is highly anisotropic. For both data sets the experimental results have been compared with the results from Monte Carlo simulations using Laget's Model including FSI. The solid lines represent the distributions calculated from the theoretical model. At $Q^2 = 0.8$ (GeV/c)$^2$ the calculated angular distributions agree well with the experimental results for $p_m = 0.2$ GeV/c and $\theta_{nq} < 40^\circ$ only. For all other kinematic settings at this momentum transfer they do not very well reproduce the experimental results for $R$. At $Q^2 = 2.1$ (GeV/c)$^2$ and at $p_{m} = $ 0.2 GeV/c the experimental results agree quite well with the theoretical calculations while larger discrepancies  exist at higher missing momenta.

\bibliography{biblography_werner}
\bibliographystyle{unsrt}

%\end{spacing}

%\end{document}

\clearpage

%\documentclass{article}
%\usepackage{epsfig}
%\setlength{\textwidth}{6.5in}
%\setlength{\oddsidemargin}{0in}
%\setlength{\evensidemargin}{0in}
%\setlength{\textheight}{9in}
%\setlength{\topmargin}{0in}
%\setlength{\headheight}{0in}
%\setlength{\headsep}{0in}
%\usepackage{mciteplus}

%\begin{document}
 %\bibliographystyle{plain}
 
% \section{Summaries of Experimental Activities}
 
\subsection[E04-007: Precision Measurements of Electroproduction of $\pi^0$ near Threshold]{E04-007: Precision Measurements of Electroproduction of $\pi^0$ near Threshold}

%\begin{center}
%\bf Precision Measurements of Electroproduction of $\pi^0$ near Threshold: A Test of Chiral QCD Dynamics
% \end{center}

\begin{center}
contributed by Richard Lindgren and Cole Smith\\
for\\
J.R.M.~Annand, D.W.~Higinbotham, R.~Lindgren, B.~Moffit, B.~Norum, V.~Nelyubin, spokespersons, \\
M. Shabestari and K. Chirapatimol, students\\
and\\
The Hall-A Collaboration\\
\end{center}
 
 \subsubsection{Introduction}
The experiment is designed to measure the electroproduction reaction $p(e,e'p)\pi^0$ of neutral pions off the proton at the lowest possible invariant mass W. Results from previous electroproduction measurements at Mainz with four-momentum transfers of $Q^2$ = 0.10 GeV/c$^2$~\cite{Distler_98} and $Q^2$ = 0.05 GeV/c$^2$~\cite{Merkel_02} were in disagreement with the $Q^2$ dependence predicted by Heavy Baryon Chiral perturbation theory (HBChPT) and also inconsistent with the predictions of the MAID model~\cite{ref:MAID}.  If the Mainz discrepancies remain unresolved, they will constitute a serious threat to the viability of Chiral Dynamics as a useful theory of low energy pion production.  Our experiment has measured absolute cross sections as precisely as possible from threshold to $\Delta W$ = 30~MeV above threshold at four-momentum transfers in the range from $Q^2$ = 0.050 GeV/c$^2$  to $Q^2$ = 0.150 GeV/c$^2$ in small steps of $\Delta Q^2$ = 0.01 GeV/c$^2$. This will cover and extend the Mainz kinematic range allowing a more sensitive test of chiral corrections to Low Energy Theorems for the S and P wave pion multipoles. In addition, the beam asymmetry was measured, which can be used to test predictions for the imaginary components of the of S wave $E_{0+}$ and  $L_{0+}$ pion multipoles, which are sensitive to unitary corrections above the $n \pi^+$ threshold.  Mainz recently repeated the electroproduction measurements and now report~\cite{Merkel_09} that the new results are more consistent with HBChPT predictions, but are in disagreement with their own previous measurements~\cite{Merkel_02}. In view of the importance of knowing whether or not HBChPT is valid in this domain, it is imperative that an independent set of measurements be reported.
  
\subsubsection{Analysis Status}
The default LHRS optics data base used in our analysis up to now requires higher order polynomials to eliminate kinematic distortions near the edges of the acceptance.  Our attempt to reduce these distortions by fitting an LH2 sieve run introduced some unphysical oscillations in the dependence on $L.tgph$. To eliminate this unphysical behavior the LHRS optics has been further refined by eliminating some of the higher order terms which was causing the over-fitting of the matrix coefficients. Although now the fits overall are more reasonable without the fine oscillations, the refit has introduced a small shift in the LHRS horizontal angle.  We find also that the hydrogen elastic and carbon elastic cross sections are uniformly lower by 5$\%$ and 10 $\%$, respectively, from expected according to previous measurements.  Although the discrepancy in the elastic yields could derive from this angle shift, it is also possible we are underestimating other corrections.  We have also studied single track efficiency corrections, which are fairly large for the high rate LH2 data and less so for the C runs.  We discovered and corrected some previous errors in our application of the Mo-Tsai radiative and straggling correction. Also to improve our understanding of the LHRS acceptance and to study the effect of multiple scattering on vertex reconstruction in the LH2 target we added the J.J. LeRose forward transport functions to our GEANT simulation of the LH2 target, target chamber exit window, LHRS entrance window and BigBite. (LHRS was not otherwise simulated except through r.m.s. smearing of the e- momentum vector according to results of previous empirical studies.) This simulation is now under study and hopefully we will reduce the hydrogen and carbon elastic yield discrepancy closer to the 5$\%$ level. Our main effort is to reduce the systematic error on the $\pi^0$ production cross section, W and  $Q^2$. Further checks are also in progress on the BigBite optics to reduce remaining reconstruction errors and improve missing mass systematics. To help speed up systematic studies a reduced data set saving only essential variables needed for the remaining analysis has been created and stored on the Galileo computer farm at UVa.

%\newpage

%\end{document}

  %E04007
\clearpage

\subsection%[E05-102: $A_x'$ and $A_z'$ asymmetries]
[E05-102:  Measurement of $A_x'$ and $A_z'$ asymmetries in the quasi-elastic
$^3\vec{\mathrm{He}}(\vec\mathrm{e},\mathrm{e}'\mathrm{d})$ reaction]{E05-102:  Measurement of $A_x'$ and $A_z'$ asymmetries in the quasi-elastic
$^3\vec{\mathrm{He}}(\vec\mathrm{e},\mathrm{e}'\mathrm{d})$ reaction}
\label{sec:e05102}

%\begin{center}
%Measurement of $A_x'$ and $A_z'$ asymmetries in the quasi-elastic
%$^3\vec{\mathrm{He}}(\vec\mathrm{e},\mathrm{e}'\mathrm{d})$ reaction
%\end{center}

\begin{center}
S.~Gilad, D.~W.~Higinbotham, W.~Korsch, S.~\v{S}irca, B.~E.~Norum (spokespersons) \\
and \\
the Hall A Collaboration.\\
Contributed by M.~Mihovilovi\v{c} and S.~\v{S}irca
\end{center}

\noindent The E05-102 experiment \cite{e05102} is devoted
to a detailed study of the
$$
^3\vec{\mathrm{He}}(\vec\mathrm{e},\mathrm{e}'\mathrm{d}) \>, \qquad
^3\vec{\mathrm{He}}(\vec\mathrm{e},\mathrm{e}'\mathrm{p})\mathrm{d} \>, \qquad
^3\vec{\mathrm{He}}(\vec\mathrm{e},\mathrm{e}'\mathrm{p})\mathrm{pn}
$$
processes at low $Q^2$.  Its main purpose is to approach
the ground-state structure of the $^3\mathrm{He}$ nucleus
by studying the missing-momentum ($p_\mathrm{miss}$) dependence
of the double-polarization (beam-target) transverse and longitudinal 
asymmetries.  The experiment has been performed in Summer 2009.
The data analysis of the
$^3\vec{\mathrm{He}}(\vec\mathrm{e},\mathrm{e}'\mathrm{d})$ channel
is now complete.  We have insufficient resolution to distinguish
the two- and three-body channels in the proton knockout processes,
but the analysis of the proton channel as a sum of these contributions
is also complete.

Since the large-acceptance spectrometer BigBite \cite{BBNIM}
was used in the experiment, the most problematic part of the data 
analysis has been the averaging of the theoretical asymmetries 
over the experimental acceptance.  Here we focus on the averaging 
procedure for the deuteron channel.  We have received the calculated 
asymmetries in
$^3\vec{\mathrm{He}}(\vec\mathrm{e},\mathrm{e}'\mathrm{d})$
from three theory groups: the Bochum/Krakow, the Hannover/Lisbon
and, most recently, the Pisa group.   Since the calculations
are numerically intensive, they have provided us only with 
calculations on a discrete grid of $35$ kinematic points distributed 
over the acceptance of the electron arm (HRS spectrometer).
Each kinematic point represents one combination of $(E',\theta_\mathrm{e})$.
For each such pair the asymmetries are calculated as functions
of $p_\mathrm{miss}$ and $\phi_\mathrm{dq}$.

\begin{figure}[hbtp]
\center{\epsfig{figure=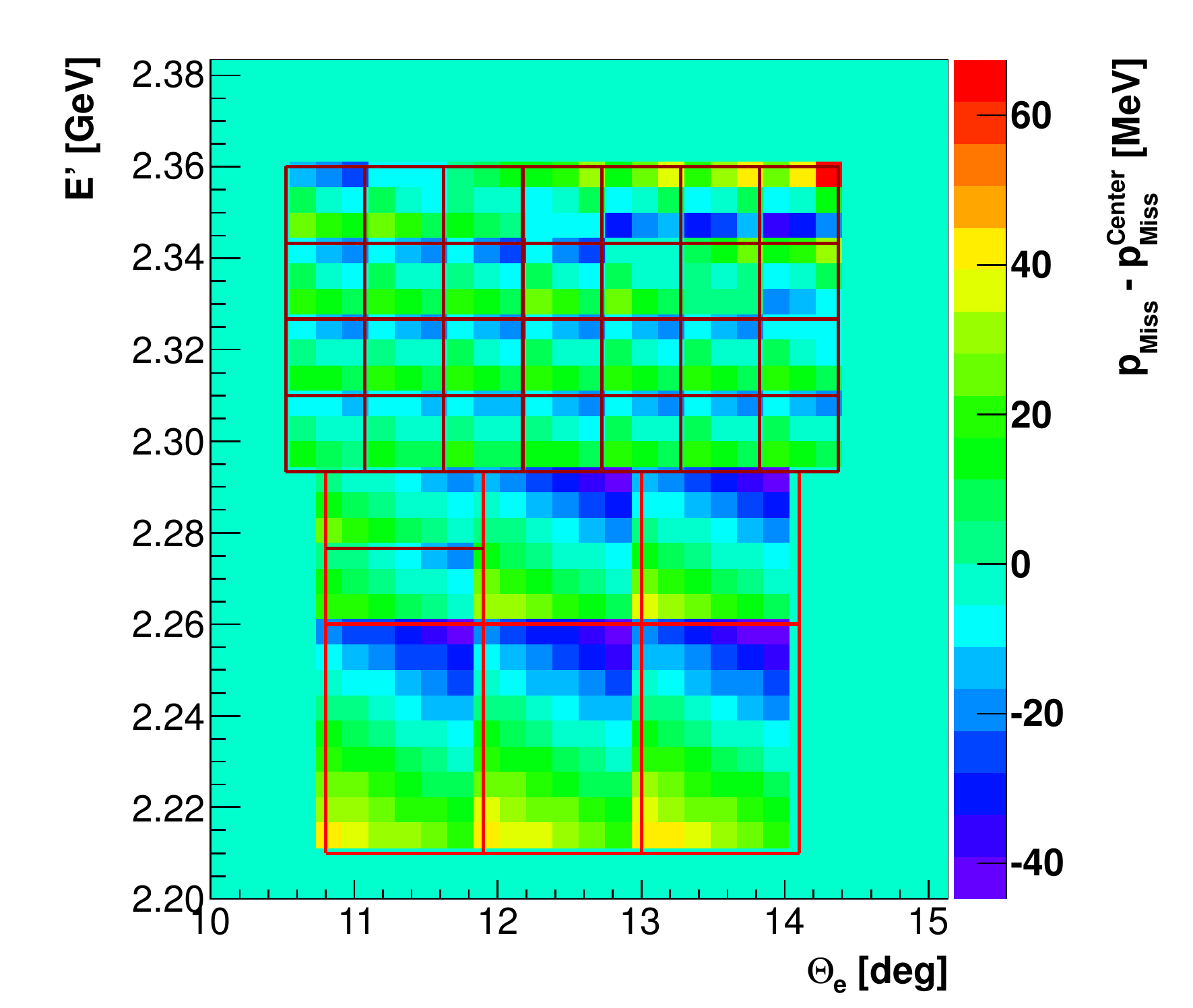,width=8.5cm}}
\vspace*{-3mm}
\caption[E05102: Deviations of $p_\mathrm{miss}$ in bins
used for theory averaging.]{Deviations of $p_\mathrm{miss}$
from the values at the bin centers (where the theoretical calculations
are available) in the $(E',\theta_\mathrm{e})$ region used 
for averaging of the theories over the experimental acceptance.}
\label{E05102-fig1}
\end{figure}

In this report we describe an important alteration of the previous
averaging procedure.  Initially, the following method was used.
From the measured data we determined all relevant kinematic variables
$E'$, $\theta_\mathrm{e}$, $p_\mathrm{miss}$, and $\phi_\mathrm{dq}$
for each event.  We then checked whether the theoretical
asymmetry is actually available for the event under consideration:
if the event was within the corresponding rectangular area
in the 35-point $(E',\theta_\mathrm{e})$ mesh, the theoretical
asymmetry {\sl at the center\/} of that rectangle was assigned
to the event.  Subsequently, we had to check whether the calculation
exists for the given $p_\mathrm{miss}$ of the actual event since
not all values of $p_\mathrm{miss}$ are available at all kinematic
points.  If the calculation existed, the event was accepted and
the theoretical asymmetry for the given set of kinematic variables
was computed.  The final theoretical asymmetry as a function of 
$p_\mathrm{miss}$ was obtained by averaging the asymmetries
over all events in each $p_\mathrm{miss}$ bin.

The advantage of this method was that the measured data and 
the computed asymmetries were compared on an equal footing,
since the experimental asymmetry was  determined by using  
only the data for which theoretical calculations exist.
But this procedure had many handicaps.  By selecting only events 
that can be furnished with the corresponding theoretical asymmetry
we have lost a huge fraction of the events.  In addition,
when calculating the theoretical asymmetry, we have always considered
the event as coming from the bin center.  This induces certain
problems because our kinematic bins are quite large. 
For example, in each bin the minimal reachable $p_\mathrm{miss}$
can be up to $40\,\mathrm{MeV}/c$ lower than the minimal 
$p_\mathrm{miss}$ available at the bin center, as shown
in Fig.~\ref{E05102-fig1}.  This is problematic 
at low $p_\mathrm{miss}$, because it forces us to throw away statistics.
Neglecting low-$p_\mathrm{miss}$ data in some bins can also lead
to variations in the final average values of the asymmetries.

Furthermore, if an event comes from near the edge of the selected bin,
the theories from the neighboring bins are almost equally valid
as the one belonging to the chosen bin: the closer one approaches the edge 
of a bin, the more important the neighboring bins become.
Hence, we need to correctly decide which theory we are going
to consider and how.

All these questions and problems have led us to develop
a new method for comparing the data to the simulation.
The main goal of this new approach is to keep as much as
possible of the measured data by avoiding unnecessary cuts.
We do not want to make the data conform to the theory but vice-versa.
This means that one needs to extend the $p_\mathrm{miss}$ range
of the calculations in order to match the larger range spanned
by the measured data.  This has been achieved by extrapolating
the calculations to smaller values of $p_\mathrm{miss}$, which is
not trivial as the $p_\mathrm{miss}$-dependence of the asymmetries 
is far from linear.  However, since typical extrapolations of
not more than approximately $20\,\mathrm{MeV}/c$ are needed,
the procedure is not expected to introduce large systematic errors.

\begin{figure}[!h]
\center{\epsfig{figure=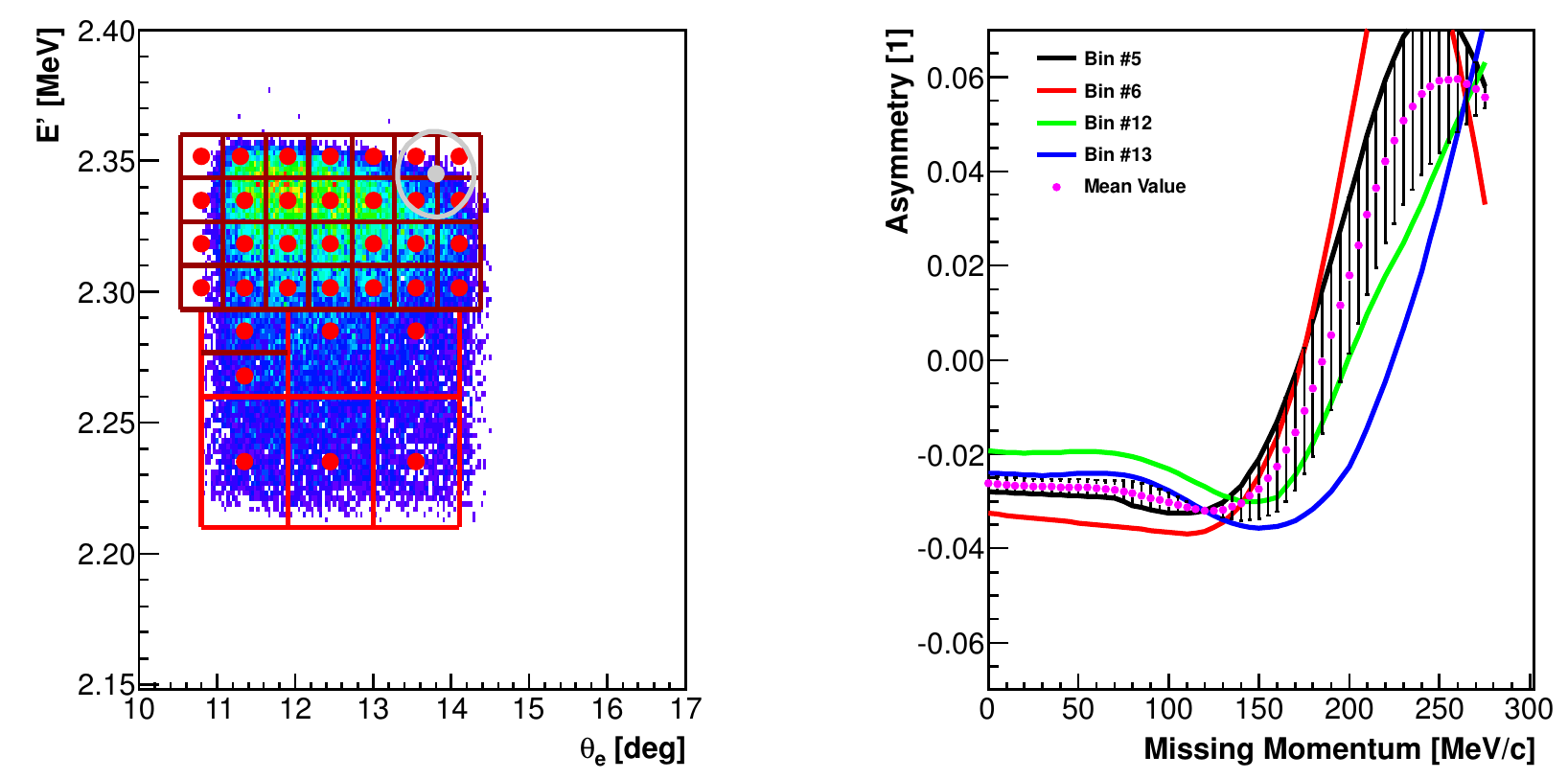,width=16.0cm}}
\vspace*{-3mm}
\caption[E05102: Averaging of the theoretical asymmetries.]
{Averaging of the theoretical asymmetries for the
$^3\vec{\mathrm{He}}(\vec\mathrm{e},\mathrm{e}'\mathrm{d})$ channel.
Left: selection of neighboring bins for a chosen event (gray point
at the intersection of four regions) and its neighborhood (gray circle)
used to weigh the corresponding theoretical predictions in the adjacent
four bins.  Right: weighted predicted asymmetries from all four bins 
encompassed by the gray circle.  The resulting averaged interpolated
asymmetry (magenta points with uncertainties corresponding
to the averaging procedure) now extends to $p_\mathrm{miss}=0$.}
\label{E05102-fig2}
\end{figure}

%\newpage

To obtain the best possible estimate for the asymmetries
in the low-$p_\mathrm{miss}$ region we defined the negative
$p_\mathrm{miss}$ axis by exploiting the fact that
$A(-p_\mathrm{miss}, \phi_\mathrm{dq}) 
= A(p_\mathrm{miss}, 180^\circ+\phi_\mathrm{dq})$.
By such mirroring of the asymmetries to the negative $p_\mathrm{miss}$-axis
the extrapolation of the asymmetries to zero turns into a more reliable
interpolation that connects the two branches of the asymmetries
at positive and negative $p_\mathrm{miss}$ across the $p_\mathrm{miss}=0$
axis.  We have assumed that the asymmetries have a modest dependence
on $p_\mathrm{miss}$ in the interpolation region, a view supported
by theoretical calculations for the bins where the asymmetries
at very low $p_\mathrm{miss}$ are available a priori.

Once the (interpolated) calculated asymmetries have been made available 
for all required $p_\mathrm{miss}$, we use each measured event in the 
following procedure.   First, we take into account not only the theoretical
value at the center of the event's bin, but also the values from
the neighboring regions in the $(E',\theta_\mathrm{e})$ plane.
We select a neighborhood around each event (see e.g. the gray circle 
surrounding an event at the intersection of four kinematic regions
in Fig.~\ref{E05102-fig2} (left)).  The size of the circle is chosen 
such that for each event the closest theoretical predictions are
selected, i.e.~the radius of the circle is approximately the same
as the linear size of the region.  Then we check which theories 
are available in the selected zone; in the example shown,
the algorithm picks the theories from all four neighboring bins.
The distance of the event to each of the considered bin centers
is calculated, and these distances are then used as weights 
in the calculation of the average of all theories available 
inside the selected zone defined by the event, i.e.~larger distances
imply smaller contributions of a particular kinematic point 
to the average asymmetry.  We also calculate the difference
from the average asymmetry to the nearest predicted asymmetry,
and take this as a measure for the systematic uncertainty 
introduced by our approach.  This uncertainty is in no way related 
to the quality of the theoretical calculations themselves; 
it is solely a consequence of our averaging procedure and the fact
that the theory is available only at discrete points.
The final result for the $p_\mathrm{miss}$-dependence 
of the asymmetry is shown in Fig.~\ref{E05102-fig2} (right)
for the 5, 6, 12, 13 bin quartet.
The whole algorithm is repeated for each event.  Finally, the asymmetries 
are accumulated and averaged in each $p_\mathrm{miss}$ bin.

\clearpage

\subsection[E06-010: Transversity]{E06-010: Measurement of Single Target-Spin Asymmetry in Semi-Inclusive Pion
Electroproduction on a Transversely Polarized $^3$He Target}\label{sec:e06010}

%\begin{center}
%Measurement of Single Target-Spin Asymmetry in Semi-Inclusive Pion
%Electroproduction on a Transversely Polarized $^3$He Target 
%\end{center}

\begin{center}
J.-P.~Chen, E.~Cisbani, H. Gao, X. Jiang, J.-C. Peng  co-spokespersons, \\
and \\
the Hall A Collaboration.\\
\end{center}

\begin{center}
contributed by K. Allada, X. Yan, and Y.X. Zhao
\end{center}
\subsubsection{Introduction}
Experiment E06010 (Transversity)~\cite{e06010} was conducted in Hall A
from Oct. 2008 to Feb. 2009 using a longitudinally polarized beam and
transversely polarized $^3{He}$ target. The beam energy was 5.9
GeV. The primary goal of the experiment was to extract Collins and
Sivers moments in the semi-inclusive deep inelastic (SIDIS) reaction,
$^3{\textrm{He}^{\uparrow}}$(e, e'$\pi^{\pm}$)X. The beam helicity was
flipped at 30 Hz and the target spin direction was flipped every 20
minutes. One can either perform target single spin
asymmetry (TSSA) study by summing the two beam helicity states to
achieve unpolarized beam, or do beam-target double spin asymmetry
(DSA) study. The BigBite spectrometer was set at $30^{o}$ on
beam-right to detect scattered electron with momenta from 0.6-2.5
GeV. The left HRS \cite{hrs} was set at $16^{o}$ on beam-left to
detect the produced hadrons ($\pi^{\pm}$, $K^{\pm}$ and 
proton) with a fixed momentum of 2.35 GeV. 

There were four types of physics triggers (see Table
\ref{trigger_table}). The coincidence trigger (T5) was used for the
primary analysis, which included SSA and DSA analysis to obtain the
Collins/Sivers moments~\cite{xinpaper} and A$_{LT}$
moments~\cite{jinpaper} for charged pions, respectively. These two
results were published in ~\cite{xinpaper,jinpaper}. The following
are some of the recent progresses in E06-010 physics analysis:

\begin{itemize}
\item The results of SSA in the inclusive hadron production reaction, $e +
^3{\textrm{He}^{\uparrow}}\rightarrow hX$ ($h = \pi^{\pm}$, K$^{\pm}$,
  p), have been submitted for a publication~\cite{kalyanpaper}. This
  analysis was performed by Kalyan Allada (MIT) and independently
  checked by Yuxiang Zhao (USTC, China).
\item The final analysis of pretzelosity asymmetry for charged pions,
  extracted from the TSSA data in the SIDIS reaction
  ${^3\textrm{He}^{\uparrow}}$(e, e'$\pi^{\pm}$)X, is finished. The
  results were submitted for a publication. This analysis was
  performed by Yi Zhang (Lanzhou University, China)
\item Results of Collins and Sivers moments for charged Kaons in
  SIDIS reaction ${^3\textrm{He}^{\uparrow}}$(e, e'K$^{\pm}$)X are
  final, and a draft for publication is being prepared. This analysis was
  performed by  Youcai Wang (UIUC) and independently
  checked by Yuxiang Zhao (USTC, China).
\item The preliminary results of DSA in inclusive hadrons production reaction, $e +
^3{\textrm{He}^{\uparrow}}\rightarrow hX$, ($h =
  \pi^{\pm}$, K$^{\pm}$, p) from HRS data were obtained. The
  analysis is currently focused on producing final results for this
  channel. This analysis is being performed by Yuxiang Zhao (USTC, China).
\item Analysis of unpolarized SIDIS cross-section to
  access the Boer-Mulders function and to study the $x-z$
  factorization. This work is being performed by Xuefei Yan (Duke
  University). 
\end{itemize}
In this report we will discuss some of the on-going analysis from the
list above. 

%___________________trigger type table______________________________________________
\begin{table}[h]
\begin{center}
\begin{tabular}{c|c} \hline \hline
Trigger type & Description \\ \hline
1            & Low threshold on BigBite lead-glass \\ \hline
%2            & BigBite gas Cherenkov singles        \\ \hline
3            & Left HRS singles (S1 .and. S2)      \\ \hline
%4            & Left HRS efficiency                 \\ \hline
5            & Coincidence between BigBite and Left HRS (T1 .and. T3)
\\ \hline
6            & High threshold on BigBite lead-glass \\ \hline \hline

\end{tabular}
\\
\caption{Summary of physics triggers in E06010} \label{trigger_table}
\end{center}
\end{table}
%_______________trigger type table ends______________________________________________

\subsubsection{SSA in Inclusive Hadron Production}
\label{ssa_inc_hadron}
Our primary results were focused on the SIDIS reaction to study
transverse momentum dependent parton distribution functions (TMDs). In
addition to SIDIS reaction one can study inclusive hadron
production reaction where the scattered electron remains
undetected. We obtained the SSA from $e +
^3{\textrm{He}^{\uparrow}}\rightarrow hX$  reaction using the HRS
data for $\pi^{\pm}$, K$^{\pm}$ and protons. 

The target SSA is defined as,
\begin{equation}
A_{UT}(x_F,p_T) =
\frac{1}{P}\frac{d\sigma^{\uparrow}-d\sigma^{\downarrow}}{d\sigma^{\uparrow}+d\sigma^{\downarrow}}\textrm{sin}\phi_S
= A_N\textrm{sin}\phi_S,
\label{eqn:ssa_def}
\end{equation}
where $d\sigma^{\uparrow(\downarrow)}$ is the differential cross-section
in the target ``up''(``down'') state, $P$ is the target
polarization, and $\phi_S$ is the angle between the scattering plane
and the nucleon spin vector.
\begin{figure}[ht]
\begin{center}
\includegraphics[width=60mm]{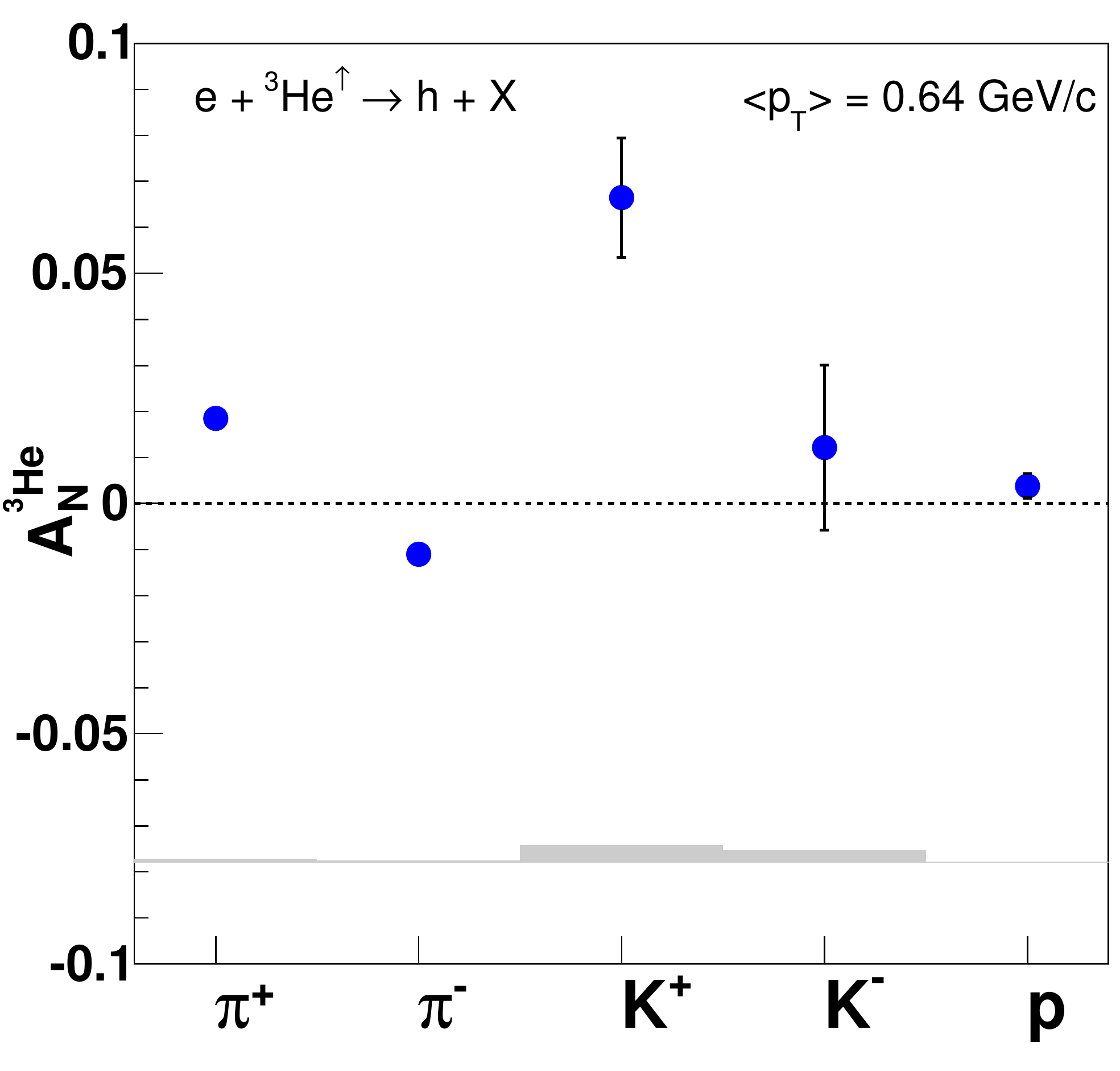}
\caption[E06-010: Transverse SSA in inclusive hadron ($\pi^{\pm}$, K$^{\pm}$, p) production on a $^3$He target.]{\label{fig:He3_vert} Inclusive SSA results on a $^3$He target
  for $\pi^{\pm}$, K$^{\pm}$ and protons in the vertical target spin
  configuration ($\phi_S= \pm 90^{\circ}$). The error bars on the
  points represents the statistical uncertainty. The grey band shows
  the magnitude of the overall systematic uncertainty for each hadron
  channel.}
\end{center}
\end{figure}

\begin{figure}[ht]
\begin{center}
\includegraphics[width=50mm]{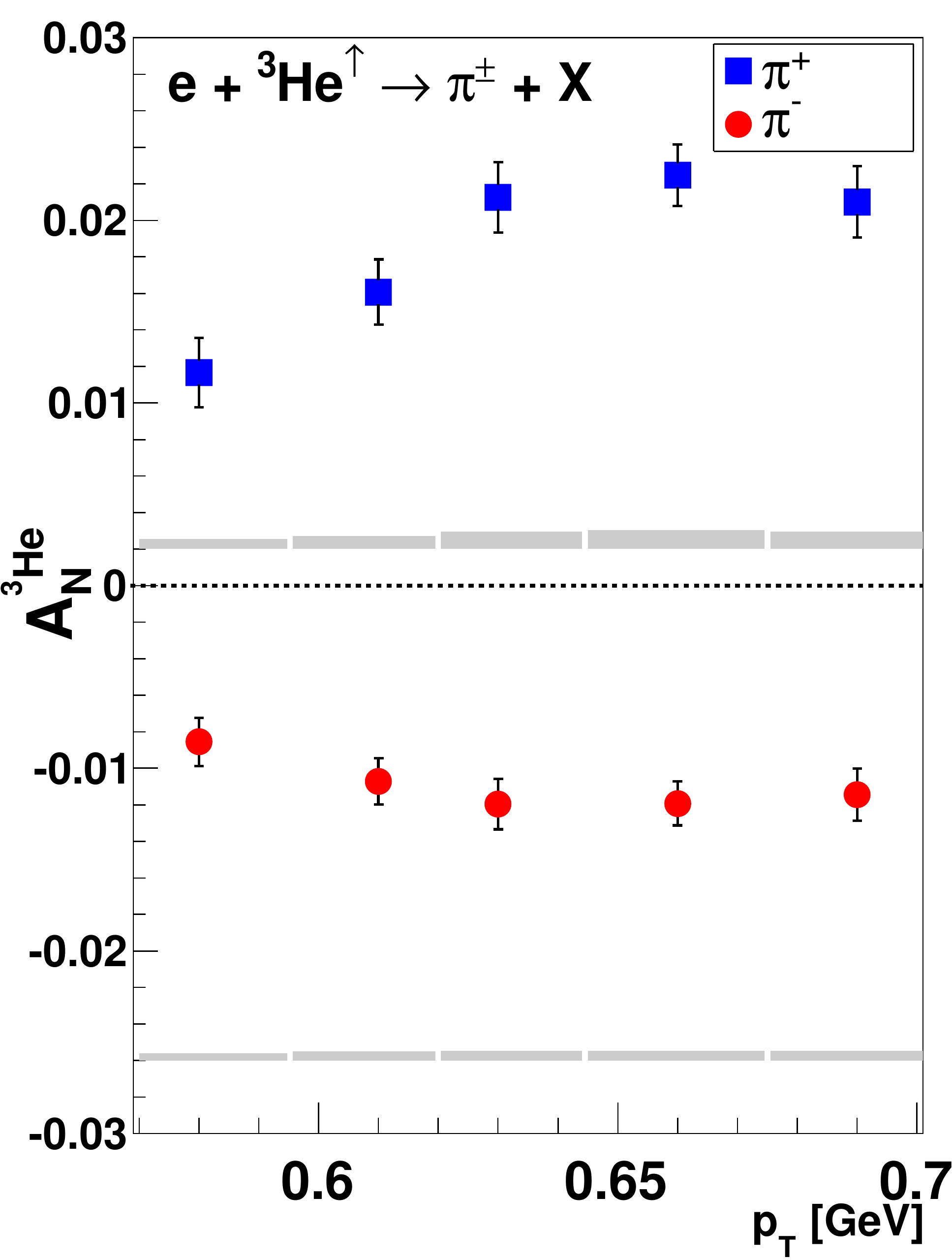}
\includegraphics[width=50mm]{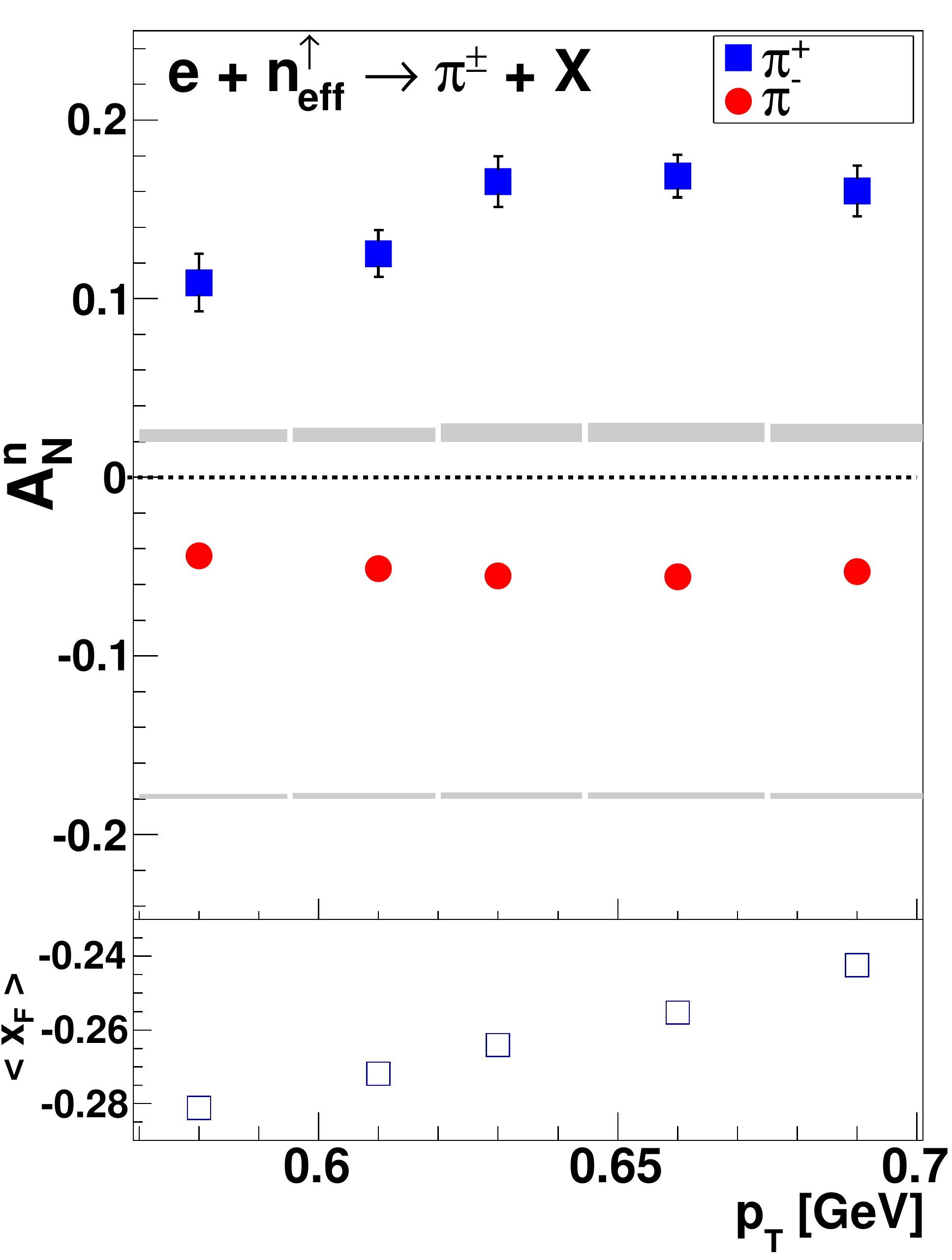}
\caption[E06-010: Transverse SSA (A$_N$) of inclusive pions as function of $p_T$ for 3He target and extracted for neutron.]{\label{fig:pt_asy} Left: $A_N$ results on a $^3$He target
  for the $\pi^{\pm}$ channel as a function of $p_T$. Right: $A_N$ results
  on a neutron target extracted from the measured $^3$He
  asymmetries. The solid band on the bottom of each panel shows the
  magnitude of the systematic uncertainty for each momentum bin. The lower
  plot shows the $x_F$ and $p_T$ correlation in this measurement.}
\end{center}
\end{figure}

The final $^3$He asymmetry results are shown for different hadron
species in Fig.~\ref{fig:He3_vert}. The center-of-mass energy was $\sqrt{s}$=3.45
GeV, and the average $p_T$ for this data was 0.64 GeV/c. For the high
statistics pion sample the SSA was divided into five bins in $p_T$. The SSAs for
charged pions as a function of $p_T$ for a $^3$He target are shown in
Fig.~\ref{fig:pt_asy} (left). A first attempt to extract the neutron
SSA from the $^3$He was done using effective polarization
approach. The final results for $A^n_N$ for charged pions on an
effective neutron target are shown in Fig.~\ref{fig:pt_asy}
(right)~\cite{kalyanpaper}. 

\subsubsection{Pretzelosity Asymmetry}
There are three leading-twist terms that appear in the SIDIS cross-section
($\sigma_{UT}$) obtained from an unpolarized beam and a transversely
polarized target. The first two are Collins and Sivers terms which
were studied in our earlier analysis~\cite{xinpaper,jinpaper}. The
third term is known as pretzelosity ($h^{\perp}_{1T}$), one of the
least known TMDs. In a class of relativistic quark
models~\cite{avakian, lorce}, pretzelosity can be  expressed as the
difference between the helicity and the transversity, and can
be intuitively related to orbital angular momentum of quarks.

\begin{figure}[H]%[ht]
\begin{center}
\includegraphics[width=100mm]{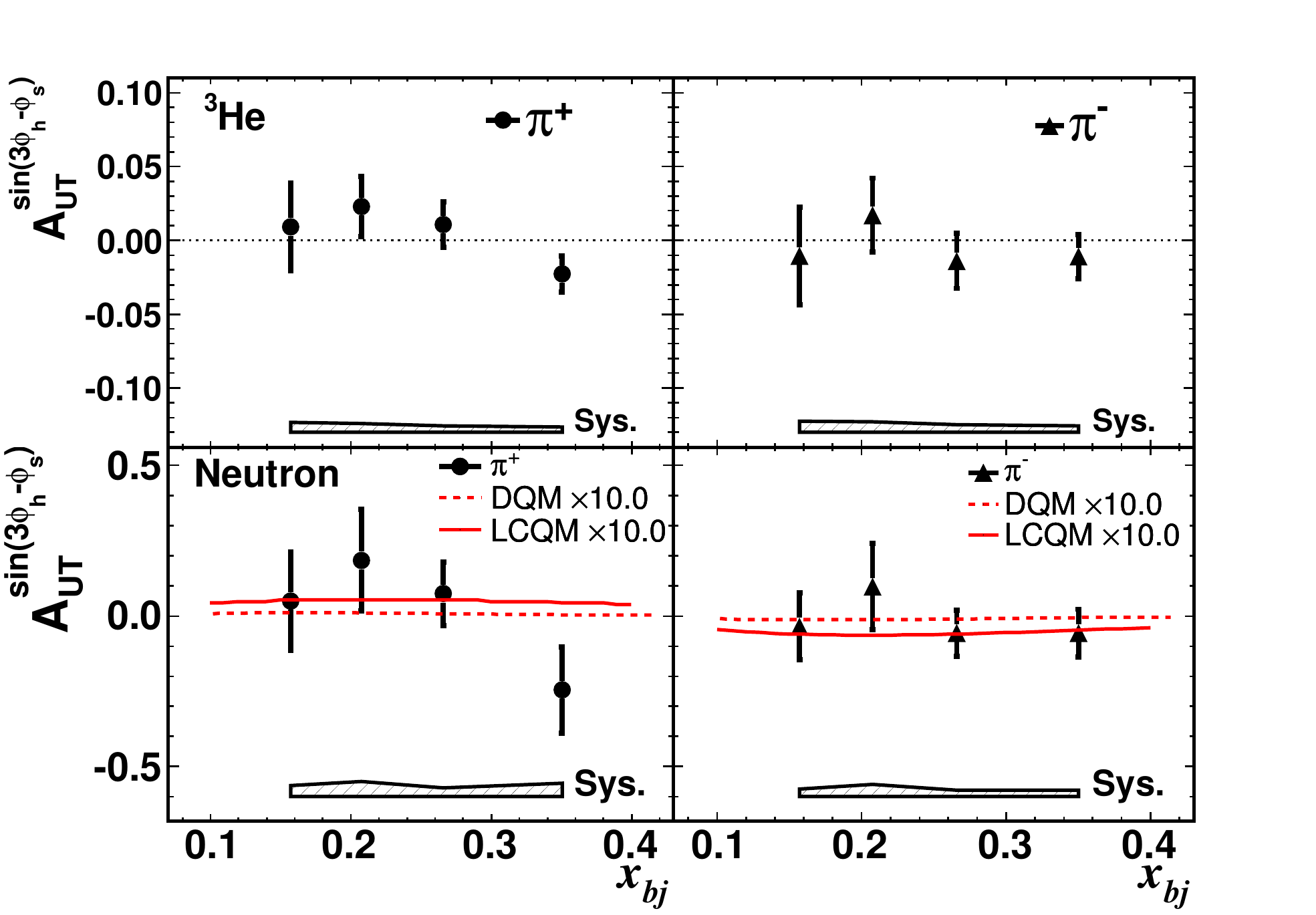}
\end{center}
\caption[E06-010: The pretzelosity asymmetries of pions for 3He target and extracted for neutron.]{ The extracted pretzelosity asymmetries on $^3$He nuclei
(top panels) and on the neutron (bottom panels) are shown
together with uncertainty bands for both $\pi^+$ and $\pi^-$
electroproduction.} \label{pretz_plot}
\end{figure}

The SSA data from SIDIS reaction ${^3\textrm{He}^{\uparrow}}$(e,
e'$\pi^{\pm}$)X was binned in $\phi_S$ and $\phi_h$, the spin and
hadron angle, respectively. The pretzelosity asymmetry was obtained by
performing a 2D fitting in the $\phi_S$ and $\phi_S$ using three terms (Collins, Sivers and
Pretzelosity) in the fitting function. The extracted pretzelosity
asymmetries on $^3$He target for charged pions are shown in
Fig.~\ref{pretz_plot}. To extract the pretzelosity asymmetry on the
neutron, the effective polarization method was used. The results of
the extracted pretzelosity moment on the neutron are shown in the
bottom two panels of Fig.\ref{pretz_plot} and are compared with
various models.
%____________PretzelosityResultPlot___________________________
%_____________________________________________________

\subsubsection{SSA of Kaon in SIDIS}
The HERMES experiment~\cite{HERMESPAPERkaon} observed that the Collins
effect for $K^{+}$ is larger than that for $\pi^{+}$ on proton target,
whereas for $K^{-}$ it is small and consistent with
zero. The Sivers effect reported by the HERMES experiment shows that the
$K^{+}$ asymmetry is larger than $K^{-}$ asymmetry on proton
target. It is also important to note that the $K^{\pm}$
Collins and Sivers moments from the COMPASS experiment on deuteron
target are consistent with zero, suggesting large cancellations between
contributions from proton and neutron. Our results from polarized
$^3{He}$ data will provide crucial independent information on Kaon
Collins and Sivers moments.

In this section, we report the preliminary results of single-spin
asymmetries of charged kaons produced in semi-inclusive deep inelastic
scattering of electrons from a transversely polarized $^3{He}$
target. Both the Collins and the Sivers moments for $K^{+}$ and
$K^{-}$ are extracted over the kinematic range of
0.05$<$x$<$0.5.

\begin{figure}[H]%[ht]
\begin{center}
\includegraphics[width=100mm]{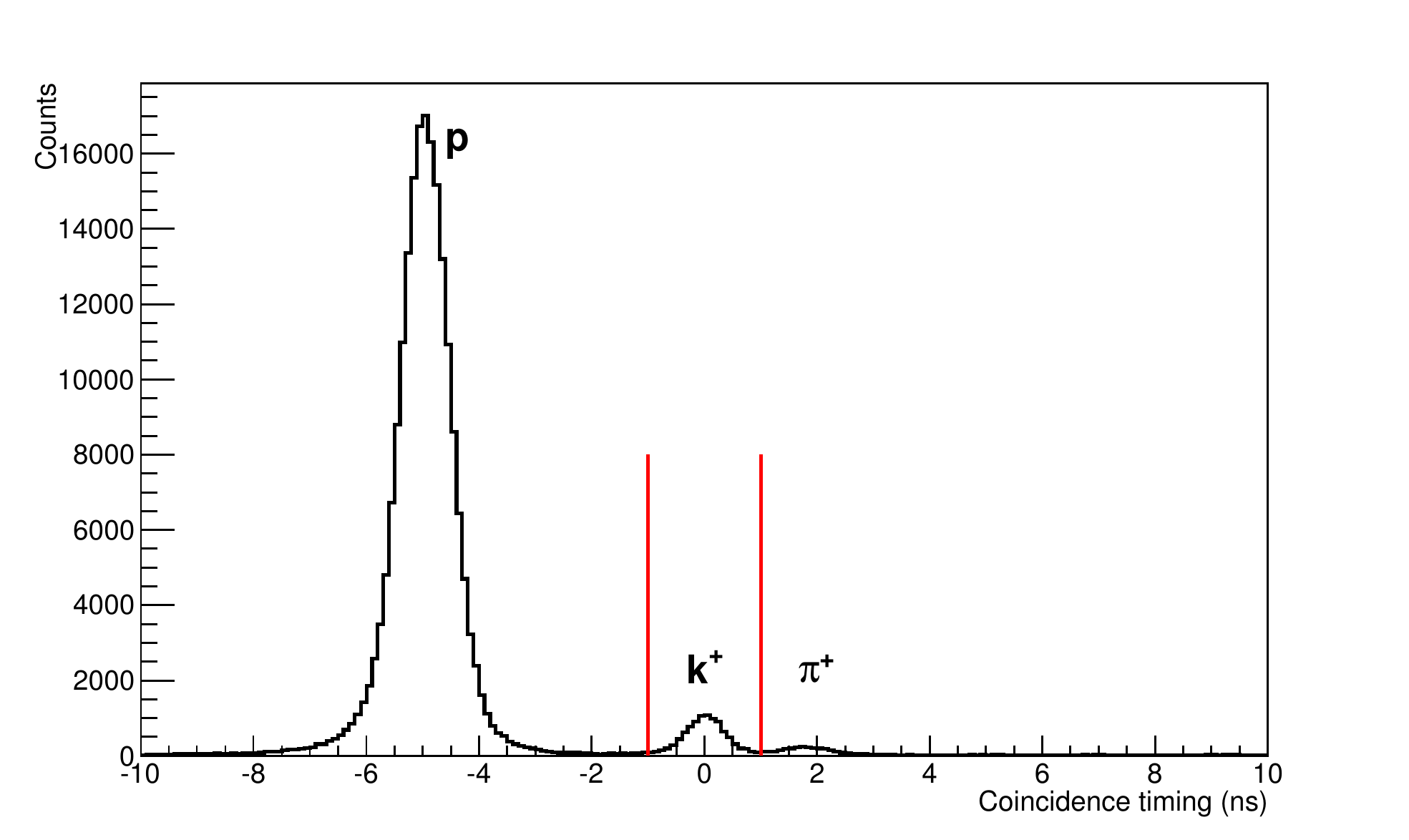}
\end{center}
\caption[E06-010: Coincidence ToF spectrum between the BigBite and the HRS in E06-010 experiment.]{ Coincidence ToF spectrum between BigBite and HRS with a veto
cut on A1 detector to suppress the pion background. The red lines shows
 the Kaon selection cuts.}
 \label{CTOFkaonPlot}
\end{figure}

The electron identification was achieved by using cuts on the BigBite
pre-shower energy $E_{ps}$ and the ratio E/p of the total pre-shower and
shower energy to the reconstructed momentum. After all the cuts,
$\pi^{-}$ contamination was controlled to be less than 1\%. 
Left HRS was configured for hadron detection. In addition to three
Cherenkov detectors ($CO_{2}$ gas Cherenkov detector, Aerogel
detector (A1) and RICH), coincidence time of flight(CTOF) between HRS and
BigBite was also used to perform the hadron PID. During E06010, RICH
performance was not optimal, its efficiency was very low and kaon
yield itself is relatively small, so kaon identification was relied on
CTOF, and RICH was used as a cross-check. Figure \ref{CTOFkaonPlot}
shows the CTOF spectrum with a veto on A1 to suppress the pion
background. After combining all the HRS cuts and CTOF
cut, $\pi^{+}$($\pi^{-}$) contamination in $K^{+}$($K^{-}$) sample was
controlled to be less than 2\%(5\%). Random contamination in
$K^{+}$($k^{-}$) sample is less than 4\%(1\%). Total number of
$K^{+}$($K^{-}$) after PID cuts are 9481(1797). 
%_____________CTOFkaonPlot_____________________________
%______________________________________________________

\begin{figure}[H]%[ht]
\begin{center}
\includegraphics[width=120mm]{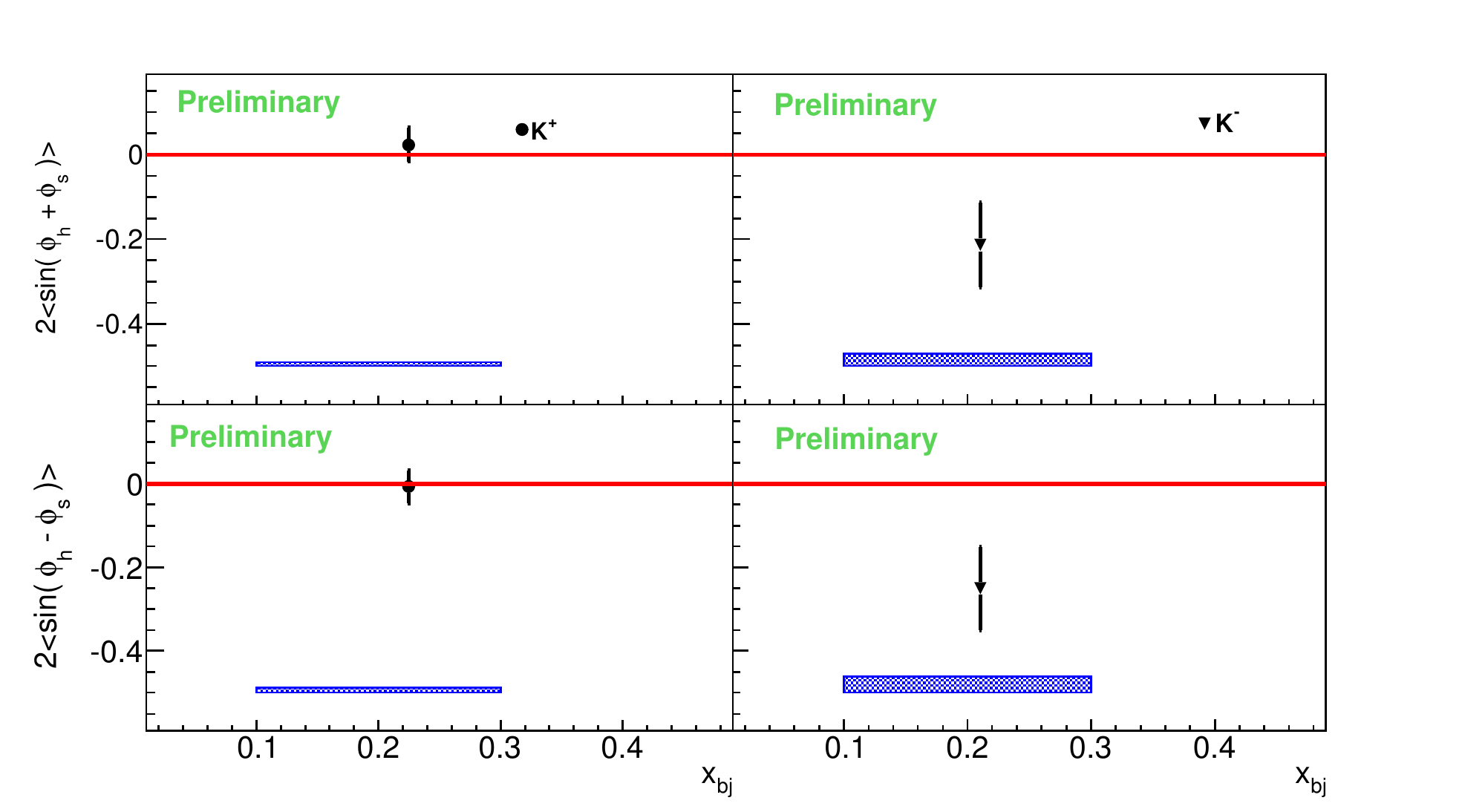}
\end{center}
\caption[E06-010: Preliminary results of Collins and Sivers moments on $^3$He for kaon electroproduction.]{ Preliminary results of Collins and Sivers moments on
  $^3{He}$ for kaon electro-production. } \label{KaonResultPlot} 
\end{figure}

\paragraph{Preliminary Results}
Due to the low statistics for kaons, the data was averaged over full
range of $x_{bj}$. The Collins and Sivers moments were extracted simultaneously by
using Maximum Likelihood Estimation(MLE) \cite{MLEnote}, the
preliminary results are shown in figure \ref{KaonResultPlot}. $K^{+}$
Collins and Sivers asymmetries are consistent with zero within error
bar, while $k^{-}$ Collins and Sivers asymmetries favor negative
sign. 
%____________KaonResultPlot___________________________
%_____________________________________________________
\subsubsection{Unpolarized SIDIS Cross-Section}

Most of the data analysis effort for experiment E06010 so far has
been focused on single-spin asymmetry and double-spin asymmetry to
extract information on various TMDs such as Collins, Sivers, Pretzelosity
and Transversal Helicity (g$_{1T} ^q$) \cite{xinpaper,jinpaper,yipaper}. Recently
we started a new analysis on unpolarized differential cross-section
from E06010 in order to access the Boer-Mulders function, which describes
the distribution of transversely polarized quarks in an unpolarized
nucleon. Moreover, we plan to study the $x-z$ factorization in SIDIS
using unpolarized cross-section.

Unpolarized differential cross-section obtained from an unpolarized
beam and an unpolarized target, is equivalent to a proper combination
of differential cross-sections with different beam and target polarizations
extracted from experiment E06010. An extraction of differential cross-section
from experiment E06010 requires further study of the BigBite acceptance
model in the existing SIMC-Transversity simulation. The BigBite acceptance
is being studied by comparing the elastic H$_2$ calibration runs
at 1.23 and 2.4 GeV, with the SIMC-Transversity simulation combined
with a well-developed elastic electron-proton-scattering event generator
with radiative effects included.

After the acceptance thoroughly studied and SIMC-Transversity simulation
updated, the unpolarized differential cross-section of SIDIS will
be extracted, from which the Boer-Mulders function can be obtained.

%__________Add reference here_________________________________________

%
% Here's how to do the references.  We will be using the APS style.
%
%
%\bibliographystyle{unsrt}
%\bibliography{references}

\clearpage

%\documentclass{article}
%===================================================================================================
% packages 
%\usepackage{epsfig}
%\usepackage{subfigure}
%\usepackage{graphicx}
%\usepackage{epstopdf}
%\usepackage{amsmath}
%\usepackage{amstext}
%\usepackage{amsfonts}
%\usepackage{amssymb}
%===================================================================================================
% graphics path 
\graphicspath{{./figs/}}
%===================================================================================================
% dimension options  
\setlength{\textwidth}{6.5in}
\setlength{\oddsidemargin}{0in}
\setlength{\evensidemargin}{0in}
\setlength{\textheight}{9in}
\setlength{\topmargin}{0in}
\setlength{\headheight}{0in}
\setlength{\headsep}{0in}
%===================================================================================================
% commands 
\newcommand{\ExpID}{E06-014}
\newcommand{\Fig}[1]{(Fig.~\ref{fig:#1})}
\newcommand{\Figure}[1]{Figure~\ref{fig:#1}}
\newcommand{\SubFig}[1]{\subref{fig:#1}}
\newcommand{\Sect}[1]{(Sect.~\ref{sect:#1})}
\newcommand{\Equation}[1]{Equation~\ref{eq:#1}}
\newcommand{\Table}[1]{Table~\ref{tab:#1}}
\newcommand{\HeliumThree}{\ensuremath{^3}He}
\newcommand{\PolHeThree}{\ensuremath{^3\vec{\textrm{He}}}}
\newcommand{\NTwo}{\ensuremath{\textrm{N}_{2}}}
\newcommand{\Es}{\ensuremath{E_s}}
\newcommand{\Ep}{\ensuremath{E_p}}
\newcommand{\dtwon}{\ensuremath{\textrm{d}_{\textrm{2}}^{\textrm{n}}}}
\newcommand{\gOne}{\ensuremath{g_{1}}}
\newcommand{\gTwo}{\ensuremath{g_{2}}}
\newcommand{\AOne}{\ensuremath{A_{1}}}
\newcommand{\AOneN}{\ensuremath{A_{1}^{n}}}
\newcommand{\AOneHeThree}{\ensuremath{A_{1}^{^{3}\textrm{He}}}}
\newcommand{\SigRad}{\ensuremath{\sigma_{\textrm{rad}}}}
%===================================================================================================
% miscellaneous options 
\setcounter{secnumdepth}{5}
%===================================================================================================

%\begin{document}

%===================================================================================================
% header 
%===================================================================================================
\subsection[E06-014: A Precision Measurement of $d_2^n$: Probing the Lorentz Color Force]{E06-014: A Precision Measurement of $d_2^n$: Probing the Lorentz Color Force} \label{sect:e06014}
%===================================================================================================
%\begin{center}
 %  \bf{A Precision Measurement of $d_2^n$: Probing the Lorentz Color Force}
%\end{center}

\begin{center}
   S. Choi, X. Jiang, Z.-E. Meziani, B. Sawatzky, spokespersons, \\
   and                                                           \\
   the $d_2^n$ and Hall A Collaborations.                        \\
   Contributed by D.~Flay.
\end{center}
%===================================================================================================
% physics motivation 
%===================================================================================================
\subsubsection{Physics Motivation} \label{sect:phys_mot}
\paragraph{$d_2^n$: Quark-Gluon Correlations in the Nucleon}
%===================================================================================================

To date, extensive work has been done investigating the spin structure function $g_1$ within the 
context of the Feynman parton model and pQCD.  However, far less is known about the $g_2$ structure 
function.  It is known to contain quark-gluon correlations.  It follows from a spin-flip Compton 
amplitude and may be written as:

\begin{equation} \label{eq:g_2}
g_2 \left( x, Q^2 \right) = g_2^{WW}\left( x, Q^2 \right) + \bar{g}_2\left( x, Q^2 \right), 
\end{equation}

\noindent where $g_2^{WW}$ is the Wandzura-Wilczek term, which may be expressed entirely in terms 
of $g_1$~\cite{ww}:

\begin{equation} \label{eq:g_2ww}
   g_2^{WW}\left( x, Q^2 \right) = - g_1 \left( x, Q^2 \right) 
                                   + \int_x^1 \frac{g_1 \left( y, Q^2 \right)}{y} dy. 
\end{equation}

\noindent The second term is given as:

\begin{equation} \label{eq:g_2bar}
   \bar{g}_2\left( x, Q^2 \right) = - \int_x^1 \frac{1}{y}\frac{\partial}{\partial y}  
                                    \left[ \frac{m_q}{M} h_T \left( y, Q^2 \right) 
                                    + \xi \left( y, Q^2 \right) \right] dy,  
\end{equation} 

\noindent where $h_T$ is the transverse polarization density, and $\xi$ is a term arising 
from quark-gluon correlations.  Here, $h_T$ is suppressed by the ratio of the quark mass 
$m_q$ to the target mass $M$.  Therefore, a measurement of $\bar{g}_2$ provides access to 
quark-gluon interactions inside the nucleon~\cite{rlj}. \*

Additionally, a measurement of both $g_1$ and $g_2$ allows for the determination of the 
quantity $d_2^n$, which is formed as the second moment of a linear combination of $g_1$ and $g_2$: 

\begin{equation} 
d_2^n \left( Q^2 \right) = \int_0^1 x^2
                         \left[ 2g_1^n \left( x,Q^2 \right) + 3g_2^n \left( x, Q^2 \right)\right]dx 
                         = 3 \int_0^1 x^2 \bar{g}_2^n \left( x,Q^2 \right) dx. \label{eq:d2n} 
\end{equation}

\noindent $d_2^n$ also appears as a matrix element of a twist-3 operator in the operator 
product expansion~\cite{fil_ji}:

\begin{equation} \label{eq:d2pps}
   \langle P,S \mid \bar{\psi}_q \left( 0 \right) gG^{+y}\left( 0 \right) {\gamma}^{+} \psi_q \left( 0 \right) 
               \mid P,S \rangle = 2MP^{+}P^{+}S^{x}d_2^n,
\end{equation}

\noindent where $G^{+y} = \frac{1}{\sqrt{2}} \left( B^{x} - E^{y} \right)$.  We see from 
Equations~\ref{eq:g_2bar}--\ref{eq:d2pps} that $d_2^n$ is a twist-3 matrix element that 
measures quark-gluon interactions.
 
Recent work has shown~\cite{mb_1,mb_2} that at high $Q^2$, $d_2^n$ is seen as a color 
Lorentz force averaged over the volume of the nucleon. This is given by the expression 
for the transverse (color) force on the active quark immediately following its interaction 
with a virtual photon:

\begin{equation}
   F^{y} \left( 0 \right) \equiv - \frac{\sqrt{2}}{2P^{+}} 
                                 \langle P,S \mid \bar{\psi}_q \left( 0 \right) g G^{+y} 
                                 \left( 0 \right) {\gamma}^{+} \psi_q \left( 0 \right) \mid P,S \rangle 
                          =      - \frac{1}{2}M^2 d_2^n. \label{eq:tlcf}
\end{equation}

\noindent This theoretical interpretation reveals how $g_2$ and subsequently $d_2^n$ will 
allow us to examine the color interactions of the constituents inside the nucleon.

\begin{figure}[hbt]
 \begin{centering}
   \includegraphics[scale=0.5]{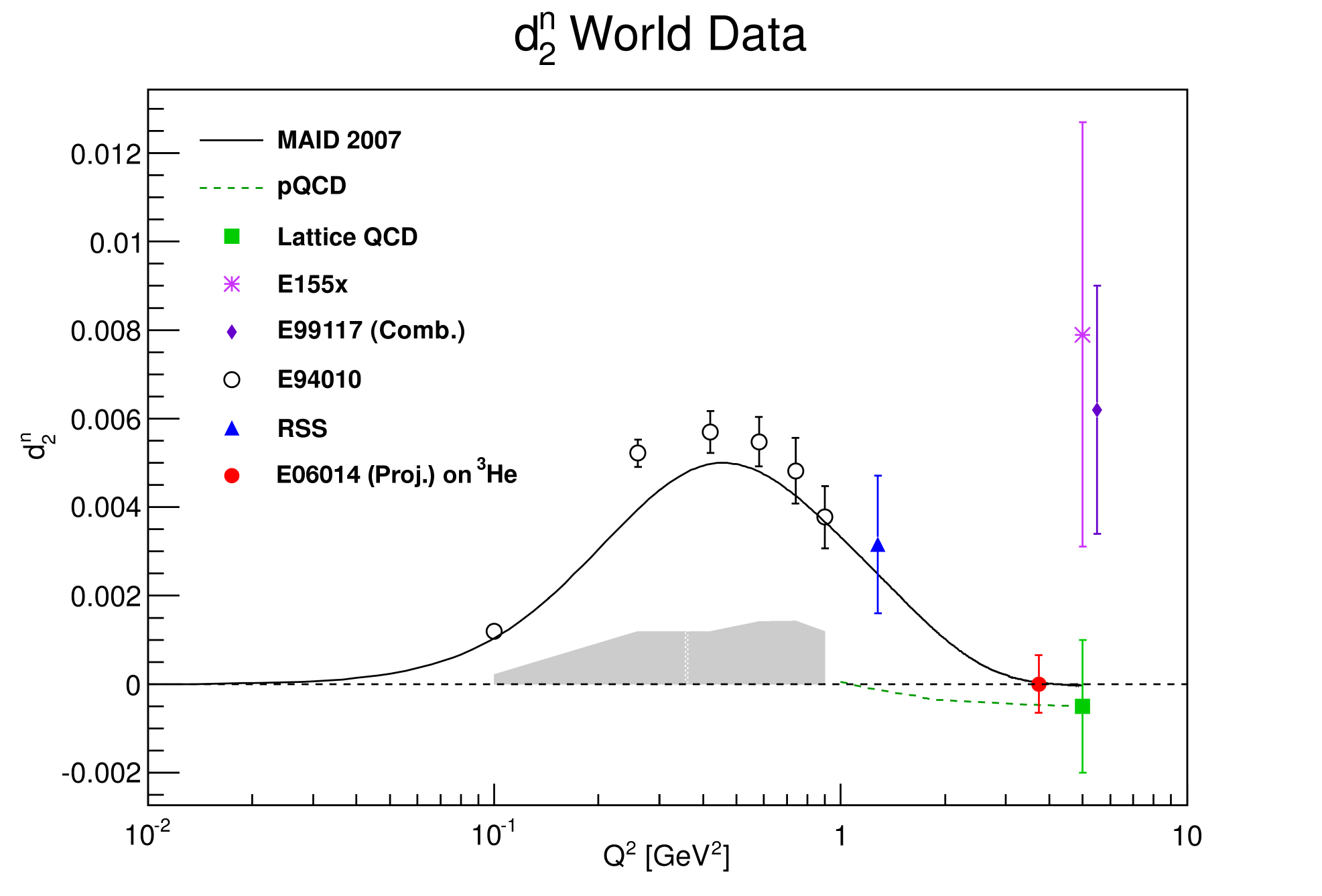} 
   \caption[World data for $d_2^n$ as a function of $Q^2$.]{\dtwon{} as a function of $Q^2$.  All the data shown with the exception of the SLAC E155x data are dominated by resonance
            contributions. E06-014 data will be dominated by the deep inelastic scattering (DIS) contribution. The projected error 
            from E06-014~\cite{PAC29} is shown, along with the lattice QCD result~\cite{LQCD}. The dashed green curve shows the 
            pQCD evolution from the lattice point~\cite{pQCDEvol} based on the calculations of~\cite{ShuryakAndVain,JiAndChou}. 
            Data from JLab experiments E94-010~\cite{E94010_Amerian,E94010_Slifer} and RSS~\cite{RSS} are included in the plot. 
            For comparison to the resonance contribution, a MAID model~\cite{MAID} is plotted. Also plotted is the total $d_2$ 
            from SLAC experiment E155x~\cite{E155x}, which consists of DIS data distributed over a large range in $Q^{2}$.}
   \label{fig:d2n_current}
 \end{centering}
\end{figure}

While bag and soliton model calculations of $d_2$ for the neutron yield numerical values 
consistent with those of lattice QCD, current experimental data differ by roughly two 
standard deviations (see the highest $Q^2$ data in \Figure{d2n_current}).  One of the 
goals of our experiment is to improve the experimental error on the value of $d_2^n$ by 
a factor of four.  It subsequently provides a benchmark test of lattice QCD calculations, 
shown in \Figure{d2n_current}.  

%===================================================================================================
\paragraph{$A_1$: The Virtual Photon-Nucleon Asymmetry}
%===================================================================================================

\begin{figure}
      \centering
      \subfigure[$A_1^n$]{
         \centering
	 \includegraphics[scale=0.50]{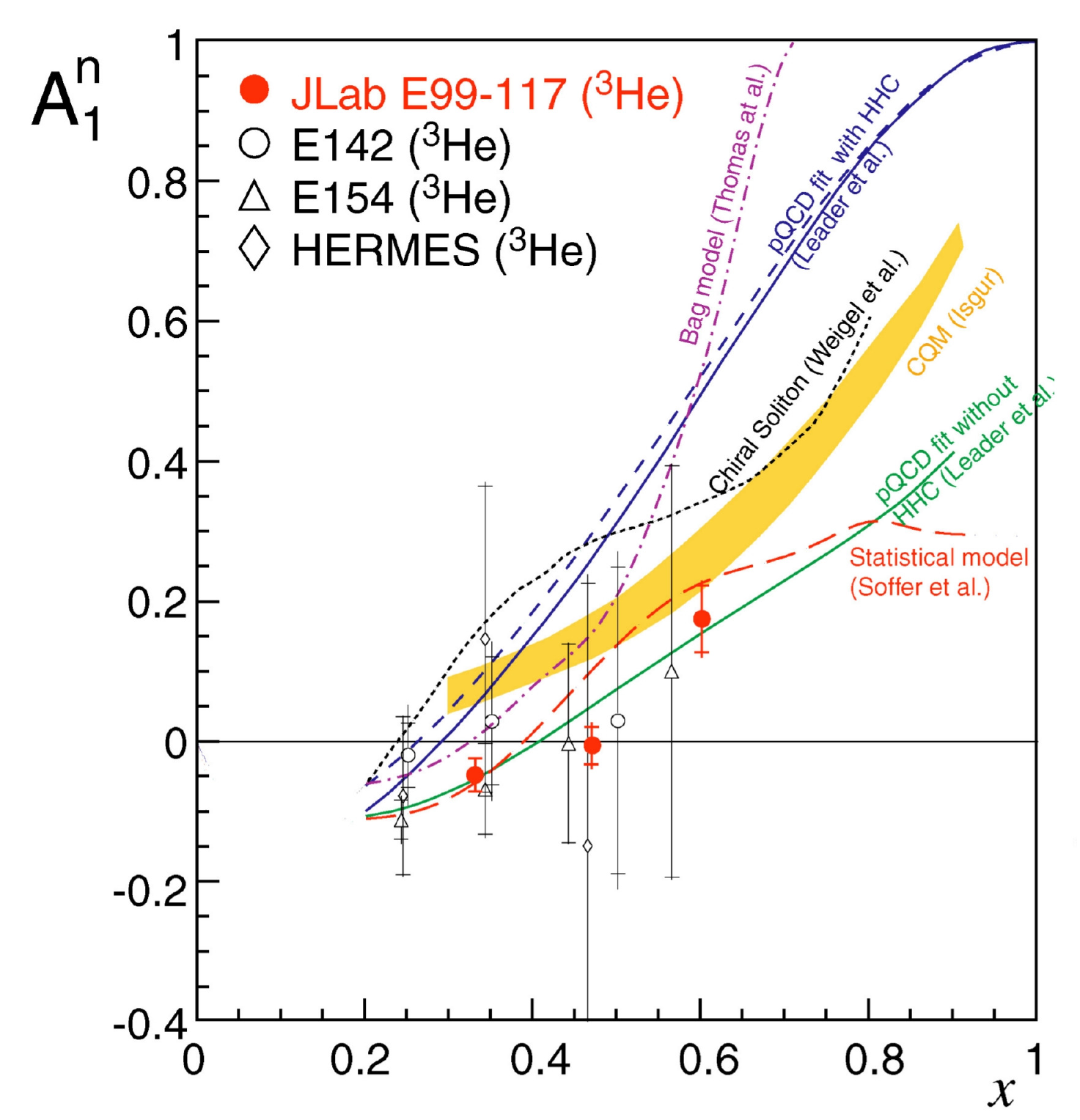}
         \label{fig:A1n_current}
      }
      \subfigure[$\Delta d/d$ and $\Delta u/u$]{
         \centering
	 \includegraphics[scale=0.80]{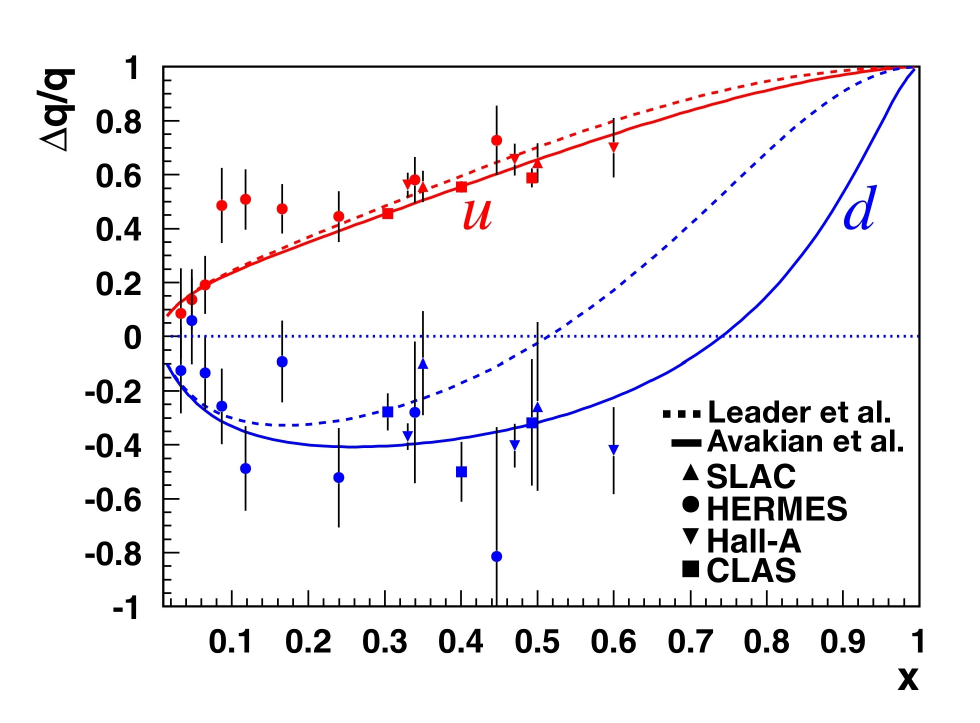}
         \label{fig:delta_q_current}
      }
      \caption[Current data for $A_1^n$, $\Delta d/d$ and $\Delta u/u$.]{Current data for $A_1^n$, $\Delta d/d$ and $\Delta u/u$. 
               \SubFig{A1n_current}: The current world data for the neutron $A_1$ from SLAC E143~\cite{E143} and E154~\cite{E154}  
               and HERMES~\cite{HERMES}, along with JLab E99-117~\cite{E99117PRC}.  Also shown are CQM models and various 
               pQCD models. Figure adapted from~\cite{E99117PRL}. 
               \SubFig{delta_q_current}: $\Delta q/q$ for the up ($u$) and down ($d$) quarks.
               The dashed curves represent a prediction from Leader {\it et al.}~\cite{pQCDHHC}, while the solid curves 
               show calculations by Avakian {\it et al.}~\cite{Avakian}. The data shown are from HERMES, SLAC and JLab.
               Figure adapted from~\cite{Avakian}.}
      \label{fig:A1_current}
   \end{figure}

Another quantity of interest is the virtual photon-nucleon longitudinal spin asymmetry $A_1$.  It provides insight into the 
quark structure of the nucleon and can be defined as: 

   \begin{equation}
      A_1\left( x, Q^2 \right) \equiv \frac{\sigma_{1/2} - \sigma_{3/2}}{\sigma_{1/2} + \sigma_{3/2}}, 
   \end{equation}
 
\noindent where the subscript 1/2 (3/2) gives the projection of the total spin of the virtual photon-nucleon system
along the virtual photon direction corresponding to the nucleon's spin anti-parallel (parallel) to the virtual photon. 
Constituent quark models (CQM) and pQCD models predict $A_1$ to be large and positive at large $x$.  \Figure{A1n_current} 
shows the current world data compared to these models.  It is seen that the CQM (yellow band~\cite{CQM}) describes the 
trend of the data reasonably well.  The pQCD parameterization with hadron helicity conservation (dark blue curve~\cite{pQCDHHC})---assuming 
quark orbital angular momentum to be zero---does not describe the data adequately.  However, the pQCD model allowing for quark 
orbital angular momentum to be non-zero (green curve~\cite{pQCDNoHHC}) describes the data well, pointing perhaps to the 
importance of quark orbital angular momentum in the spin structure of the nucleon.  

Combining $A_1^n$ data measured on a polarized effective neutron target with $A_1^p$ data measured on a
polarized proton target allows access to $\Delta u/u$ and $\Delta d/d$. Recent results from Hall A~\cite{E99117PRC} and 
from CLAS~\cite{CLAS} showed a significant deviation of $\Delta d/d$ from the pQCD predictions, which have that ratio 
approaching 1 in the limit of $x \rightarrow 1$ \Fig{delta_q_current}. As part of the 12 GeV program, two approved experiments 
(one in Hall A~\cite{E1206122} and one in Hall C~\cite{E1210101}) will extend the accuracy and $x$ range of this measurement, but a 
measurement of $A_1^n$ at the kinematics of this experiment (E06-014) will provide valuable support (or refutation) 
of prior JLab results, while producing additional input for theoretical models in advance of the coming experiments at 12 GeV.

%===================================================================================================
% the experiment 
%===================================================================================================
\subsubsection{The Experiment} \label{sect:exp}
%===================================================================================================

\begin{figure}[hbt]
  \begin{centering}
    \includegraphics[scale=0.5]{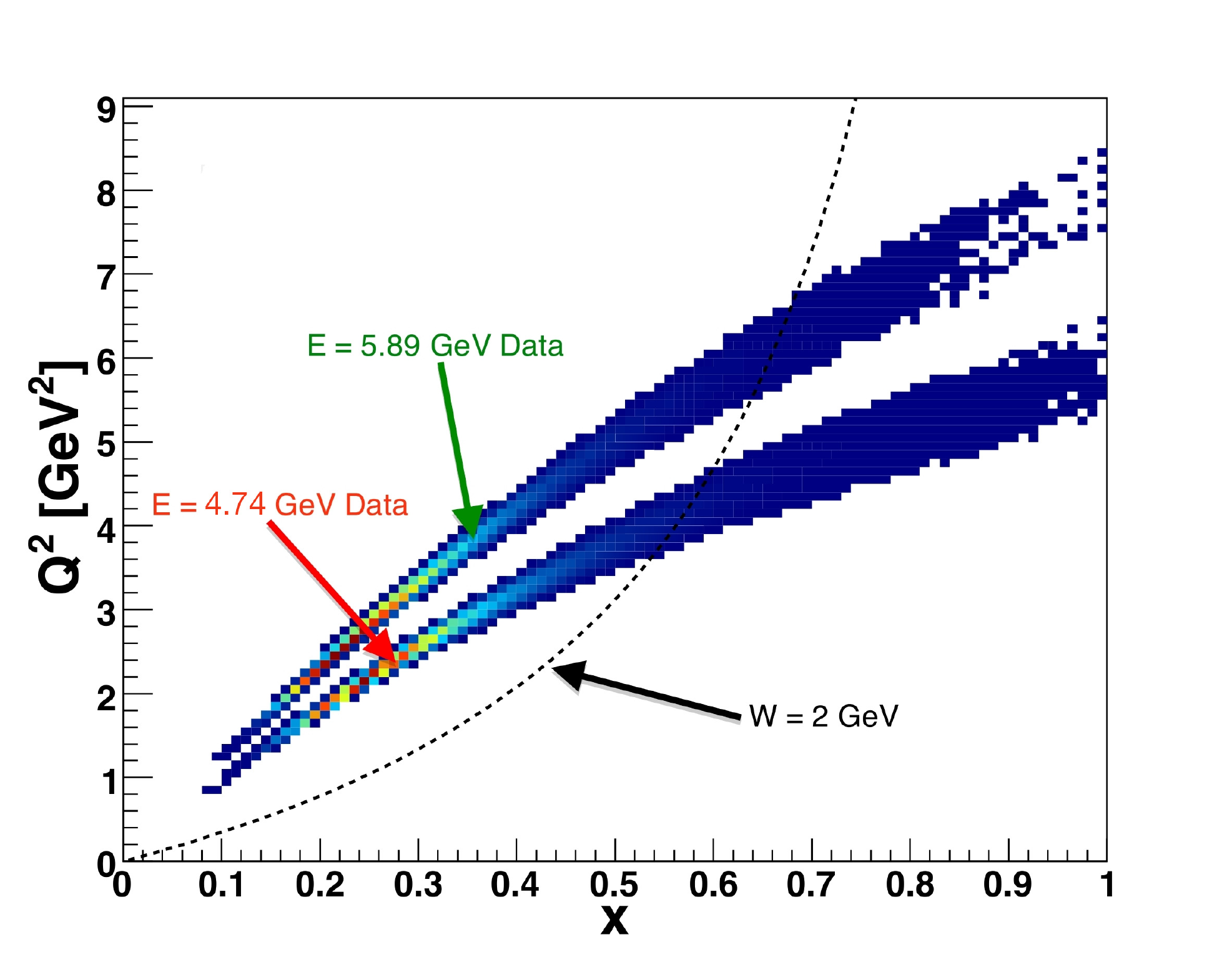}
    \caption[The E06-014 kinematic coverage in $Q^2$ and $x$.]{The E06-014 kinematic coverage in $Q^2$ and $x$. The lower band is the 4.74 GeV data set and the upper
             band is the 5.89 GeV data set. The black dashed line shows $W = 2$ GeV. The regions to the left and right of this line 
             correspond to DIS and resonance kinematics, respectively.
             }
    \label{fig:coverage}
  \end{centering}
 \end{figure}

The experiment ran in Hall A of Jefferson Lab from February to March of 2009, with two beam energies of 
$E = 4.74$ and $5.89$ $\textrm{GeV}$, covering the resonance and deep inelastic valence quark regions, characterized 
by $0.2 \leq x \leq 0.7$ and $2 \textrm{ GeV}^2 \leq Q^2 \leq 6$ $\textrm{GeV}^2$. The coverage in the $x$ and $Q^2$ plane is 
shown in \Figure{coverage}. \*

In order to measure $d_2^n$, we scattered a longitudinally polarized electron beam off of a \PolHeThree{} target in two polarization
configurations -- longitudinal and transverse.  \PolHeThree{} serves as an effective polarized neutron target since roughly $86\%$ of the 
polarization is carried by the neutron.  This is due to the two protons in the nucleus being primarily bound in a spin singlet 
state~\cite{jlf,fb_awt_ira}. \*

We measured the unpolarized cross section $\sigma_0$ and the double-spin asymmetries $A_{\parallel}$ and $A_{\perp}$.  The cross section 
was measured by the Left High-Resolution Spectrometer (LHRS), while the asymmetries were measured by the BigBite Spectrometer.  
The LHRS and BigBite were oriented at scattering angles of $\theta = 45^{\circ}$ to the left and right of the beamline, 
respectively.  

Expressing the structure functions entirely in terms of these experimental quantities, we have the expression for $d_2$:

\begin{equation}
   d_2 = \int_0^1 \frac{MQ^2}{4{\alpha}^2} \frac{x^2 y^2}{\left( 1 - y\right) \left( 2 - y\right)}{\sigma}_0
         \left[ \left( 3 \frac{1 + \left( 1 - y \right)\cos \theta}{\left( 1 - y \right)\sin \theta} + 
                \frac{4}{y} \tan \left( \theta/2 \right)\right)A_{\perp} + \left( \frac{4}{y} - 3 \right)A_{\parallel} \right] dx,
\end{equation}

\noindent where $x = Q^{2}/2M\nu$, $\nu = E - E'$ is the energy transferred to the target, $E'$ is the scattered 
electron energy, and $y = \nu/E$ is the fractional energy transferred to the target.  The asymmetries are given by: 

\begin{eqnarray*}
A_{\parallel} = \frac{ N^{\downarrow \Uparrow} - N^{\uparrow \Uparrow}}{ N^{\downarrow \Uparrow} + N^{\uparrow \Uparrow} } 
\quad \textrm{and} \quad
A_{\perp} = \frac{ N^{\downarrow \Rightarrow} - N^{\uparrow \Rightarrow}}{ N^{\downarrow \Rightarrow} + N^{\uparrow \Rightarrow} }, 
\end{eqnarray*}

\noindent where $N$ is the number of electron counts measured for a given configuration of beam helicity (single arrows) and target 
spin direction (double arrows). 

While $d_2$ was the main focus of the experiment, the measurement of the asymmetries allowed for the extraction of $A_1$: 

        \begin{equation}
           A_1 = \frac{1}{D\left( 1 + \eta\xi \right)}A_{\parallel} - \frac{\eta}{d\left( 1 + \eta\xi \right)}A_{\perp},  
        \end{equation} 

\noindent where $D$, $\eta$, $\xi$ and $d$ are kinematic factors~\cite{Anselmino}.

%===================================================================================================
% analysis progress
%===================================================================================================
% summary of previous work 
%===================================================================================================
\subsubsection{Data Analysis Progress}
\paragraph{Summary of Completed Work}
%===================================================================================================

Nearly all of the analyses for E06-014 have been completed, including detector calibrations for both 
the LHRS and the BigBite spectrometer~\cite{2010Report} and various background studies for the 
spectrometers relating to nitrogen dilution in the target and pair-produced electrons~\cite{2011Report,2012Report}.  

The experiment used a polarized electron beam at energies of 4.74 and 5.89 GeV. The polarization 
of the electron beam was measured independently through Compton and M\o ller scattering, and the
analysis of these measurements revealed a beam polarization of $\sim$ 72\%~\cite{2012Report}. 

Knowledge of the target polarization is crucial when performing a double-spin asymmetry experiment. 
E06-014 used the standard Hall A polarized \HeliumThree{} target with two holding field directions: 
longitudinal and transverse in plane, with respect to the electron beam direction.  The target 
polarization was extracted through electron paramagnetic resonance (EPR). The longitudinal polarization 
was cross checked using nuclear magnetic resonance (NMR) measurements.  During the running of the 
experiment, the polarization of the target was $\sim$ 50\%~\cite{2012Report}. 
 
%===================================================================================================
% LHRS 
\paragraph{Unpolarized Cross Sections} \label{sect:uxs}
%===================================================================================================

The LHRS was used to measure the unpolarized cross section.  The analysis for the extraction of the 
experimental cross section, \SigRad, for the E = 4.74 GeV and 5.89 GeV data sets is shown in~\cite{2011Report}. 

%===================================================================================================
\paragraph{Unpolarized Cross Section Radiative Corrections} \label{sect:rad_cor_xs}
%===================================================================================================

Electrons lose energy due to interactions with material.  This includes the material
before and after the target, and the target material itself.  These interactions will alter
the electron's \emph{true} incident energy and also its \emph{true} scattered energy.
This ultimately results in a different cross section than the true value.  These effects
are characterized by ionization (or Landau straggling) and bremsstrahlung.  There are also
higher-order processes at the interaction vertex that must also be considered.  Collectively,
the removal of these effects is called \emph{radiative corrections}.

A first correction that must be done \emph{before} carrying out the radiative corrections
is to subtract the elastic radiative tail, since it is long and affects all states of
higher invariant mass $W$~\cite{MT}.  For these kinematics, the elastic tail is
small and affects the lowest bins in scattered electron energy \Ep{} at the
$\lesssim$ 1\% level. The elastic tail was computed using the ROSETAIL code~\cite{Rosetail}. 
The model used for the elastic \HeliumThree{} form factors was from Amroun~\cite{Amroun}.

The \HeliumThree{} quasi-elastic tail, however, is much larger.  The quasi-elastic
radiative tail was computed by utilizing an appropriate model of the \HeliumThree{}
quasi-elastic cross section~\cite{QFS} and applying radiative effects~\cite{Stein}.
The tail was then subtracted from the data.  The model was checked against existing
quasi-elastic \HeliumThree{} data covering a broad range of kinematics.

In considering the effects mentioned above, the \emph{measured} cross section is realized in
terms of a triple integral:

\begin{equation} \label{eqn:triple-int}
   \SigRad\left(\Es,\Ep \right) = \int_0^T \frac{dt}{T} \int_{\Es^{min}}^{\Es} d\Es' \int_{\Ep}^{\Ep^{max}}d\Ep' 
                         I\left( \Es,\Es',t \right) \sigma_r\left(\Es',\Ep'\right) I\left( \Ep,\Ep',T-t \right),
\end{equation}

\noindent where \SigRad{} is the measured (radiated) cross section, $\sigma_r$ is the
\emph{internally}-radiated cross section. \Es{} is the incident electron energy, \Ep{} is the
scattered electron energy.  $I\left(E_0,E,t\right)$ is the probability of finding an
electron with incident energy $E_0$ that has undergone bremsstrahlung with energy $E$
at a depth $t$ inside a material~\cite{MT,Stein}.

\begin{figure}
      \centering
      \subfigure[\Es{} = 4.74 GeV]{
         \centering
         \includegraphics[scale=0.35]{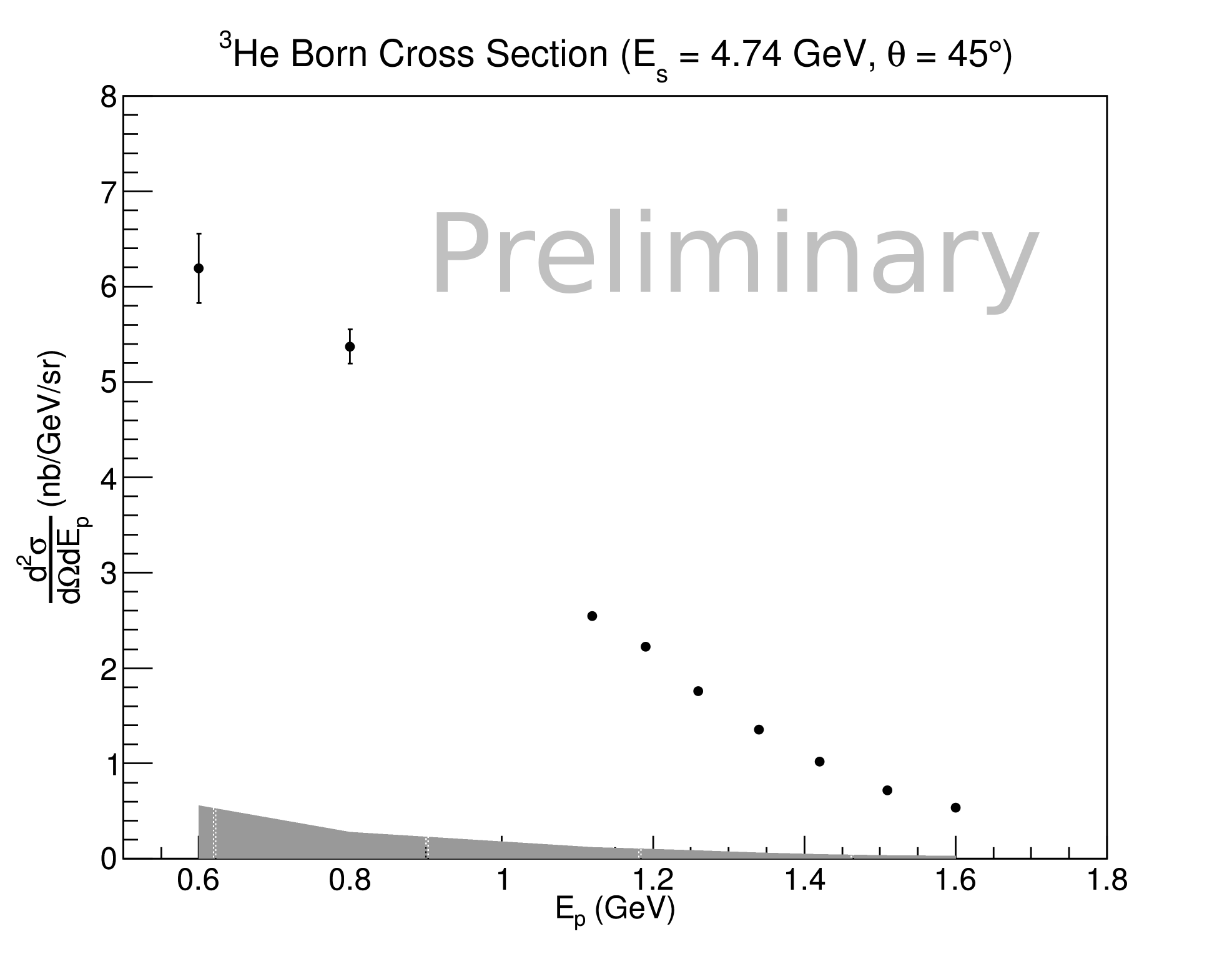}
         \label{fig:xs_4}
      }
      \subfigure[\Es{} = 5.89 GeV]{
         \centering
         \includegraphics[scale=0.35]{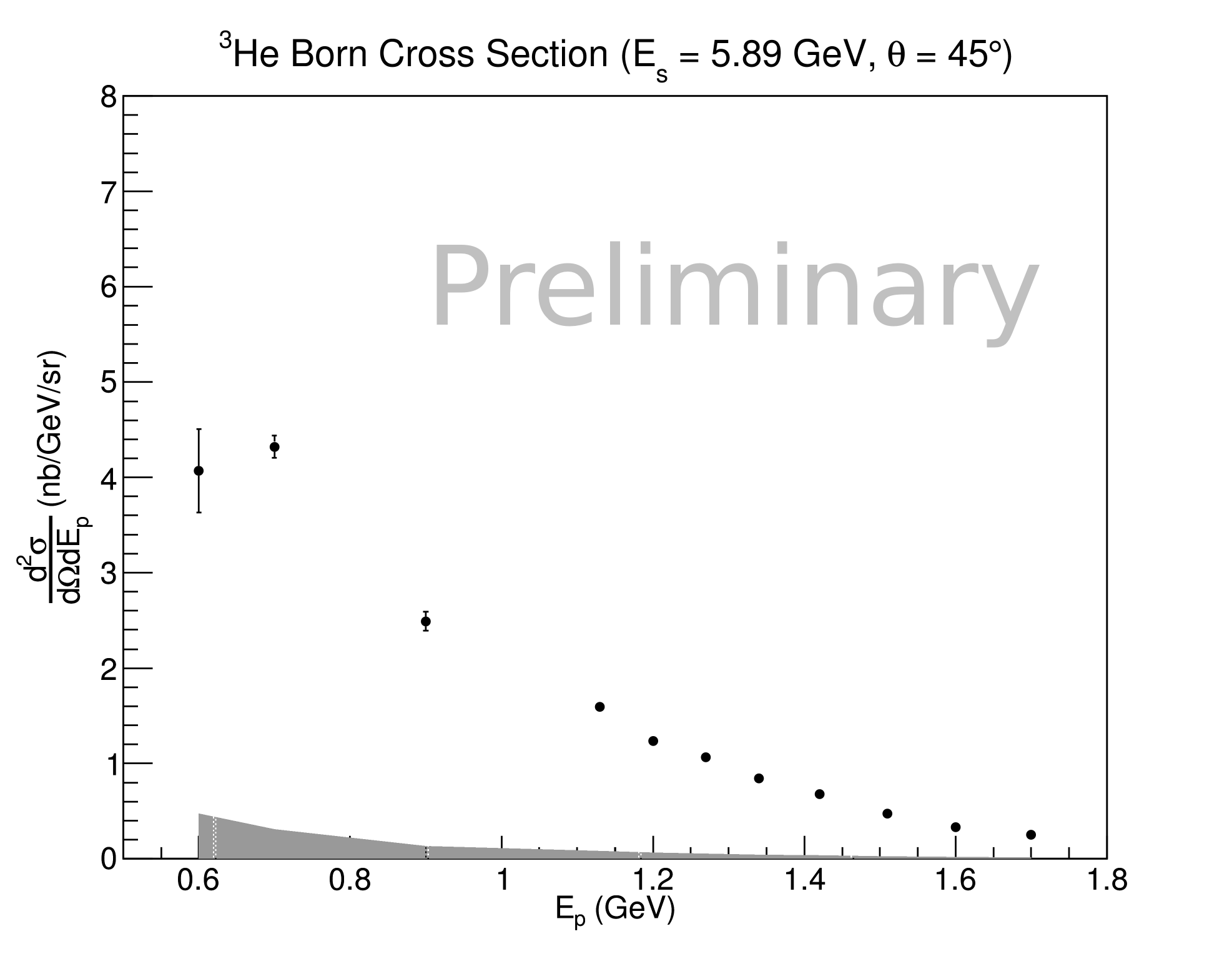}
         \label{fig:xs_5}
      }
      \caption[The  \HeliumThree{} Born cross sections.]{The  \HeliumThree{} Born cross sections.  The error bars indicate the 
               statistical error, while the band indicates the systematic error.
               \SubFig{xs_4}: \Es{} = 4.74 GeV data; \SubFig{xs_5}: \Es{} = 5.89 GeV data.}
     \label{fig:BornXS}
   \end{figure}

In order to \emph{unfold} the Born cross section, an iterative procedure is carried
out in RADCOR~\cite{RadCor}.  It amounts to calculating:

\begin{equation} \label{eqn:unfold} 
   \sigma_b^i = \frac{1}{\textrm{C}}\left[ \sigma_{\textrm{rad}} 
              - \int \left( \ldots \right)dE_{s}' 
              - \int \left( \ldots \right)dE_{p}'\right],  
\end{equation}

\noindent where C and the two integrals are defined in Equation IV.2 in~\cite{MT}.
$\sigma_b^i$ is the Born cross section obtained for the $i^{\textrm{th}}$ iteration of
the code, \SigRad{} is the radiated cross section to be corrected.  $\sigma_b^i$ is then
re-inserted into equation for the next iteration. It was found that the calculation
converges within the first 3--4 iterations.  \Figure{BornXS} shows the resulting Born cross sections.

In E06-014, we took data for only two \Es{} values of 4.74 GeV and 5.89 GeV.  However,
we need enough data to properly calculate the integrals above.  Therefore, we used a
suitable cross section model~\cite{Bosted} to fill in the rest of the phase space for
each data set.

%===============================================================================
\paragraph{Unpolarized Cross Section Systematic Errors} \label{sect:XSErr}
%===============================================================================

There are a number of contributions to the systematic errors on the cross section
calculation~\cite{2012Report}.  We will focus our discussion on the radiative corrections.

The systematic errors corresponding to the radiative corrections include the
elastic and quasi-elastic tail subtraction, material thicknesses in the electron's
path, and dependence on the input model used for the radiative correction calculations.

The systematic error of subtracting the elastic tail from the data is $\ll$ 1\%, determined
by considering different models for the elastic \HeliumThree{} form factors.

In a similar fashion as the elastic tail, the systematic effect of the subtraction
of the quasi-elastic tail was determined by considering different quasi-elastic
cross section models to compute the tail.  We found that the error is $\approx$ 5--6\%
for the lowest bin in \Ep, and falls to $\approx$ 1\% for all other bins for which we have data.

To determine the error related to the material thicknesses in the electron's path,
we varied the thicknesses in our calculations by up to 10\%, and saw a change in
our resulting Born cross section of $\lesssim$ 1.5\%.

The error corresponding to the input model used in the radiative correction procedure
was determined by using different models.  The resulting Born cross section changed
by at most $\approx$ 5\% for the lowest bin in \Ep{} and dropped to $\lesssim$ 1\%
for all other bins.

%===================================================================================================
\paragraph{The Double-Spin Asymmetries}
%===================================================================================================

The BigBite spectrometer was used to measure the parallel and perpendicular double-spin asymmetries 
between longitudinally polarized electrons and a longitudinally or transversely polarized \HeliumThree{} 
target.  These asymmetries were then corrected for imperfect beam and target polarizations.  Corrections
were also made for dilution effects due to the presence of \NTwo{} in the target~\cite{Kom}, and 
contamination due to pions and pair-produced electrons.  The full details of these analyses may be found 
in~\cite{2011Report,2012Report}.   

%===================================================================================================
\paragraph{Asymmetry Radiative Corrections}
%===================================================================================================

\begin{figure}
      \centering
      \includegraphics[scale=0.6]{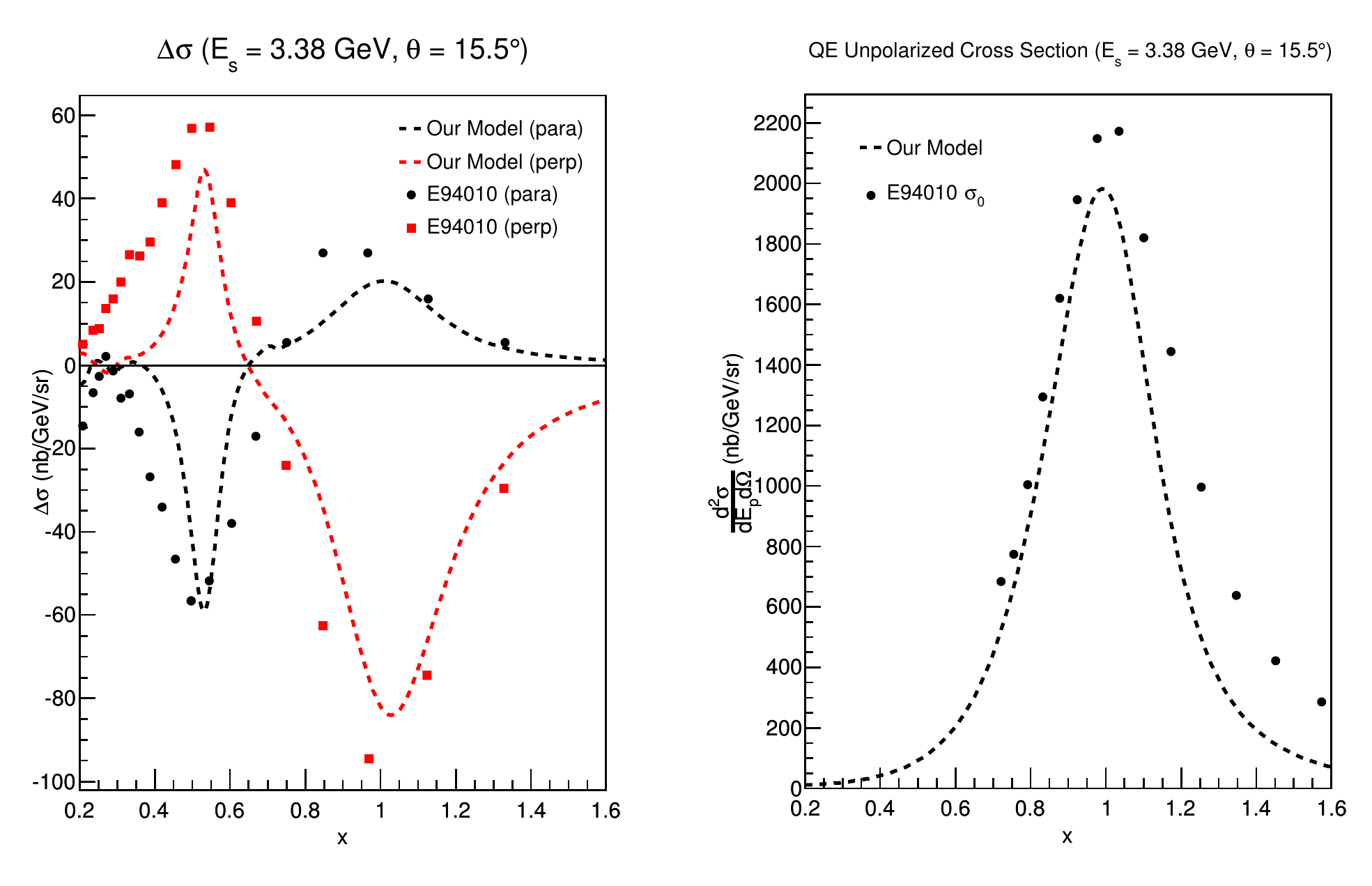}
      \caption[Our model of $\Delta \sigma_{\parallel,\perp}$ as compared to 
               JLab E94-010 data.]{Our model of $\Delta \sigma_{\parallel,\perp}$ as compared to 
               JLab E94-010 data.  Our model consists of combining 
               a smeared version of P.~Bosted's nucleon form factors to describe the 
               quasi-elastic region and the MAID model for the resonance region.  
               The DIS region is modeled using the DSSV PDFs. 
               }
     \label{fig:delta-sig-model}
   \end{figure}

To compute the radiative corrections for asymmetries, we utilized the radiative
correction code RADCOR mentioned in \ref{sect:rad_cor_xs}.  To do this, we carried out
the corrections on polarized cross section differences, $\Delta\sigma$, related to asymmetries by:

\begin{equation}
   \Delta\sigma_{\parallel,\perp}^{r} = 2\sigma_{0}^{r}A_{\parallel,\perp}^{r}, \label{eqn:delta_sig}
\end{equation}

\noindent where $A_{\parallel,\perp}$ indicates a radiated asymmetry where the target
is polarized either parallel ($\parallel$) or perpendicular ($\perp$) with
respect to the incident electron beam momentum.  The unpolarized cross section
is $\sigma_0^{r}$, where the $r$ indicates that radiative effects have been applied.
We used the F1F209~\cite{Bosted} model for the unpolarized cross section. 
After the data have been converted to polarized cross section differences, they are
imported into the RADCOR code in a similar fashion as was done for the unpolarized
cross sections.  The difference here, however, is that a model for the polarized
cross section differences is needed to complete the integrals mentioned in
Equation~\ref{eqn:unfold}.  This model consists of three components describing
different kinematic regions: DIS, the quasi-elastic region, and the resonance region. 
\noindent The model used for the DIS region was the DSSV global analysis parton
distribution function (PDF) model~\cite{DSSV}, which describes world data quite
well in our kinematic region of interest.  For the quasi-elastic region, we utilized
P.~Bosted's nucleon form factors~\cite{PBostedNucleonFF}, smeared by a quasi-elastic
scaling function~\cite{Amaro} to simulate the nuclear effects of \HeliumThree.  Putting
together the nucleon form factors and the smearing function yields a quasi-elastic $\Delta\sigma$
which fits world data well. For the resonance region, we used the MAID model~\cite{MAID}, 
which adequately describes world data.  Putting the DIS, quasi-elastic and resonance 
contributions together, we built up an appropriate $\Delta\sigma$ that describes the physics 
to a reasonable level, an example of which is shown in Figure~\ref{fig:delta-sig-model} 
where we compare our model to JLab E94-010 data~\cite{E94010_Amerian,E94010_Slifer}.  
In the radiative correction procedure, the quasi-elastic tail was not subtracted first, 
but rather included in the integration.  The elastic tail was found to be very small
and was not subtracted. 

\begin{figure}
      \centering
      \subfigure[E = 4.74 GeV]{
         \centering
         \includegraphics[scale=0.5]{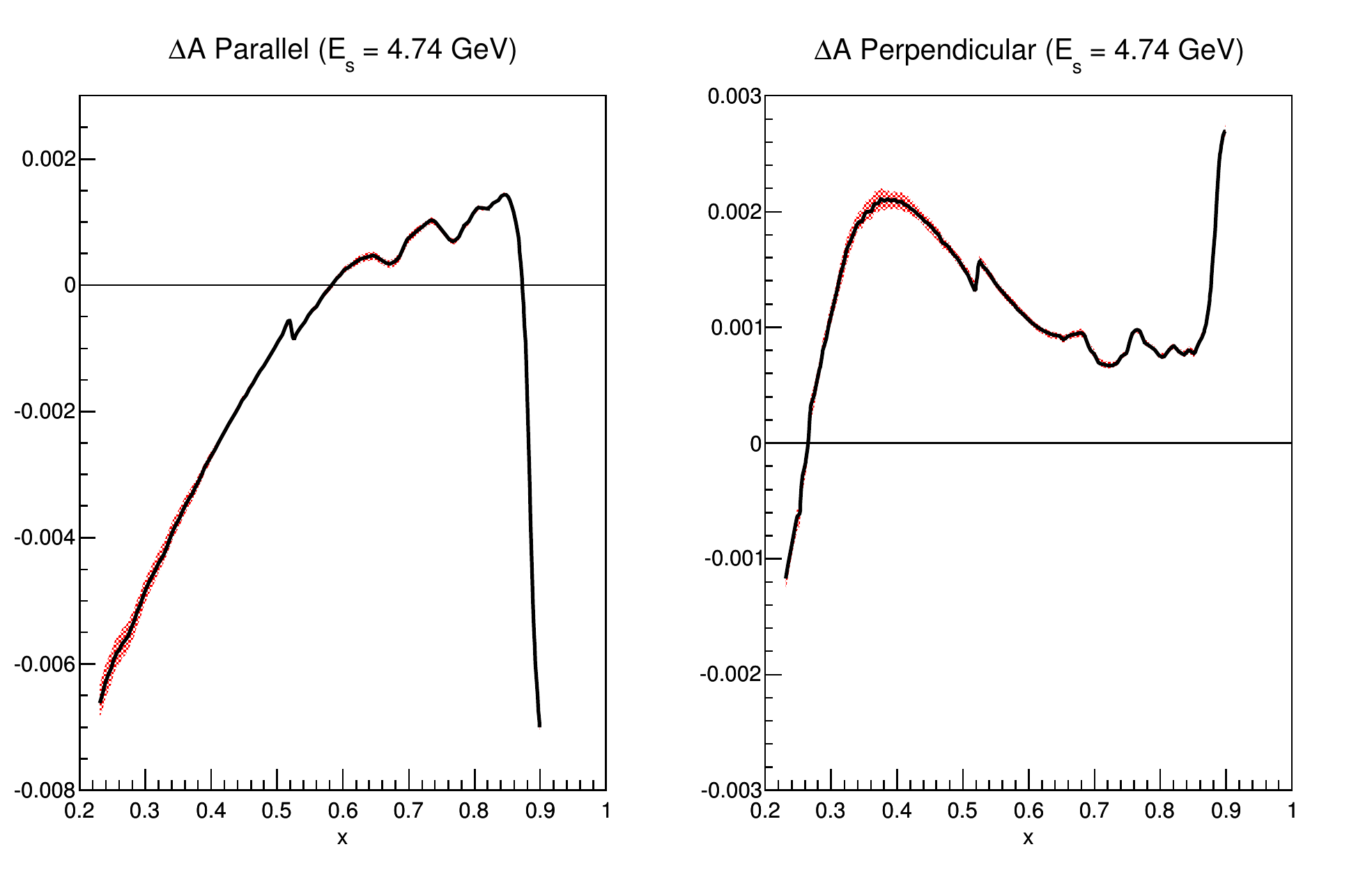}
         \label{fig:asym-rc-size_4}
      }
      \subfigure[E = 5.89 GeV]{
         \centering
         \includegraphics[scale=0.5]{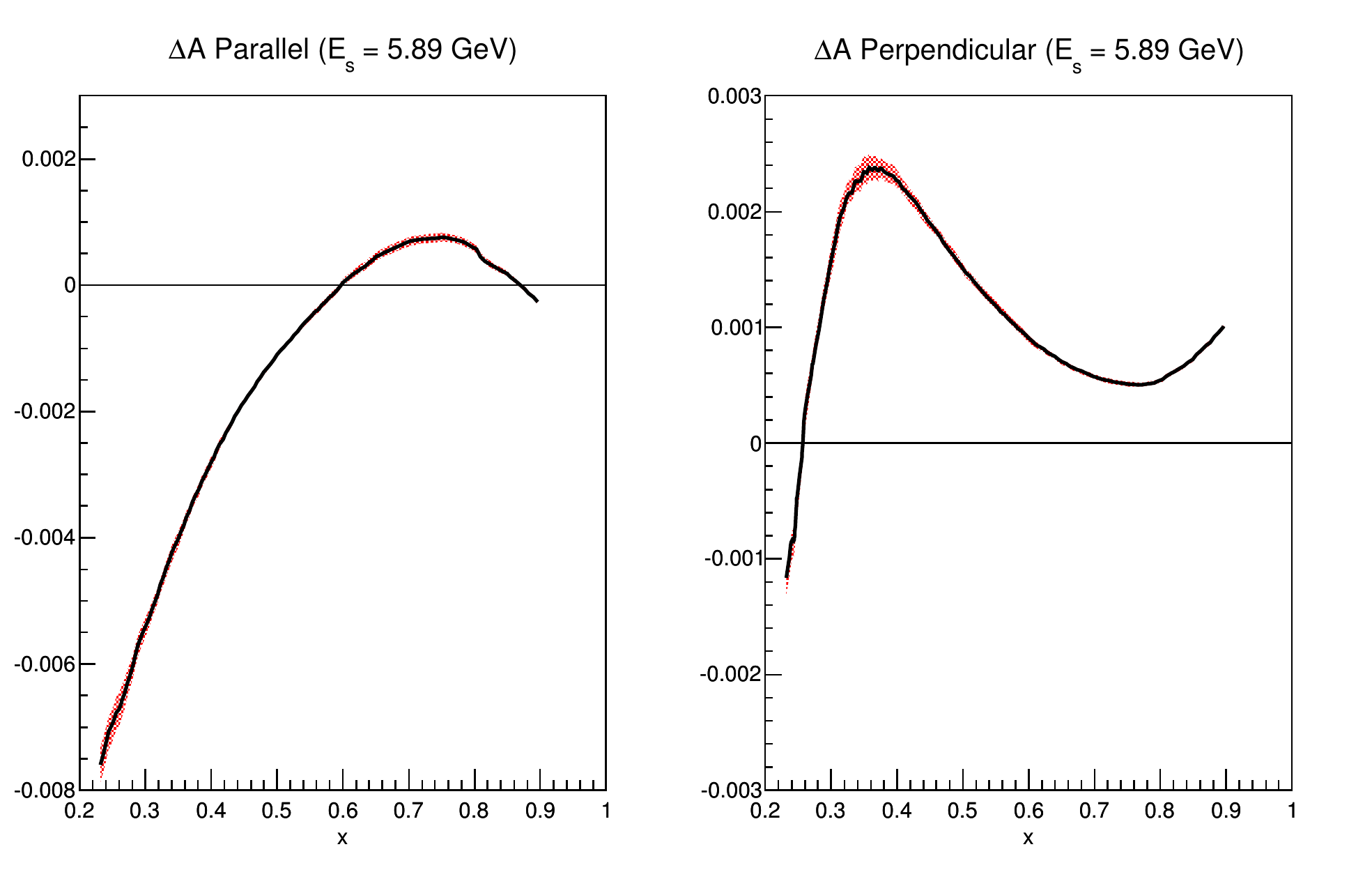}
         \label{fig:asym-rc-size_5}
      }
      \caption[The size of the radiative correction on the asymmetries.]{The size of the radiative correction on the asymmetries, defined as $\Delta A = A_{b} - A_{r}$.
               The red band indicates the systematic error associated with the correction, see Section~\ref{sect:asym-rc-syst}. 
               The $\Delta A$ shown in these plots are applied to the data as an additive correction to 
               obtain the Born asymmetry.     
               \SubFig{asym-rc-size_4}: E = 4.74 GeV data; 
               \SubFig{asym-rc-size_5}: E = 5.89 GeV data.}
      \label{fig:asym-rc-size}
   \end{figure}

To minimize statistical fluctuations in the radiative corrections, the corrections were
done to a model of our data set.  After obtaining the Born $\Delta\sigma$ from RADCOR,
the corresponding asymmetry was obtained by inverting Equation~\ref{eqn:delta_sig} (but using
the {\it Born} $\sigma_0$) to find $A$.  Then, the size of the radiative correction at
the asymmetry level was determined as:

\begin{equation}
   \Delta A = A_{b} - A_{r},  
\end{equation}

\noindent where $A_{b}$ is the Born asymmetry and $A_{r}$ is the radiated asymmetry.
This $\Delta A$ was applied to our data for both the parallel and perpendicular cases
as an additive correction.  The size of the radiative correction as a function of $x$ 
is shown in Figure~\ref{fig:asym-rc-size}.  The red band indicates the systematic error, which is
discussed in Section~\ref{sect:asym-rc-syst}.  

\begin{figure}
      \centering
      \subfigure[Parallel Asymmetries]{
         \centering
         \includegraphics[scale=0.35]{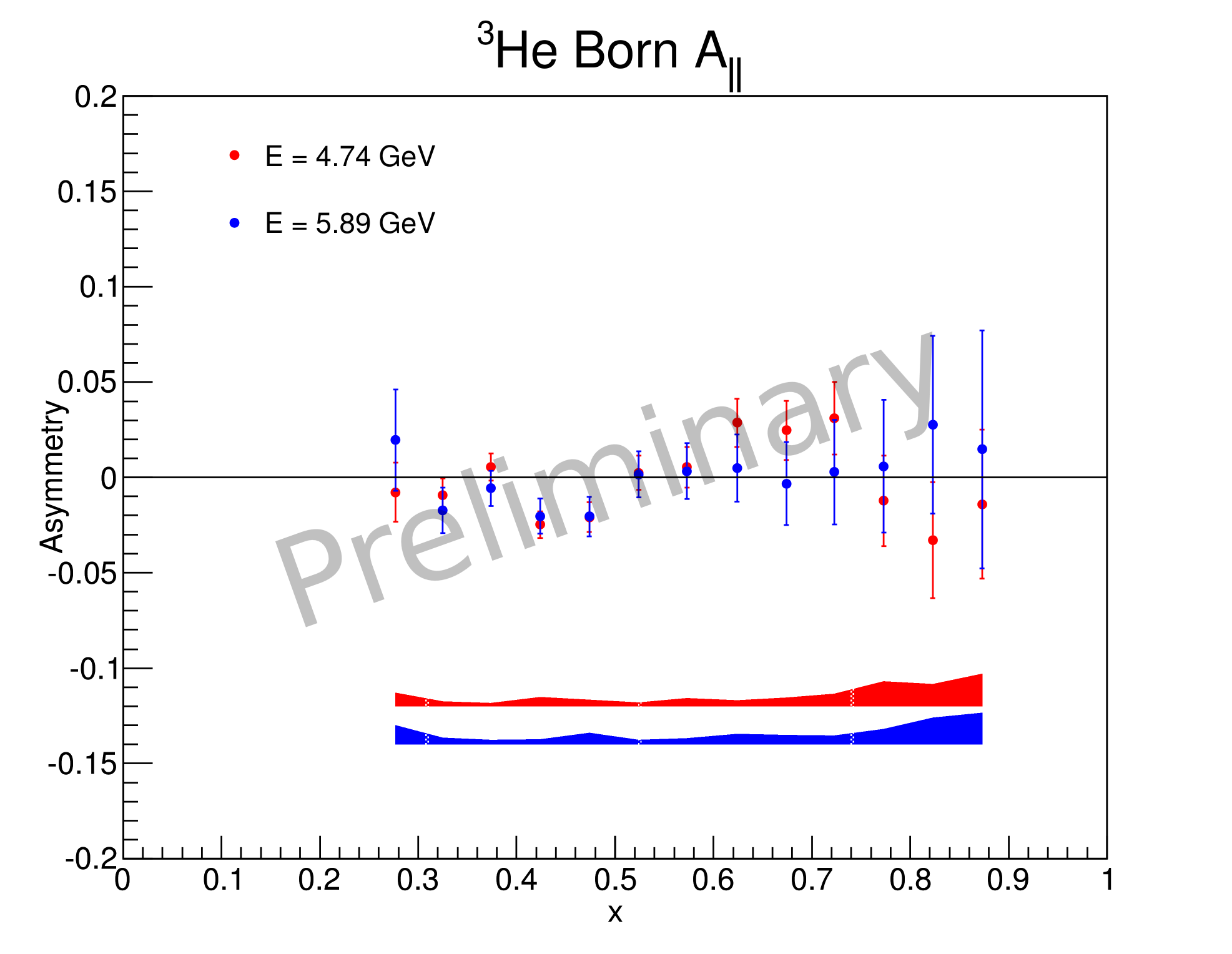}
         \label{fig:para-born-asym}
      }
      \subfigure[Perpendicular Asymmetries]{
         \centering
         \includegraphics[scale=0.35]{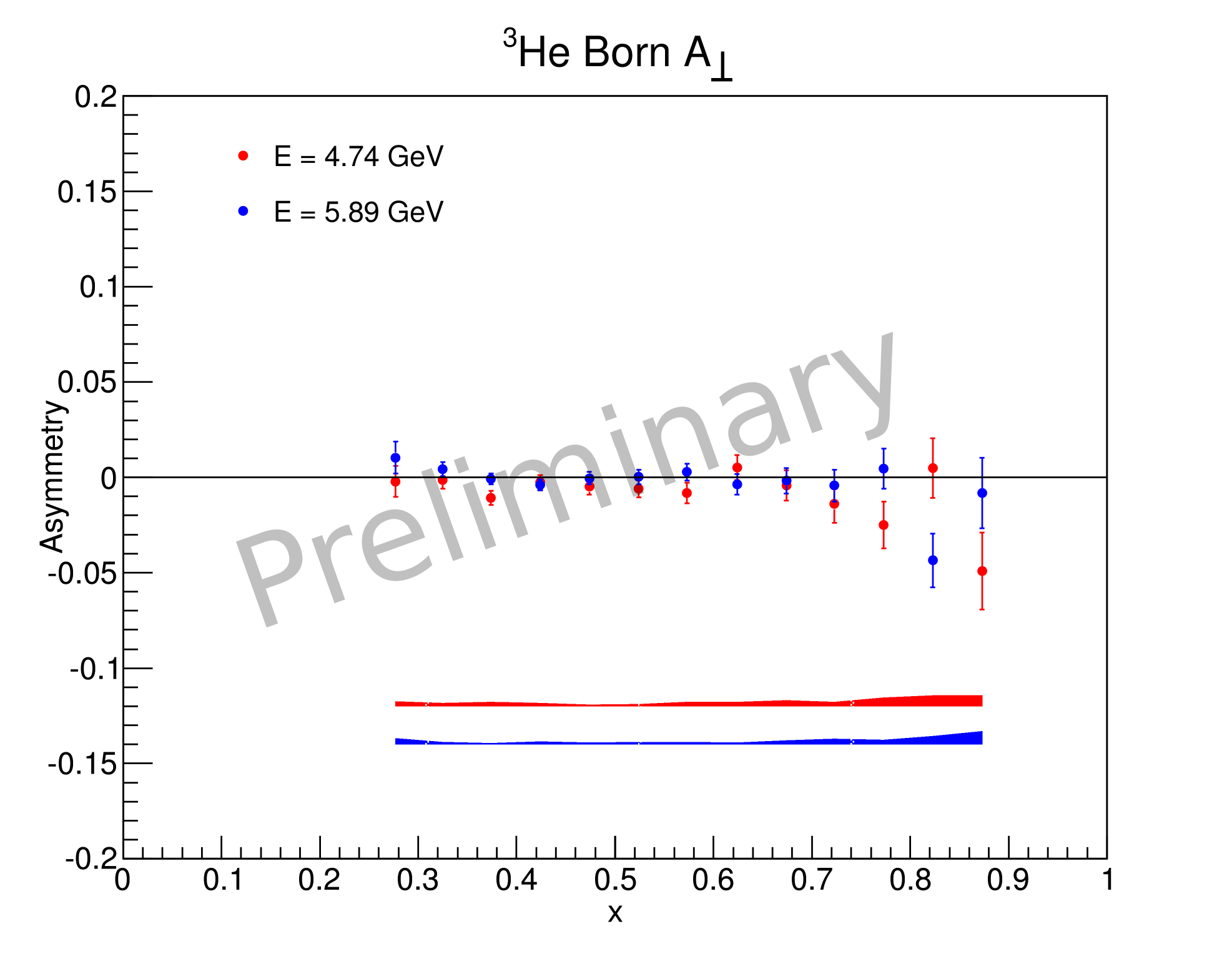}
         \label{fig:perp-born-asym}
      }
      \caption[The Born asymmetries for E = 4.74 GeV and E = 5.89 GeV.]{The Born asymmetries for E = 4.74 GeV (red) and E = 5.89 GeV (blue).
               The error bars indicate the statistical errors, while the colored
               bands show the systematic errors.      
               \SubFig{para-born-asym}: parallel asymmetries; 
               \SubFig{perp-born-asym}: perpendicular asymmetries.}
      \label{fig:born-asym}
   \end{figure}

The Born asymmetries for our data and their systematic errors are shown in \Figure{born-asym}.
The error bars indicate the statistical errors, while the colored bands indicate the systematic
errors, which were obtained by varying all of the inputs needed\footnote{Such quantities include 
the electron cuts, the nitrogen dilution factor, beam and target polarizations, and pion and 
pair production contamination factors.} to extract the asymmetries within reasonable limits and 
observing the change in the asymmetry.   

%===================================================================================================
\paragraph{Asymmetry Radiative Correction Systematic Errors} \label{sect:asym-rc-syst}
%===================================================================================================

To investigate the systematic errors for the radiative corrections, there are two main contributions 
to consider: material thicknesses and model dependence.

To address the first consideration, the thicknesses, they were changed by $\pm$ 10\% and the result 
was compared to the unmodified result.  The change was found to be $\lesssim$ 1.5\%, similar to what 
was seen for the unpolarized cross sections.

The model dependence of the radiative corrections was determined as follows:
the input spectra to the integrals were varied {\it at random} by $\pm$ 10\% for
30 trials, and the size of the correction changed by $\lesssim$ 5\%.

%===================================================================================================
% physics results 
\subsubsection{Preliminary Physics Results}
\paragraph{The Virtual Photon-Nucleon Asymmetry} \label{sect:A1_analysis}
%===================================================================================================

\begin{figure}
      \centering
      \includegraphics[scale=0.6]{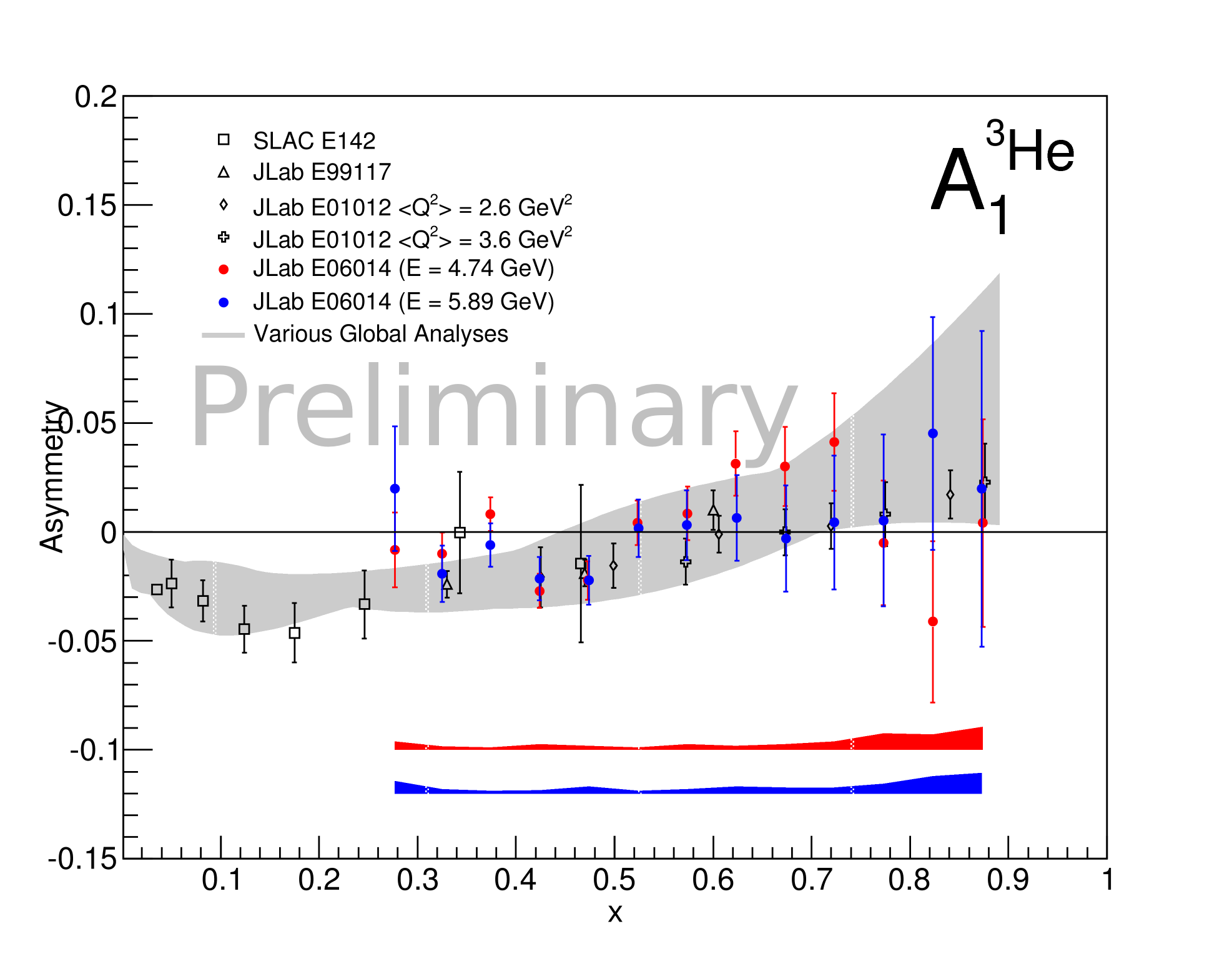}
      \caption[Preliminary \AOneHeThree{} for E = 4.74 and 5.89 GeV compared to world data and selected models]{\AOneHeThree{} compared to the world data from SLAC E142~\cite{E142} and JLab 
               E01-012~\cite{E01012} and E99-117~\cite{E99117PRC}. The error bars on the world data
               indicate the in-quadrature sum of statistical and systematic errors, while 
               the error bars on our data are statistical only.  The colored bands at the 
               bottom of the plot show the systematic errors.  The gray band shows an envelope 
               of various global analyses~\cite{pQCDHHC,DSSV,Gamberg,Soffer,Stratmann,CJ12,CTEQ}.
               }
      \label{fig:A1He3}
   \end{figure}

\Figure{A1He3} shows the preliminary results for \AOneHeThree{} at $E$ = 4.74 and 5.89 GeV, respectively.  
Also shown are world data from SLAC E142~\cite{E142} and JLab E01-012~\cite{E01012} and E99-117~\cite{E99117PRC}.  
The red (blue) data points indicate our E = 4.74 GeV (E = 5.89 GeV) data.  The error bars on the world
data are the in-quadrature sum of the statistical and systematic errors, while the error bars on 
our data are statistical only.  The colored bands at the bottom of the plot indicate the systematic 
errors.  The systematic errors were determined by varying all of the inputs to the computation of 
\AOneHeThree{} to reasonable levels and observing the change in the asymmetry.  The gray band represents 
various global analyses~\cite{pQCDHHC,DSSV,Gamberg,Soffer,Stratmann,CJ12,CTEQ}. The data from this experiment are consistent 
with the world data across a wide range in $x$, despite the larger error bars in the resonance region,
which corresponds to $x > 0.519$ $(0.623)$ for $E$ = 4.74 GeV (5.89 GeV).    

%\newpage 

%===================================================================================================
\paragraph{The Spin Structure Functions} \label{sect:spin_struc_func}
%===================================================================================================

En route to extracting $d_2^n$, the spin structure functions \gOne{} and \gTwo{} can be obtained 
according to: 

\begin{eqnarray}
  g_1 &=& \frac{MQ^2}{4{\alpha}^2}\frac{2y}{\left( 1 - y\right) \left( 2 - y\right)}\sigma_0 
          \left[ A_{\parallel} + \tan \left(\theta/2 \right)A_{\perp}\right] \\
  g_2 &=& \frac{MQ^2}{4{\alpha}^2}\frac{y^2}{\left( 1 - y\right) \left( 2 - y\right)}\sigma_0 
          \left[ - A_{\parallel} + \frac{1 + \left( 1 - y \right)\cos \theta}
          {\left( 1 - y \right)\sin \theta}A_{\perp}\right], 
\end{eqnarray}

\begin{figure}
      \centering
      \includegraphics[scale=0.6]{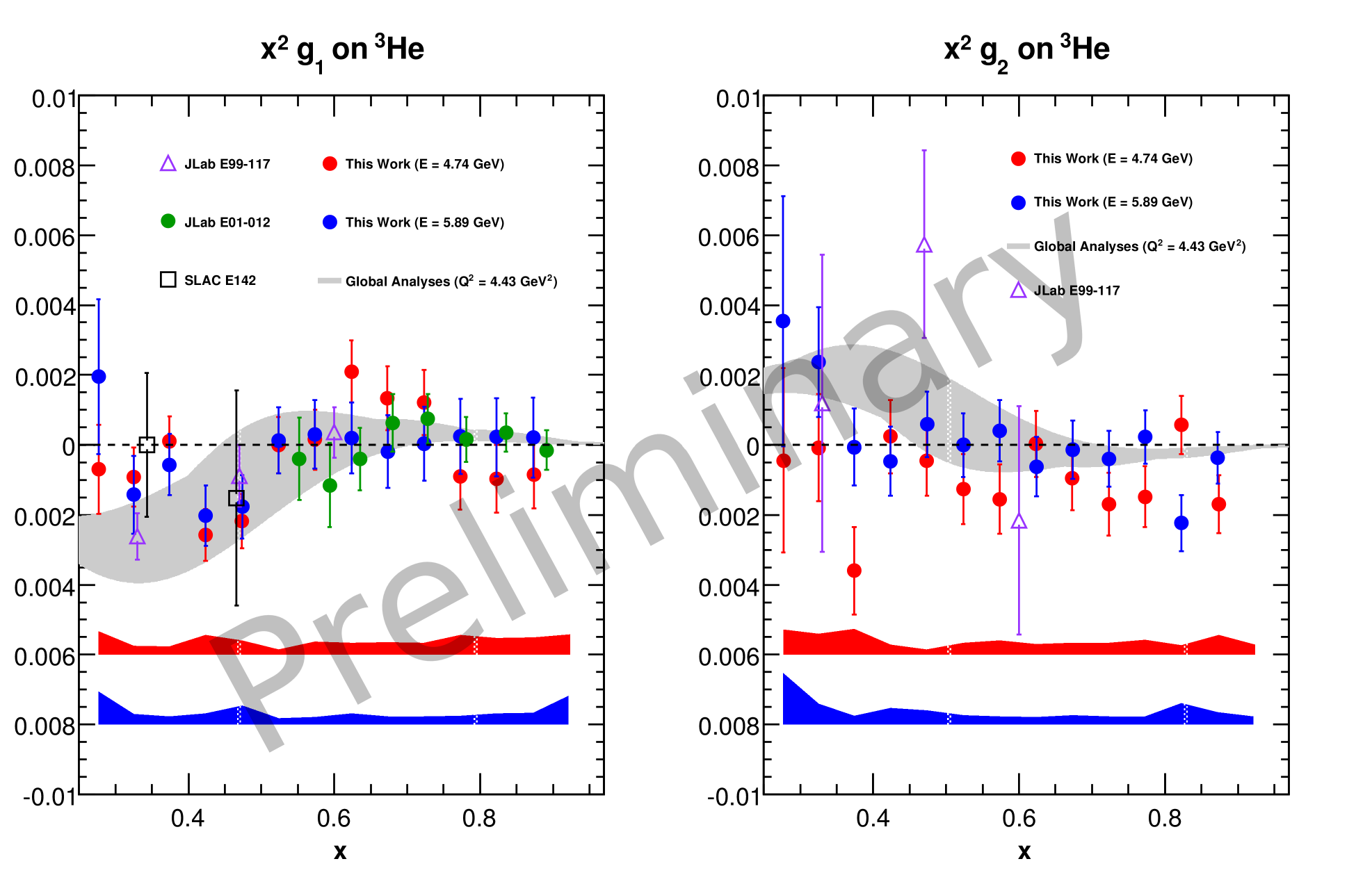}
      \caption[Preliminary $g_{1}$ and $g_{2}$ on $^{3}$He for E = 4.74 and 5.89 GeV compared to world data and various models.]{Preliminary results for the spin structure functions \gOne{} and \gTwo{} on a \HeliumThree{} target
               for E = 4.74 GeV (red) and 5.89 GeV (blue) compared to the world data~\cite{E142,E01012,E99117PRC}
               and various global analyses ~\cite{pQCDHHC,DSSV,Gamberg,Soffer,Stratmann} represented by the gray band. 
               The error bars on our data are statistical only; the colored bands at the bottom of the plot indicate the 
               systematic errors.
      }
      \label{fig:pol_sf}
   \end{figure}

The preliminary results for \gOne$^{^{\textrm{3}}\textrm{He}}$ and \gTwo$^{^{\textrm{3}}\textrm{He}}$ are shown
in \Figure{pol_sf}, which compares the data to various models represented by the gray band~\cite{pQCDHHC,DSSV,Gamberg,Soffer,Stratmann} 
and the world data.  The systematic errors on our data were obtained by varying all of the inputs 
needed to compute $g_1$ and $g_2$ to reasonable levels and observing the change in the result. 

%===================================================================================================
\subsubsection{Current and Future Work}
%===================================================================================================

At present, we are working on finalizing our analysis to evaluate $d_2^{n}$ and $A_1^n$. 
Additionally, from $d_2^n$ we can extract the color electric and magnetic forces~\cite{fil_ji,mb_1,mb_2}. 
From our $A_1^n$ data, we can also perform a flavor decomposition to obtain the quantities $\Delta u/u$
and $\Delta d/d$.  

% Additionally, we are preparing our results for publication. 

%===================================================================================================
% references 
%\clearpage

%\end{document}

\clearpage

\subsection{E07-007 and E08-025: Deeply Virtual Compton Scattering}
\label{sec:e07007-e08025}

%\begin{center}
%Deeply Virtual Compton Scattering
%\end{center}

\begin{center}
P. Bertin, A. Camsonne, C. Hyde, M. Mazouz, C. Mu\~noz Camacho and J. Roche, spokespersons, \\
and \\
the Hall A Collaboration.\\
contributed by C.~Mu\~noz~Camacho.
\end{center}

\subsubsection{Introduction}\label{sec:dvcsintro}

   Deeply Virtual Compton Scattering (DVCS) refers to the electroproduction of photons in the Deep Inelastic Scattering (DIS) kinematics: $ep\to ep\gamma$. In the Bjorken regime, this reaction is sensitive to the Generalized Parton Distributions (GPDs) of the nucleon. In the photon electroproduction, the DVCS amplitude interferes with the so-called Bethe-Heitler (BH) process, where the photon is radiated by the electron, instead of the proton. Previous measurements of DVCS cross sections, both on the proton~\cite{dvcsref1} and the deuteron~\cite{dvcsref2}, showed a significant contribution of the pure DVCS$^2$ term, comparable to the BH$\cdot$DVCS interference.

   The goal of DVCS experiments E07-007 and E08-025 is to perform a Rosenbluth-like separation of the pure DVCS$^2$ and the BH$\cdot$DVCS interference terms from the photon electroproduction cross section. The cross section was measured at 3 different values of $Q^2$ with an LH2 target (E07-007) and at one $Q^2$ with an LD2 target (E08-025), all at constant $x_B=0.36$. Each measurement was performed at two different incident beam energies $E_b$.

A secondary goal of the experiments is to perform an L/T separation of the $\pi^0$ electroproduction cross section, from both a proton and deuteron target. The only measurements of $\pi^0$ electroproduction cross sections in Jefferson Lab kinematics are still unseparated~\cite{dvcsref3,dvcsref4}. For the proton target, $\sigma_L$ and $\sigma_T$ will be measured as a function of $Q^2$ within a limited range: 1.5 to 2.0~GeV$^2$.

\subsubsection{Status of the analysis}\label{sec:dvcsana}

Calibrations of all subsystems are completed and final physics analysis has started for all different channels and kinematic settings.

\begin{figure}[hbt]
\center{\epsfig{figure=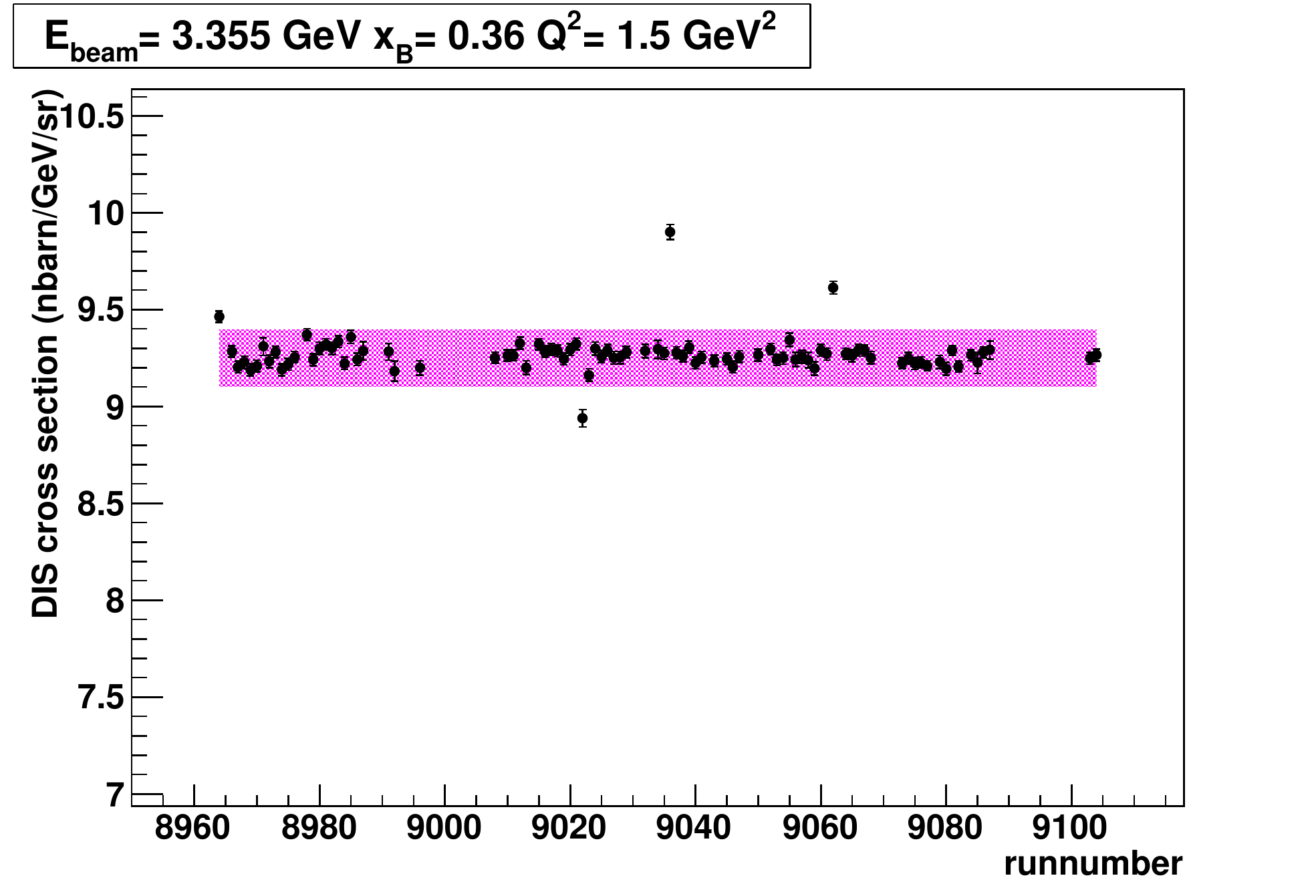,width=0.5\linewidth}}
\caption[DIS cross section as a function of run number.]{DIS cross section as a function of the run number for the kinematic setting with $Q^2=1.5$~GeV$^2$ and $E_b=3.356$~GeV. The magenta band shows the stability of the results to 1.6\% once 4 particular runs with identified problems are removed from the analysis.}
\label{fig:dis}
\end{figure}

Stability of the data and global normalization has been checked by computing the DIS cross section in a run-by-run basis. Figure~\ref{fig:dis} shows, for one of the kinematic settings ($Q^2=1.5$~GeV$^2$ and $E_b=3.356$~GeV), the DIS cross section measured as a function of the run number. The stability of the results is below 2\%, once a few runs with identified problems are removed.

Table~\ref{tab:dis} shows a summary of the DIS results for all our kinematics. Results are compared to the ones obtained using the parametrization of the structure functions~\cite{dvcsref5}. Also, the stability of our results is presented in the last column.

\begin{table}[ht]
\caption[DIS results for all kinematics.]{DIS cross section measured for each kinematic setting. The second column shows the result using the parametrization of the structure functions~\cite{dvcsref5} and the third column shows our measurement. The relative difference between these two values is also presented. Finally, the last column represents the stability of our results as a function of time (run number).}
\begin{center}
\begin{tabular}{|c||c|c|c|c|}
\hline
Kinematic setting & $\frac{d\sigma^{TMC}}{d\Omega dE}$ & $\frac{d\sigma^{TMC}_{exp}}{d\Omega dE}$  & Relative difference& Stability\\
&(nb/GeV/sr) & (nb/GeV/sr) & (\%) & (\%)\\
\hline 
\hline
$Q^2=1.50$~GeV$^2$, $E_b=3.356$~GeV & 9 & 9.26 & +2.8 & 1.6 \\ \hline
$Q^2=1.50$~GeV$^2$, $E_b=5.552$~GeV & 55.2 & 53.3 & -3.4 & 1.3  \\ \hline
$Q^2=1.75$~GeV$^2$, $E_b=4.454$~GeV & 13.14 & 13.14 & 0 & 2 \\ \hline
$Q^2=1.75$~GeV$^2$, $E_b=5.552$~GeV & 28.93 & 27.9 & -3.4 & 1.3 \\ \hline
$Q^2=2.00$~GeV$^2$, $E_b=4.454$~GeV & 6.6 & 6.9 & +4.5 & 4 \\ \hline
$Q^2=2.00$~GeV$^2$, $E_b=5.552$~GeV & 15.93 & 15.26 & -4 & 2.2 \\ \hline
\end{tabular}
\label{tab:dis}
\end{center}
\end{table}

\begin{figure}[hbt]
\center{\epsfig{figure=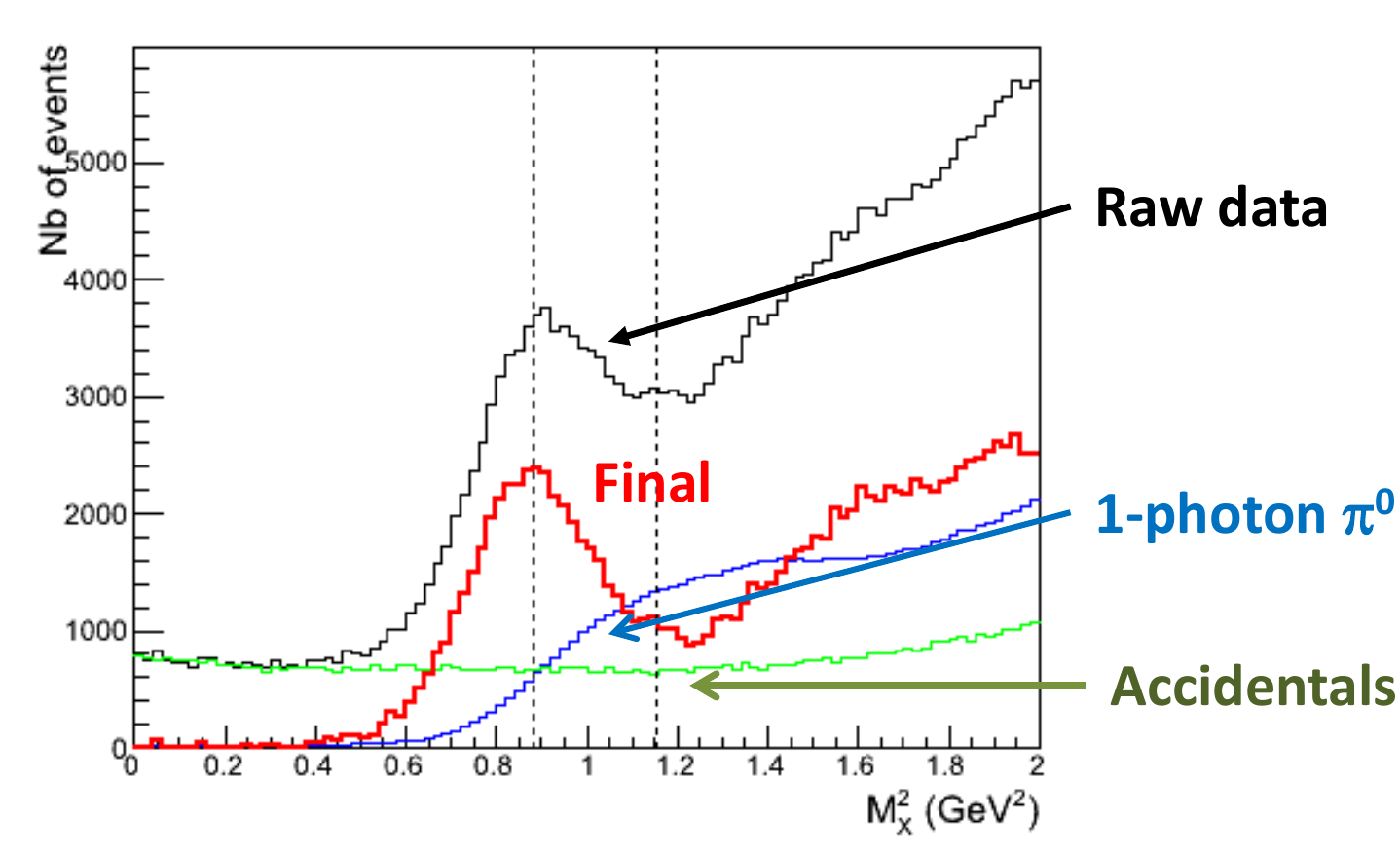,width=0.6\linewidth}}
%\center{\epsfig{figure=e07007-mm.eps,width=0.6\linewidth,bb=177 170 593 412, clip}}
\caption[E07-007: DVCS off the proton missing mass squared.]{DVCS $ep\to e\gamma X$ missing mass squared $M_X^2$ off the proton, $Q^2=1.5$~GeV$²$ and $E_b=3.356$~GeV. The black histogram shows the raw data and the red is the final sample, once accidentals (green) and the contamination from 1-photon $\pi^0$s (blue) are subtracted.}
\label{fig:mmproton}
\end{figure}

All photon data have been processed and number of counts for DVCS off the proton and the deuteron are available for all of the kinematic settings. Figure~\ref{fig:mmproton} shows the 
$ep\to e\gamma X$ missing mass squared $M_X^2$ off the proton for the kinematic setting $Q^2=1.5$~GeV$^2$ and $E_b=3.356$~GeV. Neutral pion decays that yield only one photon in the DVCS calorimeter are subtracted from the raw data using the sample of events where the 2 photons of the decay are detected. This contribution is shown in blue in Figure~\ref{fig:mmproton}.

\begin{figure}[hbt]
\center{\epsfig{figure=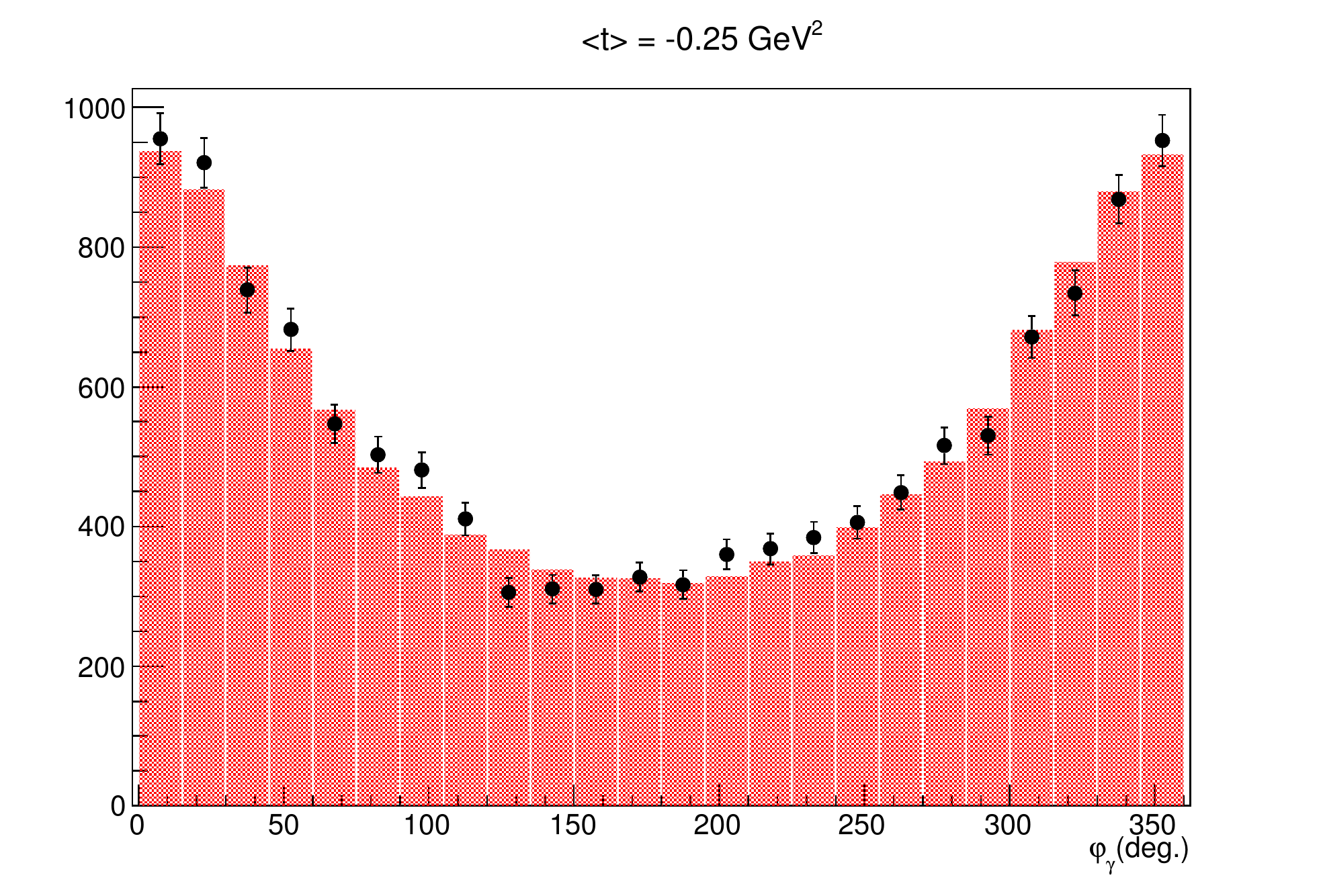,width=0.5\linewidth}}
\caption[E07-007: Final DVCS counts and comparison with Monte Carlo.]{Final number of DVCS counts for $Q^2=2.0$~GeV$^2$, $E_b=5.552$~GeV and $t=-0.25$~GeV$^2$ (black points) as a function of the angle $\varphi_\gamma$ between the leptonic and hadronic planes. The red histogram shows the estimate obtained from a Monte Carlo simulation of the experimental setup.}
\label{fig:counts}
\end{figure}

The DVCS cross section is extracted using a Monte Carlo simulation of the experimental setup and integrating the known kinematic factors of the different cross-section terms over each experimental bin. Thus, the average value of the Compton Form Factors (GPD integrals) are accurately computed in each bin, regardless of the rapid variation of the cross section within the bins. Figure~\ref{fig:counts} shows, for one particular bin in the momentum transfer $t$ of the kinematic setting $Q^2=2.0$~GeV$^2$ and $E_b=5.552$~GeV, the experimental number of counts as a function of the angle $\varphi_\gamma$ between the leptonic and hadronic planes, together with the estimate obtained from the Monte Carlo simulation.

\begin{figure}[hbt]
\center{\epsfig{figure=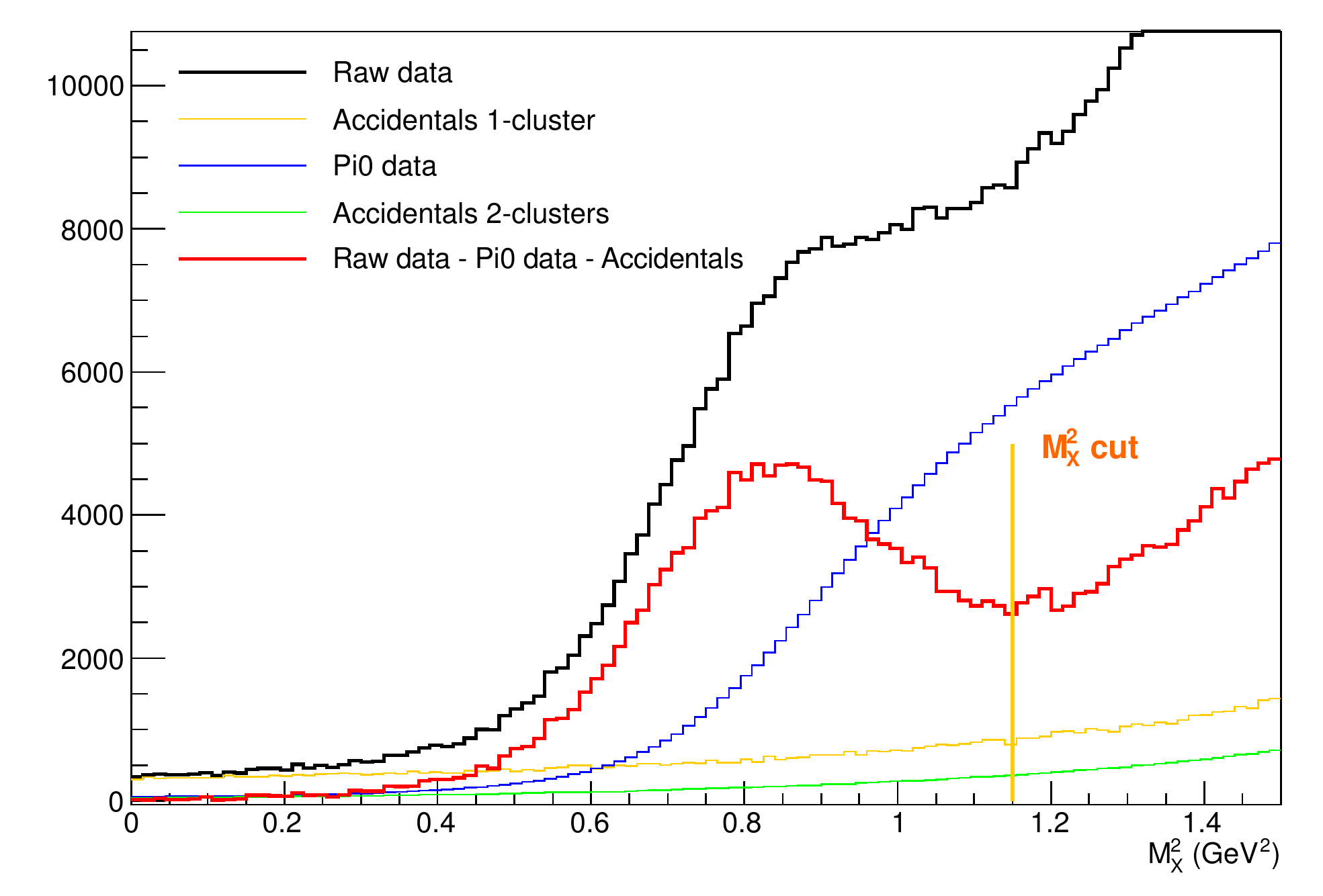,width=0.5\linewidth}\epsfig{figure=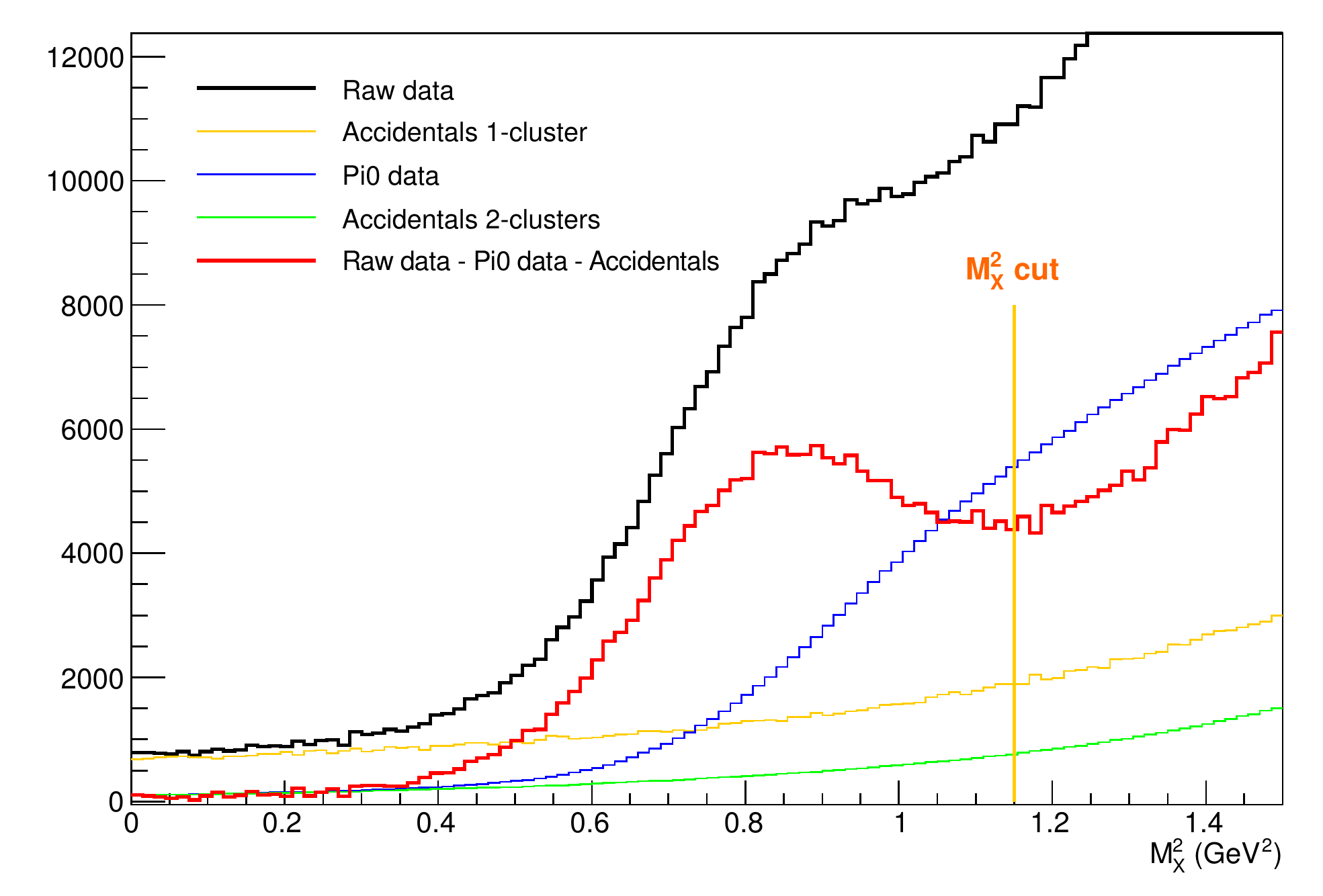,width=0.5\linewidth}}
\caption[E07-025: DVCS off the deuteron missing mass squared.]{DVCS $ep\to e\gamma X$ missing mass squared $M_X^2$, $Q^2=1.75$~GeV$²$ and $E_b=5.552$~GeV. The black histogram shows the raw data and the red is the final sample, once accidentals (green and yellow) and the contamination from 1-photon $\pi^0$s (blue) are subtracted. The left plot shows data off a LH2 target and the right plot corresponds to an LD2 target.}
\label{fig:mmdeuteron}
\end{figure}

DVCS events from the LD2 target are also in advanced stage of analysis. Measurements of DVCS off the neutron relies on the measurements with an LD2 and the subtraction of the proton contribution obtained using an LH2 target. Figure~\ref{fig:mmdeuteron} shows the $ep\to e\gamma X$ missing mass squared for the same kinematics with both an LH2 (left) and LD2 (right) target. Accidental and $\pi^0$ contamination are subtracted from the raw data (black) in order to obtain the final DVCS sample (red). The vertical line at $M_X^2=(M_n+m_\pi)^2=1.15$~GeV$^2$ shows the cut applied to ensure exclusivity.

%
% Here's how to do the references.  We will be using the APS style.
%

\clearpage

%\documentclass{article}
%\usepackage{epsfig}
%\setlength{\textwidth}{6.5in}
%\setlength{\oddsidemargin}{0in}
%\setlength{\evensidemargin}{0in}
%\setlength{\textheight}{9in}
%\setlength{\topmargin}{0in}
%\setlength{\headheight}{0in}
%\setlength{\headsep}{0in}

%\begin{document}

%\section{Summaries of Experimental Activities}

\subsection{E08-005: Target Single-Spin Asymmetry $A_y^0$ in the Quasi-Elastic $^3$He$^{\uparrow}$($e,e'n$) Reaction}
\label{sec:quasi-elastic}

\begin{center}
T. Averett, D. Higinbotham, V. Sulkosky, (Spokespersons) \\
and \\
the Hall A Collaboration.\\
contributed by E. Long (University of New Hampshire).
\end{center}

\subsubsection{Progress of $^3$He($e,e'n$) $A_y^0$}\label{sec:details}
Progress has been made on the $^3\mathrm{He}^{\uparrow}$($\vec{e},e'n$) target single-spin asymmetry, $A_y^0$, for experiment E08-005, where the target was polarized in the vertical direction, transverse to the beam. In plane wave impulse approximation (PWIA), this asymmetry is exactly zero. Any non-zero measurement indicates higher-order effects, such as final state interactions (FSI) and meson exchange currents (MEC). 

\begin{figure}[hbt]
\center{\epsfig{figure=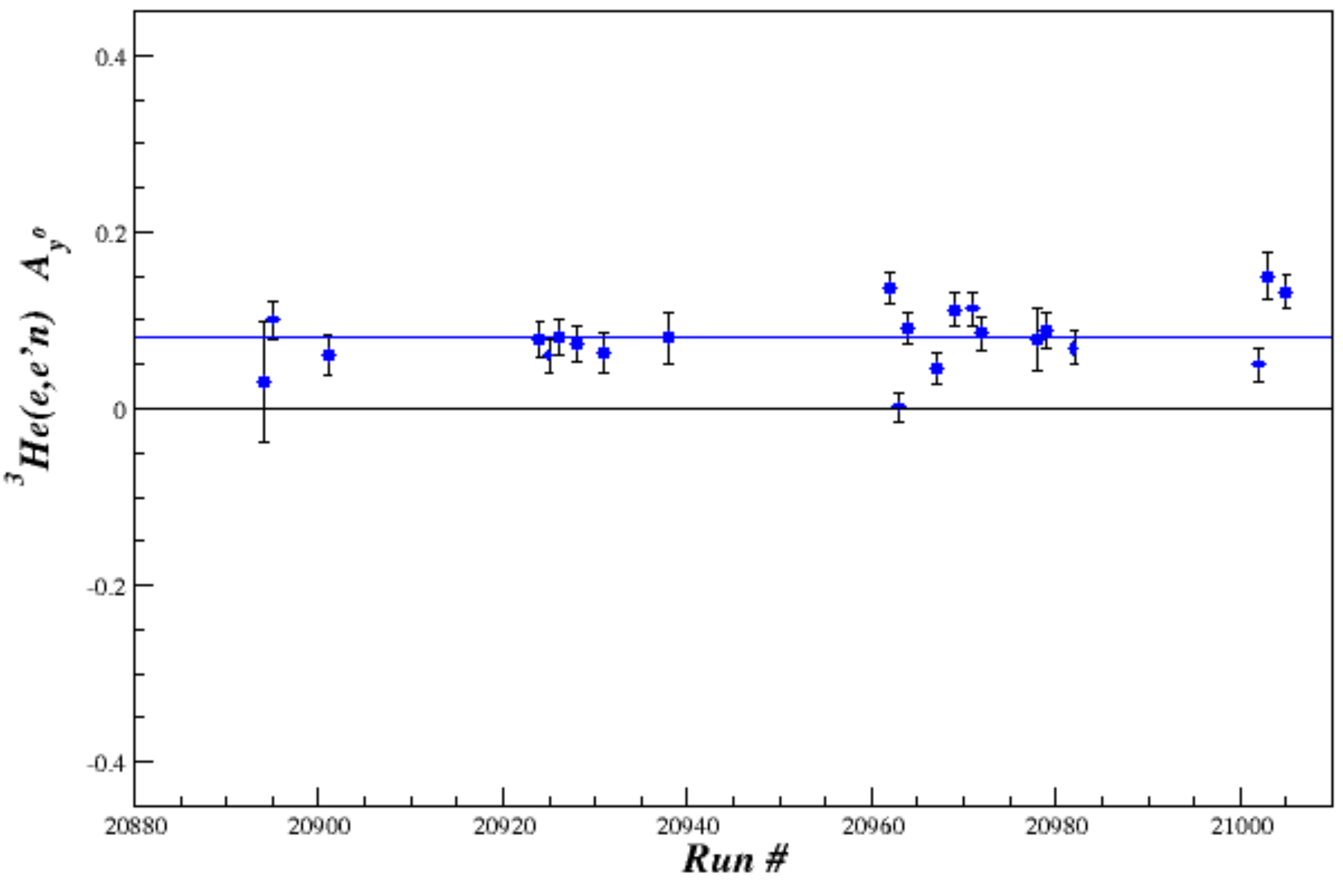,width=8cm}}
\caption{$A_y^0$ calculated on a run-by-run basis at $Q^2=0.46 \mathrm{~(GeV}/c)^2$.}
\label{runbyrun}
\end{figure}

Data from the RHRS were used to isolate the quasi-elastically scattered electrons using standard kinematic cuts. Neutrons were identified using the Hall A Neutron Detector. New analysis was conducted this past year to examine run-by-run fluctuations of $A_y^0$, as illustrated in Fig. \ref{runbyrun} for $Q^2=0.46 \mathrm{~(GeV}/c)^2$.

\begin{figure}[hbt]
\center{a)\epsfig{figure=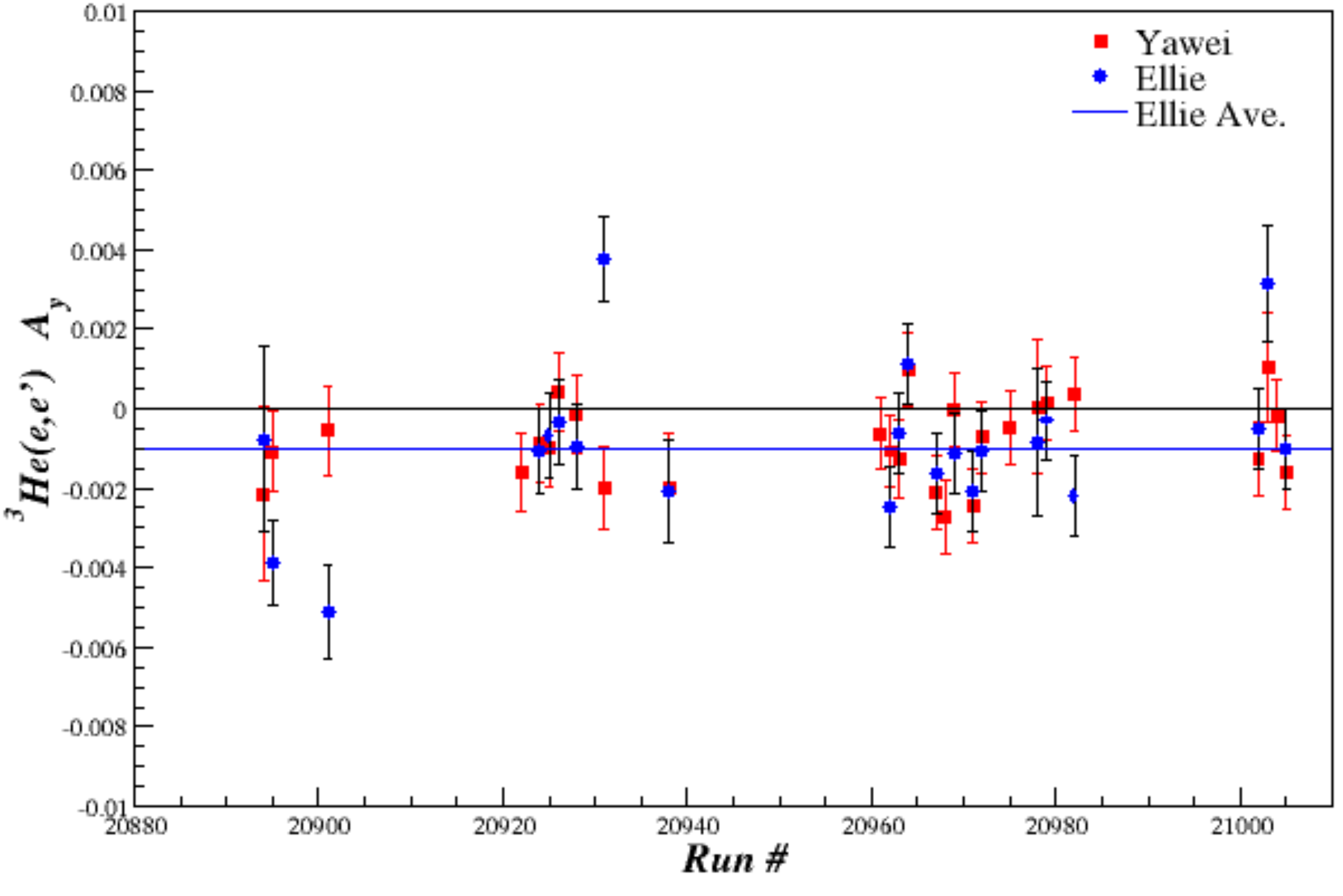,width=8cm} b)\epsfig{figure=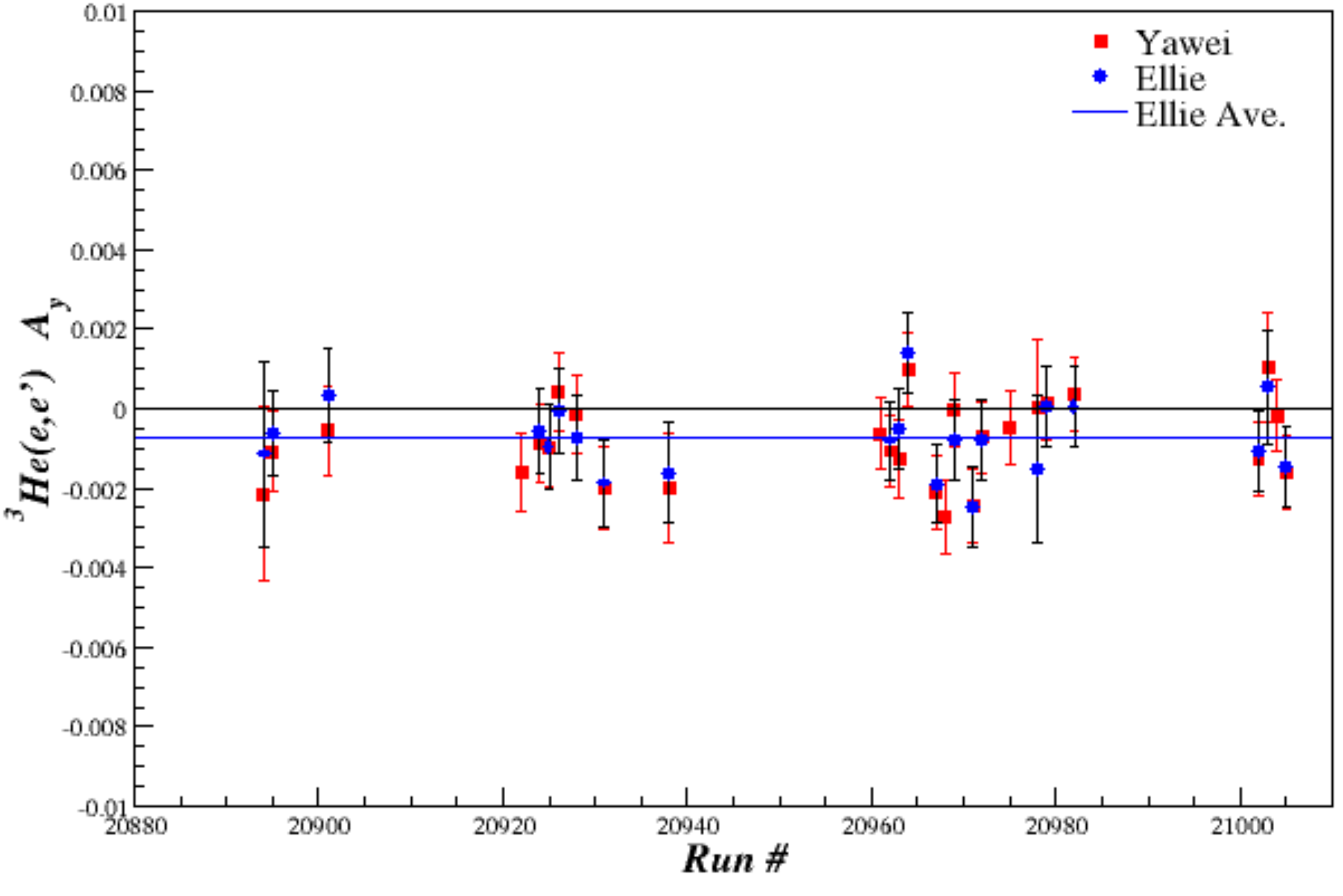,width=8cm}}
\caption{The $^3$He$^{\uparrow}(e,e')$ target single-spin asymmetry was compared to Y. Zhang's analysis of the same dataset. Calculations were done using a) the original analysis script and b) the corrected analysis script, which accounted for minor corrections to the charge and live-time calculations.}
\label{eeprime}
\end{figure}

The analysis was compared directly to Y. Zhang's $^3$He$^{\uparrow}(e,e')$ target single-spin asymmetry by removing the neutron cuts. Through this comparison, it was found that there were minor corrections to the calculation of the charge and the live-time that were included into the $A_y^0$ analysis. This discrepancy, and fix, are shown in Fig. \ref{eeprime} for $Q^2=0.46 \mathrm{~(GeV}/c)^2$.

\begin{figure}[hbt]
\center{a)\epsfig{figure=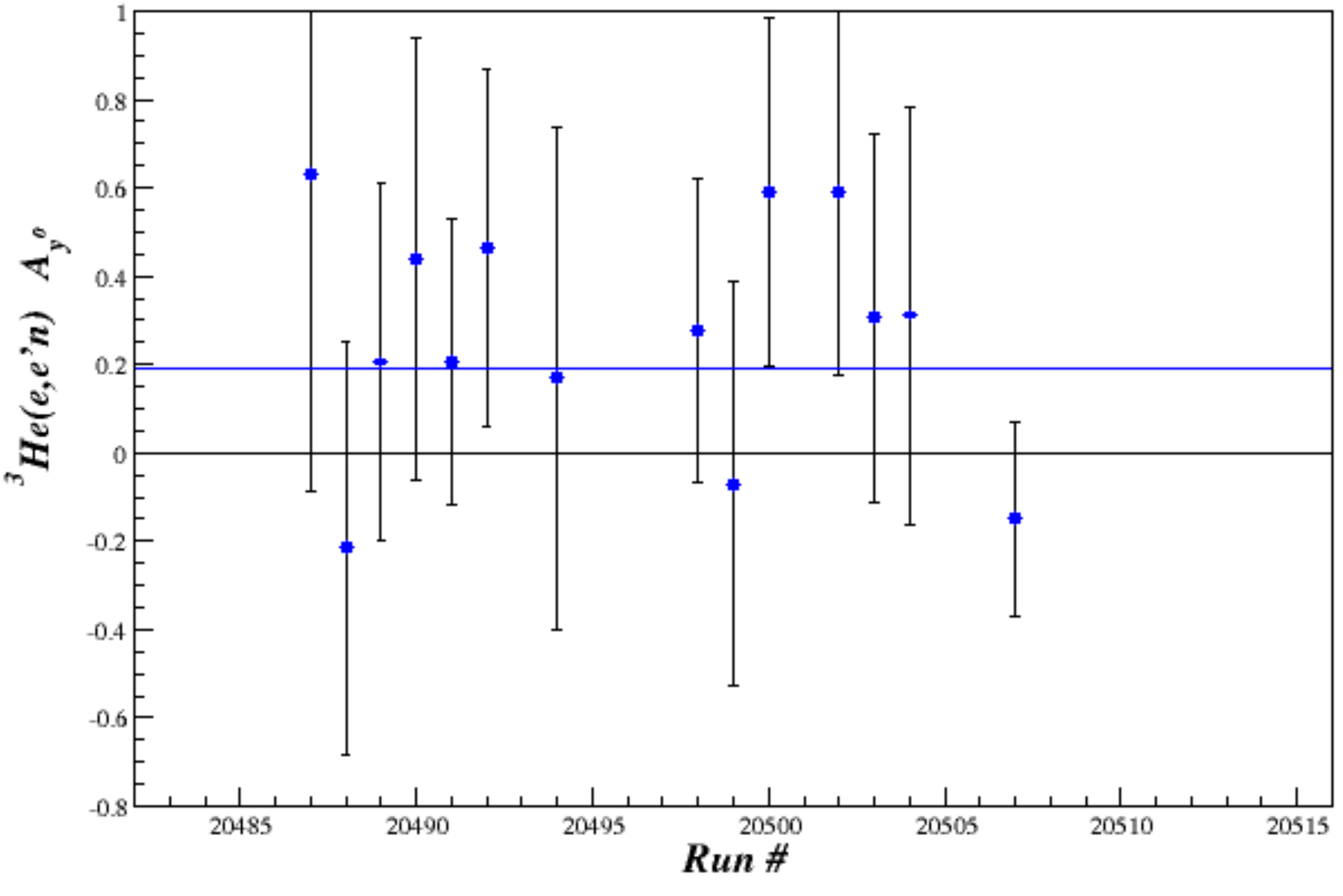,width=8cm} b)\epsfig{figure=new_plot_q2_05_ay0.eps,width=8cm} \\
c)\epsfig{figure=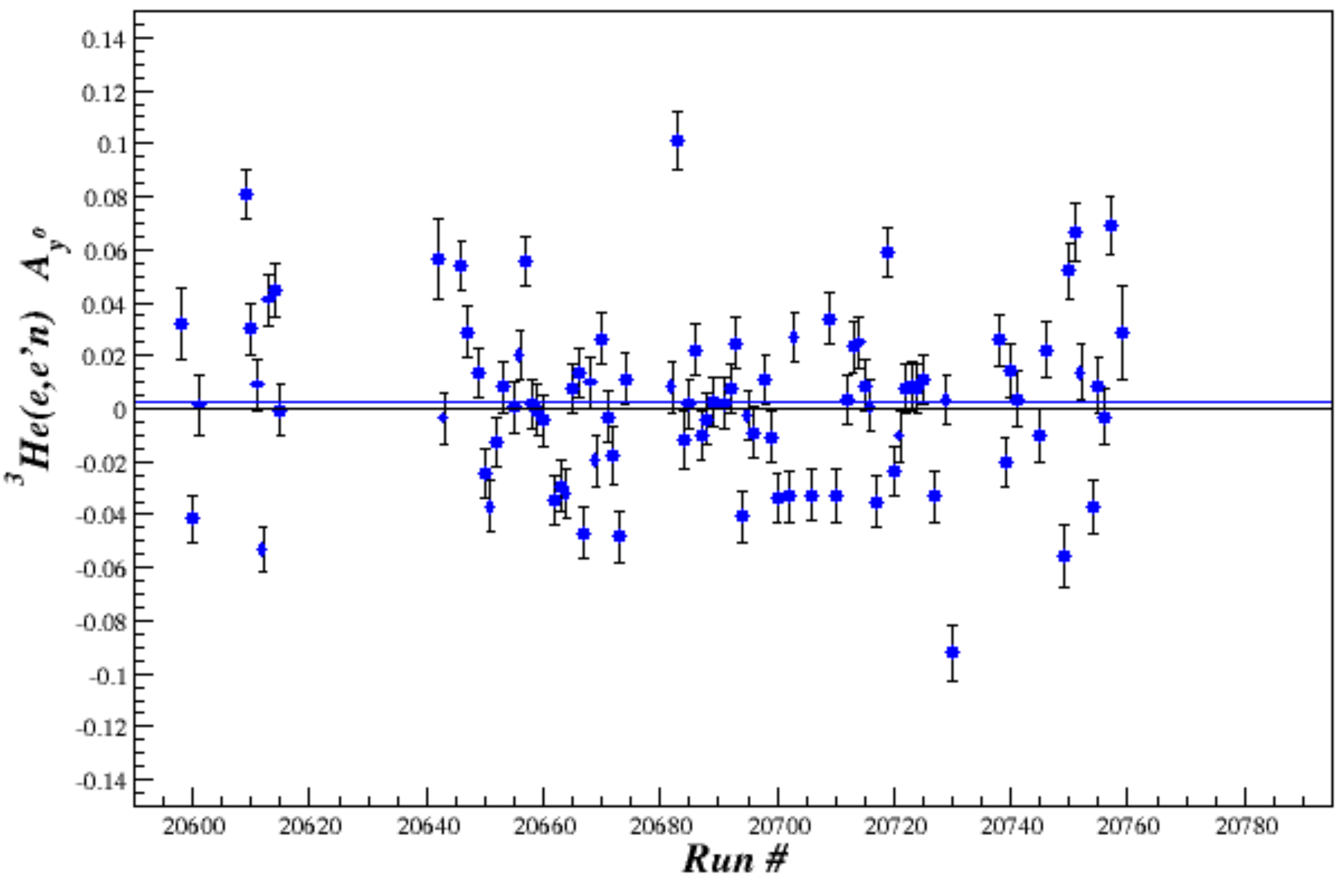,width=8cm}}
\caption{Run-by-run analysis of $A_y^0$ for $Q^2$ of a) 0.13 (GeV/$c)^2$, b) 0.46 (GeV/$c)^2$, and c) 0.95 (GeV/$c)^2$}
\label{een}
\end{figure}

However, these corrections did not account for the fluctuations seen when the neutron cuts were applied, shown in Fig. \ref{een}. The fluctuations were included as a systematic uncertainty of the neutron cuts using the $\chi^2$ scaling method described in the Review of Particle Physics. In this method, a scaling factor $S$ is defined by
\begin{equation}
	S=\left[ \chi^2/dof \right] ^{1/2}.
\end{equation}
This factor is then applied to the total uncertainty. The contribution due to only these fluctuations is then separated into a systematic contribution and included in the total uncertainty analysis.

A paper summarizing the final results has been prepared and is circulating the collaboration for comments. We expect to submit the final results to Physics Review Letters in early 2014.

%\end{document}

\clearpage

%\documentclass[12pt,a4paper,oneside]{article}
%\usepackage{graphicx}
%\usepackage{geometry}
%\usepackage{rotfloat}

%\begin{document}
%\setcounter{section}{1}
%\setcounter{subsection}{1}

\subsection{E08-007-II: G$^p_E$ at Low Q$^2$}
\begin{center}
M. Friedman, for the E08-007 collaboration
\end{center}

\subsubsection{Motivation}

The measurement of the proton form factor ratio at low $Q^2$ is important for several reasons. 
First, the form factors are fundamental properties
of the nucleon that should be measured well to test our understanding of the nucleon. 
Second, although theory generally indicates that the 
form factors vary smoothly with $Q^2$, there are an unsatisfyingly large number 
of theoretical calculations, fits, and data points that suggests this might not be the case, and
that there might be narrow structures in the form factors. Experiment E08-007-II is
good enough to either confirm or refute existing suggestions of few percent structures
in the form factors, or in the form factor ratio. Third, it has become apparent that the
existing uncertainties in the form factors are among the leading contributions to uncertainties 
in determining other physics quantities, such as the nucleon Zemach radius, the
strange form factors determined in parity violation, and the generalized parton distributions 
determined in deeply virtual Compton scattering. The improvement possible with this 
measurements is substantial. The proton electric and magnetic "radii" are also directly 
related to the form factor slope at $Q^2=0$:
\begin{equation}
\left<r^{2}_{E/M}\right>=-\frac{6}{G_{E/M}(0)}\left(\frac{dG_{E/M}(Q^2)}{dQ^2}\right)_{Q^2=0}.
\end{equation}
Recent results from muonic-hydrogen lamb shift measurements \cite{lamb} suggest a significantly smaller charge radius for the proton
than the established values, and precise measurement of the form factors at very low $Q^2$ may help to resolve this 
discrepancy. 

\subsubsection{The Experiment}

E08-007-II was run in parallel with E08-027. Details about the experiment can be found in the experiment section of E08-027.
All E08-007-II runs were taken with a magnetic field of 5T. The kinematics are listed in Table \ref{tab:results}. 

\subsubsection{Experimental Progress}
Details about optics, helicity and other calibrations are listed in the experimental progress section of E08-027. The target polarization
analysis is completed, with relative uncertainties of $\sim$2\%-3\%. These uncertainties are still under investigation. 

A preliminary extraction of the raw data has been completed, with the available calibrations, for the entire data set. Analysis of the data with the available 
optics is complete, and the physical asymmetries are listed in Table \ref{tab:results}.

The proton data in this experiment is diluted by elastic and quasi-elastic scattering off nitrogen (in the ammonia) and helium (cooling liquid).
While elastic cross section data for $^4$He is available, we have only a single experiment for $^{14}$N elastic cross section in the relevant $Q^2$ range. Neither 
$^4$He nor $^{14}$N has experimental quasi-elastic cross section data. In order to estimate our dilution we are working in parallel on three techniques: A. Using $^{12}$C
dilution runs and scaling the magnitude by the elastic peaks. B. Writing MC simulations, based on phenomenological cross section
models, and comparing to experimental data. C. Analyzing unpublished $^{14}$N quasi-elastic cross section data.

\begin{figure}
\centering
\includegraphics[width=100mm]{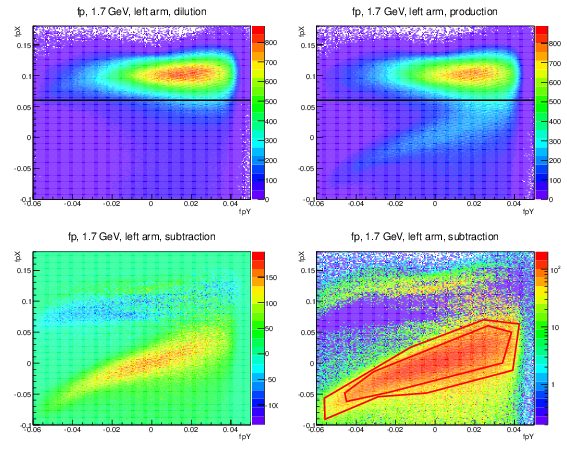}
\caption{\label{fig:dilution}Background subtraction procedure, 1.7 GeV left arm data. Top-left is the dilution run, and top-right is the production run. 
The black line represents the fpX limit used for normalization. The separation between hydrogen elastic peak and other events is not clear enough. 
The bottom-left is the subtraction after normalizing, and the bottom-right is the same in log scale. Two different cuts are show, and used for 
consistency purposes}
\end{figure}

\begin{figure}
\centering
\includegraphics[width=100mm]{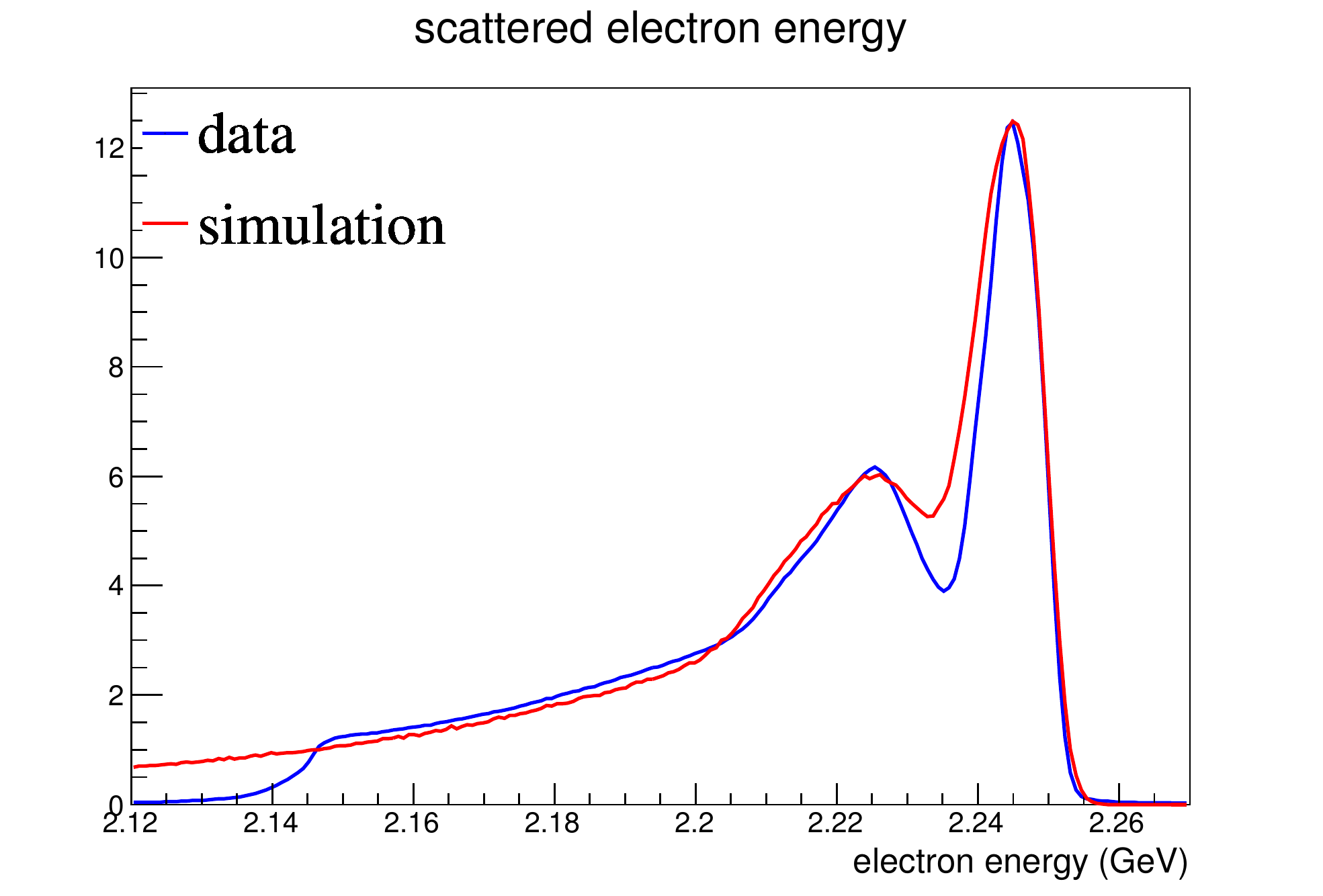}
\caption{\label{fig:simulations}Monte-Carlo simulation of production data at 2.2 GeV (left arm), compared to experimental data. 
Agreement is not yet good enough for dilution analysis.}
\end{figure}

At present, the dilution analysis is based on carbon data with results shown in Table \ref{tab:results}. An example extraction
is shown in Fig. \ref{fig:dilution}.
MC simulations are not yet sufficiently good for extracting experimental data. (See Fig. \ref{fig:simulations} for example). When optics calibrations are done,  
we will be able to improve the transport calculations from the target to the focal plane and hopefully improve our results. 

\begin{figure}
\centering
\includegraphics[width=100mm]{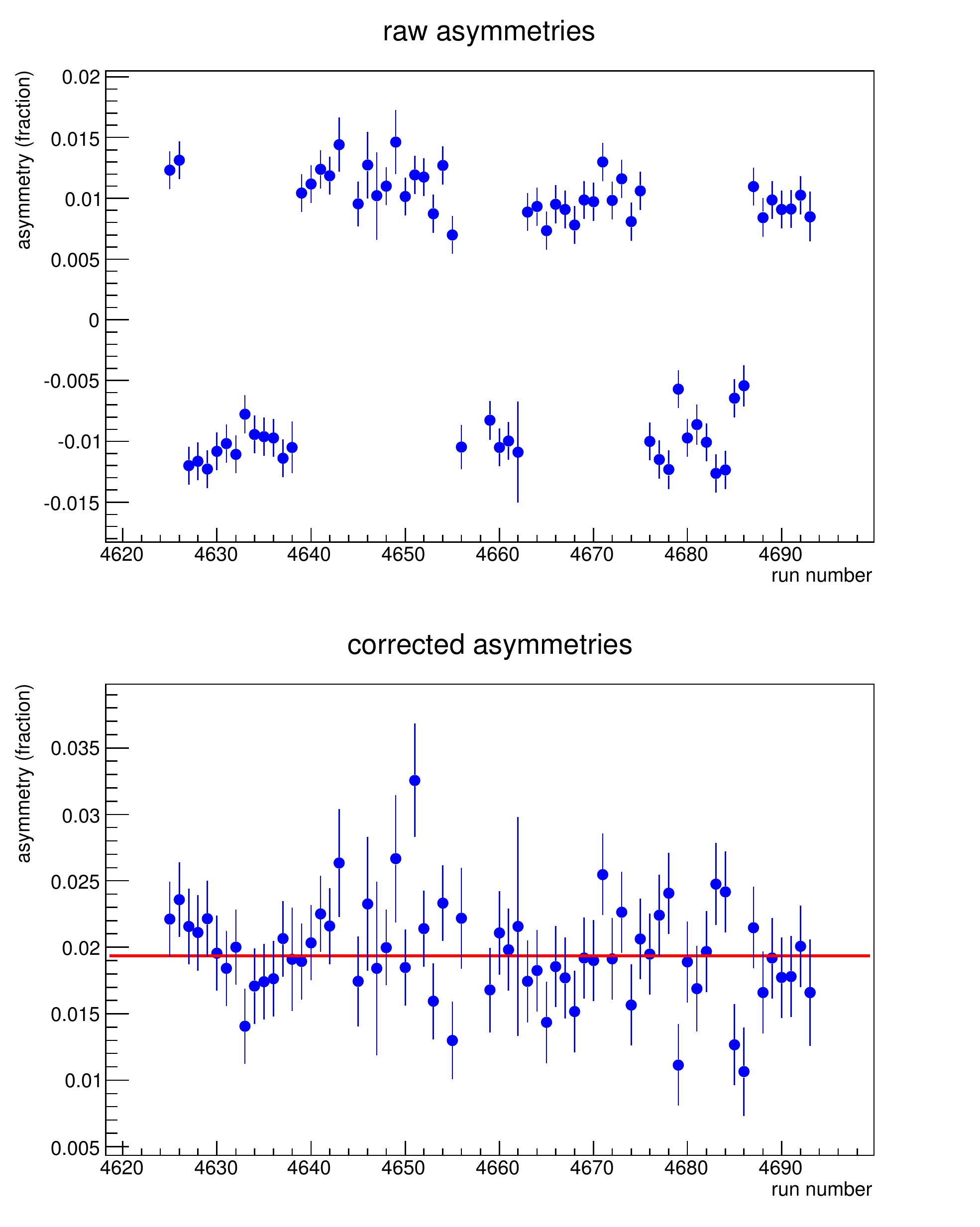}
\caption{\label{fig:asymmetries}An example for physical asymmetry extraction. This data is for the left arm, 1.7 GeV runs. 
The uncertainty is extracted by the fit, and the $\chi^2$ values are listed in Table \ref{tab:results}.}
\end{figure}

After background analysis is done, we are able to extract physical asymmetries for each kinematic setup and binning. Asymmetries are calculated for
each run separately, and the mean value and its uncertainty are extracted by fitting to a constant number. Fig. \ref{fig:asymmetries} 
shows an example of such extraction. Table \ref{tab:results} summarizes our preliminary results. These should be taken with caution, since calibrations
and dilution analysis are not yet final. 

\begin{sidewaystable}[H]
\centering
\begin{tabular}{|lccc|cccc|cccc|}
\hline
\multicolumn{4}{|c|}{bin      }                & \multicolumn{4}{|c|}{cut 1    }                          & \multicolumn{4}{|c|}{cut 2    }                          \\
\hline 
arm   & $E_e$ & $Q^2$ range  & $Q^2$ value     & dilution & A              &  $\Delta$A/A &  $\chi^2$/ndf & dilution & A               & $\Delta$A/A &  $\chi^2$/ndf \\ 
      & (GeV) & (GeV$^2$)    & (GeV$^2$)       &          & (\%)           &     (\%)     &               &          & (\%)            &     (\%)    &               \\
\hline
left  & 2.2   &  0.045-0.080 & 0.057$\pm$0.008 & 0.74     & 3.03$\pm$0.046 & 1.52         & 1.57          & 0.68     & 2.96$\pm$0.042  &  1.42        & 1.53 \\
right & 2.2   &  0.056-0.082 & 0.065$\pm$0.005 & 0.67     & 3.39$\pm$0.059 & 1.74         & 0.85          & 0.59     & 3.41$\pm$0.058  &  1.70        & 1.21 \\
left  & 2.2   &  0.028-0.050 & 0.037$\pm$0.006 & 0.75     & 1.56$\pm$0.021 & 1.35         & 1.20          & 0.70     & 1.48$\pm$0.021  &  1.42        & 1.34 \\
right & 2.2   &  0.038-0.064 & 0.047$\pm$0.006 & 0.71     & 1.93$\pm$0.029 & 1.50         & 1.60          & 0.66     & 1.74$\pm$0.029  &  1.67        & 1.39 \\
left  & 1.7   &  0.020-0.045 & 0.028$\pm$0.006 & 0.71     & 1.93$\pm$0.038 & 1.97         & 1.25          & 0.66     & 1.95$\pm$0.035  &  1.79        & 0.96 \\
right & 1.7   &  0.031-0.050 & 0.037$\pm$0.004 & 0.78     & 2.17$\pm$0.071 & 3.27         & 0.79          & 0.73     & 2.20$\pm$0.055  &  2.50        & 0.87 \\
left  & 1.7   &  0.017-0.027 & 0.020$\pm$0.003 & 0.54     & 1.24$\pm$0.071 & 5.87         & 0.90          & 0.48     & 1.18$\pm$0.066  &  5.59        & 0.82 \\
right & 1.7   &  0.023-0.033 & 0.027$\pm$0.003 & 0.67     & 1.68$\pm$0.056 & 3.33         & 1.23          & 0.64     & 1.53$\pm$0.047  &  3.07        & 1.38 \\
left  & 1.1   &  0.009-0.020 & 0.012$\pm$0.0027 & 0.26     & 1.78$\pm$0.060 & 3.37        & 0.79          & 0.23     & 1.72$\pm$0.052  &  3.02        & 0.80 \\
right & 1.1   &  0.010-0.022 & 0.014$\pm$0.0026 & 0.18     & 2.33$\pm$0.120 & 5.15        & 0.74          & 0.15     & 2.78$\pm$0.097  &  3.49        & 0.70 \\
\hline 
\end{tabular}
\caption{\label{tab:results}Preliminary asymmetries for the GEp experiment. All numbers should be taken with caution. See text for details.}
\end{sidewaystable}

%\end{document}
 % 
\clearpage

%\documentclass{article}
%\usepackage{epsfig}
%\setlength{\textwidth}{6.5in}
%\setlength{\oddsidemargin}{0in}
%\setlength{\evensidemargin}{0in}
%\setlength{\textheight}{9in}
%\setlength{\topmargin}{0in}
%\setlength{\headheight}{0in}
%\setlength{\headsep}{0in}

%\begin{document}

%\section{Summaries of Experimental Activities}

\subsection{E08-009: $^4He(e,e'p)^3H$ at $x_b=1.24$}
\label{sec:e08009}

%\begin{center}
%$^4He(e,e'p)^3H$ at $x_b=1.24$
%\end{center}

\begin{center}
A. Saha, D. Higinbotham, F. Benmokhtar, S. Gilad, and K. Aniol, spokespersons\\
and \\
Students: S. Iqbal(CSULA) and N. See(CSULA)\\
and\\
the Hall A Collaboration.\\
contributed by K. Aniol.
\end{center}

\subsubsection{Experimental Conditions}\label{sec:conditions}
The data were taken in collaboration with the SRC(E07-006) measurement during April 13 and April 14,
2011 for 16 hours of running. Our measurements provide the low missing momenta data taken at the
0.153 GeV/c and 0.353 GeV/c kinematic settings, which complement the high missing momenta data of the
SRC experiment. A 20 cm long cryogenic $^4He$ target at 20K and 10 atmosphere provided a thickness
of about $8\times 10^{22}/cm^2$. The electron beam energy was 4.454 GeV. This is the first measurement
of $^4He(e,e'p)X$ at this value of $x_b=1.24$. The M.S. thesis by Sophia Iqbal~\cite{sana} gives greater detail
on the data's analysis.

\subsubsection{Motivation}\label{sec:motivation}
A theoretical description of $^4He(e,e'p)X$ is critical for understanding nuclear structure. In particular, one
must be able to include many body forces in the theory. The reaction we measured actually includes
multiple exit channels, that is, X = $^3H$, $n+^2H$ and n+n+p nuclear and nucleonic channels. At the
beam energy used here meson production also contributes to X. Our first goal is to compare
the data for $^4He(e,e'p)^3H$ to theoretical calculations provided by the Madrid group~\cite{madrid}.   
The missing energy spectra also reveal a broad peak attributed~\cite{fatiha} to the absorption of the virtual photon
on a pair of nucleons.

\subsubsection{$E_{miss}$ Spectra}\label{sec:emiss}

\begin{figure}[hbt]
\center{\epsfig{figure=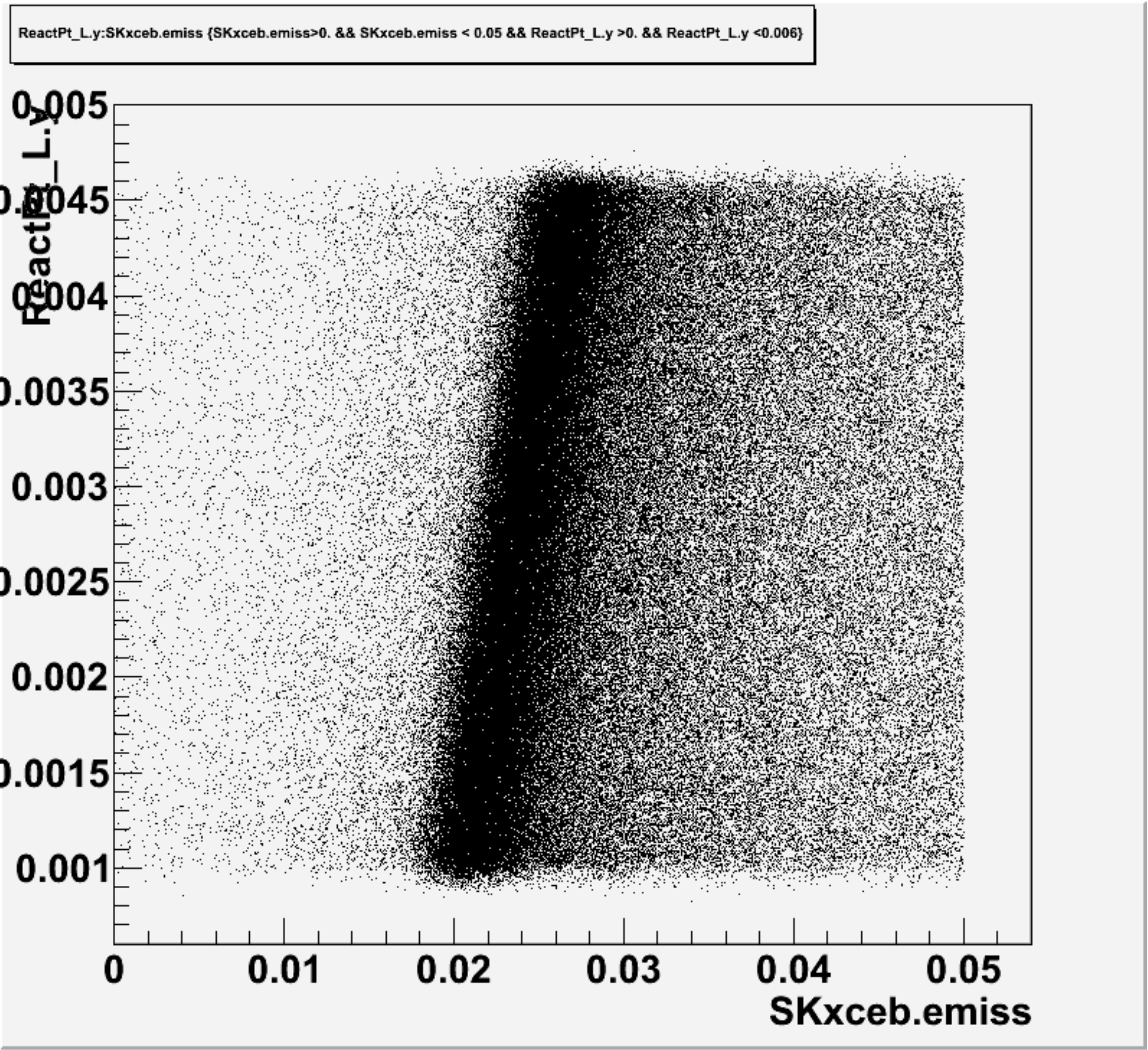,width=9.0cm}}
\caption[E08009: $E_{miss}$ defined by SKxceb.emiss vs $y_{tgt}$ for the $0.153 GeV/c$ kinematic setting..]{$E_{miss}$ defined by SKxceb.emiss vs $y_{tgt}$ for the $0.153 GeV/c$ kinematic setting.}
\label{fig:SKemiss}
\end{figure}

The $E_{miss}$ spectrum calculated from the Hall A analyzer variable SKxceb.emiss shows a strong
dependence on the \verb|ReactPt_L.y| value as seen in figure~\ref{fig:SKemiss}. We made a linear fit in the
root analysis script correcting for the slope using:\\

 TString \verb|ec_emiss|("SKxceb.emiss - 1.7255*(\verb|ReactPt_L.y| - 0.0006)");\\

\begin{figure}[hbt]
\center{\epsfig{figure=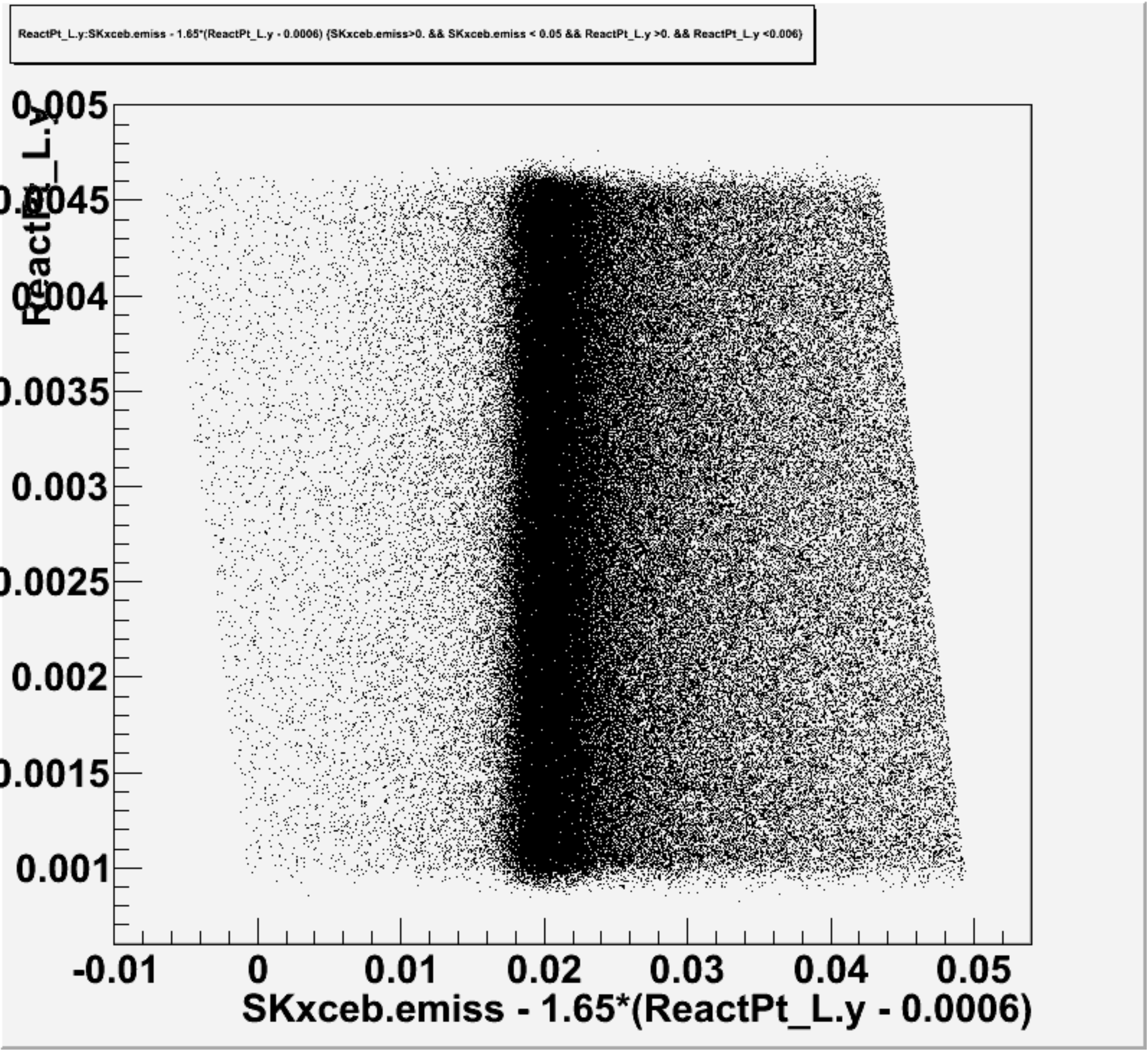,width=9.0cm}}
\caption[E08009: $E_{miss}$ corrected for the $y_{tgt}$ slope for the $0.153 GeV/c$ kinematic setting.]{$E_{miss}$ corrected for the $y_{tgt}$ slope for the $0.153 GeV/c$ kinematic setting.}
\label{fig:ecemiss}
\end{figure}

The two dimensional plot after applying this correction is shown in figure~\ref{fig:ecemiss}.

\begin{figure}[hbt]
\center{\epsfig{figure=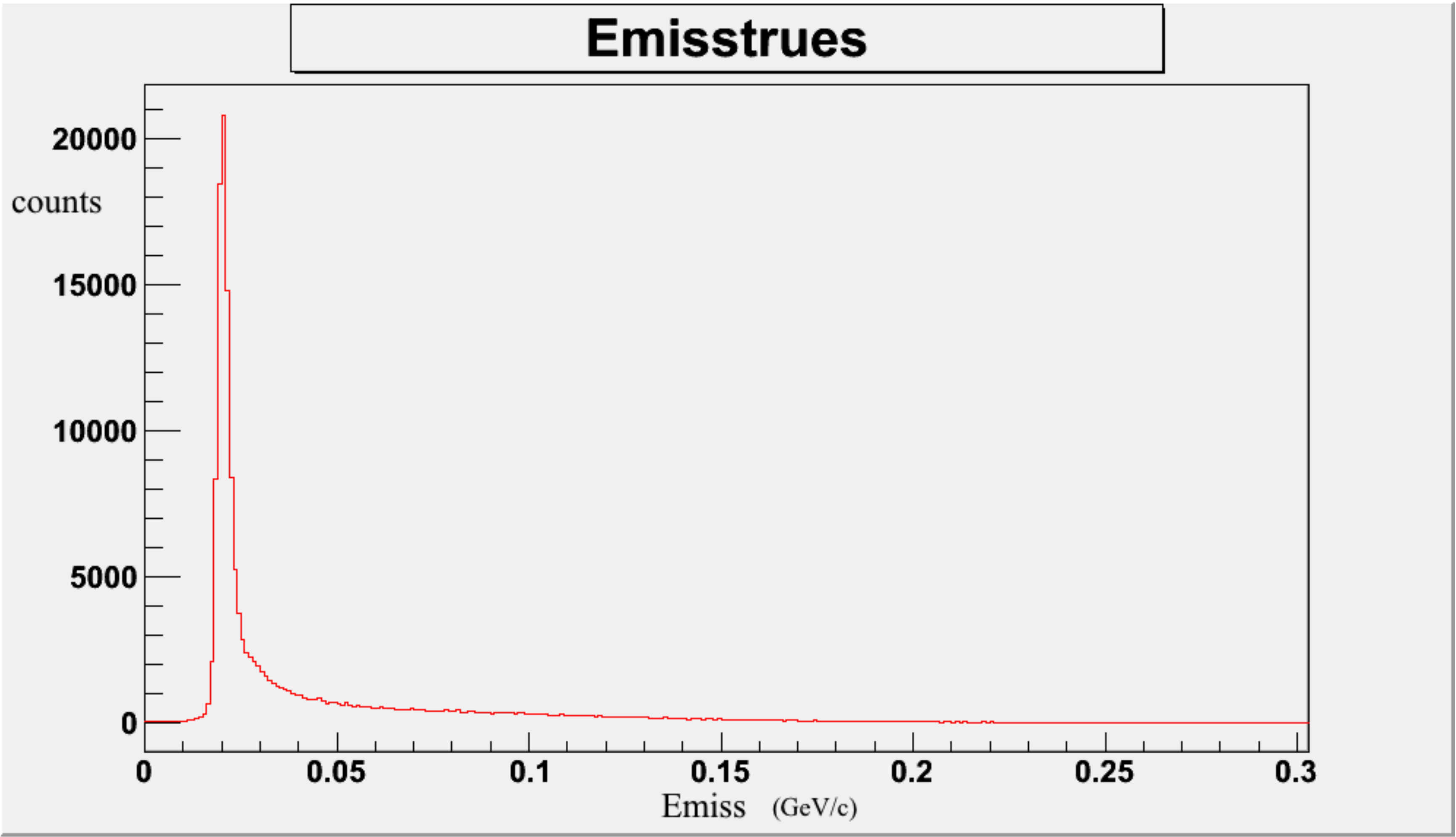,width=9.0cm}}
\caption[E08009: $E_{miss}$ for the $0.153 GeV/c$ kinematic setting.]{$E_{miss}$ for the $0.153 GeV/c$ kinematic setting.} 
\label{fig:fig1}
\end{figure}

\begin{figure}[hbt]
\center{\epsfig{figure=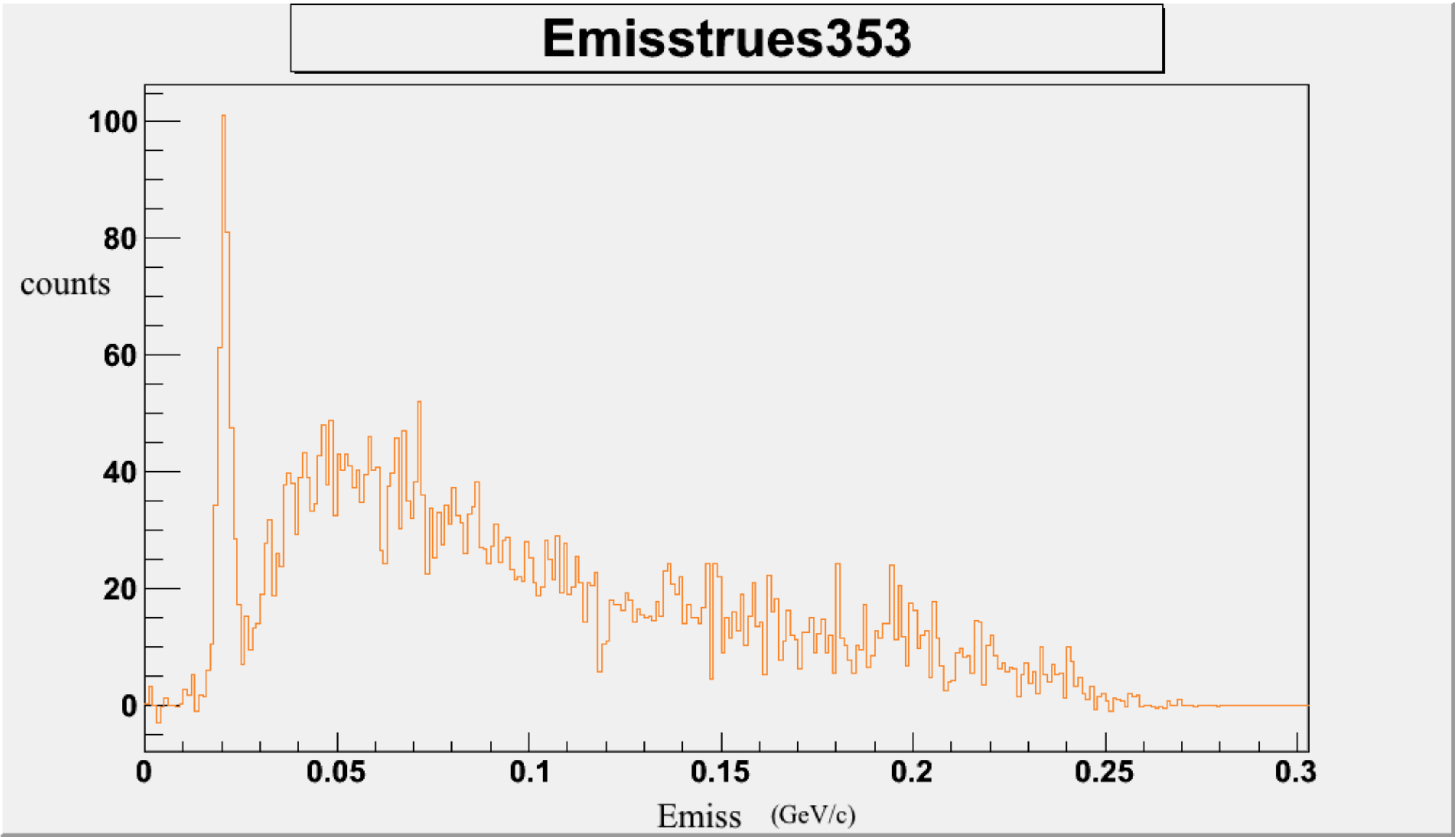,width=9.0cm}}
\caption[E08009: $E_{miss}$ for the $0.353 GeV/c$ kinematic setting.]{$E_{miss}$ for the $0.353 GeV/c$ kinematic setting.}
\label{fig:fig2}
\end{figure}

\begin{figure}[hbt]
\center{\epsfig{figure=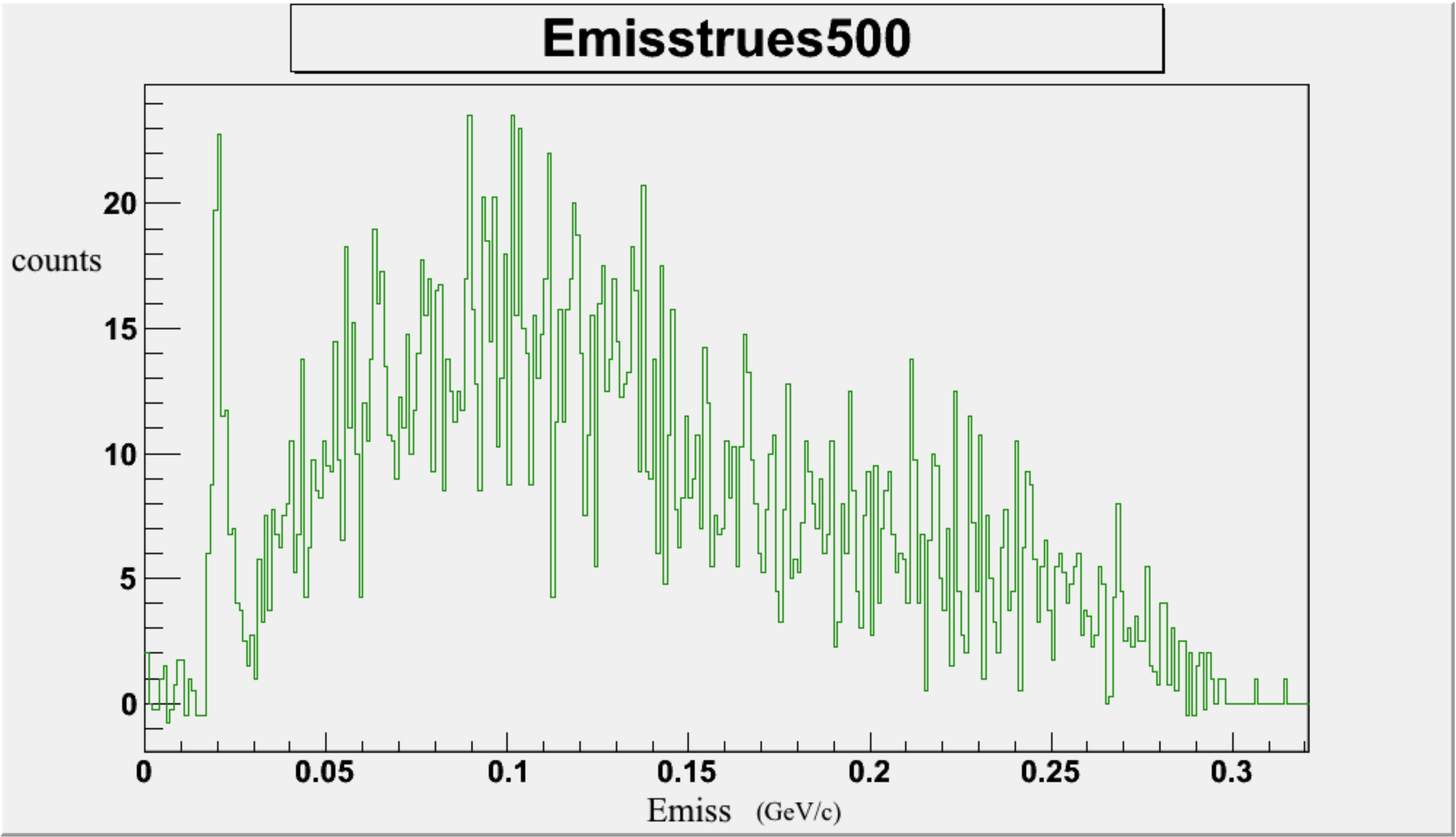,width=9.0cm}}
\caption[E08009: $E_{miss}$ for the $0.500 GeV/c$ kinematic setting.]{$E_{miss}$ for the $0.500 GeV/c$ kinematic setting.}
\label{fig:fig3}
\end{figure}

The $E_{miss}$ spectra with the $y_{tgt}$ corrections applied are shown in figures~\ref{fig:fig1},
~\ref{fig:fig2},~\ref{fig:fig3}.

\subsubsection{Simulations}\label{sec:simulations}
During the root replay of the GEANT simulations the electron momenta and proton momenta were subjected
to Gaussian broadening to match the FHWM of the strong triton peak seen in the $0.153 GeV/c$ data.

\begin{verbatim}
eexs = eexs*gRandom.Gaus(mean,sig1);
eeys=eeys*gRandom.Gaus(mean,sig1);
eezs=eezs*gRandom.Gaus(mean,sig1);

ppxs=ppxs*gRandom.Gaus(mean,sig1);
ppys=ppys*gRandom.Gaus(mean,sig1);
ppzs=ppzs*gRandom.Gaus(mean,sig1);
\end{verbatim}

\begin{figure}[hbt]
%\center{\epsfig{figure=emiss_1e-4_495e-4_sig.eps,width=9.0cm}}
\includegraphics[width=0.45\textwidth]{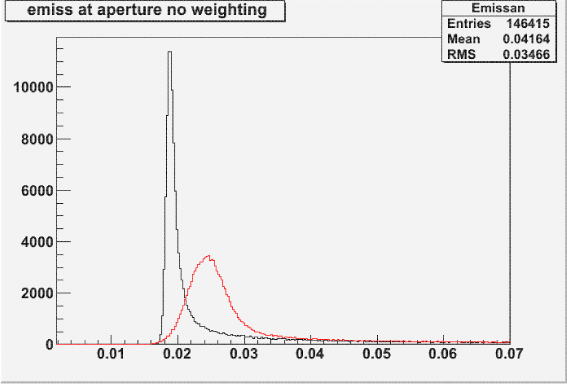}
\caption[E08009: GEANT simulations, Gaussian broadened, of the
 triton missing energy spectrum. The black curve is for sigma = $1.\times 10^{-4}$ and the red curve
is offset to match the data with sigma = $6.53 \times 10^{-4}$.]{GEANT simulations, Gaussian broadened, of the
 triton missing energy spectrum. The black curve is for sigma = $1.\times 10^{-4}$ and the red curve
is offset to match the data with sigma = $6.53 \times 10^{-4}$.}
\label{fig:broad}
\end{figure}

\begin{figure}[hbt]
\center{\epsfig{figure=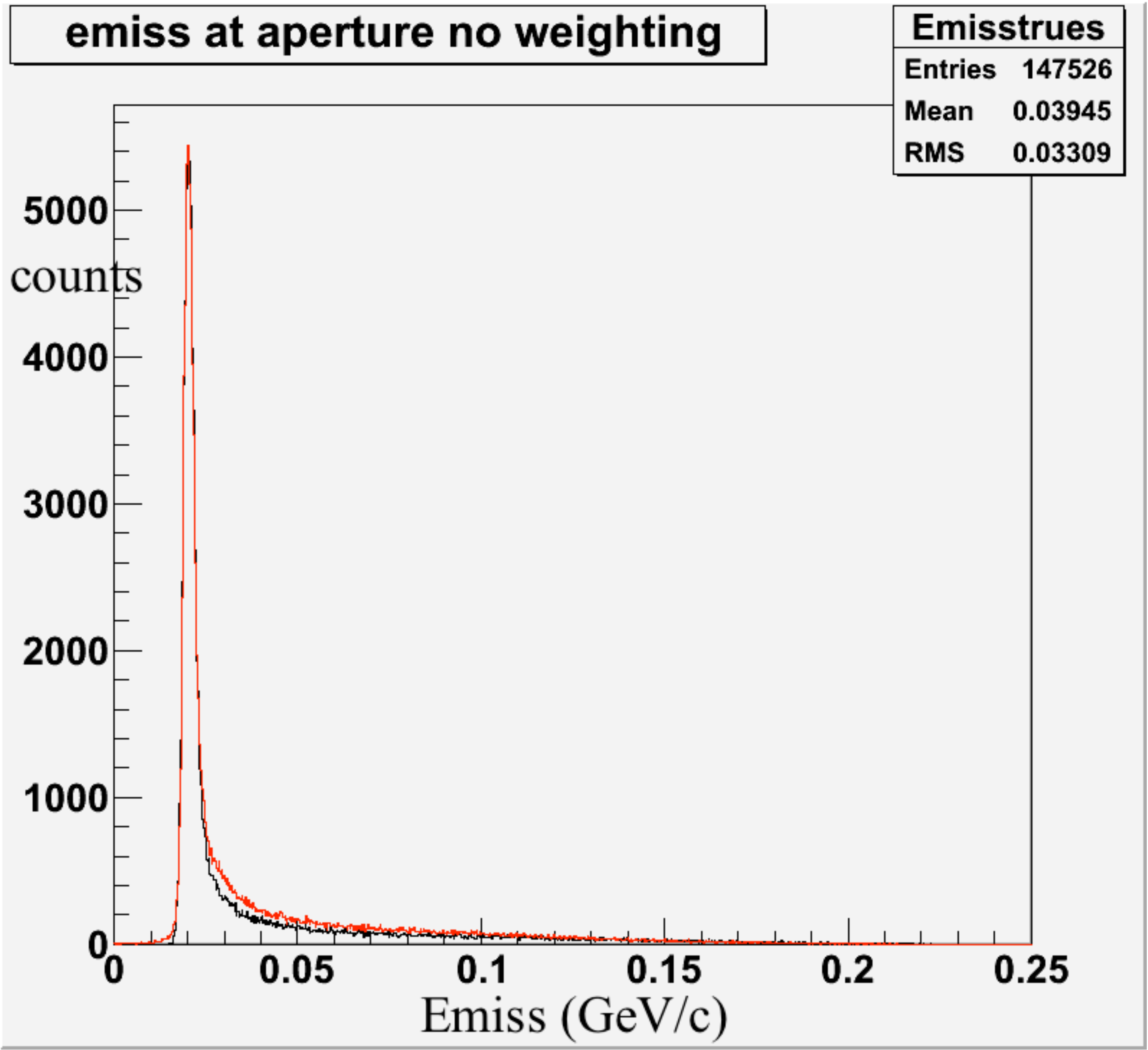,width=9.0cm}}
\caption[E08009: GEANT simulation(black curve) and data(red curve) for
the $0.153 GeV/c$ kinematic setting.]{GEANT simulation(black curve) and data(red curve) for
the $0.153 GeV/c$ kinematic setting.}
\label{fig:153sim}
\end{figure}

\begin{figure}[hbt]
\center{\epsfig{figure=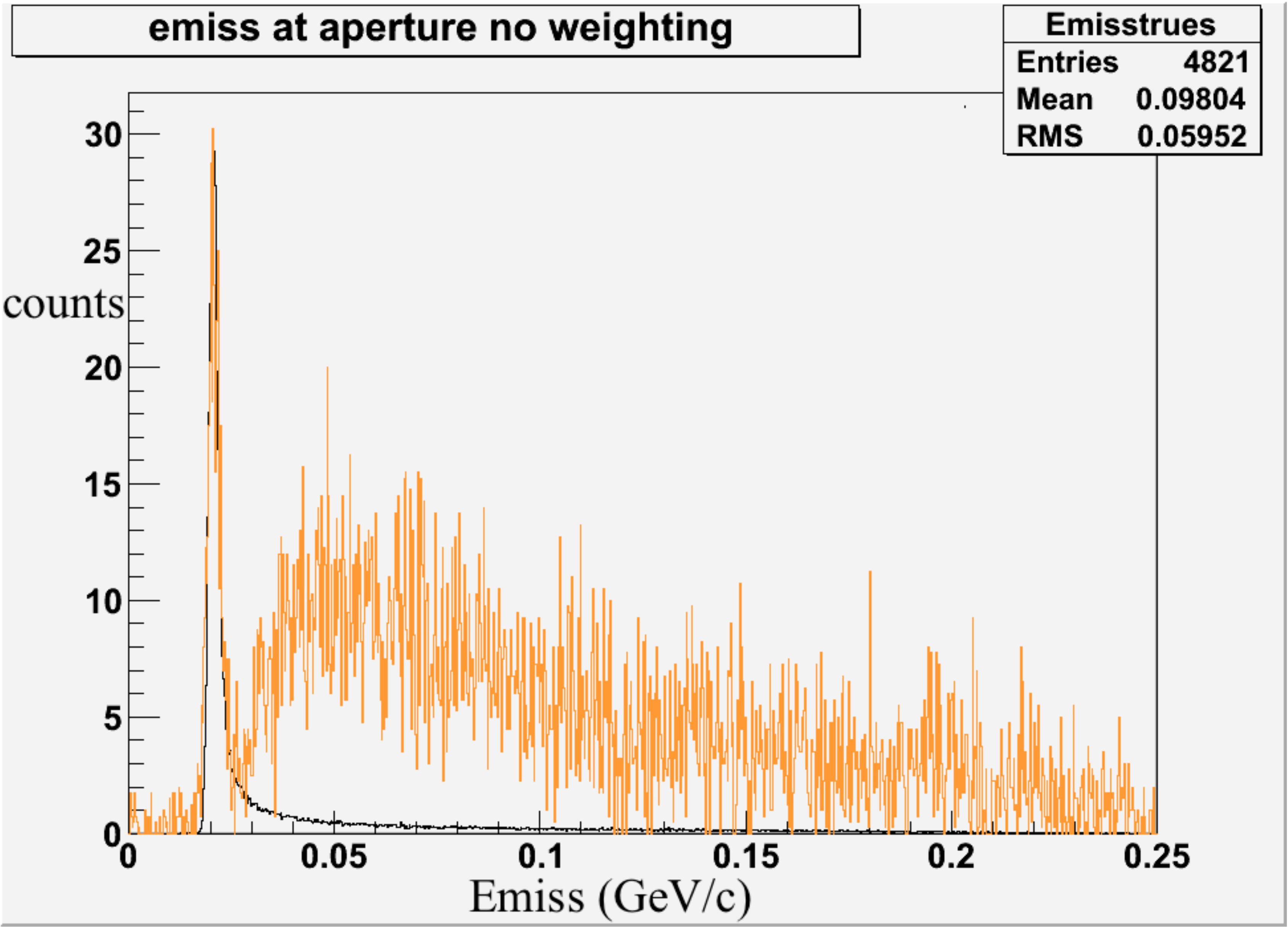,width=9.0cm}}
\caption[E08009: GEANT simulation(black curve) and data(orange curve) for
the $0.353 GeV/c$ kinematic setting.]{GEANT simulation(black curve) and data(orange curve) for
the $0.353 GeV/c$ kinematic setting.}
\label{fig:353sim}
\end{figure}

A comparison between the simulated $E_{miss}$ triton spectrum using sig1 = $1.\times 10^{-4}$ and
sig1 = $6.53 \times 10^{-4}$ is shown in figure~\ref{fig:broad}. 
The data and the simulated triton peaks are shown in figures~\ref{fig:153sim},~\ref{fig:353sim}.

The wide momentum acceptances of the HRSs' allows for a broad missing
momentum acceptance. In the simulation each point within the 
spectrometers' apertures has an equal probability of being a target
for a vertex electron or proton. The 3 body kinematical and geometrical
limitations are correctly determined by GEANT. We thus define the
missing momentum acceptance factor, $f(p_m)$, for a bin of missing
momentum centered around $p_m$ as

\begin{equation}
f(p_m)=\frac{n(p_m)}{\sum n(p_m)}.
\label{eq:fpm}
\end{equation}

\begin{figure}[hbt]
\center{\epsfig{figure=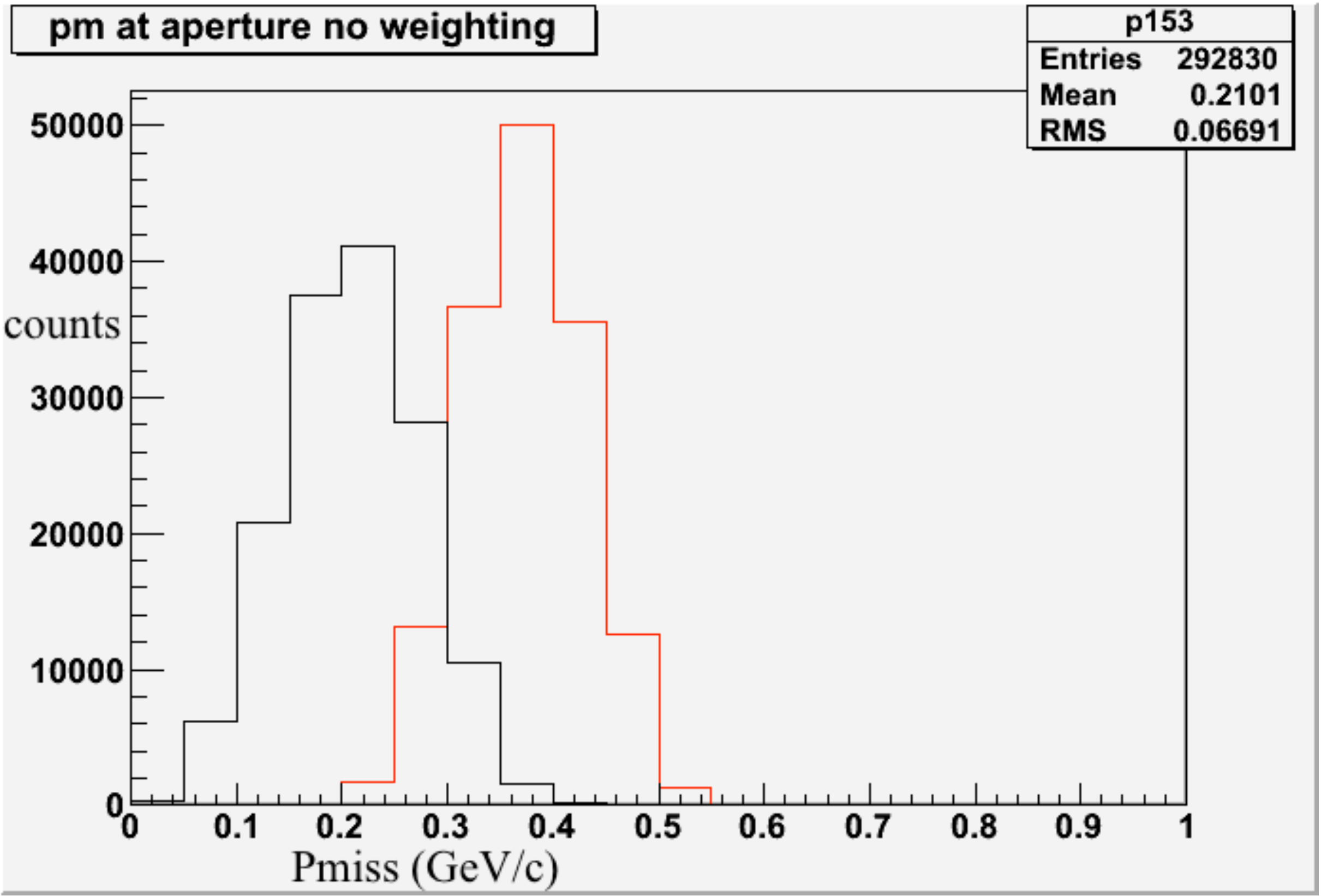,width=9.0cm}}
\caption[E08009: GEANT missing momentum spectra for
$0.153GeV/c$(black) and $0.353GeV/c$(red) as determined at the spectrometer
apertures.]{GEANT missing momentum spectra for
$0.153GeV/c$(black) and $0.353GeV/c$(red) as determined at the spectrometer
apertures.}
\label{fig:153-353-pm}
\end{figure}

Where $n(p_m)$ is the number of triton events in the missing momentum
bin centered on $p_m$ and $\sum n(p_m)$ is the total number of triton
events over all missing momenta for the particular kinematic setting.
This factor, $f(p_m)$ is used in the cross section determinations
for each missing momentum bin. An example of the simulation prediction
for the missing momenta spectra is shown in figure~\ref{fig:153-353-pm}.\\

The simulation also allows us to determine the fraction of triton
events which fall outside of the window we place around the triton
peak in the data. This loss of tritons is due to radiative processes
and multiple scattering. A more extended discussion of the simulation
is in reference~\cite{simulation}.

\subsubsection{SRC target behavior}\label{sec:srctgt}

The SRC target presented challenges for cross section measurements. We are grateful to
Patricia Solvignon and Zhihong Ye for sharing their results about their \verb|ztgt| analysis
of this target and especially to Silviu Covrig for providing CFD calculations~\cite{silviu}.

\begin{figure}[hbt]
\center{\epsfig{figure=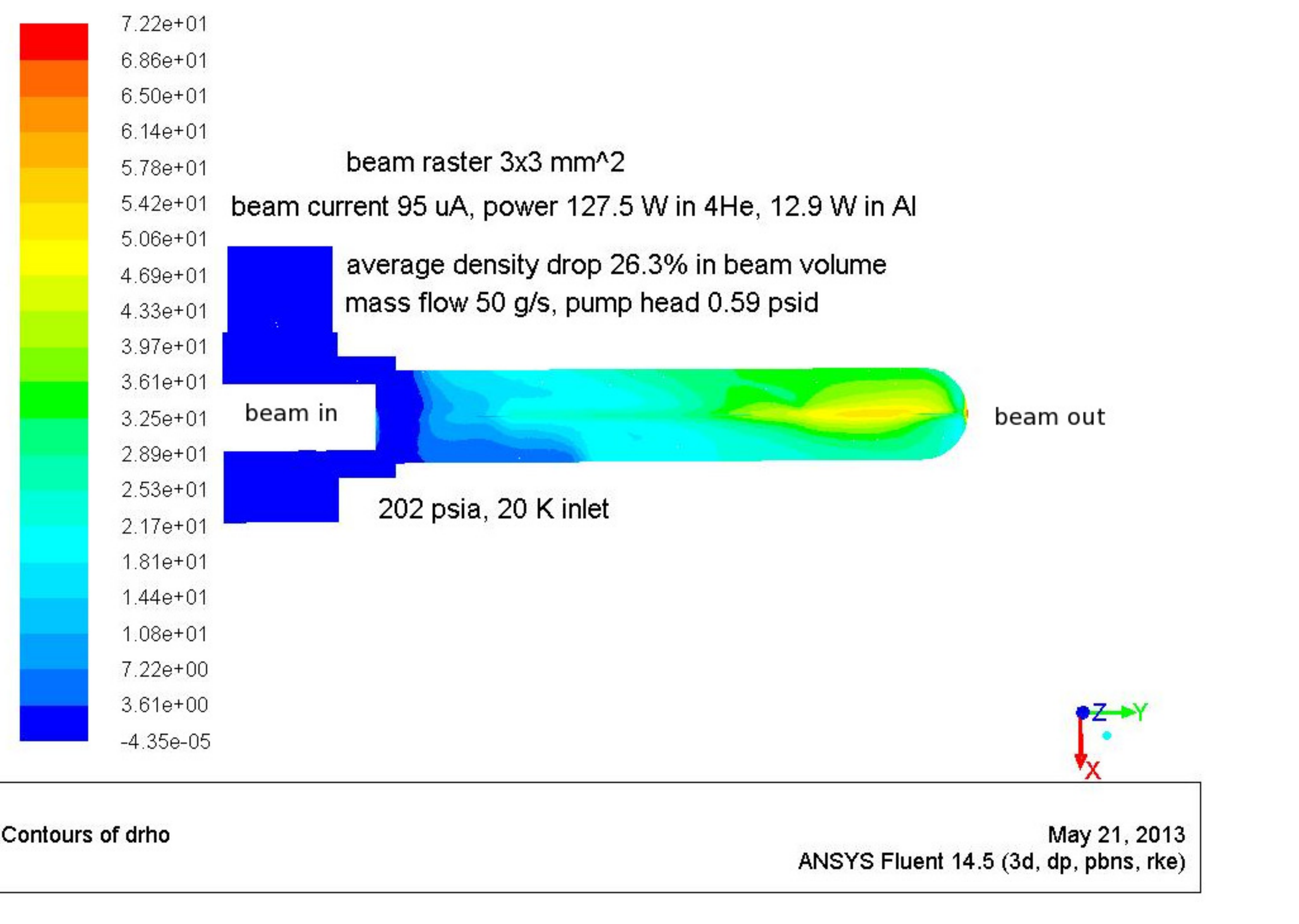,width=9.0cm}}
\caption[E08009: Computational Fluid Dynamic calculation~\cite{silviu} 
for the SRC target. drho is the loss in percent of density from the input fluid.]{Computational Fluid Dynamic calculation~\cite{silviu} 
for the SRC target. drho is the loss in percent of density from the input fluid.}
\label{fig:cfd}
\end{figure}

The SRC target container is an aluminum can of length 20cm. Cryogenic $^4He$ enters and
exits at the upstream end of the target. There is no outlet for the fluid at the
downstream end of the can. A diagram of the target from a CFD calculation for
a 95$\mu A$ electron beam is shown in figure~\ref{fig:cfd}.

\begin{figure}[hbt]
\center{\epsfig{figure=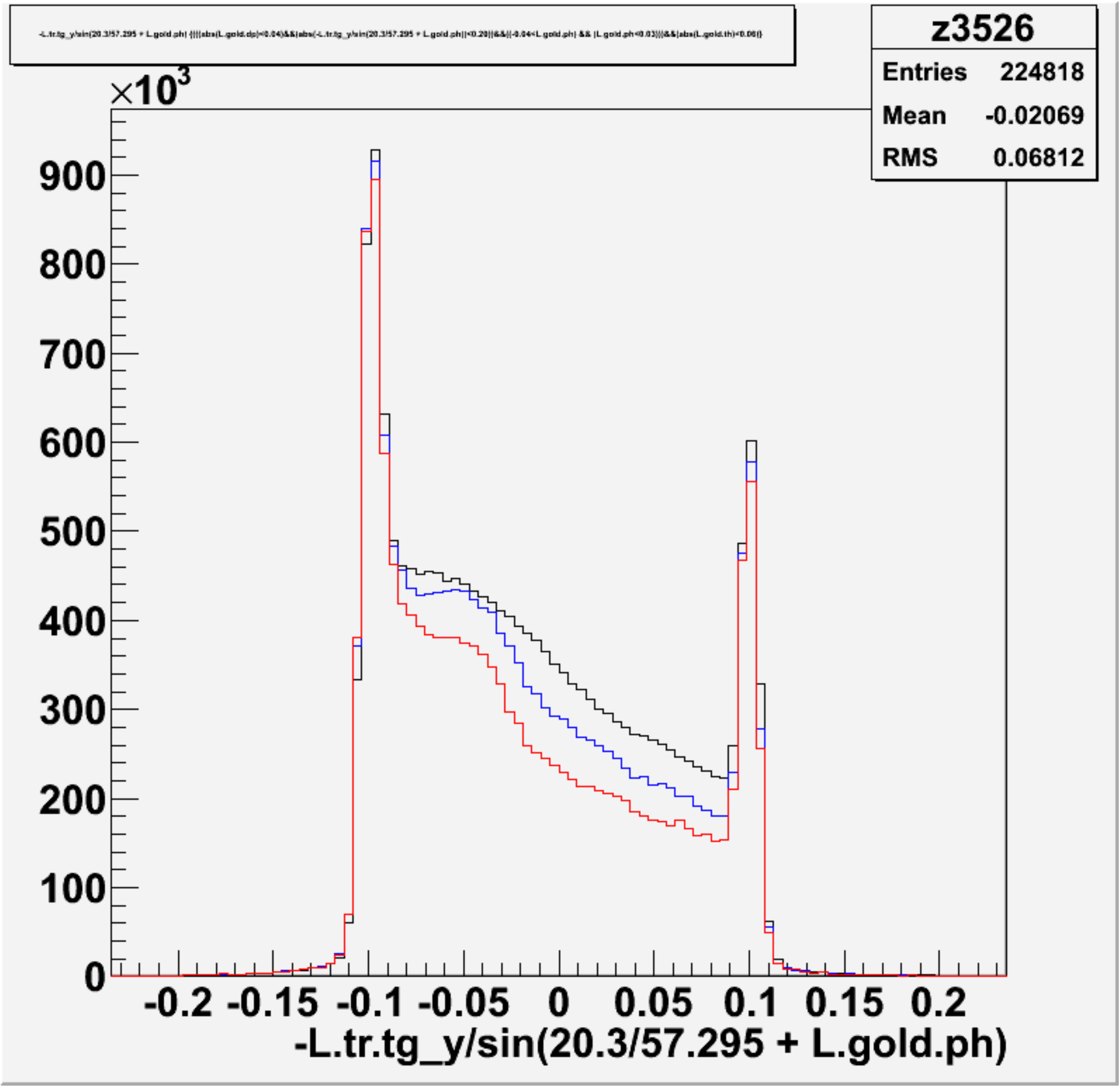,width=9.0cm}}
\caption[E08009: Normalized counts per Coulomb(vertical
 axis) along the beam's path for 4 different beam currents.]{Normalized counts per Coulomb(vertical
 axis) along the beam's path for 4 different beam currents. $4\mu A$(black),
$47\mu A$(blue), $60\mu A$(red). The horizontal axis is along z in meters.
The aluminum end caps are seen as sharp spikes at $\pm0.1m$.}
\label{fig:ztgtL}
\end{figure}

The issue with the SRC target, which inspired extensive  study, was the non monotonic
decrease in event rate as a function of distance between the beam in and 
beam out points along the z axis seen in figure~\ref{fig:ztgtL}. The first concern
was that the beam was hitting some structures besides the $^4He$ target. However,
there were no other indications of a misdirected beam or beam halo hitting
anything but the cryogenic target. The shape of the ztgtL distribution was
very stable. The beam currents used for the short runs of E08009, $60\mu A$, $47\mu A$,
 and the long SRC runs at $4\mu A$ show no significant difference in
ztgtL shape except for the decrease in count rate.\\

Although the CFD calculations do not match exactly with the ztgtL distributions
seen in the experiment, they do show that the ztgtL distribution will not
be a smooth function along the beam axis. Perhaps with sufficient tuning
of the parameters or models used in the CFD calculations a match can be
made with experiment. The SRC target is a challenge computationally. We
were able to extract the target density, $\rho(z,I)$ from a combination
of the measure ztgtL distributions and the $4\mu A$ predictions from CFD.
More details of the SRC target analysis are found in references~\cite{sana} and
\cite{srctgt}.

\subsubsection{Cross Section Results and Theory}\label{sec:xsect}

\begin{figure}[hbt]
\center{\epsfig{figure=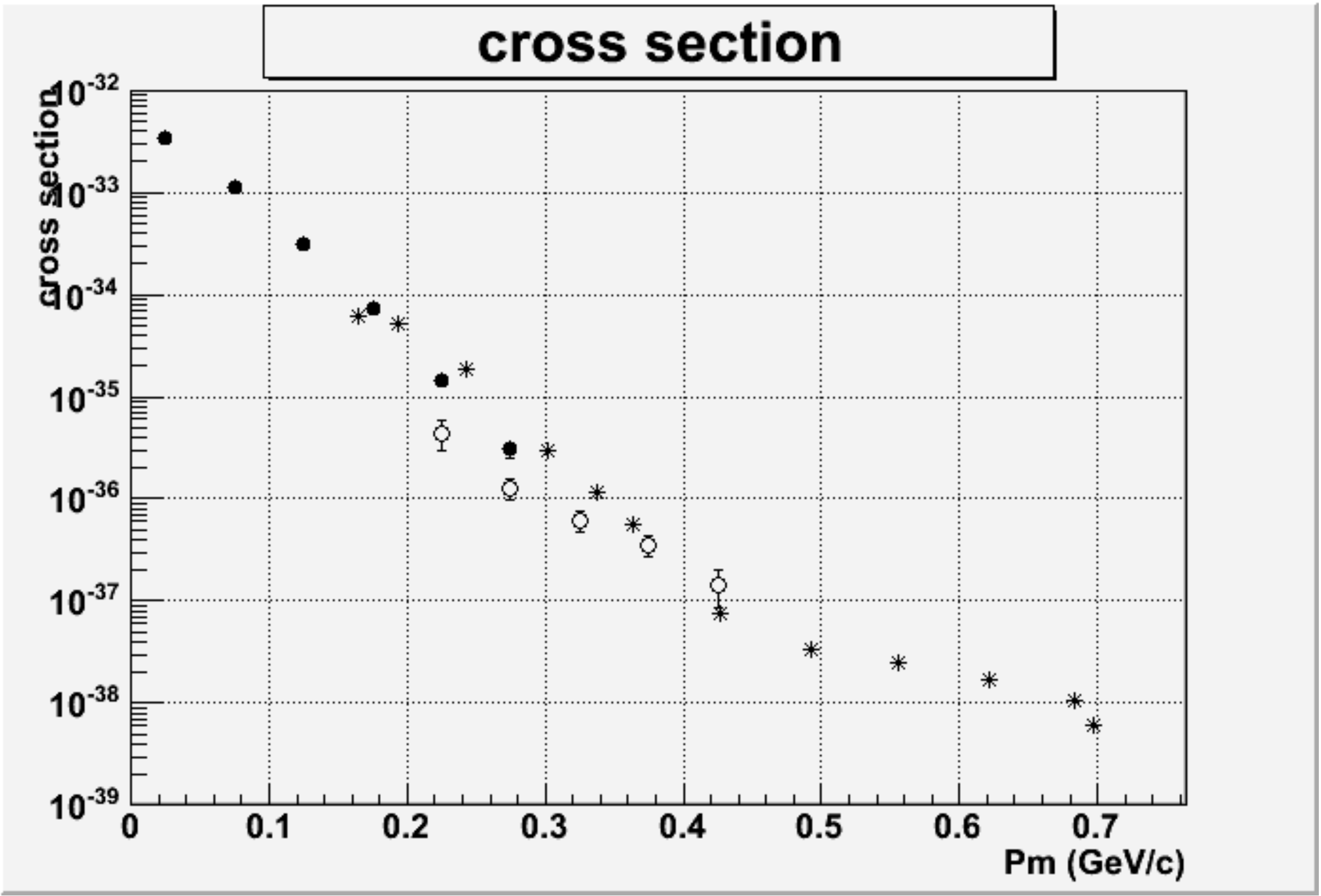,width=9.0cm}}
\caption[E08009: Cross sections for $0.153 GeV/c$(closed dots), 
$0.353GeV/c$(open dots) and Madrid Theory(asterisks)~\cite{madrid}.]{Cross sections for $0.153 GeV/c$(closed dots), 
$0.353GeV/c$(open dots) and Madrid Theory(asterisks)~\cite{madrid}. The units are
$\frac{cm^2}{MeV*sR^2}$. The first two theory values are underestimates
because the theory values we have are only valid above $0.150 GeV/c$, but the
simulations and data continue below $0.150GeV/c$.}
\label{fig:xsect}
\end{figure}

Our data and theory for the cross section for $^4He(e,e'p)^3H$, 
$\frac{d\sigma}{dE d\Omega_e d\Omega_p}$, is shown in figure~\ref{fig:xsect}.
A comparison between theory and data for proton angles between 35 degrees and
47 degrees yields

\begin{equation}
\frac{\Sigma \sigma_{data}(\theta_p) sin(\theta_p) d\theta_p}{\Sigma \sigma_{theory}(\theta_p) sin(\theta_p) d\theta_p} \le 0.68.
\label{eq:ratio}
\end{equation}

Tables containing the cross sections for data and theory are found in the thesis~\cite{sana}.

%
% Here's how to do the references.  We will be using the APS style.
%

%\end{document}

\clearpage

%\documentclass{article}
%\usepackage{epsfig}
%\setlength{\textwidth}{6.5in}
%\setlength{\oddsidemargin}{0in}
%\setlength{\evensidemargin}{0in}
%\setlength{\textheight}{9in}
%\setlength{\topmargin}{0in}
%\setlength{\headheight}{0in}
%\setlength{\headsep}{0in}

%\begin{document}

%\section{Summaries of Experimental Activities}

\subsection[E08-010: Coulomb quadrupole amplitude in N $\rightarrow\Delta$ at
low Q$^2$]{E08-010: Measurement of the Coulomb quadrupole amplitude at the $\gamma^*p\rightarrow \Delta(1232)$
in the low momentum transfer region}
\label{sec:e08010}

%\begin{center}
%{\bf Measurement of the Coulomb quadrupole amplitude at the $\gamma^*p\rightarrow \Delta(1232)$\\
%in the low momentum transfer region}
%\end{center}

\begin{center}
S.~Gilad, D.~W.~Higinbotham, A.~Sarty and N.~F.~Sparveris, spokespersons, \\
\vskip 0.3cm
Graduate students: D.~Anez (St. Mary's), A.~Blomberg (Temple) \\
\vskip 0.3cm
and the Hall A Collaboration.\\
\vskip 0.3cm
contributed by N.F.~Sparveris
\end{center}

\subsubsection{Introduction}\label{sec:introduction}

Hadrons are composite systems with complex quark-gluon and meson
cloud dynamics that  give rise to non-spherical components in their
wave-function which in a classical limit and at large wavelengths
will correspond to a "deformation". In recent years an extensive
experimental and theoretical effort has been focused on identifying
and understanding the origin of possible non-spherical amplitudes in
the nucleon wave-function
\cite{Ru75,is82,pho2,pho1,frol,pos01,merve,bart,Buuren,spaprl,kelly,spamami,stave,elsner,joo1,St08,ungaro,villano,aznau,spaepja,dina,sato,dmt00,kama,mai00,multi,said,pv,rev3}.
The spectroscopic quadrupole moment provides the most reliable and
interpretable measurement of these components; for the proton, the
only stable hadron, it vanishes identically because of its spin 1/2
nature. As a result, the presence of resonant quadrupole amplitudes
in the $N\rightarrow \Delta$ transition has emerged as the
definitive experimental signature of non spherical amplitudes.
Spin-parity selection rules in the $\gamma^* N\rightarrow \Delta$
transition allow only magnetic dipole (M1) and electric quadrupole
(E2) or Coulomb quadrupole (C2) photon absorption multipoles (or the
corresponding pion production multipoles $M^{3/2}_{1+},
E^{3/2}_{1+}$ and $S^{3/2}_{1+}$ ($L^{3/2}_{1+}$) respectively) to
contribute. The ratios CMR $= Re(S^{3/2}_{1+}/M^{3/2}_{1+})$ and EMR
$= Re(E^{3/2}_{1+}/M^{3/2}_{1+})$ are routinely used to present the
relative magnitude of the amplitudes of interest. Non-vanishing
resonant quadrupole amplitudes will signify the presence of
non-spherical components in either the proton or in the
$\Delta^{+}(1232)$, or more likely at both; moreover, their $Q^2$
evolution is expected to provide insight into the mechanism that
generate them.

In the constituent-quark picture of hadrons, the non-spherical
amplitudes are a consequence of  the non-central, color-hyperfine
interaction among quarks \cite{Gl79}. However, it has been shown
that this mechanism only provides a small fraction of the observed
quadrupole signal at low momentum transfers, with the magnitudes of
this effect for the predicted E2 and C2 amplitudes \cite{is82} being
at least an  order of magnitude too small to explain the
experimental results and with the dominant M1 matrix element being
$\simeq$ 30\% low \cite{is82}. A likely cause of these dynamical
shortcomings is that the quark model does not respect chiral
symmetry, whose spontaneous breaking leads to strong emission of
virtual pions (Nambu-Goldstone bosons) \cite{rev3}. These couple to
nucleons as $\vec \sigma \cdot \vec p$, where $\vec \sigma$ is the
nucleon spin, and $\vec p$ is the pion momentum. The coupling is
strong in the p-wave and mixes in nonzero angular-momentum
components. Based on this, it is physically reasonable to expect
that the pionic contributions increase the M1 and dominate the E2
and C2 transition matrix elements in the low-$Q^2$ (large distance)
domain. This was first indicated by adding pionic effects to quark
models \cite{Lu97}, subsequently shown in pion cloud model
calculations \cite{sato,kama}, and recently demonstrated in chiral
effective field theory calculations \cite{pv,Ga06}. Our current
understanding of the nucleon suggests that at low-$Q^2$ (large
distance) the pionic cloud effect dominates while at high-$Q^2$
(short distance) intra-quark forces dominate.

\begin{figure}[t]
\begin{center}
\epsfig{file=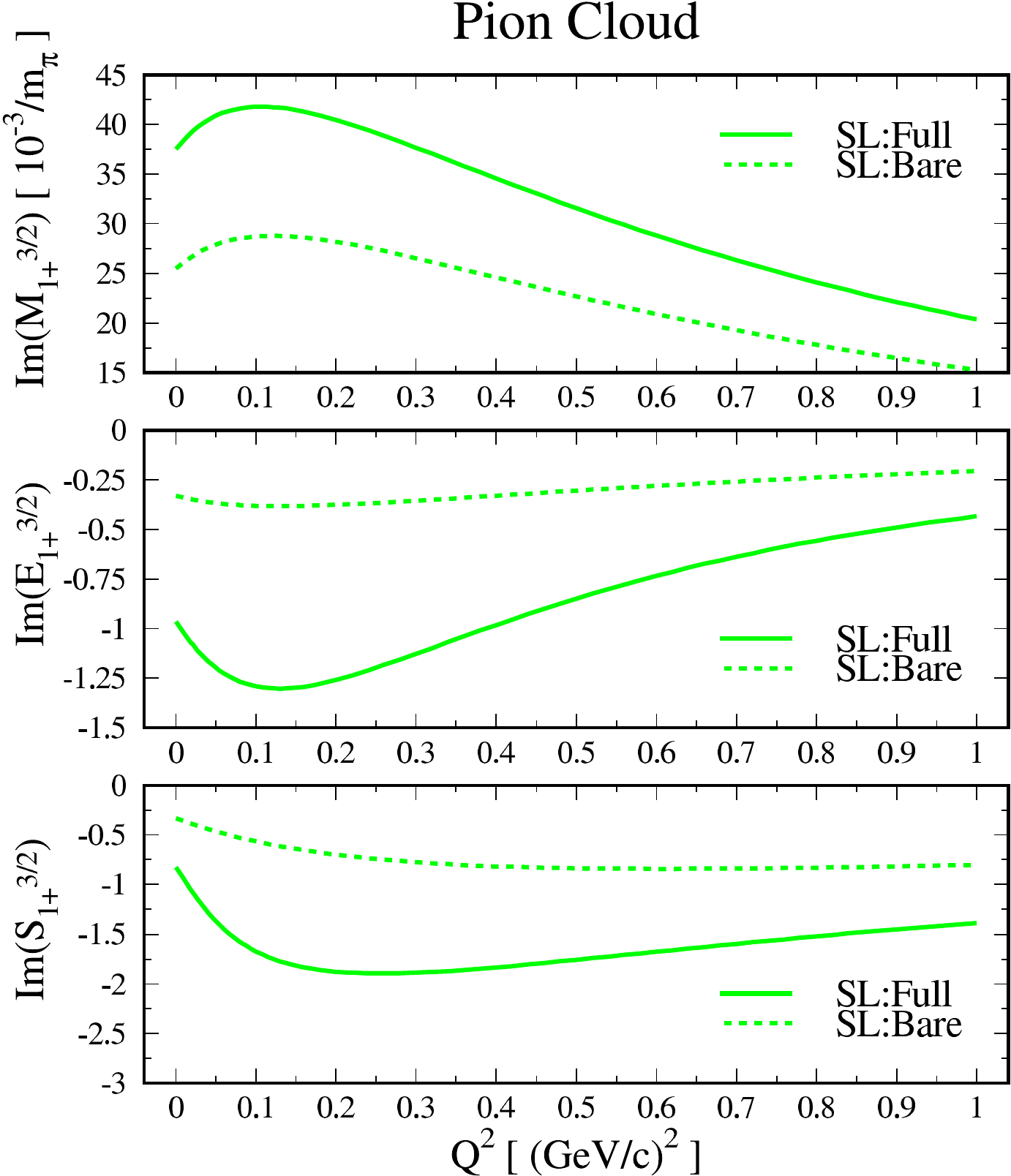,width=7.0cm}
\end{center}
\caption{\label{fig:pioncloud} The effect of the pionic cloud to the
resonant amplitudes as predicted by the Sato-Lee calculation
\cite{sato}. Solid line includes the pion cloud contribution while
the dashed line neglects the pion cloud effect.}
\end{figure}

Recent high precision experimental results
\cite{pho2,pho1,frol,pos01,merve,bart,Buuren,spaprl,kelly,spamami,stave,joo1,St08,ungaro,villano,aznau,spaepja}
are in reasonable agreement with predictions of models suggesting
the presence of non-spherical amplitudes and in strong disagreement
with all nucleon models that assume sphericity for the proton and
the $\Delta$. With the existence of these components well
established, recent investigations have focused on understanding the
various mechanisms that could generate them. Dynamical reaction
models with pion cloud effects \cite{sato}, \cite{dmt00} bridge the
constituent quark models gap and are in qualitative agreement with
the $Q^2$ evolution of the experimental data. These models calculate
the resonant channels from dynamical equations; they account for the
virtual pion cloud contribution dynamically but have an empirical
parametrization of the inner (quark) core contribution which gives
them some flexibility in the observables. They find that a large
fraction of the quadrupole multipole strength arises due to the
pionic cloud with the effect reaching a maximum value in the region
$Q^2=0.10~ (GeV/c)^2$ (see Fig.~\ref{fig:pioncloud}). Results from
effective field theoretical (chiral) calculations \cite{pv,Ga06},
solidly based on QCD, can also successfully account for the
magnitude of the effects giving further credence to the dominance of
the meson cloud effect in the low $Q^2$ region. Recent results from
lattice QCD \cite{dina} are also of special interest since they are
for the first time accurate enough to allow a comparison to
experiment. The chirally extrapolated \cite{pv} values of CMR and
EMR are found to be nonzero and negative in the low $Q^2$ region, in
qualitative agreement with the experimental results, thus linking
the experimental evidence for the non-spherical amplitudes directly
to QCD while highlighting the importance of future lattice
calculations using lighter quark masses and further refining the
chiral extrapolation procedure.

\begin{figure}[h]
\vspace*{-0.1in} \centering
\begin{tabular}{cc}
\includegraphics[width=8.0cm]{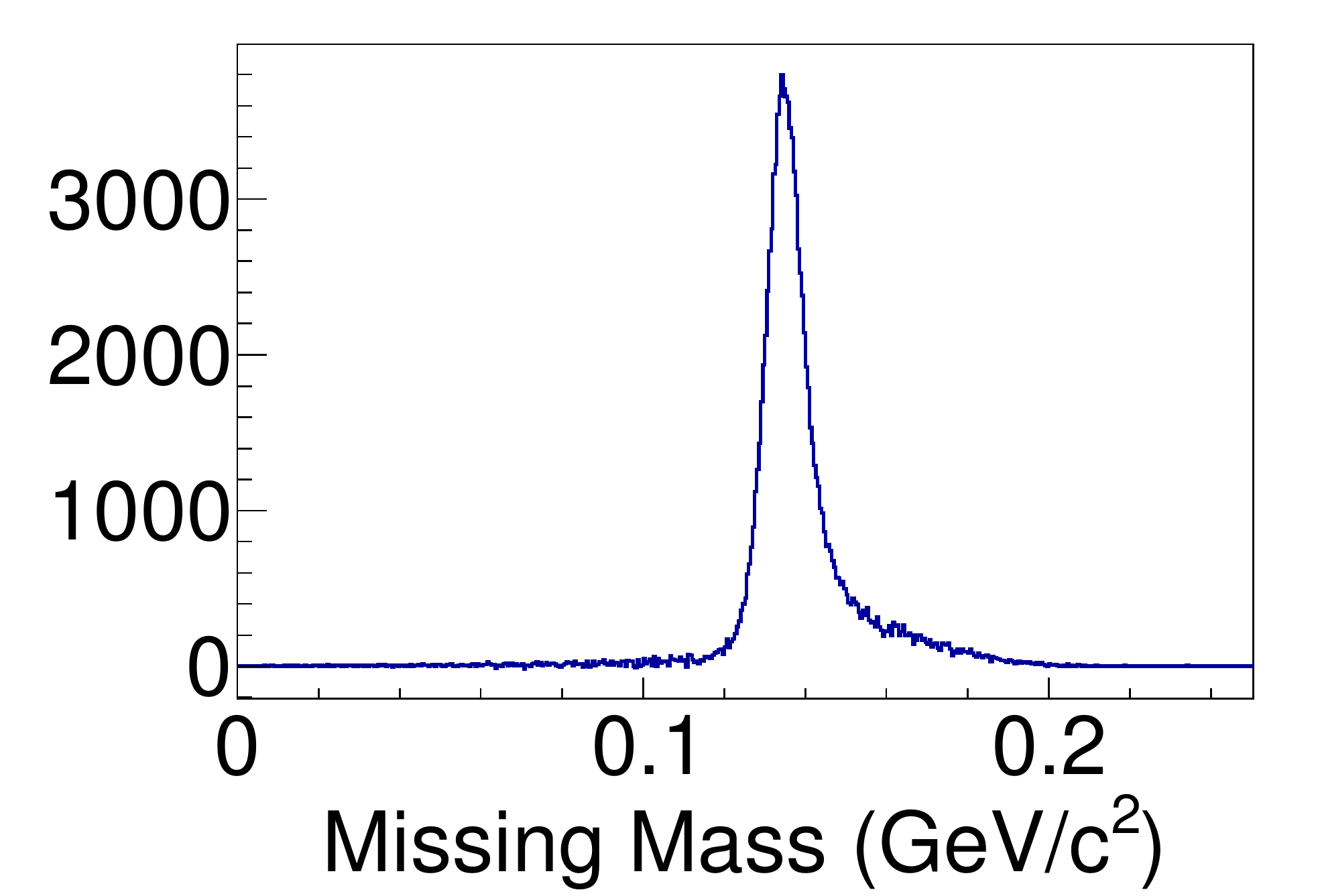}
&
\includegraphics[width=8.0cm,height=5.4cm]{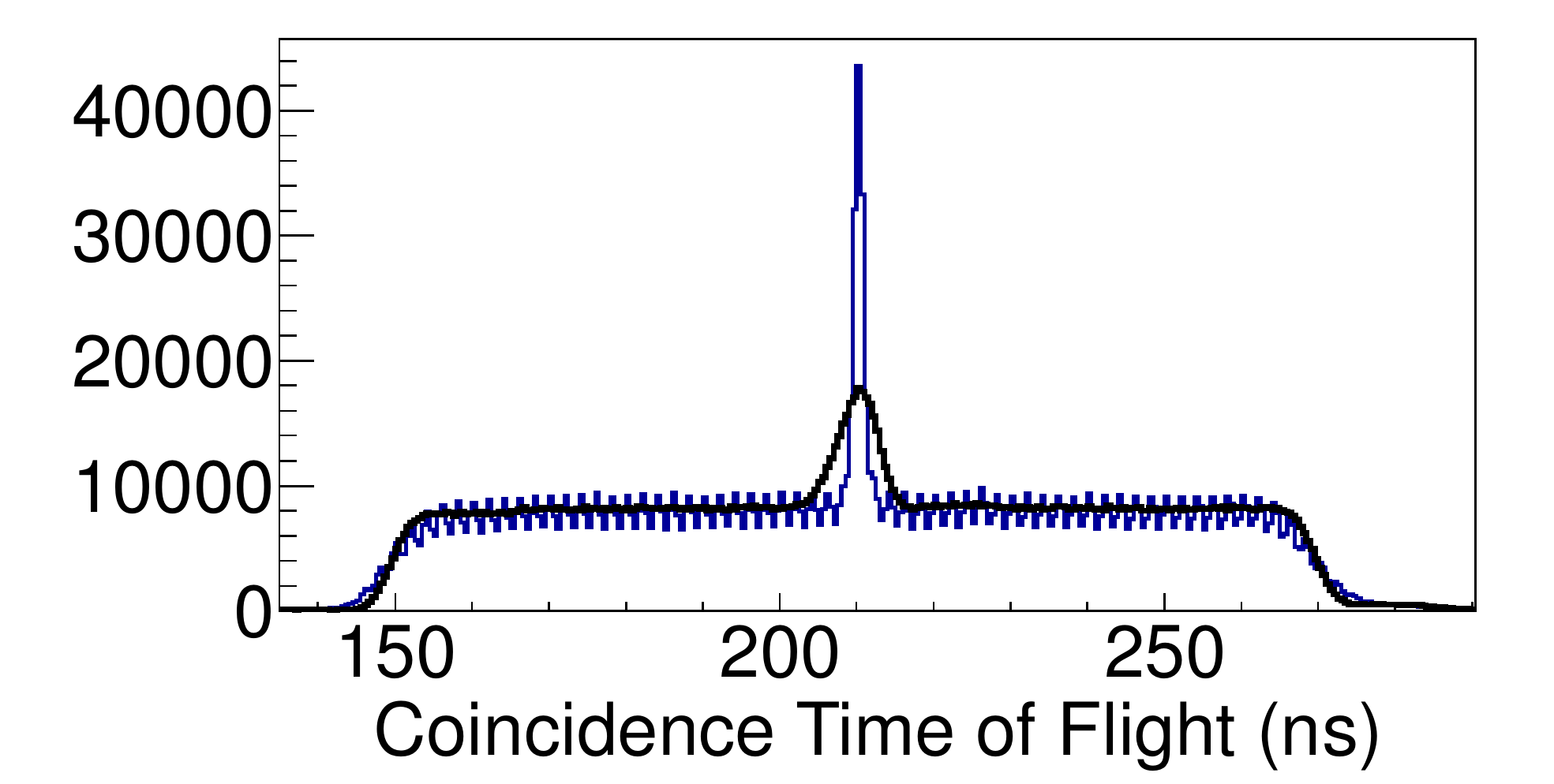}
\end{tabular}
\linespread{0.5} \caption{\label{fig:calib} Left panel: The missing
mass spectrum for the reconstructed (undetected) pion after the
subtraction of accidentals. Right panel: Raw and corrected
coincidence Time of Flight spectrum; an excellent timing resolution
of 1.6~ns has been achieved after the ToF corrections.}
\end{figure}

\subsubsection{The Experiment}\label{sec:experiment}

\begin{figure}[h]
\begin{center}
\epsfig{file=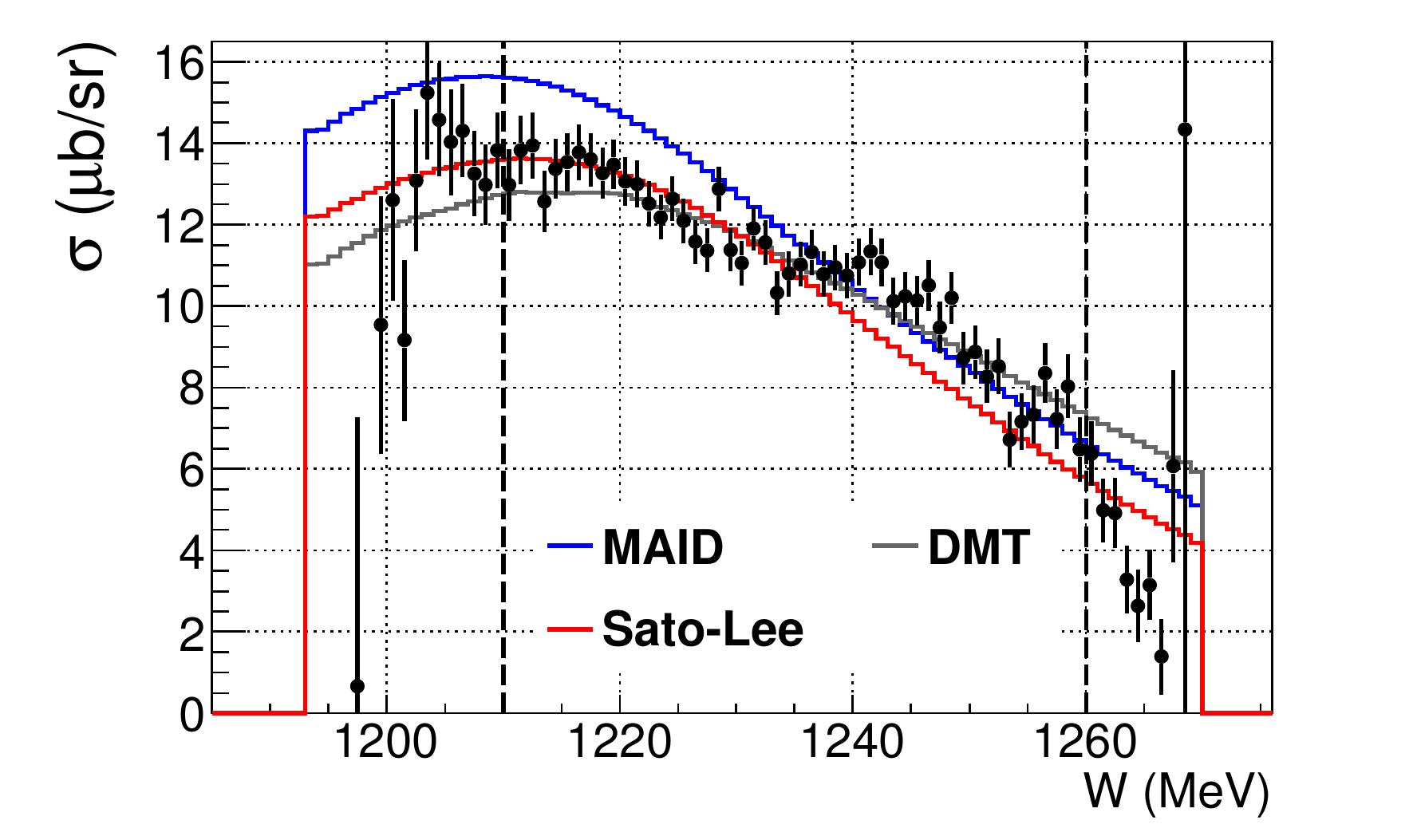,width=12.0cm}
\end{center}
\caption{The measured cross section is presented for the parallel
cross section measurement at $Q^2=0.125~(GeV/c)^2$.}
\label{fig:wxsec}
\end{figure}

\begin{figure}[h]
\begin{center}
\epsfig{file=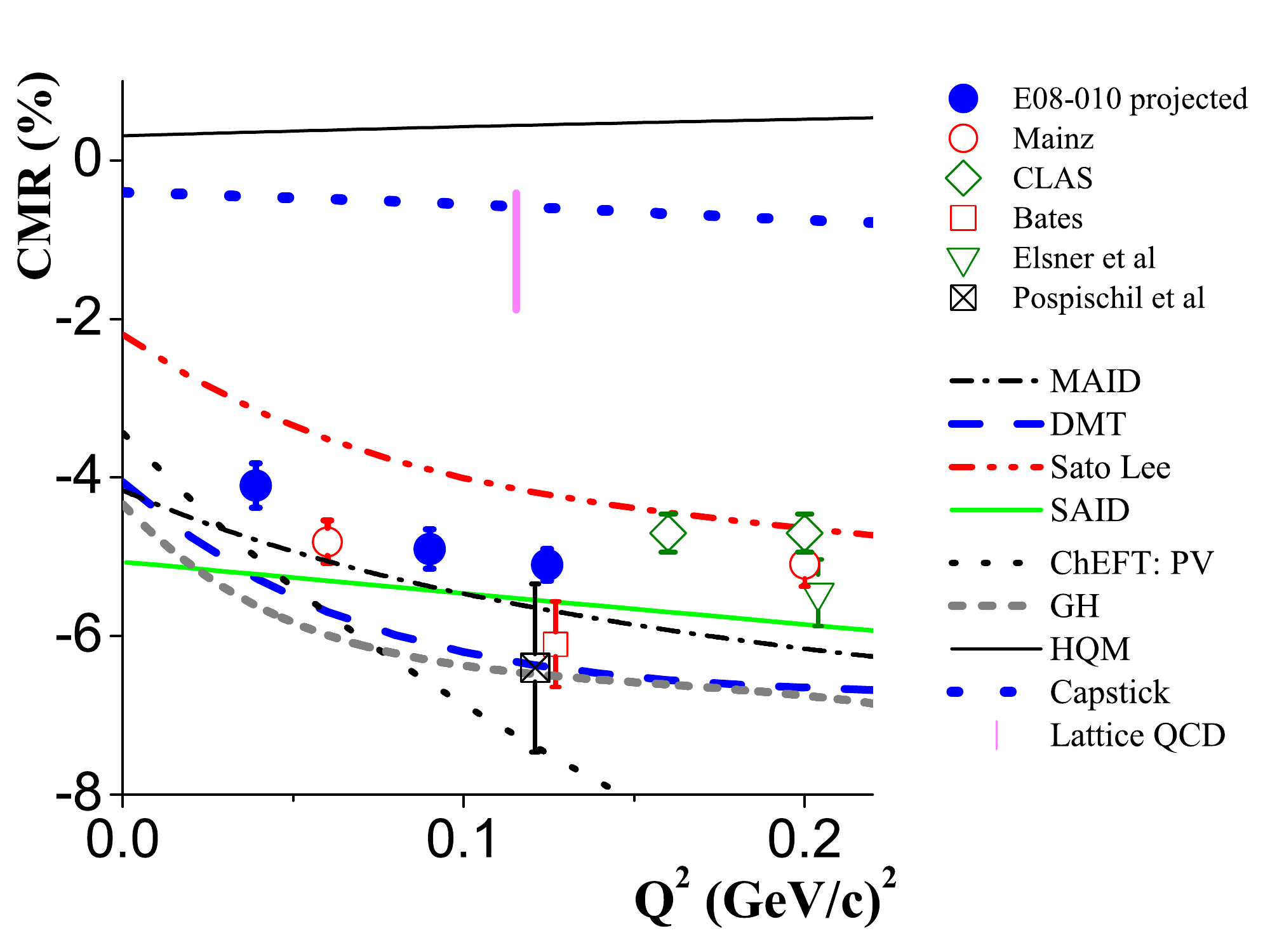,width=12.0cm}
\end{center}
\caption{The CMR at the low momentum transfer region. The projected
E08-010 uncertainties are presented along with the world data and
the theoretical model predictions} \label{fig:cmr}
\end{figure}

The E08-010 experiment aim to explore the low momentum transfer
region at the nucleon - $\Delta(1232)$ transition, where the pionic
cloud effects are expected to dominate. The experiment ran in
February and March of 2011 and achieved all the quantitative and
qualitative goals of the experiment proposal. High precision
measurements of the $p(e,e^\prime p)\pi^{\circ}$  excitation channel
were provided. The two High Resolution Spectrometers were utilized
to detect in coincidence electrons and protons respectively while
the 6 cm and 15 cm liquid hydrogen targets and an electron beam of
$E_{o}=~1.15~GeV$ at $75~\mu A$ were used throughout the experiment.
High precision measurements were conducted in the
$Q^2=0.04~(GeV/c)^2$ to $0.13~(GeV/c)^2$ range. The experiment will
offer results of unprecedented precision in the low momentum
transfer region and will extend the knowledge of the Coulomb
quadrupole amplitude lower in momentum transfer. Furthermore these
measurements will resolve observed discrepancies between
measurements of other labs. Two parallel analysis efforts are
currently in progress, by Temple University and St. Mary's
University, in order to provide important cross checks throughout all
the steps of the analysis. The analysis stage involving calibrations
has been completed. In Fig.~\ref{fig:calib} the Missing Mass
spectrum (after background subtraction), corresponding to the
undetected pion, as well as the corrected time of flight spectrum
are presented. An excellent timing resolution of 1.6 ns has been
achieved. Currently the effort has moved on to the kinematical phase
space analysis, the extraction of the spectrometer cross sections
and the extraction of the resonant amplitudes. In
Fig.~\ref{fig:wxsec} the measured cross section is presented for the
parallel cross section measurement at the highest momentum transfer
kinematics of this experiment. The cross section measurements will
allow for the extraction of CMR with unprecedented precision at the
low momentum transfer region. The projected uncertainties for the
CMR are presented in Fig.~\ref{fig:cmr}. The new results will allow
an in depth exploration of the nucleon dynamics, offering a very
precise signature of the pion cloud, and will provide strong
constraints to modern theoretical calculations that will in turn
allow for a more complete understanding of the nucleon structure.

%\end{document}

\clearpage

\newcommand{\pvdisqsqI}{1.121}
\newcommand{\pvdisqsqII}{1.925}

%\begin{document}

%\section{Summaries of Experimental Activities}

\subsection{E08-011: $\vec e-^2$H Parity Violating Deep Inelastic Scattering (PVDIS) at CEBAF 6 GeV}
\label{sec:e08011}

%\begin{center}
%$\vec e-^2$H Parity Violating Deep Inelastic Scattering (PVDIS) at CEBAF 6 GeV
%\end{center}

\begin{center}
R. Michaels, P.E. Reimer, X. Zheng, spokespersons, \\
K. Pan, D. Wang, graduate students, \\
and \\
the Hall A Collaboration.\\
contributed by R. Michaels
\end{center}

In the past year, our collaboration has published three journal
articles to document the results of 
our 2009 run \cite{e08011:ref1,e08011:ref2,e08011:ref3}.
The DAQ, trigger, detectors, PID, and deadtime corrections 
were documented in detail
in ref \cite{e08011:ref1}.  The measurements in the resonance
region were published in \cite{e08011:ref2} and compared to 
models; this was the first parity-violating data covering the
full resonance region, and the result showed that quark-hadron duality
holds true, at the several percent level, 
for the electroweak structure functions.
Recently, a paper on the main DIS results was 
accepted by Nature \cite{e08011:ref3}.
A fourth and probably final archival paper with many other details 
is expected to be finished in 2014.  

As a reminder, the experiment measured the parity violating asymmetry 
of $\vec e-^2$H deep inelastic scattering (PVDIS).  A simple 
formula which neglects sea quarks and radiative corrections is

\begin{eqnarray}
 A_{PV}^{DIS} &\approx& {{3 G_FQ^2}\over{10\sqrt{2}\pi\alpha}}
   \left[\big(2 C_{1u} - C_{1d}\big) + Y_3 \big(2 C_{2u} - C_{2d}\big)\right]
\end{eqnarray}
where $G_F$ is the Fermi constant, $\alpha$ is the fine structure constant, 
$Q^2$ is the square of the four-momentum transfer (we ran at
$Q^2 = 1.1$ and $1.9$ (GeV/$c)^2$), and $Y_3$ is a kinematic function.  
The constants ``$C_{(1u),(1d),(2u),(2d)}$'' 
are products of the fundamental electroweak
coupling constants for up ($u$) and down ($d$) quarks;
in the Standard model they can be written
as simple functions of the Weinberg angle. 
Consider the first-order Feynmann diagram for the process.  The 
parity violation arises either from axial coupling at the electron
vertex (the $C_1$ terms) or at the quark vertex (the $C_2$ terms).
For details, see ref \cite{e08011:ref3}.
While Qweak \cite{qweak:expt} and other experiments provide
precise constraints on the $C_1$ terms, the unique contribution
of the e08011 measurement is an extraction of the $C_2$ terms.
We have improved the precision of the vector-electron 
axial-vector-quark (VA) interaction term combination
$2C_{2u}-C_{2d}$ over the world value 
\cite{Prescott:1978tm,Prescott:1979dh} 
by a factor of five. 

The main experimental challenges were the high event rate (600 kHz),
the pion contamination (controlled at the $< 4 \times 10^{-4}$ level),
the deadtime uncertainty ( $< 0.4$\% ), and the beam polarimetry
[$dP/P = (1.2 - 1.8)$\%] \cite{e08011:ref1,e08011:ref2,e08011:ref3}.
The success of the experiment bodes well for the 12 GeV
DIS parity program, which has similar, though significantly more
challenging, experimental issues and goals.

%\end{document}

\clearpage

%\documentclass{article}
%\usepackage{epsfig}
%\setlength{\textwidth}{6.5in}
%\setlength{\oddsidemargin}{0in}
%\setlength{\evensidemargin}{0in}
%\setlength{\textheight}{9in}
%\setlength{\topmargin}{0in}
%\setlength{\headheight}{0in}
%\setlength{\headsep}{0in}

%\begin{document}

%\section{Summaries of Experimental Activities}

\subsection{E08-014: The $x>2$ experiment}
\label{sec:e08014}

%\begin{center}
%The $x>2$ experiment
%\end{center}

\begin{center}
J. Arrington, D. Day, D. Higinbotham and P. Solvignon, spokespersons, \\
and \\
the Hall A Collaboration.\\
      contributed by P. Solvignon and Zhihong Ye.
\end{center}

\subsubsection{Motivations}\label{sec:e08014-motivation}
The shell model has been partially successful in describing many features of nuclei such as 
the structure and energies of the nuclear excited states. However, about 30-40\% of the 
nucleonic strength predicted by the shell model to be in shells below the Fermi level is not 
seen in the experimental data~\cite{Lapikas}. This missing strength is thought to be due to 
the nucleon-nucleon (NN) interaction at short distances and the fact that the close packing 
of nucleons in nuclei results in a significant probability of overlapping nucleon wavefunctions. 
These overlapping nucleons belong to a short range correlated cluster and exhibit high momenta, 
well above the Fermi momentum in the nucleus~\cite{Frankfurt1}. 

Short-range correlations (SRC) are now well accepted as a key ingredient in the formulation of 
realistic nuclear wave functions. This means that the experimental characterization of SRC is 
crucial to the development of accurate nuclear structure calculations. Recent results from JLab 
experiment E01-015~\cite{Shneor} confirmed the overwhelming dominance of the proton-neutron pairs 
in two-nucleon SRCs. These two-nucleon knockout experiments are very sensitive to the isospin 
structure as they are able to measure both pp and pn correlations. However, they have also to 
deal with potentially large final state interactions which plague coincidence measurements at 
high missing momentum.

Although inclusive scattering is typically isospin-blind, isospin sensitivity, also 
called ``tensor dominance'', can be identified through a careful choice of complementary targets. 
Isospin-independent and isospin-dependent models predict 25\% differences in the cross-section 
ratios of the two medium-weight nuclei, $^{48}$Ca and $^{40}$Ca. E08-014 complements two-nucleon 
knockout experiments, for which other physical processes make it difficult to extract a 
model-independent and precise quantitative measure of the isospin asymmetry. Further insights 
will be obtained after the JLab 12GeV upgrade from the use of two light mirror nuclei $^3$He and 
$^3$H~\cite{prop2}; in addition to further enhanced isospin sensitivity in these light nuclei, 
realistic theoretical calculations can be performed where the nucleon-nucleon potential components 
and their amplitudes can be separated. 
%\begin{figure}[hbt]
%\center{\epsfig{figure=ratios_he3_clas_us_1.eps,width=9.0cm}}
%\caption[E08014_mot: Results from Hall C E02-019.]{Results from Hall C experiment E02-019~\cite{Nadia}}
%\label{fig:e02019}
%\end{figure}

At $x>2$, the cross-sections from nuclei heavier than deuterium are expected to be dominated by 
three-nucleon short-range correlations (3N-SRCs). Results from Hall C experiment 
E02-019~\cite{Nadia} show a discrepancy with the CLAS results~\cite{egiyan05} in the $x>2$ region, 
while being in very good agreement the 2N-SRC region. E02-019 is at 
higher $Q^2$ than CLAS, and this is consistent with the hint of possible $Q^2$ dependence in the 
CLAS results (see figure 3 of the original proposal~\cite{prop}). These new data and observations 
make our measurement decisive in the effort to map precisely the 3N-SRC region and resolve this 
new issue. E08-014 will also be the first measurement of isospin dependence of 3N-SRC. The amplitude 
and properties of SRCs have important implications not only for the structure of the neutron stars 
and their cooling process~\cite{Frankfurt2} but also in the search for neutrino oscillation~\cite{Martini}.

\subsubsection{Analysis status}\label{sec:e08014-details}
JLab experiment E08-014 ran in April-May 2011. This experiment aims at mapping the 2N and 3N-SRC 
scaling behaviors. It also provides the first test of the SRC isospin dependence in inclusive 
electron-nucleus scattering by using two Calcium isotopes. This experiment used the standard Hall A 
high resolution spectrometers configured for electron detection.

The calibrations of all beam diagnostic elements, spectrometer optics and detectors are done and all 
efficiencies have been evaluated. This past year the analysis efforts were directed on the target density 
study, complicated by the non-uniformed flow of coolant along the cells.  The cooling flow was covering 
only the upstream half of the cell and therefore the gas/liquid density in this part of the cell was higher 
than in the downstream half. Also, as expected, this density gradient between the upstream and downstream 
parts of the cell increases with the beam current. This is mostly an issue for the determination of the target 
luminosity and the radiative corrections. Data were taken at a range of currents and extrapolated 
to zero current to determine the overall density change due to beam heating, while the z-dependence 
of the density as observed from the data is replicated in the simulation to account for the varying 
density. The cross section are generated versus the variable $x_{bj}$ with each bin being an average over t
he target length, i.e. $x_{bj}$ does not dependent on the vertex position. However the radiative corrections 
depend on the location of the reaction. The density distribution along the cell was extracted by fitting the 
vertex spectra of each cell and then fed into the Monte Carlo simulation to determine the radiative correction 
factor related to the vertex position.
				   
%\begin{figure}[hbt]
%\center{\epsfig{figure=H2_Boiling_VZ.eps,width=8.0cm}
%\epsfig{figure=He3_Boiling_VZ.eps,width=8.0cm}}
%\caption{Target spectra along the beam direction for $^2$H (left plot) and for $^3$He (right plot).}
%\label{fig:bump}
%\end{figure}

\begin{figure}[hbt]
\center{\epsfig{figure=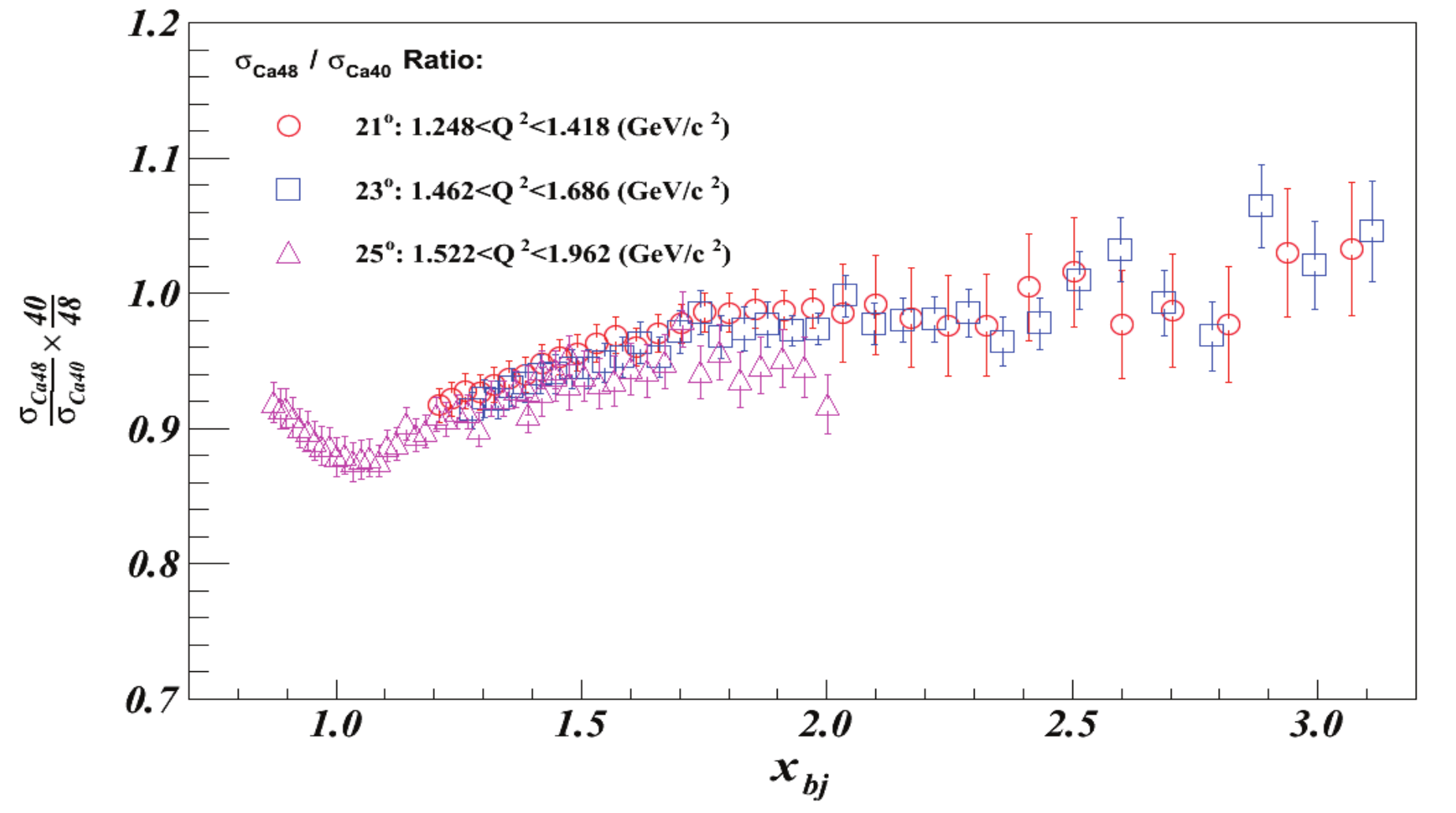,width=10.0cm}}
\caption{Preliminary results (statistical errors only) on the isospin ratio, i.e. per-nucleon cross section ratio $^{48}$Ca/$^{40}$Ca. Theoretical calculation~\cite{Vanhalst} predicts this ratio to be one in the 2N-SRC region.}
\label{fig:xs}
\end{figure}

The analysis of E08-014 is in its final stage for the isospin study  with the data on $^{40}$Ca and $^{48}$Ca.
The iteration of the cross section model (Hall C XEM model) has been carried on in order to fit our kinematical 
region. Corrections for the Coulomb distortion~\cite{Solvignon} have to be added to the analysis. Then the 
radiative corrections and cross section model iteration procedure will be finalized. Preliminary results on the 
isospin ratio are shown in Fig.~\ref{fig:xs}. Recent theoretical work predicts that the probability to find a np pair is
the same in $^{40}$Ca and in $^{48}$Ca~\cite{Vanhalst}.

For the 2N-SRC and 3N-SRC measurements, special care needs to be taken for these data because of the 
non-uniformity of the $^2$H, $^3$He and $^4$He target density.

%
% Here's how to do the references.  We will be using the APS style.
%

%\end{document}

\clearpage

%\documentclass{article}
%\usepackage{epsfig}
%\setlength{\textwidth}{6.5in}
%\setlength{\oddsidemargin}{0in}
%\setlength{\evensidemargin}{0in}
%\setlength{\textheight}{9in}
%\setlength{\topmargin}{0in}
%\setlength{\headheight}{0in}
%\setlength{\headsep}{0in}

%\begin{document}

%\section{Summaries of Experimental Activities}

\subsection{E08-027:  $g_2^p$}
\label{sec:e08027}

%\begin{center}

%\end{center}

\begin{center}
A. Camsonne, J.P. Chen, D. Crabb, K. Slifer, spokespersons, \\
and \\
the Hall A Collaboration.\\
contributed by M. Cummings
\end{center}

\subsubsection{Motivation}

The deviation of the nucleon's spin dependent properties from point like behavior in inclusive electron scattering can be described by the spin structure functions (SSF) $g_1$ and $g_2$.  While $g_1$ can be expressed in terms of quark distribution functions, $g_2$ contains contributions from higher order interactions, and so has no simple interpretation in the quark-parton model.  Measurements of $g_2$ for the proton, specifically at low to moderate Q$^2$, are scarce; currently, the lowest momentum transfer investigated is 1.3 GeV$^2$, by the RSS collaboration~\cite{Wesselmann:2006mw}.

The data from this experiment will provide insight on several outstanding physics puzzles, such as why Chiral Perturbation Theory ($\chi$PT) calculations fail to predict the behavior of the longitudinally-transverse spin polarizability ($\delta_{LT}$) for the neutron~\cite{Amarian:2004yf}; a surprising outcome as $\delta_{LT}$ is seen as a good test of QCD dynamics due to its insensitivity to the delta resonance~\cite{Bernard:2002bs,Kao:2002cp,Bernard:2012xu}.  Additionally, they will provide a test of the Burkhardt-Cottingham Sum rule, which says that the integral of $g_2$ over the Bjorken scaling variable $x$ is zero.  This sum rule has been tested for the neutron, but the lack of data for $g_2^p$ leaves this sum rule largely untested for the proton.  Furthermore, a lack of knowledge of the SSF at low $Q^2$ is a limiting factor of QED calculations of bound-state systems, such as the hydrogen atom.  The energy levels of the hydrogen atom can be measured to very high accuracy, to the point where the leading uncertainty of the corresponding QED calculations comes from the finite size of the nucleon as characterized by the SSF and elastic form factors.  Finally, recent results from PSI~\cite{Pohl:2010zz} for the proton charge radius $\langle R_p\rangle$ via measurements of the lamb shift in muonic hydrogen suggest a discrepancy from the value obtained from elastic electron-proton scattering.  The leading uncertainty in these calculations comes from differing values of the Zemach radius, determined from integrals of the SSF and elastic form factors.  

\begin{figure}[htdp]
\begin{center}
\includegraphics[width=0.45\textwidth]{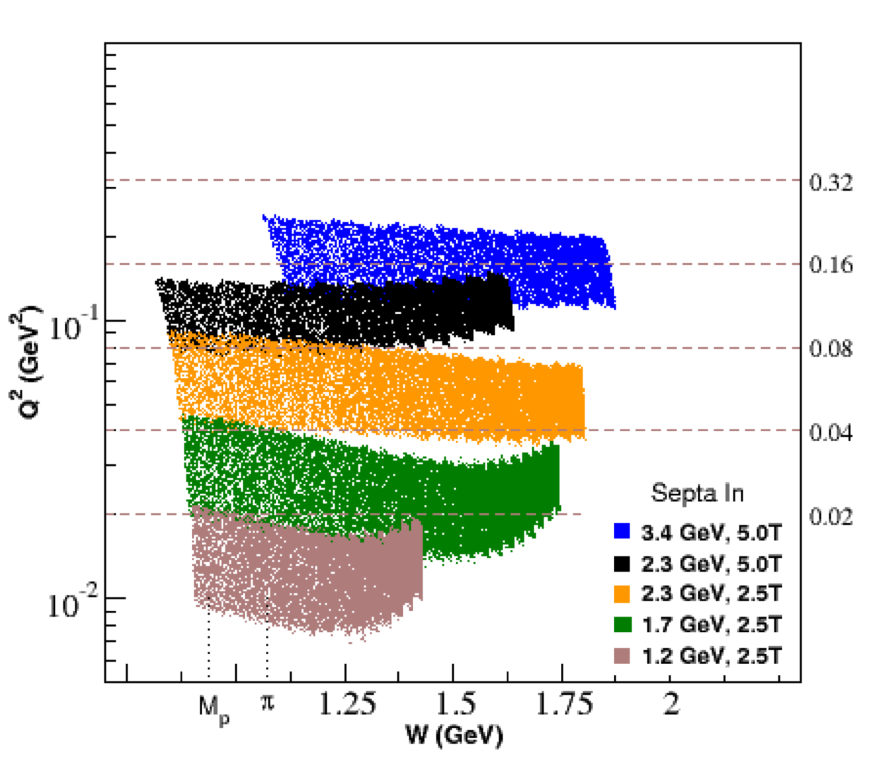}
\caption{ Achieved kinematic coverage during the experimental run period.  The vertical axis on the right hand side is the extrapolation to constant Q$^2$.  }
\label{fig:kin}
\end{center}
\end{figure}

\subsubsection{The Experiment}
\begin{figure}
\begin{center}
\includegraphics[width=0.50\textwidth]{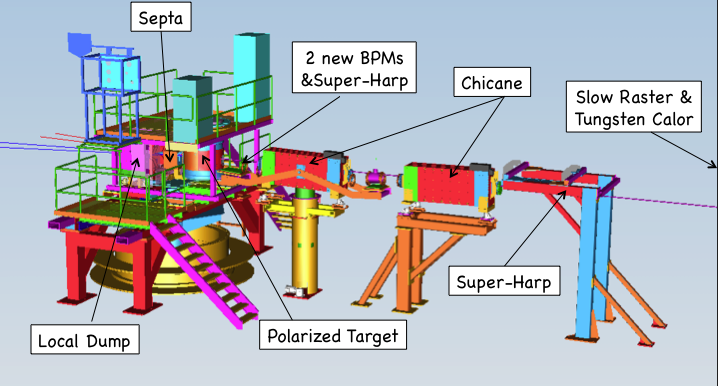}
\caption{Installation of the $g_2^p$ experiment in Hall A.  The third arm detector was located on the left-hand side of the bottom target platform.  }
\label{fig:plat}
\end{center}
\end{figure}

The $g_2^p$ experiment collected data successfully from March-May of 2012.  An inclusive measurement was performed in the low $Q^2$ region $0.02<Q^2<0.20~\textrm{GeV}^2$ (see figure \ref{fig:kin}) at forward angles to obtain the proton spin-dependent cross sections.  From these data the $g_2^p$ structure function will be extracted along with the longitudinally-transverse spin polarizability $\delta_{LT}$.  This experiment required a large scale installation in Hall A, as shown in figure \ref{fig:plat}.  A solid ammonia target was polarized through the process of Dynamic Nuclear Polarization (DNP).  In order to compensate for the the deflection of the beam by the large target magnetic field, a pair of chicane magnets was installed upstream of the target.  To reach the small scattering angle of 5.69$^\circ$ necessary for this kinematic range, a septum magnet was installed downstream of the target.  New beamline diagnostics (BPM and BCM) were required due to the low beam current (50-100nA) needed throughout the run to maintain the target polarization.  For certain kinematics, a local beam dump was necessary, located just downstream of the septum magnets.  Finally, a new scintillator detector, the third arm, was developed specifically for this experiment as a cross check of the product of the beam and target polarization.  The third arm was placed on the target platform to collect elastically scattered protons at large scattering angles to provide a measurement of the elastic proton asymmetry.

\subsubsection{Status of Analysis}

\begin{figure}[htdp]
\begin{center}
\includegraphics[width=0.60\textwidth]{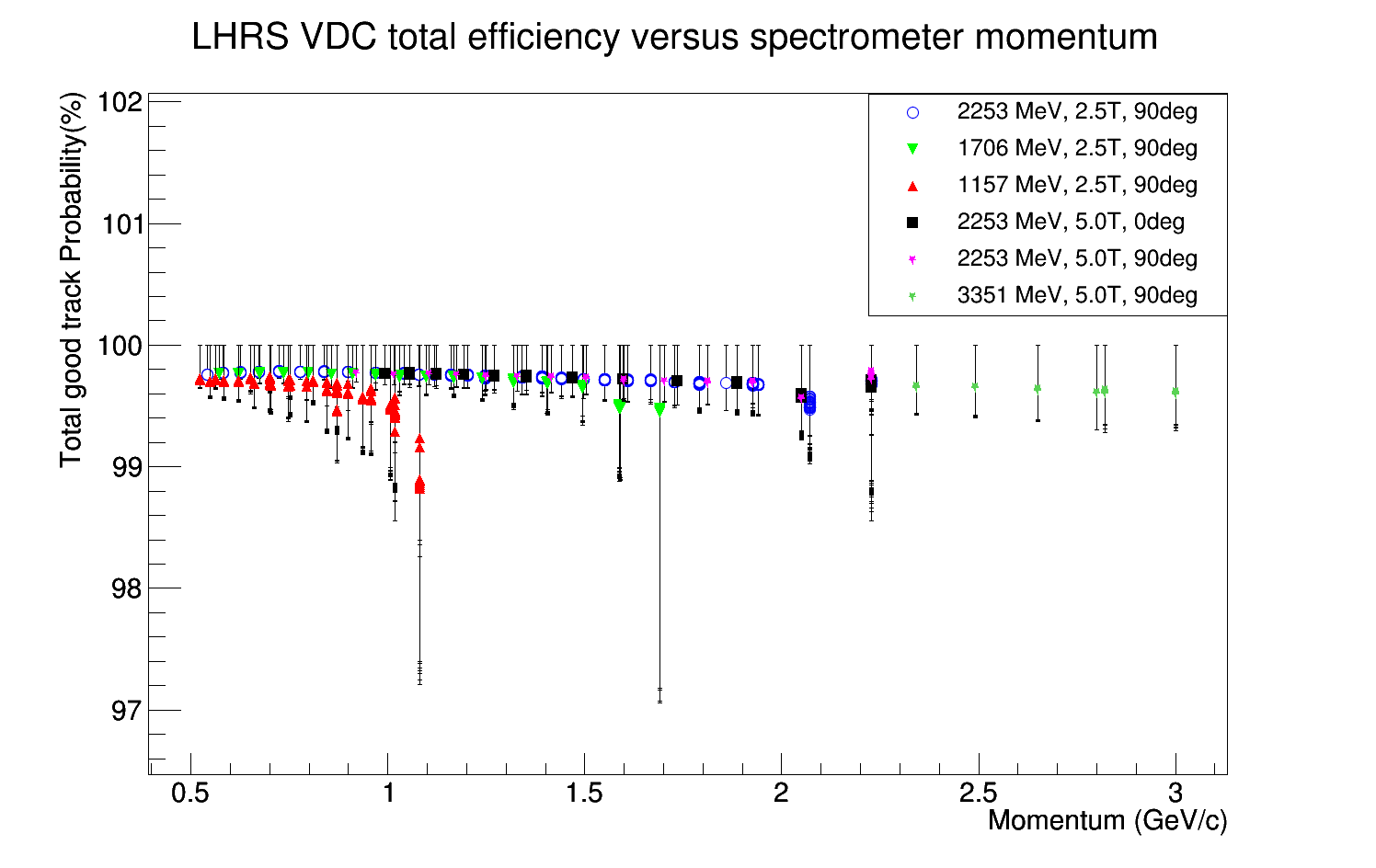}
\caption{Total VDC efficiency for all LHRS runs, after multitrack events have been accounted for.  The efficiency is $>$ 99\% for most kinematic settings. }
\label{fig:vdc_eff}
\end{center}
\end{figure}

HRS Detector efficiencies are needed as a correction to the cross section.  The VDCs provide tracking information for both arms of the HRS, which provides good position and angle reconstruction.  However, due to the high event rate, it is possible that multiple particles will pass through the drift chambers simultaneously; as many as 30\% of events can have multiple track for certain kinematic settings.  This presents a large uncertainty to the cross section if left uncorrected.  The multitrack events are carefully examined and resolved, bringing the systematic uncertainty down to below $\sim$1\% for all kinematic settings.  The total VDC efficiency can be seen in figure \ref{fig:vdc_eff}.  Efficiencies from the other spectrometer detectors, including the s1 and s2m trigger scintillators, gas Cherenkov, and lead glass calorimeters were seen to be high ($\sim$99\% or higher) throughout the run, indicating good detector performance.  PID cuts were determined to minimize the amount of residual pion contamination and maintain and overall detection efficiency of 99\%.

The optics calibration without target field has been updated to include beam positions obtained from fitting the focal plane data.   Forward and reverse transport functions between the target and focal plane, used to describe the magnetic field system without the target field, were fitted with simulation data from the SNAKE Model.  These functions have been incorporated into the $g_2^p$ simulation package to describe the trajectories of outgoing electrons.  A comparison between simulation results and optics data at the target plane is being done; the quantities match well, but more tuning to refine the comparison results is currently underway.  Additionally, efforts are being focused on optics analysis with the target field on.  Due to problems with the right-HRS septum magnet during the run, multiple calibrations will be needed to correspond to the different magnet coil configurations.

A Monte-Carlo simulation program was developed to simulate the affect of the target and septum fields in combination with the HRS.  The program is based on the Hall A Single Arm Monte-Carlo (SAMC) package and is extended to work with the target field.  The simulation setup has been tuned with our experimental configuration, including two target field settings and several different versions of the septum field model.  The programs also include several different cross section models and fittings in the elastic and resonance kinematic regions.  The package will be used in the optics calibration with target field and the spectrometer acceptance study.

Beam position information is very important for optimization of the optics.  For the straight through calibration, two harps were used to determine the position of the beam.  During the course of the run, the BPMB division (attenuation) was changed; present efforts are focused on determining a different calibration method to account for this change.  Additionally, a study is underway to understand and account for the fluctuation of the pedestal throughout the run period.  The $g_2^p$ experiment is using the same helicity scheme set up by the QWEAK experiment.  Both the helicity flip rate and the DAQ rate are high, so a new standalone package was developed to decode the helicity information under these conditions.

\begin{figure}[htdp]
\hspace{0.5cm}
\includegraphics[width=0.45\textwidth]{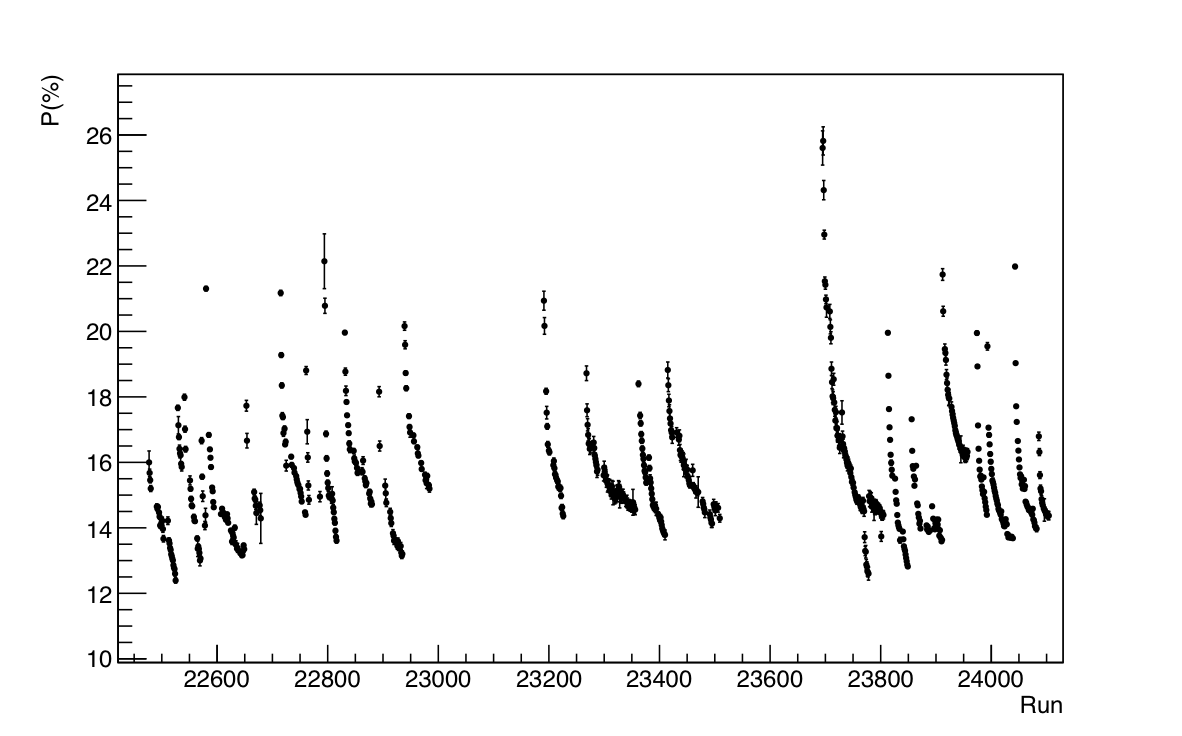}
\includegraphics[width=0.45\textwidth]{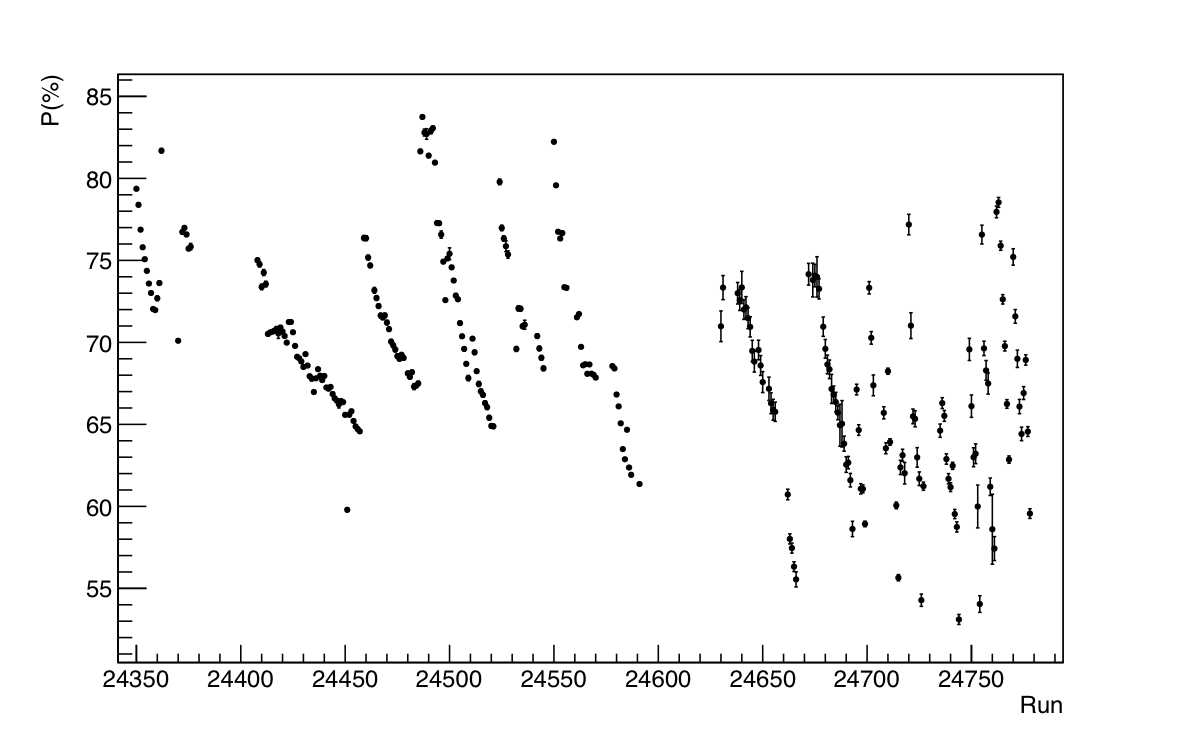}
\caption[E08-027 Preliminary Results]{\label{fig:tarpol}
Run by run polarization results for the 2.5T magnetic field setting ({\bf left}) and 5T magnetic field setting ({\bf right}).
}
\end{figure}

 A precise measurement of the target polarization is needed to extract the physics asymmetry.  During this experiment two different target fields, 2.5T and 5T, were necessary to achieve the desired kinematic range.  The calibration constants, used to convert the measured NMR signal into a useable polarization, have been calculated for all ammonia samples and applied to the data to determine the average polarization on a run-by-run basis.  The final run polarizations can be seen in figure \ref{fig:tarpol}.  An average polarization of 70\% for the 5T field setting and 15\% for the 2.5T field setting were observed.   Polarization uncertainties due to measurement precision and statistical fluctuations have been calculated to less than 5\%.

\begin{figure}[htdp]
\hspace{0.5cm}
\includegraphics[width=0.42\textwidth]{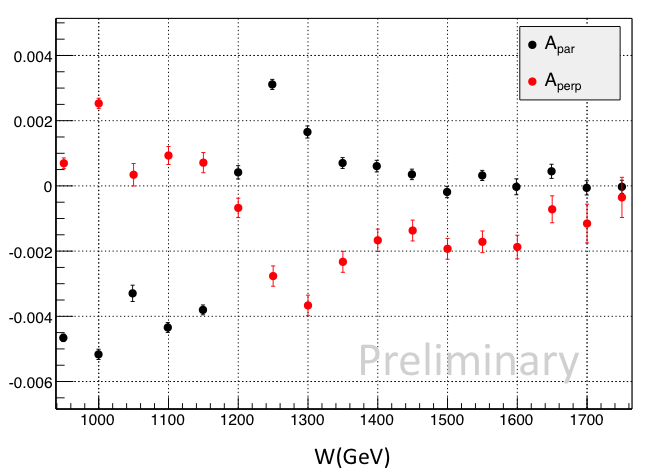}
\hspace{1cm}
\includegraphics[width=0.4\textwidth]{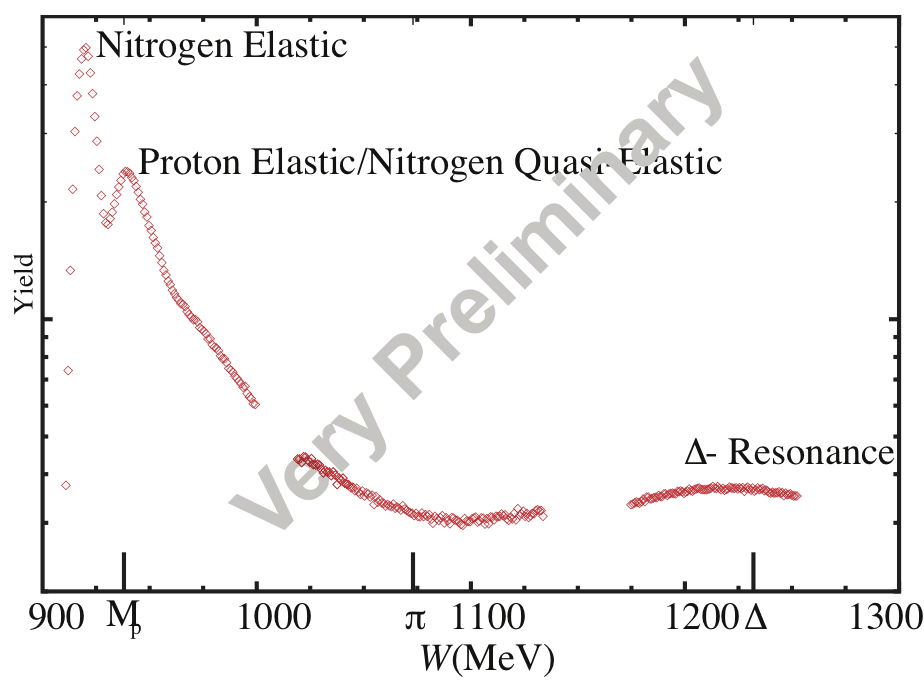}
\caption[E08-027 Preliminary Results]{\label{fig:pre_results}
{\bf Left:}  Preliminary physics asymmetries for the longitudinal and transverse magnetic field setting.
{\bf Right:} 
Preliminary yields showing the nitrogen and proton elastic peaks.
}
\end{figure}

Once beam position calibrations, optics with target field and acceptance studies have been completed, the cross section can be extracted.  Studies are underway to calculate the packing fraction and dilution factor, which will determine the percentage of electrons that were not scattered from a proton in the ammonia target.  Analysis of data taken with the third arm, which will be used as a cross check of the beam and target polarization, is also in progress.  Preliminary physics asymmetries and yields can be seen in figure \ref{fig:pre_results}.

%\end{document}
\clearpage
%\newpage

%\documentclass{article}
%\usepackage{graphicx}
%\usepackage{url}
%\setlength{\textwidth}{6.5in}
%\setlength{\oddsidemargin}{0in}
%\setlength{\evensidemargin}{0in}
%\setlength{\textheight}{9in}
%\setlength{\topmargin}{0in}
%\setlength{\headheight}{0in}
%\setlength{\headsep}{0in}

%\begin{document}

%\section{Summaries of Experimental Activities}

\subsection{The Super Bigbite Spectrometer}
\label{sec:sbs}

\begin{center}
    Progress by the Super Bigbite Collaboration
\end{center}

\begin{center}
contributed by S.~Riordan \\
for the Super Bigbite Collaboration.
\end{center}

This a summary of the progress on the DOE funded SBS Project as well as the many separate dependencies, experiments using the same equipment, and collaborative efforts.

\subsubsection{Overview}\label{sec:sbs:overview}

The Super Bigbite (SBS) project is a collection of experiments based around utilizing large-acceptance single dipole spectrometers designed to operate in a high rate environment in conjunction with the upgraded $12~\mathrm{GeV}$ CEBAF accelerator.  The types of experiments this is suited for typically require high statistics but only moderate momentum resolution and may either be coincidental electron-hadron scattering (requiring two arms) or single-arm inclusive.  The experimental program as a DOE project is formally defined by three high $Q^2$ elastic form factor measurements, specifically $G_E^p/G_M^p$, $G_E^n/G_M^n$, $G_M^n$, but the collaborative effort also includes $A_1^n$ and SIDIS using a $ {}^3\overrightarrow{\mathrm{He}}$ target.  These experiments are 
\begin{itemize}
    \item{E12-07-109, GEp}
    \item{E12-09-016, GEn}
    \item{E12-09-019, GMn}
    \item{E12-09-018, SIDIS}
    \item{E12-06-122, $A_1^n$}.
\end{itemize}

This program is addressing an important part of the Jefferson Lab's 12 GeV goals.  It is pushes the boundaries of previous high momentum-transfers and large Bjorken-$x$ with what is now kinematically accessible with the 11 GeV beam in Hall A and allows for many tests of theoretical models describing fundamental nucleon structure.  In particular, these tests aim towards understanding the nucleon in non-perturbative QCD, mapping out the non-perturbative to perturbative transition, and relating to fundamental questions about the role of angular momentum in nucleon structure.  

Over the last year there have been many developments including progress on new hardware systems and structuring of the collaboration.  More documentation over the entirety of the project can be found online~\cite{sbs:webpage}.

\subsubsection{Organization}

The SBS Collaboration has been formally organized with the adoption of a charter and consists of the overlap of collaborations of experiments using all or parts of the equipment of SBS. This includes those beyond the experiments in the formal DOE project.  The coordinating committee for the collaboration consists of two program scientists, the Hall~A leader, and a representative from each experiment, one of which acting as the chair.  This body consists of

\begin{itemize}
    \item{Gordon Cates, UVA, Program Scientist}
    \item{Evaristo Cisbani, INFN, GEp Representative}
    \item{Cynthia Keppel, JLab, Hall Leader}
    \item{Nilanga Liyanage, UVA, $A_1^n$ Representative}
    \item{Andrew Puckett, UConn, SIDIS Representative}
    \item{Seamus Riordan, UMass Amherst, GEn Representative}
    \item{Brian Quinn, CMU, GMn Representative and Chair}
    \item{Bogdan Wojtsekhowski, JLab, Principle Program Scientist}
\end{itemize}

Additionally, outside of the collaboration, in 2014 Mark Jones of Jefferson Lab will replace John LeRose as Project Manager.

\subsubsection{Instrumentation Progress}

\paragraph{48D48 Magnet}

One critical component of the SBS project is the 48D48 magnet, which serves as the magnetic element for the hadronic arm of these experiments.  In the last year, considerable progress has been made in acquiring the existing magnet, designing the iron and coil configuration, as well as simulating magnetic characteristics.   The existing magnet iron was acquired from Brookhaven in August.  A contract to construct a new power supply was awarded and new coils for the magnet have been been ordered.  Drawings for the modification of the iron yoke have also been completed.  Commissioning of the magnet and power supply is expected to begin later in 2014.

Numerous TOSCA magnetic field simulations have been performed, including analysis of the full magnetic systems of Bigbite and Super Bigbite.  These studies are ongoing.  This magnetic shielding is particularly important, as the magnet will be situated near the target areas as well as have beamline components run through it when at small angles.  Field maps from these simulations have become available and can now be included within detailed Geant4 simulations of backgrounds and counting rates.

\paragraph{GEM Detectors}

Several sets of GEM detectors are being constructed by groups at INFN and UVA and will be used in both the hadron and electron arms of these experiments.  In the last year, two full sized $50\times50~\mathrm{cm}^2$ chambers have been built at UVA and one was successfully tested at Fermilab as well as several prototype chambers which have been thoroughly tested with cosmic rays.  The UVA contract to produce the first 29 modules was awarded and orders for the GEM components have been placed.

The INFN collaboration has assembled three $40\times50~\mathrm{cm}^2$ ``pre-final'' modules and have been performing in-beam tests at DESY in magnetic fields up to $500~\mathrm{Gauss}$.  Two full chambers are expected to be produced by the end of 2014.  Orders for all GEM components and electronics have been placed.

Testing of two APV25 readout systems are being carried out, using a stand-alone SRS (scalable readout system) at UVA or the VME-based INFN MPD system.

The INFN group is also working on the development of two small silicon microstrip planes ($10\times20~\mathrm{cm}^2$, $50~\mathrm{\mu m}$ pitch) which could improve the tracking of the primary particles in the front tracker. Impact of these silicon planes on the SBS tracking is under careful evaluation.

\paragraph{Calorimetry}

There are two primary calorimetry projects required to fully realize SBS.  This includes the electromagnetic calorimeter for the GEp experiment and the hadronic calorimeter, both outside of the DOE project, which is used in the hadron arm in the coincident trigger experiments.

A primary concern for the electromagnetic calorimeter is the darkening of lead-glass blocks as they absorb radiation.  The essence of the sensitivity of the lead glass to radiation is due to the heavy metals in the glass and the low electrical conductivity, which leads to the accumulation of ions.  Two curing processes have been used in the past and are under investigation.  A UV curing process, which typically requires downtime during running is under investigation by collaborators at William and Mary, Norfolk State University, and Christopher Newport University.  Continuous UV curing is one of the goals of the group's effort, in particular finding the appropriate UV source and filters to eliminate visible light.  This group has also studied the possibility of using the sampling calorimeter technology, in the form of shashliks to avoid the problem of radiation damage. A viable solution was arrived at, but the current funding situation precludes its use at the present time.

Additionally, a thermal annealing method which operates continuously is under investigation at JLab.  The key elements of the annealing concept have been fully tested and engineering aspects of this system are being developed.  The use of low-energy but high intensity accelerators at Idaho State University is planned to irradiate blocks for final proof of the annealing methods.

The hadronic calorimeter, HCal-J, is being designed and prototyped at Carnegie Mellon University with support from INFN Catania and is heading to full construction.  A design of optical parts of the modules the wavelength shifter and light guides, has been completed as well a concept for the overall support structure.  Samples of specialized plastic scintillator were obtained to determine an optimal PPO concentration to maximize the light yield and production of this scintillator for the full detector has been funded and initiated.  A mechanical prototype module is expected to be completed this January, a complete detailed design in June, and full scale construction beginning later in 2014.

\paragraph{Polarized ${}^3\mathrm{He}$ Target}

The polarized ${}^3\mathrm{He}$ target is at the heart of the $A_1^n$, $G_E^n$, and SIDIS experiments and provides an effective polarized neutron target and has been employed at the lab for many years with beam currents up to $15~\mathrm{\mu A}$ with $60\%$ polarization.   However to reach sufficient luminosity, beam currents up to $60~\mathrm{\mu A}$ and extending the length to almost $60~\mathrm{cm}$ presents new challenges.   To meet these goals, a design that includes metal end-cap windows, a convection design, and two laser pumping chambers is being developed at UVA.  In the last year, progress has been made in prototyping this target, in particular testing different glass-to-metal seals which will still maintain high-quality polarization.  Additionally, progress has been made in developing a target chamber and holding coils with $<0.1~\mathrm{Gauss/cm}$ field gradient, which is also crucial to achieve high polarization.

\paragraph{Gas Cherenkov}

A 600 PMT gas Cherenkov detector to be used for electron identification is being developed jointly by collaborators at William and Mary, North Carolina A\&T State University, University of Glasgow, and James Madison University.  Full-scale simulations and a small-scale prototyping is completed and the detector geometry has been frozen.  Prototyping of the mirror frame and forming the mirror shape have been completed and development for an LED-based alignment system is also completed.  Fabrication and testing will occur in early 2014.  A particularly detailed Geant4 simulation has also been developed at William and Mary to study light collection efficiencies down to the PMT level.  

In additional particular efforts were made to optimize the design of the PMT array, which requires magnetic shielding.  A prototype mu-metal and iron box was produced and tested by North Carolina A\&T State University which will allow for $<10\%$ change of gain in a $30~\mathrm{Gauss}$ environment which was consistent with expectations.  James Madison University will be testing the phototubes to be used and front-end electronics are being developed by University of Glasgow.  

\paragraph{Coordinate Detector}

A scintillator coordinate detector will serve as a hodoscope to determine the electron position in front of the electromagnetic calorimeter in the GEp experiment.  The position resolution of such a detector will allow for a much cleaner identification of proton elastic events than by the calorimeter alone by exploiting the specific electron-hadron kinematic correlation.  It consists of two detector planes of $0.5\times3\times102~\mathrm{cm}^3$ horizontal scintillator bars split into left and right halves.  A set of multi-anode PMTs from FNAL were donated to Jefferson Lab in 2012 and have been tested by collaborators at Saint Mary's University in Halifax for uniformity on gain between pixels.  Front-end electronics based on the NINO discriminator chips is being developed at University of Glasgow.  Last summer a prototype module was constructed and will be used for mechanical construction tests.

\paragraph{Monte Carlo}

The Geant4 and ROOT-based SBS Monte Carlo simulation, {\verb g4sbs }, included realizing a fully detailed representation of the experiment geometry.  In particular, recent descriptions of the 48D48 magnet coil and shielding, cryotarget, and detector stacks were included.  The code is now hosted on github~\cite{sbs:github}.

%\end{document}

\clearpage

\section{Publications}

\newcommand{\etal}{{\em et~al.}}

Publications published during 2013, either in preprint or finally, based
on experiments run in Hall A of  Jefferson Lab.

\begin{enumerate}

\item{Monaghan, P. \etal, Measurement of the $^{12}$C(e,e'p)$^{11}$B Two-Body
    Breakup Reaction at High Missing Momentum Values, 
\href{http://arxiv.org/abs/1301.7027}{arxiv:1301.7027}
}

\item{Wang, D. \etal,
Measurement of the Parity-Violating Asymmetry in
                        Electron-Deuteron Scattering in the Nucleon Resonance
                        Region, 
\href{http://prl.aps.org/abstract/PRL/v111/i8/e082501}{Phys.Rev.Lett. 111 082501}
}

\item{Pomerantz, I. \etal,
Hard Two-body Photodisintegration of $^3$He,
\href{http://prl.aps.org/abstract/PRL/v110/i24/e242301}{Phys.Rev.Lett. 110 242301}
}

% march 2013
\item{Wang, Y. \etal, A MRPC prototype for SOLID-TOF in JLab,
\href{http://iopscience.iop.org/1748-0221/8/03/P03003/}{JINST 8 P03003}
}

\item{Horowitz, C.J. \etal, Electroweak Measurements of Neutron
    Densities in CREX and PREX at JLab, USA, 
\href{http://arxiv.org/abs/arxiv:1307.3572}{arxiv:1307.3572}
}

\item{Singh, J. \etal, The Development of High-Performance Alkali-Hybrid Polarized He-3 Targets for Electron Scattering
\href{http://arxiv.org/abs/arxiv:1309.4004}{arxiv:1309.4004}
}

\item{Camsonne, A. \etal,
JLab Measurement of the $^4$He Charge Form Factor at
                        Large Momentum Transfers, 
\href{http://arxiv.org/abs/1309.5297}{arxiv:1309.5297}
}

\item{Katich, J. \etal,
Measurement of the Target-Normal Single-Spin Asymmetry
                        in Deep-Inelastic Scattering from the Reaction
                        $^{3}\mathrm{He}^{\uparrow}(e,e')X$, 
\href{http://arxiv.org/abs/1311.0197}{arxiv:1311.0197}
}

%1 October 2013, 
\item{Subedi, R. \etal, % D.Wang, K.Pan, X.Deng, R.Michaels, P.E.Reimer, A.Shahinyan, B. Wojtsekhowski, X.Zheng, 
A scaler-based data acquisition system for measuring parity-violating
asymmetry in deep inelastic scattering,
\href{http://www.sciencedirect.com/science/article/pii/S016890021300627X}{
  Nucl. Instr. Meth. A 724
(2013) 90-103}
}

% 11 November 2013,
\item{Parno, D. \etal, Comparison of Modeled and Measured Performance
    of a GSO Crystal as Gamma Detector, 
\href{http://dx.doi.org/10.1016/j.nima.2013.05.154}{Nucl. Instr. Meth. A
  728, 92-96}
}

\item{Allada, K. \etal,
Single Spin Asymmetries of Inclusive Hadrons Produced in
                        Electron Scattering from a Transversely Polarized $^3$He
                        Target, 
\href{http://arxiv.org/abs/1311.1866}{arxiv:1311.1866}
}

\item{Zhang, Y. \etal,
Measurement of pretzelosity asymmetry of charged pion
                        production in Semi-Inclusive Deep Inelastic Scattering on
                        a polarized $^3$He target, 
\href{http://arxiv.org/abs/1312.3047}{arxiv:1312.3047}
}

\end{enumerate}

%\clearpage
%\newpage

\section{Theses}
\begin{enumerate}

\item{\emph{Meausrement of the Neutron Radius of $^{208}$Pb Through Parity Violation in Electron Scattering}\\	
Kiadtisak Saenboonruang\\ 
\url{https://misportal.jlab.org/ul/publications/view_pub.cfm?pub_id=12303}}

\item{\emph{Short Range Correlations in Nuclei at Large x$_{bj }$ through Inclusive Quasi-Elastic Electron Scattering}\\	
Zhihong Ye\\ 
\url{https://misportal.jlab.org/ul/publications/view_pub.cfm?pub_id=12877}}

\end{enumerate}

%\clearpage
\newpage

\section{Hall A Collaboration Member List, 2013}
\begin{multicols}{3}

{\parindent 0cm 
{\bf Argonne National Lab}\\
John Arrington\\
Paul Reimer\\
Xiaohui Zhan\\

{\bf Brookhaven National Lab}\\
Xin Qian\\

{\bf Budker Institute of Nuclear Physics}\\
Dima Nikolenko\\
Igor Rachek\\

{\bf Cairo University}\\
Hassan Ibrahim\\

%{\bf California Institute of Technology}\\

{\bf California State University}\\
Konrad A. Aniol\\
Martin B. Epstein\\
Dimitri Margaziotis\\

{\bf Carnegie Mellon University}\\
Gregg Franklin\\
Vahe Mamyan\\
Brian Quinn\\

{\bf The Catholic University of America}\\
Marco Carmignotto\\
Tanja Horn\\ 
Indra Sapkota\\

{\bf Commissariat a l’Energie Atomique - Saclay}\\
Maxime Defurne \\
Nicole d'Hose\\
Eric Fuchey\\
Franck Sabatie\\

{\bf China Institute of Atomic Energy (CIAE)}\\
Xiaomei Li\\
Shuhua Zhou\\

{\bf Christopher Newport University}\\
Ed Brash\\

{\bf College of William and Mary}\\
David S. Armstrong\\
Carlos Ayerbe Gayoso\\
Todd Averett\\
Juan Carlos Cornejo\\
Melissa Cummings\\
Wouter Deconinck\\
 Keith Griffioen\\
Joe Katich\\
Charles Perdrisat\\
Yang Wang\\ 
Huan Yao\\
Bo Zhao\\

%{\bf Dapnia/SphN}\\

{\bf Duquesne University}\\
Fatiha Benmokhtar\\

{\bf Duke University}\\
Steve Churchwell\\
Haiyan Gao\\
Calvin Howell\\
Min Huang\\
Richard Walter\\
Qiujian Ye\\

{\bf Faculte des Sciences de Monastir (Tunisia)}\\
Malek Mazouz\\

{\bf Florida International University}\\
Armando Acha\\
Werner Boeglin\\
Luminiya Coman\\
Marius Coman\\
Lei Guo\\
Hari Khanal\\
Laird Kramer\\
Pete Markowitz \\
Brian Raue\\
Jeorg Reinhold\\

{\bf Forschungszentrum Rossendorf Institut f\"ur Kern- und. Hadronenphysik}\\
Frank Dohrmann \\

%{\bf The George Washington University}\\

{\bf Gesellschaft fur Schwerionenforschung (GSI)}\\
Javier Rodriguez Vignote\\

{\bf Hampton University}\\
Eric Christy\\
Leon Cole\\
Peter Monaghan\\

{\bf Harvard University}\\
Richard Wilson\\

{\bf Hebrew University of Jerusalem}\\
Moshe Friedman\\
Aidan Kelleher\\
Guy Ron \\

{\bf Huangshan University}\\
Hai-jiang Lu\\
XinHu Yan\\

{\bf Idaho State University}\\
Mahbub Khandaker\\
Dustin McNulty\\

{\bf INFN/Bari}\\
Raffaele de Leo\\

{\bf INFN/Catania}\\
%Vincenzo Bellini\\
%Francesco Noto\\
Antonio Guisa\\
Francesco Mammolit\\
Giuseppe Russo\\
%Maria Leda Sperduto\\
Concetta Maria Sutera\\

{\bf INFN/Lecce}\\
Roberto Perrino\\

{\bf INFN/Roma}\\
Marco Capogni\\
Evaristo Cisbani\\
Francesco Cusanno\\
Fulvio De Persio\\
Alessio Del Dotto\\
Cristiano Fanelli\\
Salvatore Frullani\\
Franco Garibaldi\\
Franco Meddi\\
Guido Maria Urciuoli\\

{\bf Institute of Modern Physics, Chinese Academy of Sciences}\\
Xurong Chen\\

{\bf Institut de Physique Nucleaire - Orsay}\\
Camille Desnault\\
Alejandro Marti Jimenez-Arguello\\
Carlos Munoz Camacho\\
Rafayel Paremuzyan\\

{\bf ISN Grenoble}\\
Eric Voutier\\

{\bf James Madison University}\\
Gabriel Niculescu\\
Ioana Niculescu\\

{\bf Jefferson Lab}\\
Alexandre Camsonne\\
Larry Cardman\\
Jian-Ping Chen\\
Eugene Chudakov\\
Mark Dalton\\
Kees de Jager\\
Alexandre Deur\\
Ed Folts\\
David Gaskell\\
Javier Gomez\\
Ole Hansen\\
Douglas Higinbotham\\
Mark K. Jones\\
Thia Keppel\\
John Lerose\\
Simona Malace\\
Bert Manzlak\\
David Meekins\\
Robert Michaels\\
Bryan Moffit\\
Sirish Nanda\\
Noel Okay\\
Lubomir Pentchev\\
Yi Qiang\\
Lester Richardson\\
Yves Roblin\\
Brad Sawatzky\\
Jack Segal\\
Dennis Skopik\\
Patricia Solvignon\\
Mark Stevens\\
Riad Suleiman\\
Stephanie Tysor\\
Bogdan Wojtsekowski\\
Jixie Zhang\\

{\bf Jozef Stefan Institute}\\
Miha Mihovilovic\\
Simon Sirca\\

{\bf Kent State University}\\
Bryon Anderson\\
Mina Katramatou\\
Elena Khrosinkova\\
Richard Madey\\
Mark Manley\\
Gerassimos G. Petratos\\
Larry Selvey\\
John Watson\\

{\bf Kharkov Institute of Physics and Technology}\\
Oleksandr Glamazdin\\
Viktor Gorbenko\\
Roman Pomatsalyuk\\
Vadym Vereshchaka\\

{\bf Kharkov State University}\\
Pavel Sorokin\\

{\bf Khalifa University}\\
Issam Qattan\\

{\bf Lanzhou University}\\
Bitao Hu\\
Yi Zhang\\

{\bf Longwood University}\\
Tim Holmstrom\\
Keith Rider\\
Jeremy St. John\\
Vincent Sulkosky\\
Wolfgang Troth\\

{\bf Los Alamos Laboratory}\\
Jin Huang\\
Xiaodong Jiang\\
Ming Xiong Liu\\

{\bf LPC Clermont-Ferrand France}\\
Pierre Bertin\\
Helene Fonvielle\\

{\bf Mississippi State University}\\
Dipangkar Dutta\\
Mitra Shabestari\\
Amrendra Narayan\\
Nuruzzaman\\

{\bf Massachusetts Institute of Technology}\\
Kalyan Allada\\
Bill Bertozzi\\
Shalev Gilad\\
Navaphon ``Tai'' Muangma\\
Kai Pan\\
Cesar Fernandez Ramirez\\
Rupesh Silwal\\

{\bf Mountain View Collage}\\
Ramesh Subedi\\

{\bf Negev Nuclear Research Center}\\
Arie Beck\\
Sharon Beck\\

{\bf NIKHEF}\\
Jeff Templon\\

{\bf Norfolk State University}\\
Wendy Hinton\\
Vina Punjabi\\

{\bf North Carolina Central University}\\
Benjamin Crowe\\
Branislav (Branko) Vlahovic\\

{\bf North Carolina A\&T State University}\\
Ashot Gasparian\\

{\bf Northwestern University}\\
Ralph Segel\\

{\bf Ohio University}\\
Mongi Dlamini\\
Norman Israel\\
Paul King\\
Julie Roche\\

{\bf Old Dominion University}\\
S. Lee Allison\\
Gagik Gavalian\\
Mohamed Hafez\\
Charles Hyde\\
Kijun Park\\
Hashir Rashad \\
Larry Weinstein\\

{\bf Peterburg Nuclear Physics Institute}\\
Viacheslav Kuznetsov \\

{\bf Regina University}\\
Alexander Kozlov\\
Andrei Semenov\\

{\bf Rutgers University}\\
Lamiaa El Fassi\\
Ron Gilman\\
Gerfried Kumbartzki\\
Katherine Myers\\
Ronald Ransome\\
Yawei Zhang\\

{\bf Saint Norbert College}\\
Michael Olson\\

{\bf Seoul National University}\\
Seonho Choi\\
Byungwuek Lee\\

{\bf Smith College}\\
Piotr Decowski\\

{\bf St Mary's University}\\
Davis Anez\\
Adam Sarty\\

%{\bf Stanford Linear Accelerator}\\

{\bf Stony Brook University}\\
Rouven Essig\\

{\bf Syracuse University}\\
Zafar Ahmed\\
Richard Holmes\\
Paul A. Souder\\

{\bf Technische Universit\"at M\"unchen}\\
Jaideep Singh\\

{\bf Tel Aviv University}\\
Nathaniel Bubis\\
Or Chen\\
Igor Korover\\
Jechiel Lichtenstadt\\
Eli Piasetzky\\
Ishay Pomerantz\\
Ran Shneor\\
Israel Yaron\\

{\bf Temple University}\\
David Flay\\
Zein-Eddine Meziani\\
Michael Paolone\\
Matthew Posik\\
Nikos Sparveris\\

{\bf Tohoku University}\\
Kouichi Kino\\
Kazushige Maeda\\
Teijiro Saito\\
Tatsuo Terasawa\\
H. Tsubota\\

{\bf Tsinghua University}\\
Zhigang Xiao\\

{\bf Universidad Complutense de Madrid (UCM)}\\
Joaquin Lopez Herraiz\\
Luis Mario Fraile\\
Maria Christina Martinez Perez\\
Jose  Udias Moinelo\\

{\bf University of Connecticut}\\
Andrew Puckett\\

{\bf Universitat Pavia}\\
Sigfrido Boffi\\

{\bf University ``La Sapienza'' of Rome}\\
Cristiano Fanelli \\
Fulvio De Persio\\

{\bf University of Glasgow}\\
John Annand\\
David Hamilton\\
Dave Ireland\\
Ken Livingston\\
Dan Protopopescu\\
Guenther Rosner\\
Johan Sjoegren\\

{\bf University of Illinois}\\
Ting Chang\\
Areg Danagoulian\\
J.C. Peng\\
Mike Roedelbronn\\
Youcai Wang\\
Lindgyan Zhu\\

{\bf University of Kentucky}\\
Dan Dale\\
Tim Gorringe\\
Wolfgang Korsch\\

{\bf University of Lund}\\
Kevin Fissum\\

{\bf University of Manitoba}\\
Juliette Mammei\\

{\bf University of Maryland}\\
Elizabeth Beise\\

{\bf University of Massachusetts, Amherst}\\
Krishna S. Kumar\\
Seamus Riordan\\
Jon Wexler\\

{\bf University of New Hampshire}\\
Toby Badman\\
Trevor Bielarski\\
John Calarco\\
Bill Hersman\\
Maurik Holtrop\\
Donahy John\\
Mark   Leuschner\\
Elena Long\\
James Maxwell\\
Sarah Phillips\\
Karl Slifer\\
Timothy Smith\\
Ryan Zielinski\\

{\bf University of Regina}\\
Garth Huber\\
George Lolos\\
Zisia Papandreou \\

{\bf University of Saskatchewan}\\
Ru Igarashi\\

{\bf University of Science and Technology of China (USTC)}\\
Yi Jiang\\
Wenbiao Yan \\
Yunxiu Ye\\
Zhengguo Zhao\\
Yuxian Zhao \\
Pengjia Zhu\\

{\bf University of South Carolina}\\
Steffen Strauch\\

{\bf University of Tennessee}\\
Nadia Fomin\\

{\bf University of Virginia}\\
Khem Chirapatpimol\\
Donal Day\\
Xiaoyan Deng\\
Gordon D. Gates\\
Gu Chao\\
Charles Hanretty\\
Ge  Jin\\
Richard Lindgren\\
Jie Liu\\
Nilanga Liyanage\\
Vladimir Nelyubin\\
Blaine Norum\\
Kent Paschke\\
Peng Chao\\
Oscar Rondon\\
Kiadtisak Saenboonruang\\
William ``Al'' Tobias\\
Diancheng Wang\\
Kebin Wang\\
Zhihong Yi\\
Zhiwen Zhao\\
Xiaochao Zheng\\
Jiayao Zhou\\

{\bf University of Washington}\\
Diana Parno\\

{\bf Yamagata University}\\
Seigo Kato\\
Hiroaki Ueno\\

{\bf Yerevan Physics Institute}\\
Sergey Abrahamyan\\
Nerses Gevorgyan\\
Edik Hovhannisyan\\
Armen Ketikyan\\
Samvel Mayilyan\\
Karen Ohanyan\\
Artush Petrosyan\\
Galust Sargsyan \\
Albert Shahinyan\\
Hakob Voskanian\\

{\bf Inactive Members}\\
Mattias Anderson\\
Maud Baylac\\
Hachemi Benaoum\\
J. Berthot\\
Michel Bernard \\
Louis Bimbot\\
Tim Black\\
Alexander Borissov\\
Vincent Breton\\
Herbert Breuer\\
Etienne Burtin\\
Christian Cavata\\
George Chang\\
Nicholas Chant\\
Jean-Eric Ducret\\
Zhengwei Chai\\
Brandon Craver \\
Natalie Degrande\\
Rachel di Salvo\\
Pibero Djawotho\\
Chiranjib Dutta\\
Kim Egiyan\\
Stephanie Escoffier\\
Catherine Ferdi\\
Megan Friend \\
Robert Feuerbach\\
Mike Finn\\
Bernard Frois\\
Oliver Gayou\\
Charles Glashausser\\
Jackie Glister\\
Greg Hadcock\\
Brian Hahn\\
Harry Holmgren\\
Sebastian Incerti\\
Mauro Iodice\\
Riccardo Iommi\\
Florian Itard\\
Stephanie Jaminion\\
Steffen Jensen\\
Sudirukkuge Tharanga Jinasundera\\
Cathleen Jones\\
Lisa Kaufman\\
James D. Kellie\\
Sophie Kerhoas\\
Ameya Kolarkar\\
Norm Kolb\\
Ioannis Kominis\\
Serge Kox\\
Kevin Kramer\\
Elena Kuchina\\
Serguei Kuleshov\\
Jeff Lachniet\\
Geraud Lavessiere\\
Antonio Leone\\
David Lhuillier\\
Meihua Liang\\
Han Liu\\
Robert Lourie\\
Jacques Marroncle\\
Jacques Martino\\
Kathy McCormick\\
Justin McIntyre\\
Luis Mercado\\
Brian Milbrath\\
Wilson Miller\\
Joseph Mitchell\\
Jean  Mougey\\
Pierre Moussiegt\\
Alan Nathan\\
Damien Neyret\\
Stephane Platchkov\\
Thierry Pussieux\\
Gilles Quemener\\
Abdurahim Rakhman\\
Bodo Reitz\\
Rikki Roche\\
Philip Roos\\
David Rowntree\\
Gary Rutledge\\
Marat Rvachev\\
Arun Saha\\
Neil Thompson\\
Luminita Todor\\
Paul   Ulmer\\
Antonin Vacheret\\
Luc Van de Hoorebeke\\
Robert Van de Vyver\\
Pascal Vernin\\
Dan Watts\\
Krishni Wijesooriya\\
Hong Xiang\\
Wang Xu\\
Jingdong Yuan\\
Jianguo Zhao\\
Jingdong Zhou\\
Xiaofeng Zhu\\
Piotr Zolnierczuk\\
}

\end{multicols}

\end{document}